\documentclass[a4paper,12pt]{article}
\usepackage[utf8]{inputenc}
\usepackage{cancel}
\usepackage{ulem}
\usepackage{amsfonts}
\usepackage{amssymb}
\usepackage{graphicx}
\usepackage{amsmath}
\usepackage{enumerate}
\usepackage{mathtools}
\usepackage{subfloat}
\usepackage{color}
\usepackage{float}
\usepackage{here}
\usepackage{cite}
\usepackage{mathrsfs}
\usepackage{float,epsfig}
\usepackage{dcolumn}
\DeclareMathOperator\arccosh{arccosh}
\usepackage{tabularx, ragged2e, booktabs, caption}
\usepackage{blindtext}
\usepackage{caption} 
\usepackage{subcaption}
\usepackage{tablefootnote}
\usepackage{booktabs} % Nice tables
\usepackage[export]{adjustbox}
\usepackage{authblk}

\usepackage{ulem}

\usepackage{graphicx}% Include figure files
\usepackage{bm}% bold math
\usepackage{amsmath,amssymb,amsthm}
\usepackage[colorlinks=true,linkcolor=blue,citecolor=red]{hyperref}
\newcommand{\be}{\begin{equation}}
\newcommand{\ee}{\end{equation}}
\newcommand{\bea}{\setlength\arraycolsep{2pt} \begin{eqnarray}}
\newcommand{\eea}{\end{eqnarray}}

\def\0{{\sst{(0)}}}
\def\1{{\sst{(1)}}}
\def\2{{\sst{(2)}}}
\def\3{{\sst{(3)}}}
\def\4{{\sst{(4)}}}
\def\5{{\sst{(5)}}}
\def\6{{\sst{(6)}}}
\def\7{{\sst{(7)}}}
\def\8{{\sst{(8)}}}
\def\sst#1{{\scriptscriptstyle #1}}

\usepackage{tikz,xcolor,hyperref}

\definecolor{lime}{HTML}{A6CE39}
\DeclareRobustCommand{\orcidicon}{%
	\begin{tikzpicture}
	\draw[lime, fill=lime] (0,0) 
	circle [radius=0.16] 
	node[white] {{\fontfamily{qag}\selectfont \tiny ID}};
	\draw[white, fill=white] (-0.0625,0.095) 
	circle [radius=0.007];
	\end{tikzpicture}
	\hspace{-2mm}
}
\foreach \x in {A, ..., Z}{%
	\expandafter\xdef\csname orcid\x\endcsname{\noexpand\href{https://orcid.org/\csname orcidauthor\x\endcsname}{\noexpand\orcidicon}}
}

\title{\bf Born-Infeld-AdS black hole phase structure: Landau theory and free energy landscape approaches}

\author{
 Md Sabir Ali\orcidS{}$^1$\footnote{alimd.sabir3@gmail.com},
 Hasan  El Moumni\orcidH{}$^2$\thanks{h.elmoumni@uiz.ac.ma (Corresponding author)} ,   Jamal  Khalloufi\orcidJ{}$^{2}$\footnote{jamalkhalloufi@gmail.com },
 Karima  Masmar\orcidK{}$^{3}$\footnote{karima.masmar@gmail.com }\footnote{Authors are in alphabetical order.}
\\
{
\small $^{1}$ Department of Physical Sciences, Indian Institute of Science Education and Research Kolkata,
Mohanpur, 741246, India.\\
\small $^{2}$ LPTHE, Physics Department, Faculty of Sciences, Ibnou Zohr University, B.P 8106, Agadir, Morocco. \\
\small $^{3}$ Laboratory of High Energy Physics and Condensed Matter, Faculty of
Sciences Ain Chock, B.P 5366, HASSAN II University, Casablanca, Morocco. 
}
}
\vspace{-3.6em}

\date{}
\begin{document} 
\maketitle
\begin{abstract}

We start with a brief overview of the basic thermodynamic properties of the Born-Infeld metric in AdS spacetime. Using the concept of the enthalpy characterizing the total mass of the black hole, in our present paper, we probe the thermal phase transition structure, the dynamic and kinetic behavior of the Born-Infeld-AdS black hole. The emergence of the triple point behavior and the possible ruling out the reentrant phase transition, for a certain parametric value of the charge on the free energy landscape, we scrutinize the stochastic dynamics and the kinetic processes. We describe such processes during the black hole phase transitions in terms of the Landau functional and equivalently by the Fokker-Planck equation in the context of black hole chemistry. 

Our analysis establishes a pertinent bridge between the thermal behavior among the different states of the Van-der-Waals-like fluids and the Born-Infeld-AdS black holes phases. To visualize the direct implications of the Landau functional of the usual Van-der-Waals-like fluids, we consistently employed the generic Landau formalism for the analysis of the black hole phase transitions of the Born-Infeld-AdS black holes. We find that such investigations are worthy of study in implementing the continuous phase transition behavior during the Hawking radiation. For more details, and  in addition to the exploitation of the Landau functional, we introduce its convexity to determine its extreme points and the corresponding stable and unstable phases of the thermal black hole systems. We systematically study the behavior of the first-order and the second-order phases and look into details of their evolution during thermal transitions. Moreover, knowing that the thermal phase transitions are controlled through a stochastic process depending upon an order parameter, the dynamics during its phases are determined through the fluctuating macroscopic variables. We recall the dynamical Fokker-Planck equation to furnish the advancement of such a process in the Born-Infeld-AdS background. We mainly focus on the probability distribution of the triple point structure where the small, large, and pure thermal radiation coexist. The evolution of the initial probability indicates that there not only the initial small black hole to the final large black hole phase occurs, but also one has the equilibrium conditions established among the thermal radiations to the small black holes or the small black holes to thermal radiations and large black hole states. We also demonstrate the first passage time for the different black hole phase behaviors to determine their time scale. In addition to this, we also calculate the mean first passage time and its fluctuation using the Crank-Niclson method. Such a study has implications in the friction effects of the kinetic turnover of different black hole phases. The friction has a direct connection to the microscopic degrees of freedom. Therefore such investigation helps us to determine the interactions of the black hole micromolecules during kinetic phase transitions.

{\noindent}

\end{abstract}

\tableofcontents

%\newpage

\section{Introduction}
In theoretical high energy physics, the thermodynamics of black holes is a fascinating and promising research area in order to probe their quantum nature. It provides a possible route to develop the quantum gravitational theory when studied in the context of anti-de Sitter (AdS)/Conformal field theory (CFT). The discovery of the Hawking-Page phase transition between the AdS thermal bath and large Schwarzschild-AdS black holes triggered a flurry of thermodynamic activities in the last few decades. The black hole chemistry, i.e., the thermodynamics with a negative cosmological constant, made an open room to understand the different and new thermodynamics phenomena from the AdS/CFT perspectives. The gravitational viewpoint of the van der Waals (vdW) fluid, the reentrant phase transition of the multicomponent liquids, the thermal behavior of the triple points, the polymer phases, and the superfluidity have uplifted the status of thermodynamics of a wide range of AdS black holes. Among various solutions, the Born-Infeld black holes in AdS spacetime have been of great importance. The Born-Infeld gravity is a nonlinear generalization of the Maxwell electrodynamics calibrated by the Born-Infeld parameter. Such solutions are very important in order to probe the short-distance behavior of the electromagnetic field. From a general relativity viewpoint, we have the Born-Infeld black hole spacetime analytically when Einstein's field equations are solved. Asymptotically, the solution recovers the Reissner-Nordstr$\Ddot{o}$m black hole. In AdS spacetime in four dimensions, the solutions have some exotic thermodynamic behaviors. The reentrant phase transition behaviors which were uncommon in the study of the thermodynamics was first observed in the case of Born-Infeld AdS spacetime \cite{Gunasekaran:2012dq} in four dimensions. Such phase transition was examined in the usual thermodynamic systems exhibiting the nicotine/water mixture \cite{NarayananKumar:1994}. The reentrant phase transition (RPT) is a thermodynamic phenomenon of multicomponent liquids when the thermal system goes through multiple phases for a monotonic variation of any of the thermodynamic variables, provided the initial and the final states of the system shows the same macroscopic behaviors. After its inception for the conventional thermodynamic system, the reentrant phase transition has been reported for a wide class of AdS black hole systems \cite{Altamirano:2013ane, Altamirano:2013uqa, Altamirano:2014tva, Kubiznak:2015bya,Frassino:2014pha,Wei:2014hba,
Hennigar:2015esa,Sherkatghanad:2014hda,Hennigar:2015wxa,Xu:2019yub, Zou:2016sab, Kubiznak:2014zwa} apart from four-dimensional Born-Infeld AdS black holes. The Born-Infeld AdS black hole in dimensions greater than four does not show the reentrant phase transition\cite{Zou:2013owa}.\\

The thermal properties of AdS black holes spacetime has surpassed the core idea of black hole mechanics and landed into much richer physics that is probed through the critical points. During the small to large phase transitions the thermodynamic properties change which in turn has an effect on the dynamic and kinetic processes that indicate the evolution of thermal systems through any thermodynamic process. Such dynamical evolution is systematically analyzed through Landau's theory of free energy, denoted by $L$, sometimes called the Landau functional. The AdS black hole systems have a direct link to the boundary conformal field theory (CFT), the duality between thermal AdS black properties and the CFT is usually called the holographic duality, or the gauge/gravity duality or the AdS/CFT correspondence. During the phase transition, the system undergoes a non-equilibrium condition and the dynamic is determined through Landau theory. The Landau parameters separate out the thermal phase transition behavior mainly of the vdW type fluid or the charge AdS black holes. This way, the Landau functional has the local minimum and thereby mimics the second-order phase transition, whereas, for the first-order transition during various phases, it would correspond to a global minimum of the extended thermodynamic phases of the AdS black hole systems. \\

Recently, the idea comprising the free energy landscape has been introduced from the perspectives of connecting the thermodynamic properties, the dynamics, and the kinetic transition processes through a stochastic process $\Grave{\text{a}}$ la Fokker-Planck equation. In thermal statistical physics, the dynamical Fokker-Planck relation is employed to study the time profile of the probability density function for any generic observable. For black hole thermal systems, such notions were first explored in the context of the phase transition due to Hawking and Page and also for the massive gravity scenarios \cite{Li:2020khm}. The concept of representing the off-shell Gibbs free energy as dependant on the order parameter (e.g., the horizon radius), was the crucial identification behind such studies. The investigations of the free energy landscape are still under improvement, though people studied it for a wide range of black holes in Einstein as well as in modified gravity theories in asymptotically AdS spacetime. Soon after the original work on the dynamics of Schwarzschild-AdS spacetime on the free energy landscape proposal was posed, the conceptualization was further extended to charged AdS black hole systems \cite{Li:2020nsy}, then to charge neutral Gauss-Bonnet gravity theories \cite{Wei:2020rcd}, and its charged version in four dimensional spacetime \cite{Li:2020spm}, in determining the dynamics of triple point for a six-dimensional electrically charged Guass-Bonnet-AdS systems \cite{Wei:2021bwy}, the dark energy modified charged AdS systems \cite{Lan:2021crt}, to the black hole spacetime when there is a minimal coupling of general relativity to the nonlinear electromagnetic sources \cite{Kumara:2021hlt,Du:2021cxs}, and lately to the Euler-Heisenberg-AdS black holes \cite{Dai:2022mko}. An analysis of the dynamical evolution also considered the effects of the path integral and the instantons approaches \cite{Liu:2021lmr}, while for others the free energy landscape is extended to include the non-Markovian effects \cite{Li:2022ylz,Li:2022yti}. It has been applied to the rotating solution as well, e.g., see for similar analysis on Kerr-AdS black holes \cite{Yang:2021ljn}. The free energy landscape problem was further explored by the generalized Fokker-Planck equation \cite{Li:2022oup}. Recently, the topology on the grounds of the free energy landscape and the identification of a dominant route of the dynamic and kinetic process during the black hole phase alternation has been analyzed for the electrically charged AdS spacetime in Gauss-Bonne gravity\cite{Li:2023ppc}.    \\

Born-Infeld electrodynamics was originally used to establish a finite energy density model for the electron in the 1930s \cite{Born:1934gh}. Then, it has piqued an intense interest in recent years for a variety of reasons.
ranging from its natural appearance in open superstrings and D-branes \cite{Leigh:1989jq}, in loop calculations,  to the low energy effective action of an open superstring \cite{Fradkin:1985qd,Gibbons:2001gy}. More recently the Born-Infeld electrodynamics has been introduced in the modification of the Einstein-Hilbert action and aroused a special emphasis \cite{Cecotti:1986gb,Cataldo:1999wr,Fernando:2003tz,Delhom:2019zrb,Wang:2020ohb} from different point of views \cite{Novello:1999pg,Garcia-Salcedo:2000ujn,Tao:2017fsy,Wang:2019kxp,Jing:2020sdf,Bi:2020vcg,Miskovic:2008ck,BeltranJimenez:2021oaq,Gan:2019jac,Liang:2019dni,He:2022opa}.

In our study, we shall consider the thermal phase transition characteristics of the AdS black holes within the Born-Infeld gravitational framework, a natural generalization of the Maxwell electrodynamics in nonlinear theory. In our case, we shall study the free energy landscape of the Born-Infeld-AdS black holes using Landau theory as well as the Fokker-Planck equation. The Born-Infeld-AdS metric in four-dimensional spacetime shows the reentrant phase transition like the nicotine/water mixture as in the conventional thermal system. We shall investigate the dynamical evolution of the black holes during the thermal and also for the reentrant phase transitions. The Born-Infeld parameter has a crucial effect on the horizon size and the radius of the horizon behaves as the order parameter. The subsequent analysis for the dependence of Landau functional or the Gibbs free energy on horizon radius may serve as a one-dimensional curve. \\

The organizations of our paper are as follows. In Section~\textcolor{blue}{2}, we review some of the basic notions of the Born-Infeld-AdS black holes regarding its solutions, and the thermodynamic quantities of physical interests. The next Section~\textcolor{blue}{3}, is devoted to studying the critical points analysis and their limiting values. The formalism of the Landau functional is vividly described in Section~\textcolor{blue}{4} in order to probe the free energy dependence on the Born-Infeld parameter. In Section~\textcolor{blue}{5}, the dynamic and kinetic processes of the thermal system during the changes of its phases are probed through the equation as proposed by Fokker and Planck. We summarize and conclude the final remarks in Section~\textcolor{blue}{6}.

\section{Born-Infeld black hole in AdS spacetime}

The starting point is the action describing the 
 four-dimensional general relativity where the Born-Infeld electrodynamics is considered \cite{Dey:2004yt} :
\begin{equation}\label{1}
	\mathcal{S} = \dfrac{1}{16 \pi} \int d^4x\sqrt{-g}\left[  R-2 \Lambda + 4 b^2 \left( 1 - \sqrt{1+\dfrac{F_{\mu \nu} F^{\mu \nu}}{2 b^2}}\right)  \right]  ,
\end{equation}
in which $R$ denotes the Ricci scalar curvature, $\Lambda$ is the cosmological constant expressed as $\Lambda = -3/l^2$, where $l$ being the AdS radius and $b$ stands for the Born-Infeld parameter having the dimension of mass and has a connection to the string tension $\alpha'$ as $b = 1 / \left( 2 \pi \alpha ' \right) $ \cite{Gibbons:2001gy}. The electromagnetic tensor field $F_{\mu \nu}$ is given by $F_{\mu \nu} = \partial_{\mu} A_{\nu}-\partial_{\nu} A_{\mu}$, where $A_{\mu}$ is the four potential. The four-dimensional static geometry with spherical symmetry has the $ansatz$ 
\begin{equation}\label{2}
		ds^2  = - f(r) dt^2 + \dfrac{dr^2}{f(r)}+ r^2 d\Omega^2,
\end{equation}
where, $d\Omega$ is the line element on a unit $2-$sphere and the blackening function $f(r)$ is obtained to be \cite{Dey:2004yt,Fernando:2003tz,Cai:2004eh}
\begin{equation}\label{3}	
	f(r) = 1 + \dfrac{r^2}{l^2}-\dfrac{m}{r}+\dfrac{2 b^2 r^2}{3}\left( 1 - \sqrt{1+\dfrac{16 \pi^2 Q^2}{b^2 r^2}}\right) +\dfrac{64 \pi^2 Q^2}{3 r^2} {}_2\mathcal{F}_1\left[ \dfrac{1}{4} , \dfrac{1}{2}, \dfrac{5}{4}, - \dfrac{16 \pi^2 Q^2}{b^2 r^2} \right], 
\end{equation}
with ${}_2\mathcal{F}_1\left[ a,b,c,d \right]$ is the hypergeometric function of the second kind, the black hole mass $M$ can be expressed via the integration constant $m$ as $M = m/8\pi$. The electric charge per unit volume $\omega=4\pi$ is denoted by $Q$. The only component with spherical symmetric distribution, we consider the following non-zero component of the four potential
\begin{equation}\label{4}	
	A_t(r) = - \dfrac{4 \pi Q}{r} {}_2\mathcal{F}_1\left[ \dfrac{1}{4} , \dfrac{1}{2}, \dfrac{5}{4}, - \dfrac{16 \pi^2 Q^2}{b^2 r^2} \right]. 
\end{equation}
The limiting case of the Reissner-Nordstrom (RN)-AdS black hole \cite{Kubiznak:2012wp, Dehyadegari:2016nkd} can easily be found by taking the limit $b \to \infty$ in the metric function and the gauge potential. The Hawking temperature of Born-Infeld AdS spacetime metric is expressed as 
\begin{equation}\label{5}	
	T = \dfrac{1}{4 \pi} \left. \dfrac{\partial f(r)}{\partial r}\right| _{r=r_h} = \dfrac{1}{4 \pi r_h} + \dfrac{3 r_h}{4 \pi l^2} + \dfrac{b^2 r_h}{2\pi} \left( 1 - \sqrt{1+\dfrac{16 \pi^2 Q^2}{b^2 r^2}}\right). 
\end{equation}
in which $r_h$ is the event horizon radius, obtained as the largest positive real solution of the blackening function, $f(r)=0$. The electric potential at spatial infinity relative to the event horizon reads as 
\begin{equation}\label{6}	
	\Phi = \dfrac{\partial M}{\partial Q} = \dfrac{4 \pi Q}{r_h} {}_2\mathcal{F}_1\left[ \dfrac{1}{4} , \dfrac{1}{2}, \dfrac{5}{4}, - \dfrac{16 \pi^2 Q^2}{b^2 r_h^2} \right]. 
\end{equation}
The first law of thermodynamics associated with the Born-Infeld-AdS black hole, are obtained by defining its key ingredients, namely  entropy $S$, pressure $P$, and the volume $V$
 \cite{Gunasekaran:2012dq}
\begin{equation}\label{7}	
\begin{split}
	S &=\int \dfrac{1}{T} \dfrac{\partial M}{\partial r_h} d r_h = \dfrac{r_h^2}{4}, \quad P = - \dfrac{\Lambda}{8 \pi}, \quad V = \dfrac{\partial M}{\partial P} = \dfrac{r_h^3}{4}, \\
	\mathcal{B}& = \dfrac{\partial M}{\partial b} =  \dfrac{b r_h^3}{6 \pi} \left( 1 - \sqrt{1+\dfrac{16 \pi^2 Q^2}{b^2 r^2}}\right) + \dfrac{4 \pi Q^2}{3 b r_h}  {}_2\mathcal{F}_1\left[ \dfrac{1}{4} , \dfrac{1}{2}, \dfrac{5}{4}, - \dfrac{16 \pi^2 Q^2}{b^2 r_h^2} \right],
\end{split}
\end{equation}
the additional quantity $\mathcal{B}$ is conjugate to $b$ which is interpreted as the Born-Infeld vacuum polarization \cite{Gunasekaran:2012dq}, so that the first law and the related Smarr formula take the forms
\begin{equation}\label{8}	
	\begin{split}
		dM &= T dS + \Phi dQ + V dP + \mathcal{B}db,\\
		M& = 2 T S + \Phi Q - 2 P V - \mathcal{B} b.
	\end{split}
\end{equation}
We should keep in mind that the thermodynamic quantities $M$, $Q$, $S$, and $V$ are written per unit volume $\omega$.

\section{On Criticality of Born-Infeld-AdS black hole}

Following \cite{Dehyadegari:2017hvd}, we look into the Born-Infeld-AdS black hole critical behavior, where the black hole charge is allowed to vary but the cosmological constant remains a constant parameter. The specific heat at a constant charge is expressed as
\begin{equation}\label{9}	
C_Q = T\left(\dfrac{\partial S}{\partial T}\right) _Q,
\end{equation}
 the stability/instability during the phase transition can be found by considering the sign (positive/negative) of this quantity.

\begin{figure}[!ht]
	\begin{center}
		\centering
			\includegraphics[scale=.6]{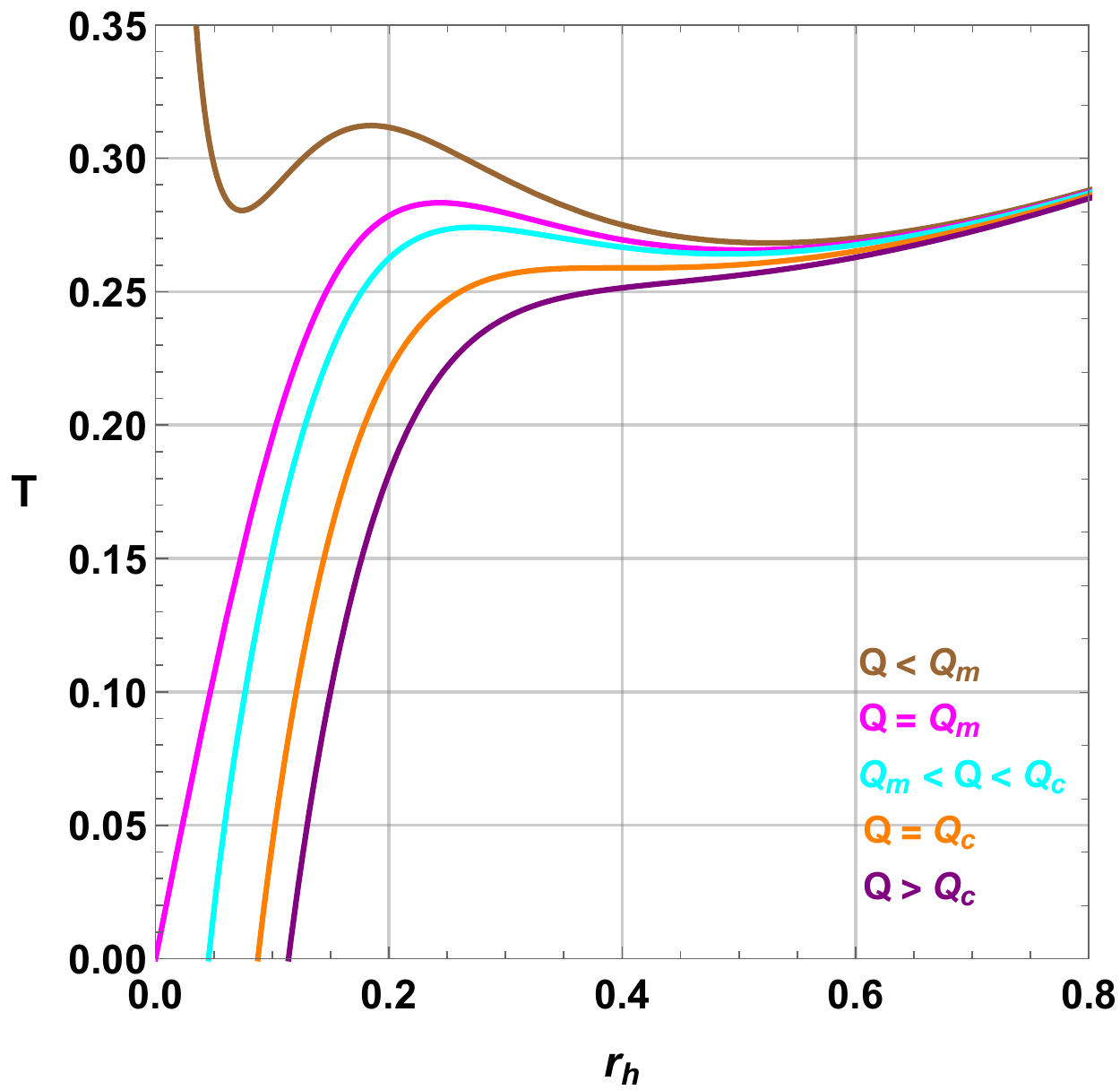} 
	\end{center}
	\caption{\footnotesize\it Temperature $T$ in terms of the event horizon radius $r_h$ for different values of charge $Q$ with $l=1$ and $b=3.5$.}
	\label{f1}
\end{figure}

\begin{figure}[ht!]
	\centering
	\begin{subfigure}[h]{0.45\textwidth}
		\centering \includegraphics[scale=.45]{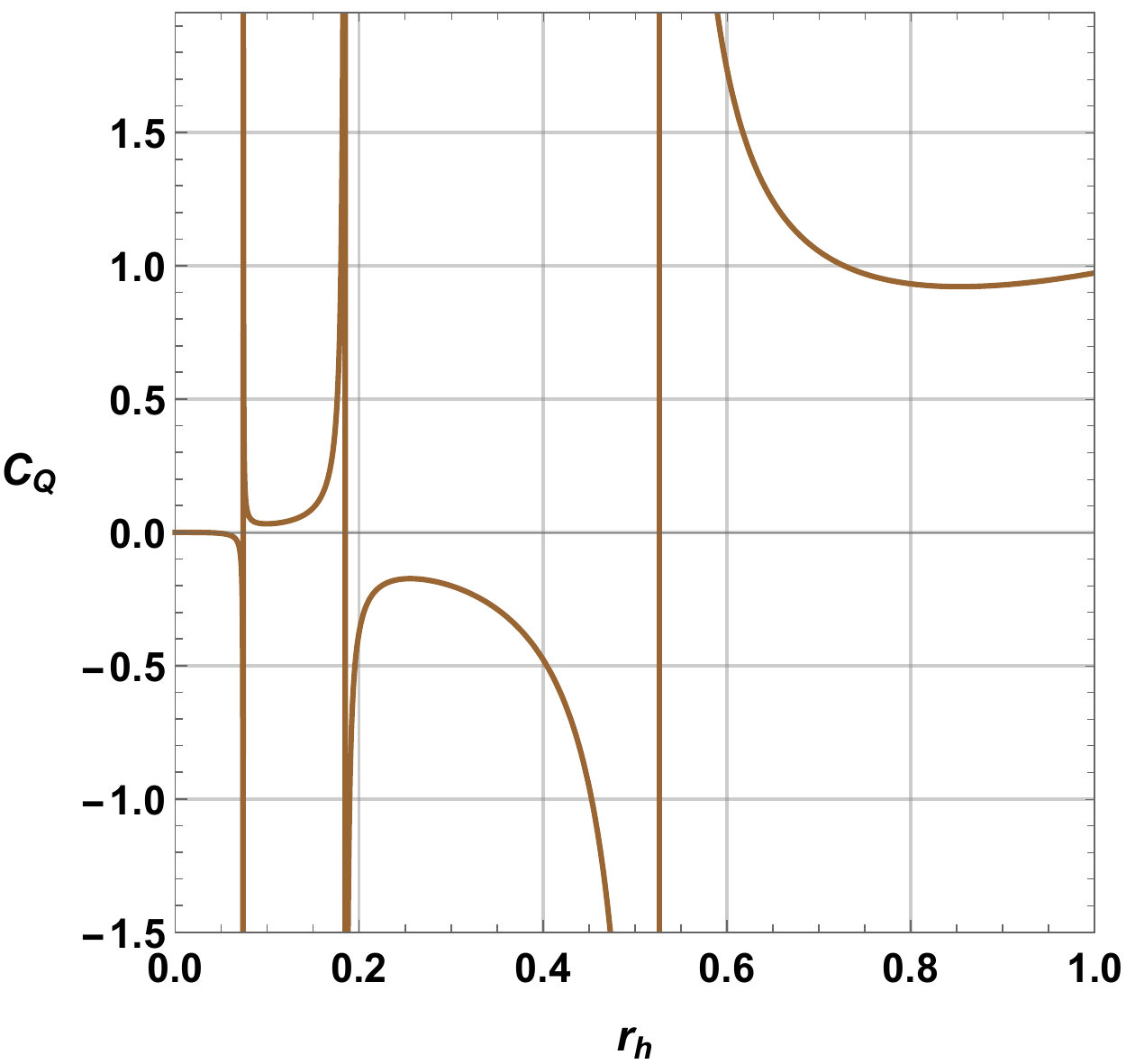}
		\caption{$Q<Q_m$}
		\label{f2_1}
	\end{subfigure}
	\hspace{1pt}	
	\begin{subfigure}[h]{0.45\textwidth}
		\centering \includegraphics[scale=.45]{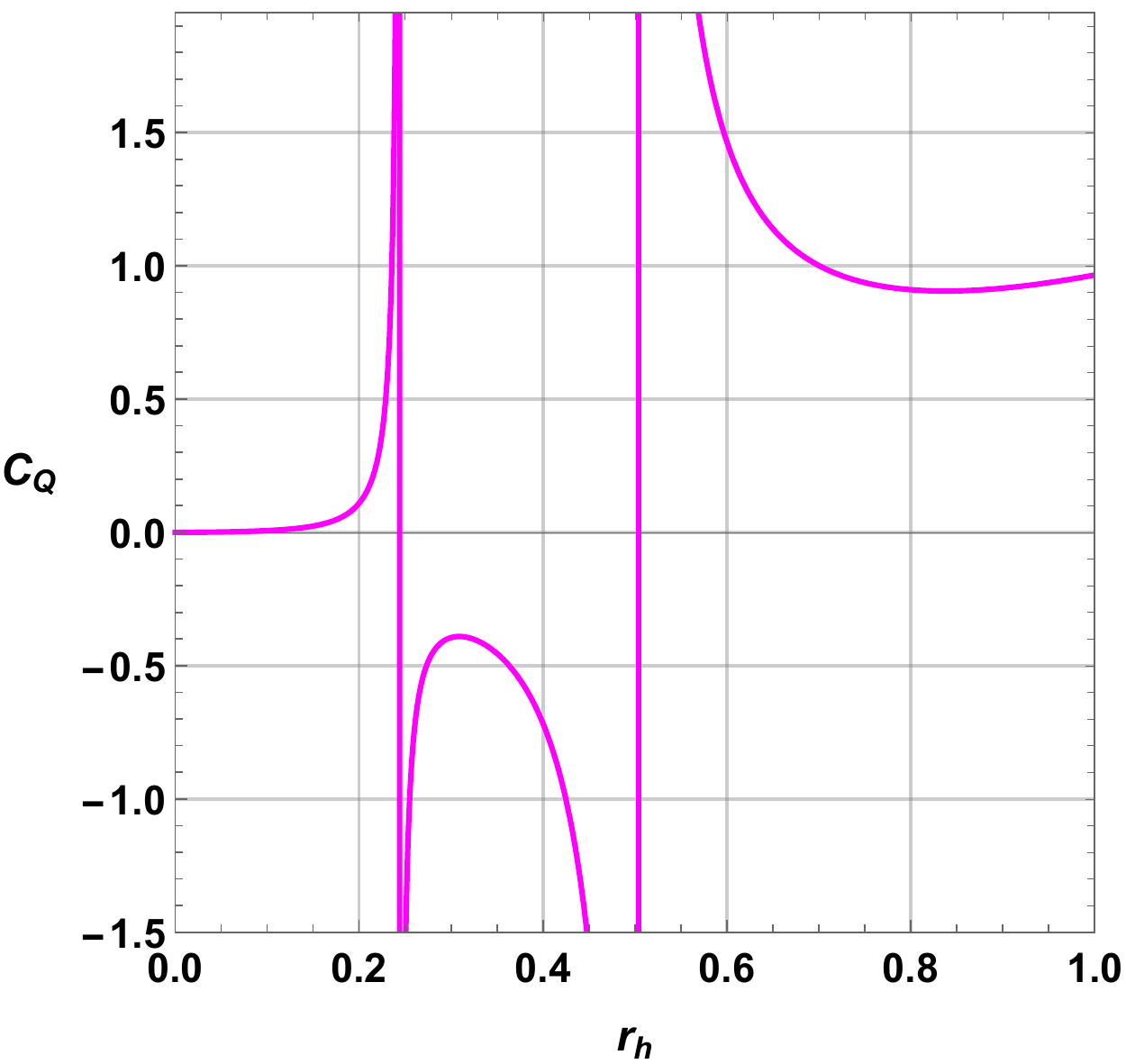}
		\caption{$Q=Q_m$}
		\label{f2_2}		
	\end{subfigure}
	\hspace{1pt}	
\begin{subfigure}[h]{0.45\textwidth}
	\centering \includegraphics[scale=.45]{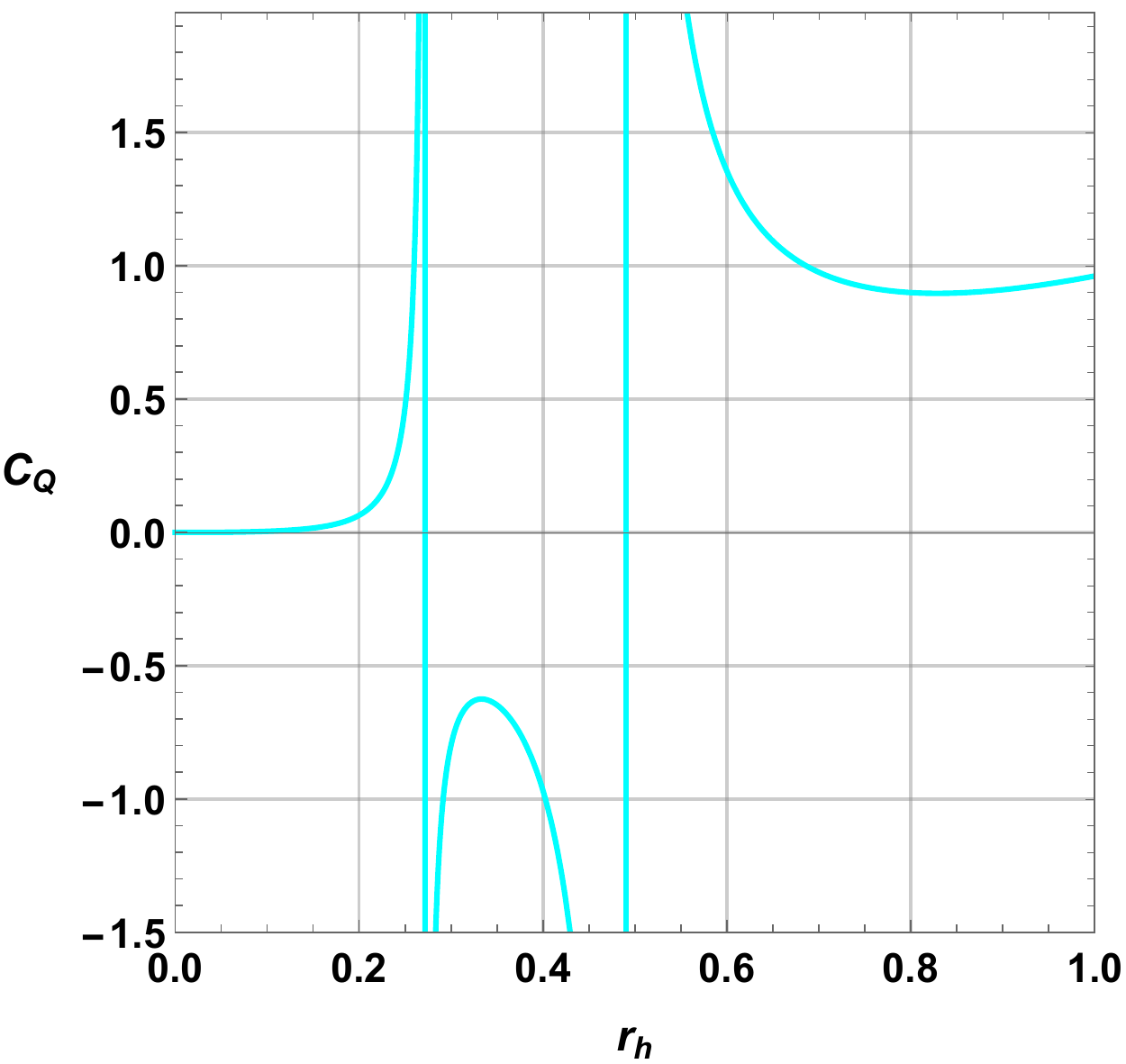}
	\caption{$Q_m<Q<Q_c$}
	\label{f2_3}	
\end{subfigure}
	\hspace{1pt}	
\begin{subfigure}[h]{0.45\textwidth}
	\centering \includegraphics[scale=.45]{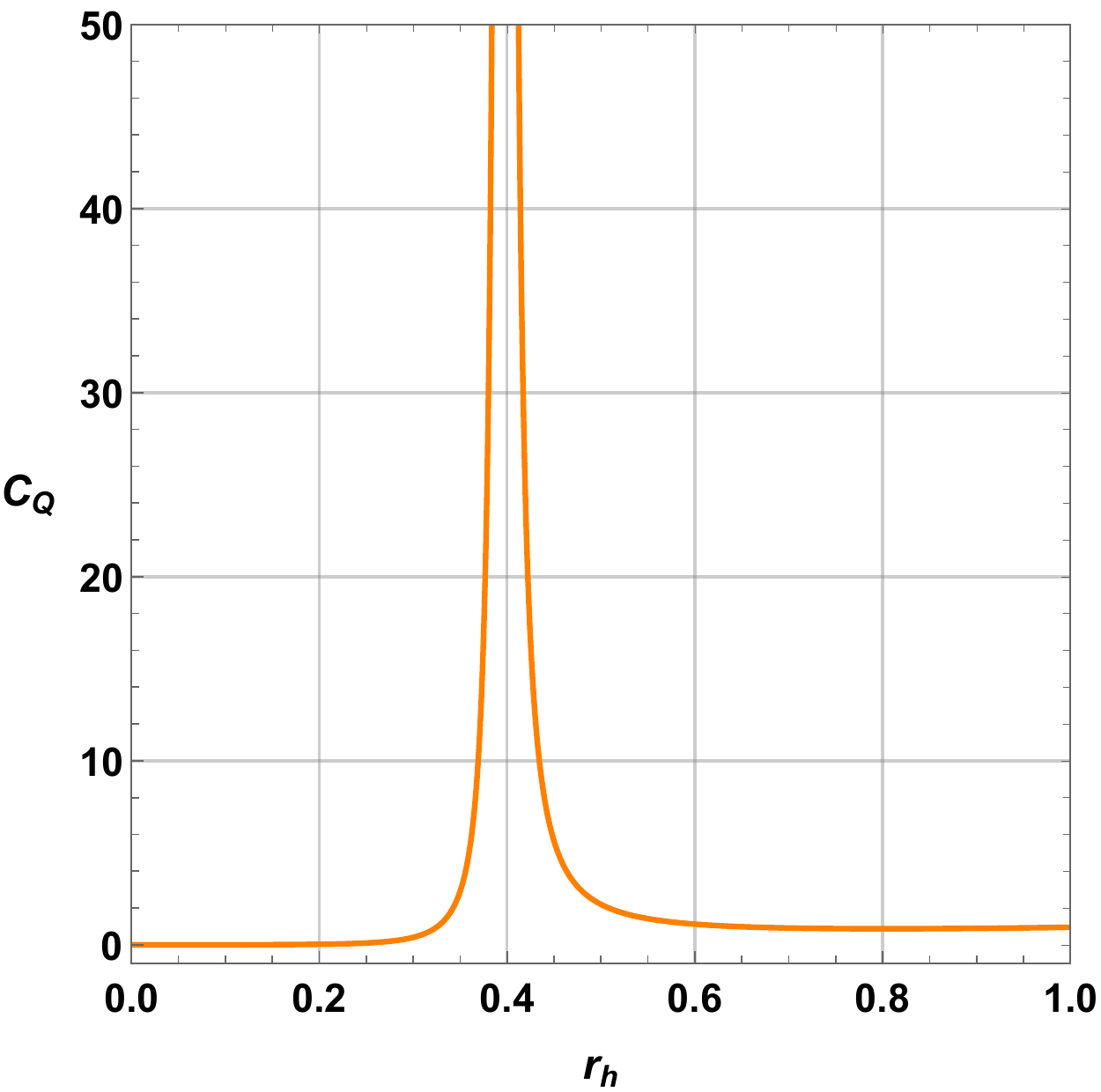}
	\caption{$Q=Q_c$}
	\label{f2_4}	
\end{subfigure}
	\hspace{1pt}	
\begin{subfigure}[h]{0.45\textwidth}
	\centering \includegraphics[scale=.45]{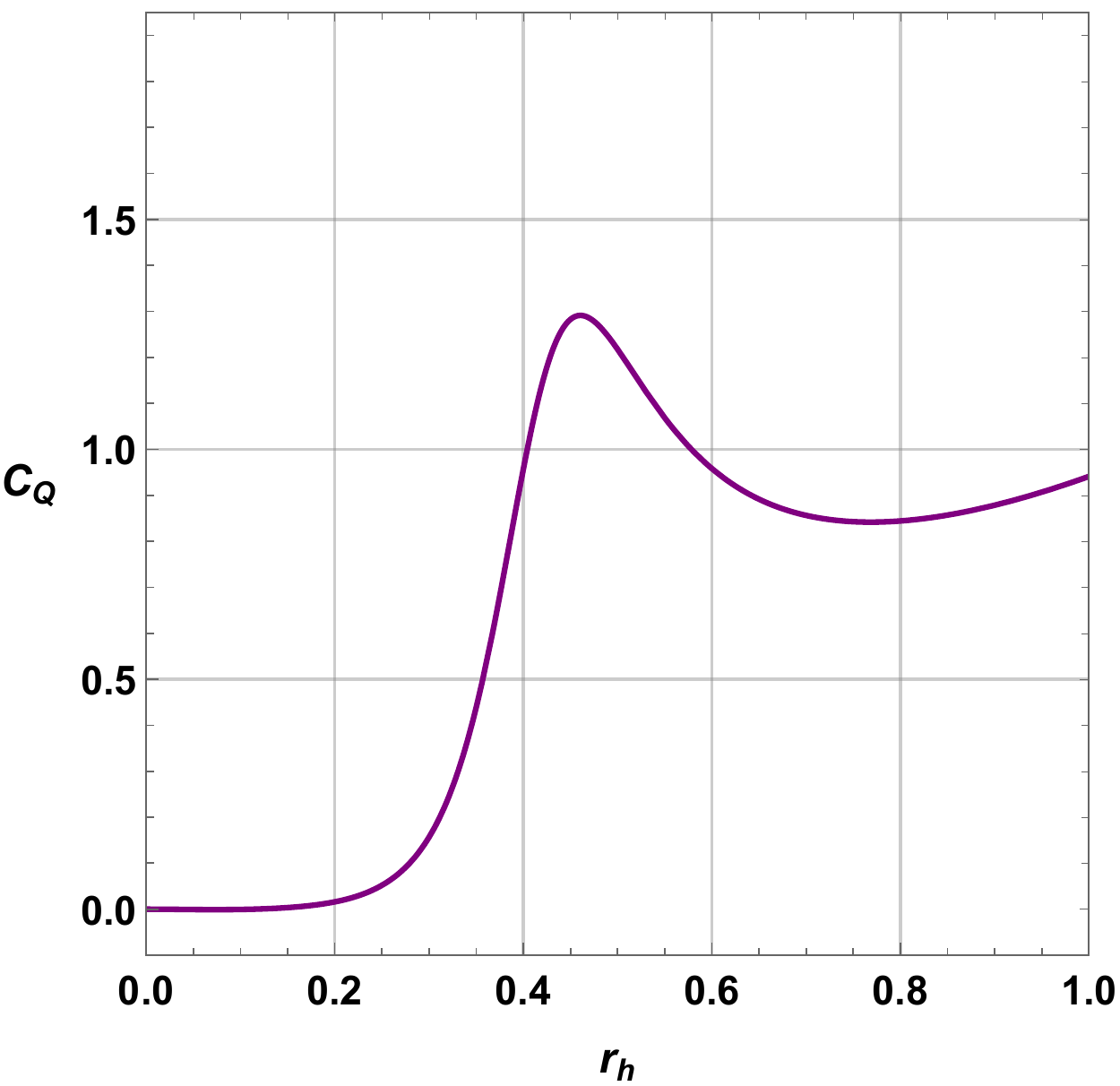}
	\caption{$Q>Q_c$}
	\label{f2_5}
\end{subfigure}
	\caption{\footnotesize\it The specific heat capacity $C_Q$ in terms of the event horizon radius $r_h$ for different values of charge $Q$ with $l=1$ and $b=3.5$.}
	\label{f2}
\end{figure}

Note that Eq.\eqref{9} is also calculated with  $l$ and $b$ taken as fixed.
To unveil the thermal behavior of the Born-Infeld-AdS spacetime, we depict the variation of the temperature $T$ and the specific heat $C_Q$ as a function of the event horizon radius $r_h$ within various values of the charge in Fig.\ref{f1} and Fig.\ref{f2} respectively. 
Obviously, one can notice that the behavior of the temperature is highly influenced by the black holes' charge for small $r_h$.
So, for small $r_h$ values, we can Taylor expand the Hawking temperature as
\begin{equation}\label{10}	
	T =\dfrac{2b}{r_h}\left( Q_m-Q \right) + \dfrac{r_h\left( 3+2 b^2l^2\right) }{4 \pi l^2}-\dfrac{b^3r_h^3}{16 \pi^2Q} +\mathcal{O}(r_h^4) ,
\end{equation}
herein $Q_m = \frac{1}{8 \pi b}$ denotes the marginal charge. Besides, the large limit $r_h$ is $\frac{3 r_h}{4 \pi l^2}$ which is independent of the charge and which explains the linearly increasing behaviour of the temperature for the large $r_h$
Depending on the value of $Q$, Born-Infeld-AdS black hole is identified as follow :
\begin{itemize}
	\item  For $Q < Q_m$, the black hole is ‘Schwarzschild-like’ (S-type). 
In the region of low temperature, black holes do not exist, much like in the Schwarzschild solution.
The largest (smallest) branch of the isocharge, as shown in Fig.\ref{f1} is connected to the huge (small) black hole, which is locally stable (unstable), as illustrated by the positive (negative) values of specific heat at the constant charge in Fig. \ref{f2_1}.
  Indeed, in the situation of $Q < Q_m$, the extremal Born-Infeld-AdS black hole can be found and therefore any charged black hole does not persist. This finding may be understood as screening effects on the electric field caused by the existence of the parameter $b$, which makes the role of charge less significant.
	
	\item When $Q \geq Q_m$, black hole is ‘Reissner-Nordstrom-like’ (RN-type). Due to the temperature crossing over from zero with decreasing $r_h$, we get an extreme black hole. 
	\item In $Q_m\leq Q < Q_c$ situation, the system exhibits a phase transition behavior of first-order between a small black hole (SBH) and a large black hole (LBH)  as one can notice from Fig.\ref{f2_2} and Fig.\ref{f2_3}.
	
	\item The situation $ Q = Q_c$, associated with Fig.\ref{f2_4}, a critical behavior appears and  phase behavior is of a second-order that occurs between a SBH and a LBH.
	\item In the last case $ Q > Q_c$, the black hole has a locally thermal stable phase for forever, and the heat capacity is positive everywhere as is revealed in Fig.\ref{f2_5}.
\end{itemize}

We now proceed to find the values of the critical points associated with the second-order phase transition with respect to the Born-Infeld-AdS black hole discussed previously.
With constant $l$ and $Q =Q_c$, we have illustrated in Fig.\ref{f1} that the critical points are determined through the inflection point as may be described by 
\begin{equation}\label{11}	
	\left. \dfrac{\partial T}{\partial r_h}\right|_{Q_c} = 0 \quad \text{and} \quad \left. \dfrac{\partial^2 T}{\partial r_h^2}\right|_{Q_c} = 0.
\end{equation} 
The criticality must have happened at the right branch of the  $T-r_h$ curve in the S-Type black hole and is thermally stable.
Then, the formulas of Eqs.\eqref{11} are expressed by considering the temperature in Eq.\eqref{5} as follows
\begin{equation}\label{12}
\begin{split}	
- 2 x^2 + \left( 1 + \dfrac{3}{2 b^2 l^2} - \dfrac{1}{2 b^2 r_c^2}\right) x + 1 &=0, \\
x^4 - \dfrac{x^2}{2} + \dfrac{x}{4 b^2 r_c^2}-\dfrac{1}{2} &= 0,
\end{split}
\end{equation}
in which, we have set  
\begin{equation}\label{13}	
	x = \left( 1+\dfrac{16 \pi ^2 Q_c^2}{b^2 r_c^4}\right) ^{-1/2},
\end{equation}
and $r_c$ denotes the critical horizon radius. To ensure the positive definiteness of the values of the critical quantities, we impose the following constraint on $x$,
\begin{equation}\label{14}	
	0 \leq x \leq 1.
\end{equation}
By the help of  Eqs.\eqref{12}, one can write the following cubic equation
\begin{equation}\label{15}	
	x^3 + p x + q = 0,
\end{equation}
where
\begin{equation}\label{16}	
	p = - \dfrac{3}{2}, \quad q = \dfrac{1}{2}\left( 1+\dfrac{3}{2 b^2 l^2}\right).
\end{equation}
Moreover, because $q$ is real valued and $p$ is negative, the cubic equation presents one or three roots which are real in nature. The existence of such three roots is controlled by $\Delta = 4 p^3 + 27 q^2\leq 0$, hence
\begin{equation}\label{17}	
	b \geq b_0 = \sqrt{\dfrac{3}{2}\left( 1 + \sqrt{2}\right)} / l \approx 1.9029/l,
\end{equation}
and their form is given by
\begin{equation}\label{18}	
	x_k = \sqrt{2} \cos\left( \dfrac{1}{3}\arccos\left[ -\dfrac{\sqrt{2}}{2}\left( 1 + \dfrac{3}{2 b^2 l^2}\right) \right] - \dfrac{2 \pi k}{3}\right), \quad k=0,1,2.
\end{equation}
It is simple to prove that the third root $x_2$ violates the condition in Eq.\eqref{14}.
Additionally, by computing the critical quantity for $x_1$, we discover that $r_c$ persists in that branch of critical isocharge where it is locally unstable for the S-type black holes. \cite{Dehyadegari:2017hvd}. Hence the only physical solution is $x_0$. For $b<b_0$ case, one real root is given by
\begin{equation}\label{19}	
	x_3 = - \sqrt{2} \cosh\left( \dfrac{1}{3} \arccosh\left[ -\dfrac{\sqrt{2}}{2}\left( 1 + \dfrac{3}{2 b^2 l^2}\right) \right] \right),
\end{equation}
which breaks the rule obtained in Eq.\eqref{14}.
Henceforth, the criticality of the Born-Infeld-AdS black holes can only be seen for  $b\geq b_0$. 
The critical quantities are revealed as soon as $x_0$ is available  
\begin{equation}\label{20}
	\begin{split}	
		r_c &=\sqrt{\dfrac{x_0}{2 b^2\left( 1 + x_0^2 - 2 x_0^4\right) }}, \\
	    Q_c &= \dfrac{1}{8 \pi b \sqrt{\left( 1+3 x_0^2-4 x_0^6\right) }},\\
	    T_c &= \dfrac{3 + 2 b^2 l^2\left( 1 + x_0 -2 x_0^3\right) }{4 \pi b l^2}\sqrt{\dfrac{x}{2\left( 1 + x_0^2-2 x_0^4\right) }}.
	\end{split}
\end{equation}
The critical charge $Q_c$ is larger than $Q_m$ where
\begin{equation}\label{21}	
b> b_1 = \sqrt{\dfrac{3}{2\left( \sqrt{6 \sqrt{3}-9}-1\right) }}/l \approx 2.8870/l.
\end{equation}
Thus for $b\geq b_1$, the criticality is the mimicker of the RN-type black hole. On the contrary, for $b_0 \leq b \leq b_1$, the criticality occurs at the right branch of $(T-r_h)_{Q_c}$ curve of the S-type black hole.

 For large values of $b$, the Tayalor expansion of the critical quantities gives rise to
\begin{equation}\label{22}
	\begin{split}	
		r_c &=\dfrac{l}{\sqrt{6}} - \dfrac{7}{24 \sqrt{6} b^2}+\mathcal{O}\left( \dfrac{1}{b^4 } \right), \\
		Q_c &= \dfrac{l}{24 \pi} - \dfrac{7}{576 \pi l b^2}+\mathcal{O}\left( \dfrac{1}{b^4 } \right),\\
		T_c &= \sqrt{\dfrac{2}{3 \pi ^2 l^2}} - \dfrac{1}{12 \pi \sqrt{6}b^2 l^3}+\mathcal{O}\left( \dfrac{1}{b^4 } \right).
	\end{split}
\end{equation}
As expected, the first terms in all the quantities of Eqs.\eqref{22} reproduce the same critical behaviors as that of the RN-AdS black holes\cite{Dehyadegari:2016nkd}.

\section{Born-Infeld AdS black hole from the Landau theory point of view}

\subsection{A brief review of Landau theory formalism}

Landau estimated a system's free energy from a perspective that displays the non-analytical nature during the phase transition and ends up capturing a significant amount of the physics. The system is defined through a global minimum of Landau free energy $L$ as a function of order parameter.  The quantity $L$ called sometimes the Landau functional which is related to the system's Gibbs free energy and has an energy dimension, but it is not the same as it   \cite{Goldenfeld, Xu:2021qyw }.

Following \cite{Xu:2021qyw}, the Landau free energy can be constructed for a general thermodynamical system  as
\begin{equation}\label{23}	
	L = \int F\left( X,T,P,Q\right) dX
\end{equation}
The thermodynamic system consisting of —temperature $T$, pressure $P$, and charge $Q$— are treated as independent parameters, whereas the parameter $X$ is regarded as an auxiliary variable.
The function $F\left( X,T,P,Q\right)$ represents various relationships that these four significant thermodynamic system parameters satisfy. 

The equation of state (EOS) describing the thermal properties are expressed as $P = f(V, T, Q)$, where $V$ is the volume which is in canonical conjugation of $P$. Based on the EOS, we may create the functional dependence of $F\left( X,T,P,Q\right)$ as
 \begin{equation}\label{24}	
 F\left( X,T,P,Q\right) = P - f(X, T,Q).
 \end{equation}
Now, one can assert that a particular thermodynamic equation describing the system's state under a particular set pertaining certain physical conditions takes the form $F\left( X,T,P,Q\right) = 0$.
The system's preferred path is the one that causes its free energy to drop to its lowest value between many distinct paths that the system can take to attain equilibrium. We incorporate an auxiliary variable $X$ which serves as an order parameter having a dimension of volume when system is heading towards the equilibrium in the isothermal, isobaric, and isocharge environments. We can discover certain actual physical thermodynamic system operations by the use of this parameter. Thus, when the functional $L$ takes the least possible value, it is considered as the most realistic state in which the system is described, the relations $F\left( X,T,P,Q\right)$ satisfied by the set $\left\lbrace X, T, P, Q\right\rbrace $: 
 \begin{equation}\label{25}	
	\dfrac{d L}{d X} = F\left( X,T,P,Q\right) = 0 \Longrightarrow X = V,
\end{equation}
The order parameter $X$ can be viewed as a volume of the system as it approaches equilibrium or, less formally, as the volume of the system in a non-equilibrium state that does not meet the system's equation of state. Although the equilibrium thermodynamic volume $V$ fulfills the system's equation of state and is the root of the function $F\left(X,T,P,Q\right) = 0$. Another benefit of creating the Landau free energy in this manner is that the convexity is connected to the thermal stability of the thermodynamical system
 \begin{equation}\label{26}	
	\delta \left. \left(   \dfrac{d L}{d X}\right)\right| _{X=V}  = - \dfrac{\partial f(V,T,Q)}{\partial V} \delta V,
\end{equation}
from which, one can notice that the extreme point is comparable to a potential well when  $\partial f(V,T,Q)/\partial V < 0$  and the corresponding state is stable, while the corresponding thermodynamic state is unstable when $\partial f(V,T,Q)/\partial V > 0$  and the extreme point is similar to a potential barrier. Hence, the $\gamma$-function is defined as 
\begin{equation}\label{27}	
	\gamma (V) = \left( 3 V\right) ^{2/3} \dfrac{\partial L(X=V)}{\partial V},
\end{equation}
which is negative when the black hole is stable and it is positive when the black hole is unstable. When $\gamma(V) = 0$, we are in presence of critical behavior. Moreover, we can interpret the zeros of $\gamma(V)$ as the fixed points by analogy with dynamical systems.

\subsection{Born-Infeld-AdS black hole thermodynamics through Landau formalism}

To construct a  phase transitions picture and classify their types, we have to probe the behavior of the thermodynamic potential associated with Born-Infeld-AdS black hole. In this sense, we recall the Gibbs free energy which is the thermodynamic potential computed from the Euclidean action with the proper boundary term in the canonical ensemble \cite{Kubiznak:2012wp}, for fixed temperature $T$, pressure $P$, and charge $Q$.
By using the Legendre transformation, one can derive the Gibbs free energy per unit volume $\omega$ as in \cite{Gunasekaran:2012dq}
\begin{equation}\label{28}	
 	 \begin{split}
	G\left( T,P,Q\right) &= M - T S\\
	& = \dfrac{1}{48 \pi r_h}\left[ 3 r_h^2 - \dfrac{3 r_h^4}{l^2} - 2 b^2 \left( 1 - \sqrt{1+\dfrac{16 \pi^2 Q^2}{b^2 r_h^4}}\right) + 128 \pi^2 Q^2 {}_2\mathcal{F}_1\left[ \dfrac{1}{4} , \dfrac{1}{2}, \dfrac{5}{4}, - \dfrac{16 \pi^2 Q^2}{b^2 r_h^2} \right] \right], 
	\end{split}
\end{equation}
in which $r_h = r_h(T, Q, P)$ and $l = l(P)$. In the following, we shall assume the charge of the Born-Infeld-AdS black hole  can vary, while the value of the pressure is fixed. Moreover, we shall take $b=3.5>b_1$, thus the critical behavior will take place in the RN-type black hole.

Depending on the value of $Q$, the phase structure of the Born-Infeld-AdS Black hole is characterized as follows:
\begin{itemize}
	\item  For $Q = 0.005 < Q_0$, we plot in  Fig.\ref{f3} the temperature as a function of horizon radius $r_h$ (Fig.\ref{f3_1}), Gibbs free energy in therms of temperature (Fig.\ref{f3_2}), Landau function $L$ in terms of the parameter $X$ for different temperatures (Fig.\ref{f3_3}), Landau function $L$ in terms of the black hole volume $V$ (Fig.\ref{f3_4}), $\gamma$-function in terms of the black hole volume $V$ F(ig.\ref{f3_5}) and on-shell Gibbs free energy $\tilde{G}$ versus the temperature (Fig.\ref{f3_6}). 
 	 \begin{figure}[!ht]
		\centering
		\begin{subfigure}[h]{0.45\textwidth}
			\centering \includegraphics[scale=.5]{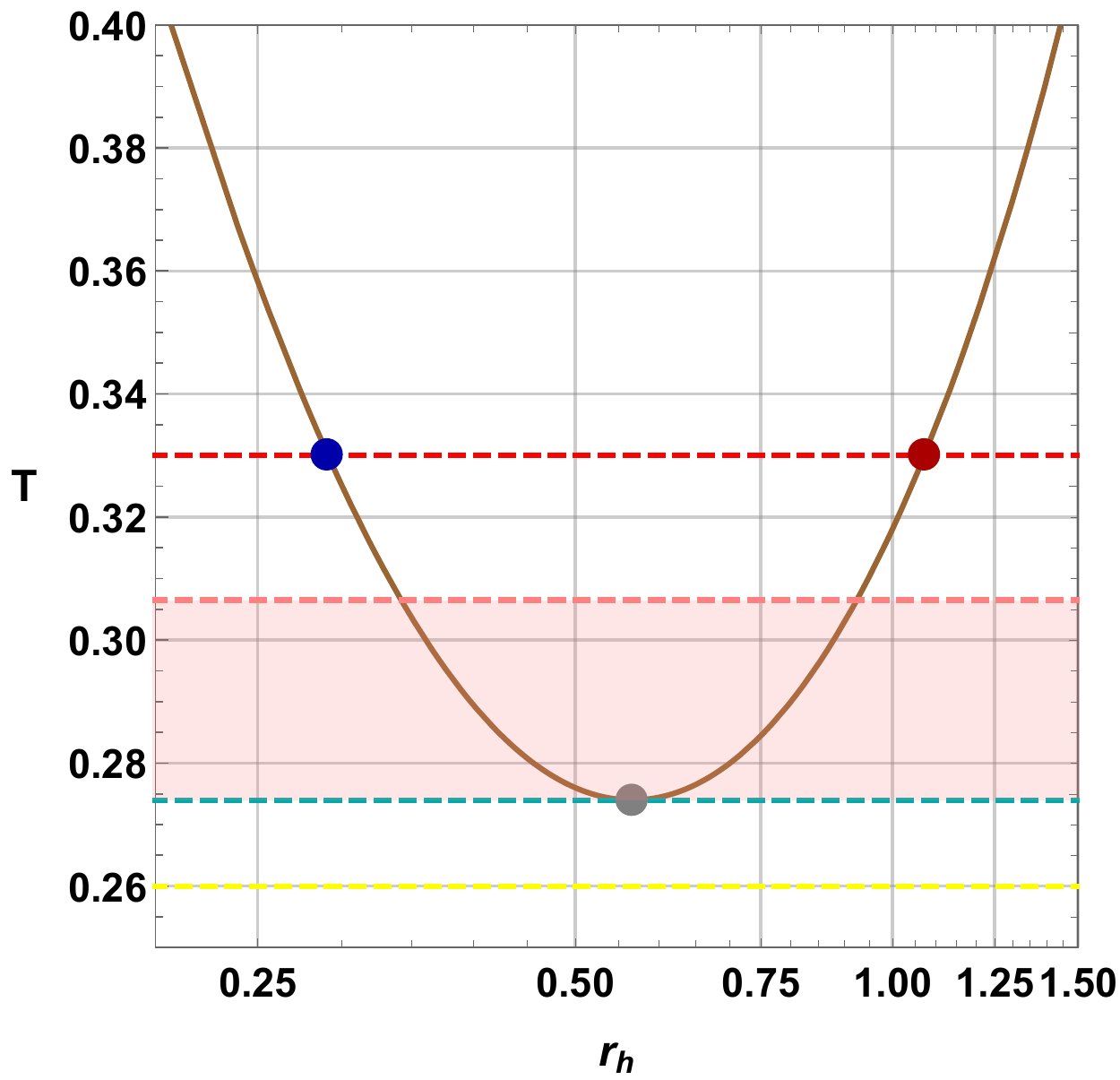}
			\caption{}
			\label{f3_1}
		\end{subfigure}
		\hspace{1pt}	
		\begin{subfigure}[h]{0.45\textwidth}
			\centering \includegraphics[scale=.5]{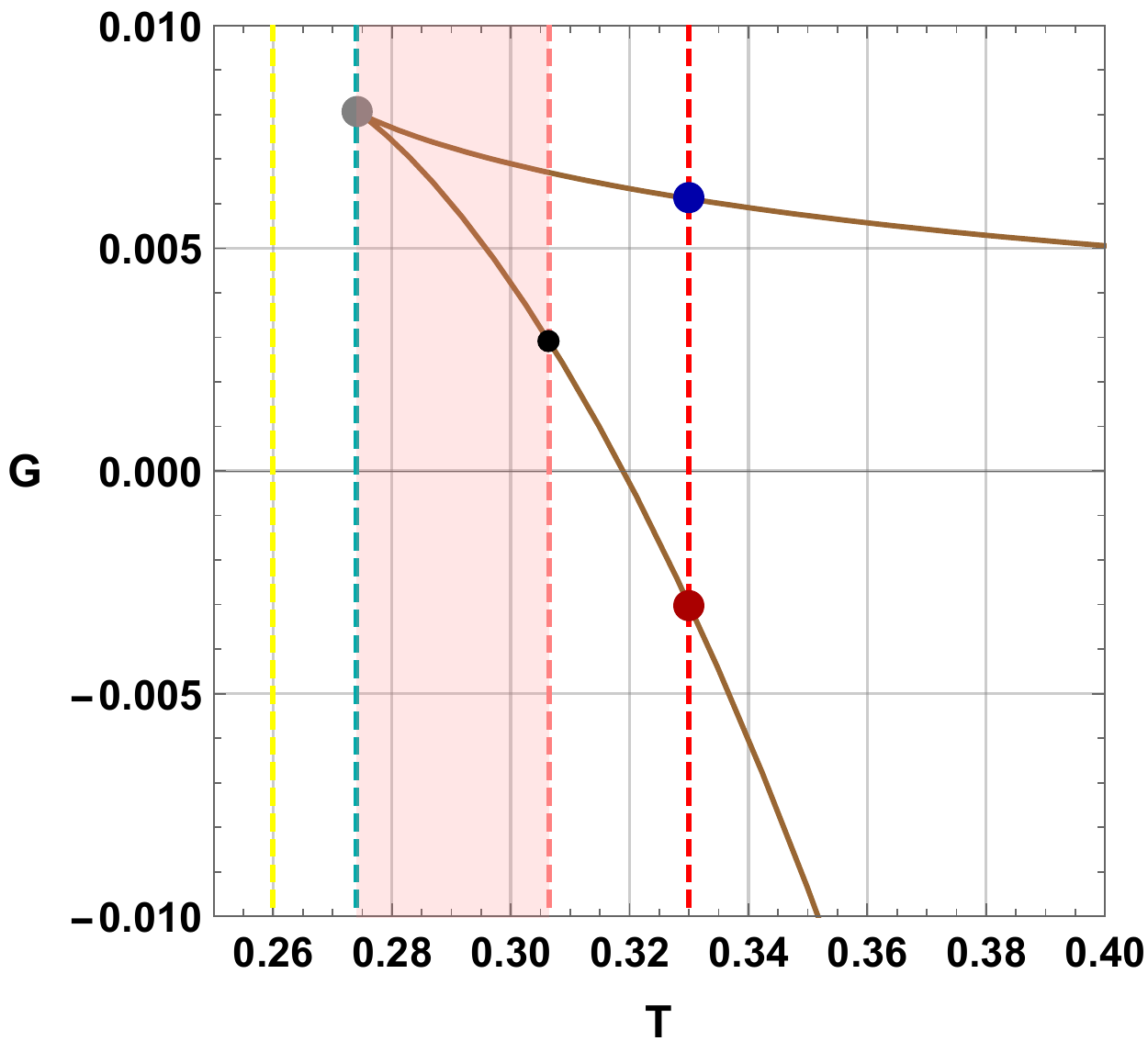}
			\caption{}
			\label{f3_2}		
		\end{subfigure}
		\hspace{1pt}	
		\begin{subfigure}[h]{0.45\textwidth}
			\centering \includegraphics[scale=.5]{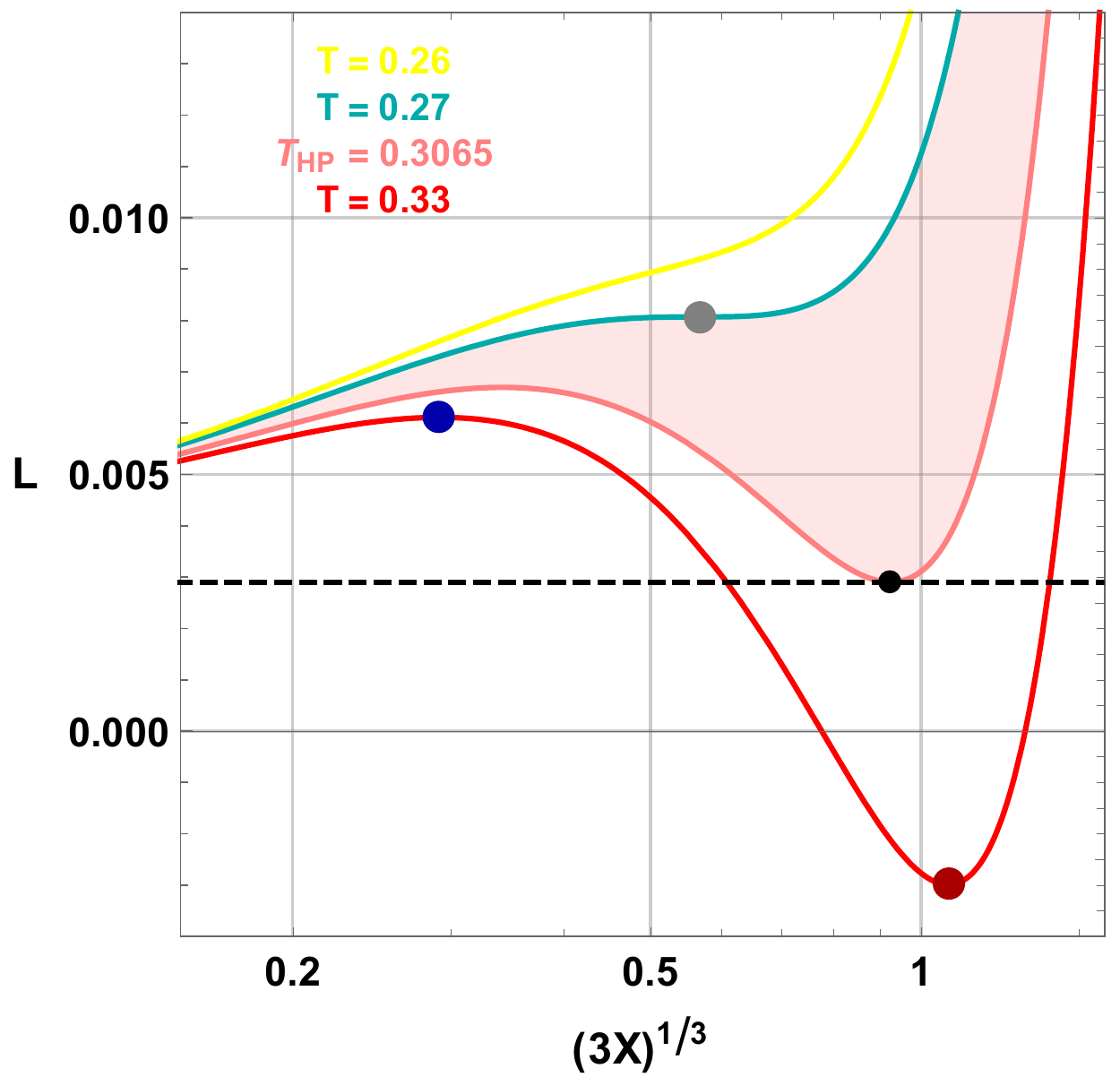}
			\caption{}
			\label{f3_3}	
		\end{subfigure}
		\hspace{1pt}	
		\begin{subfigure}[h]{0.45\textwidth}
			\centering \includegraphics[scale=.5]{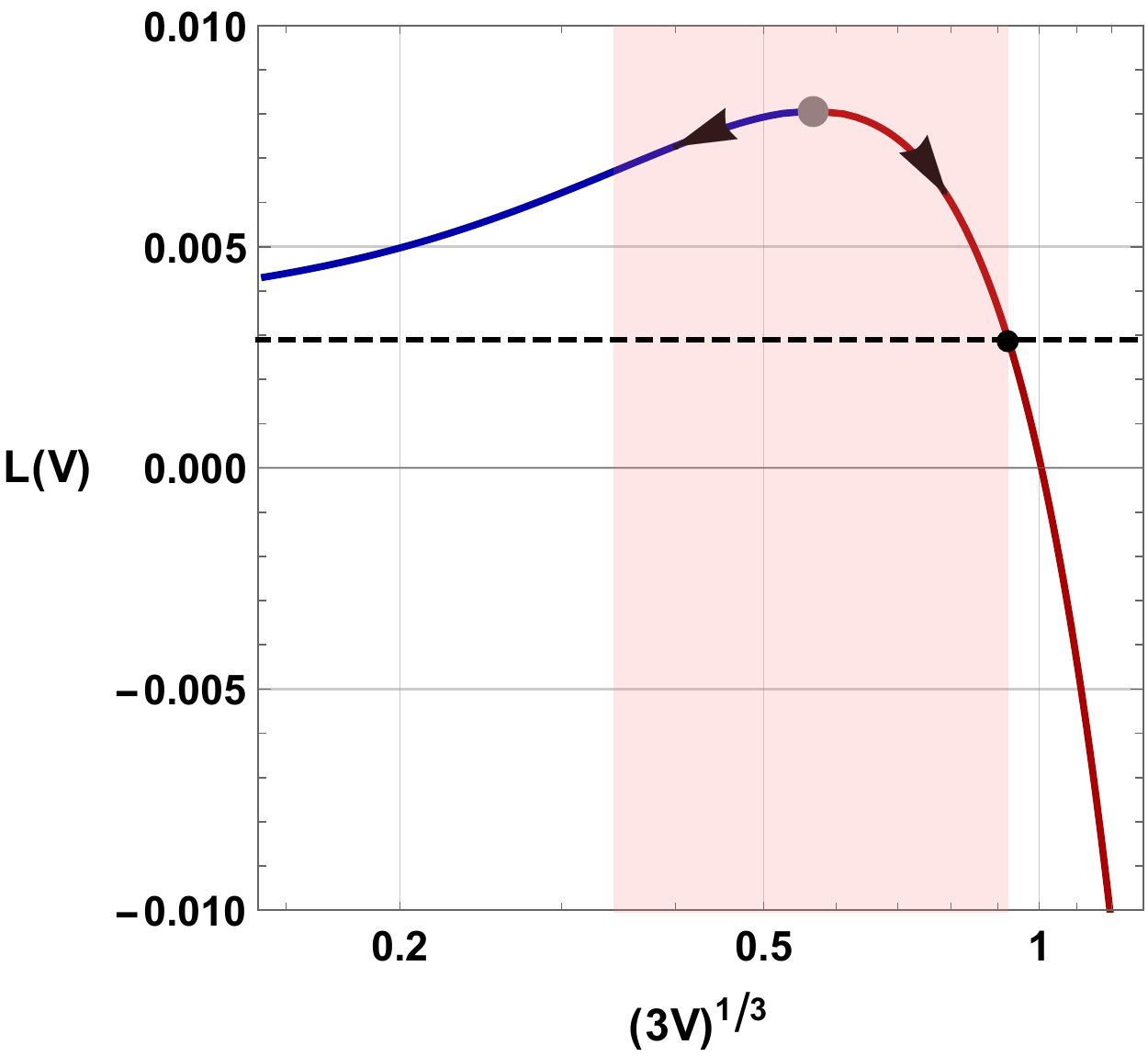}
			\caption{}
			\label{f3_4}	
		\end{subfigure}
		\hspace{1pt}	
		\begin{subfigure}[h]{0.45\textwidth}
			\centering \includegraphics[scale=.5]{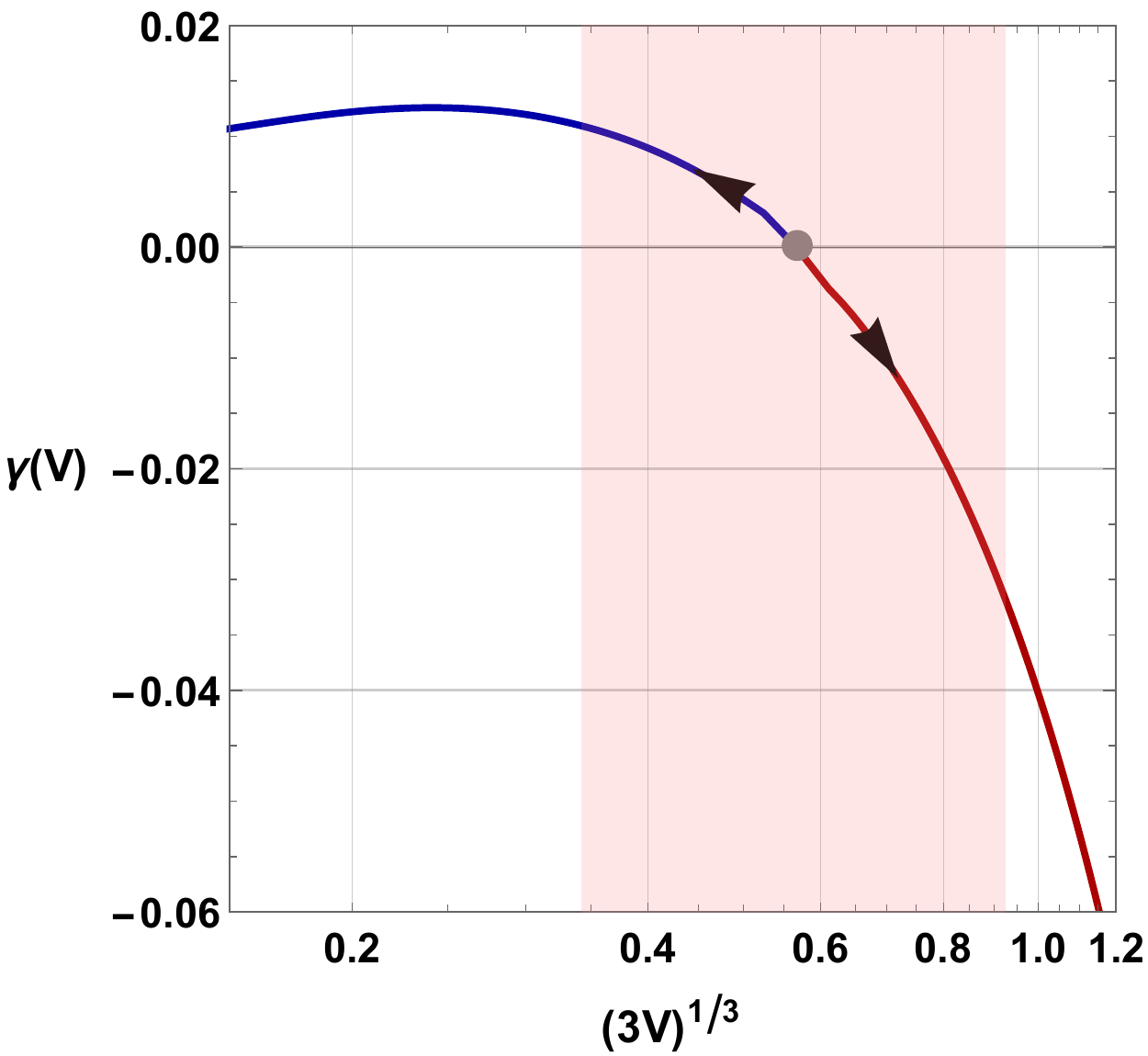}
			\caption{}
			\label{f3_5}
			
		\end{subfigure}
			\hspace{1pt}	
	\begin{subfigure}[h]{0.45\textwidth}
		\centering \includegraphics[scale=.5]{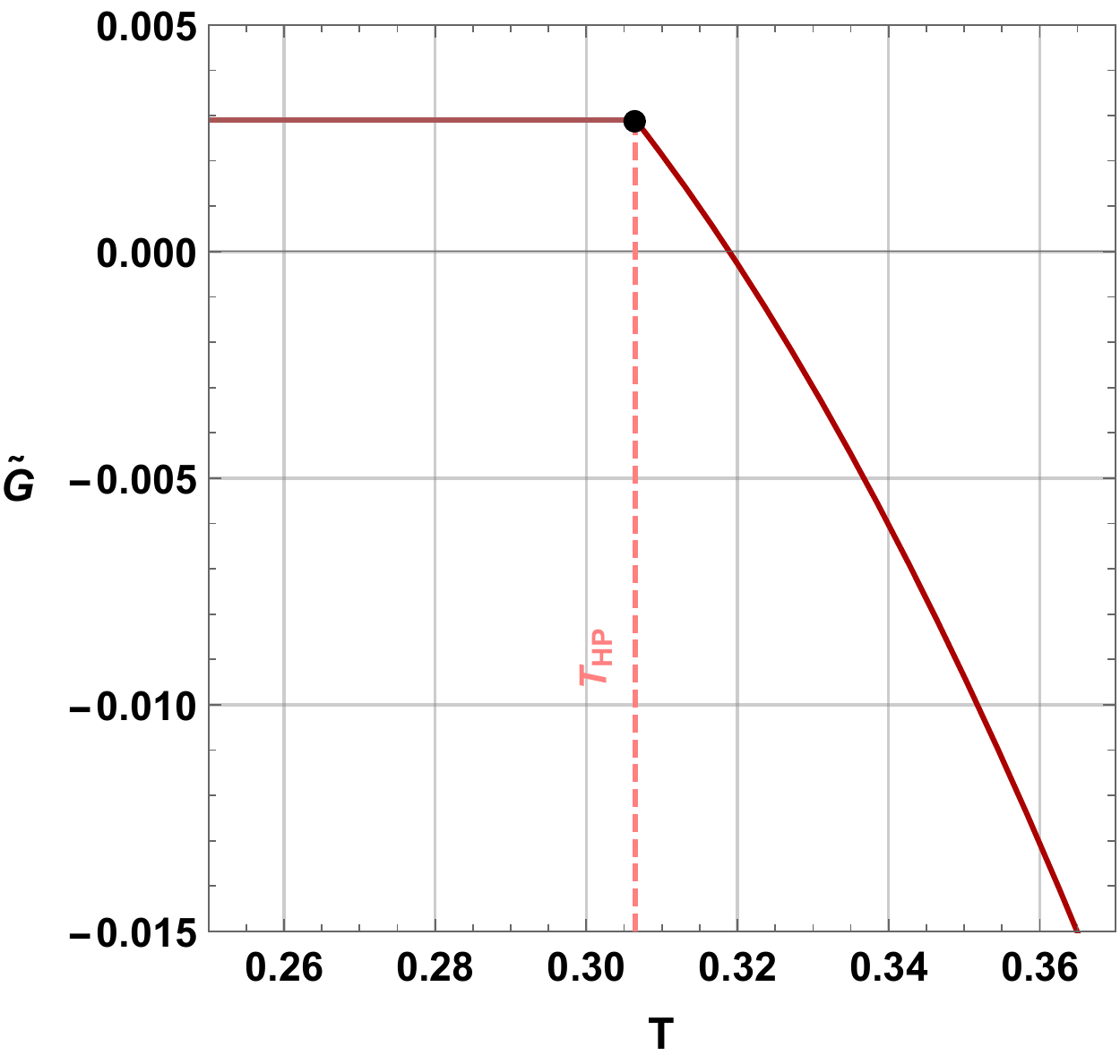}
		\caption{}
		\label{f3_6}
		
	\end{subfigure}
	
		\caption{\footnotesize \it (a) Temperature versus the event horizon radius $r_h$. (b) Gibbs free energy-temperature diagram. (c) Landau function $L$ in terms of the parameter $X$ for different temperatures. (d) Landau function $L$ in terms of the black hole volume $V$. (e)  $\gamma$-function in terms of the black hole volume $V$. (f) On-shell Gibbs free energy $\tilde{G}$ as a function of temperature $T$. The arrows indicate the evolution of the temperature and the pink region indicates where the thermal radiation phase is the global stable phase $(T<T_{HP})$ with $Q = 0.005$, $l=1$, and $b=3.5$.}
		\label{f3}
	\end{figure}
 We observe that the black hole behaves like Shwarzschild one revealing a Hawking-Page transition between the unstable small black holes phase  and stable large black hole one at $T_{HP}=0.3065$. The unstable phase corresponds to the local maximum of the Landau function (blue dot) whereas the stable phase corresponds to the local minimum (red dot). Moreover, we notice that the Landau function is a decreasing function in terms of the black hole volume when the system is stable (red curve) and it is an increasing function when the system is unstable (blue curve). Thus the $\gamma$ function is negative in the large black holes phase and positive in the small black holes one. The zero of $\gamma$-function is associated with an unstable fixed point indicating that the temperature is (locally) minimal. We can interpret these results as the large black hole becoming larger and more stable when it gets hotter, whereas the small hole becomes smaller when it gets hotter because of evaporation. The pink region indicates the zone where the temperature is less than Hawking-Page one (pink curves) and only the thermal radiation phase is globally stable.

	\item  Herein, all previous diagrams are reproduced but for $Q = Q_0 = 0.00922$ in Fig.\ref{f4}.
 
 %the temperature as a function of horizon radius $r_h$ (Fig.\ref{f4_1}), Gibbs free energy as a function of temperature Fig.\ref{f4_2}, Landau function $L$ in terms of the parameter $X$ for different temperatures (Fig.\ref{f4_3}), Landau function $L$ in terms of the black hole volume $V$ Fig.\ref{f4_6},  $\gamma$-function in terms of the black hole volume $V$ (Fig.\ref{f4_5})  and on-shell Gibbs free energy $\tilde{G}$ as a function of temperature (Fig.\ref{f4_6}). 
\begin{figure}[!ht]
	\centering
	\begin{subfigure}[h]{0.45\textwidth}
		\centering \includegraphics[scale=.5]{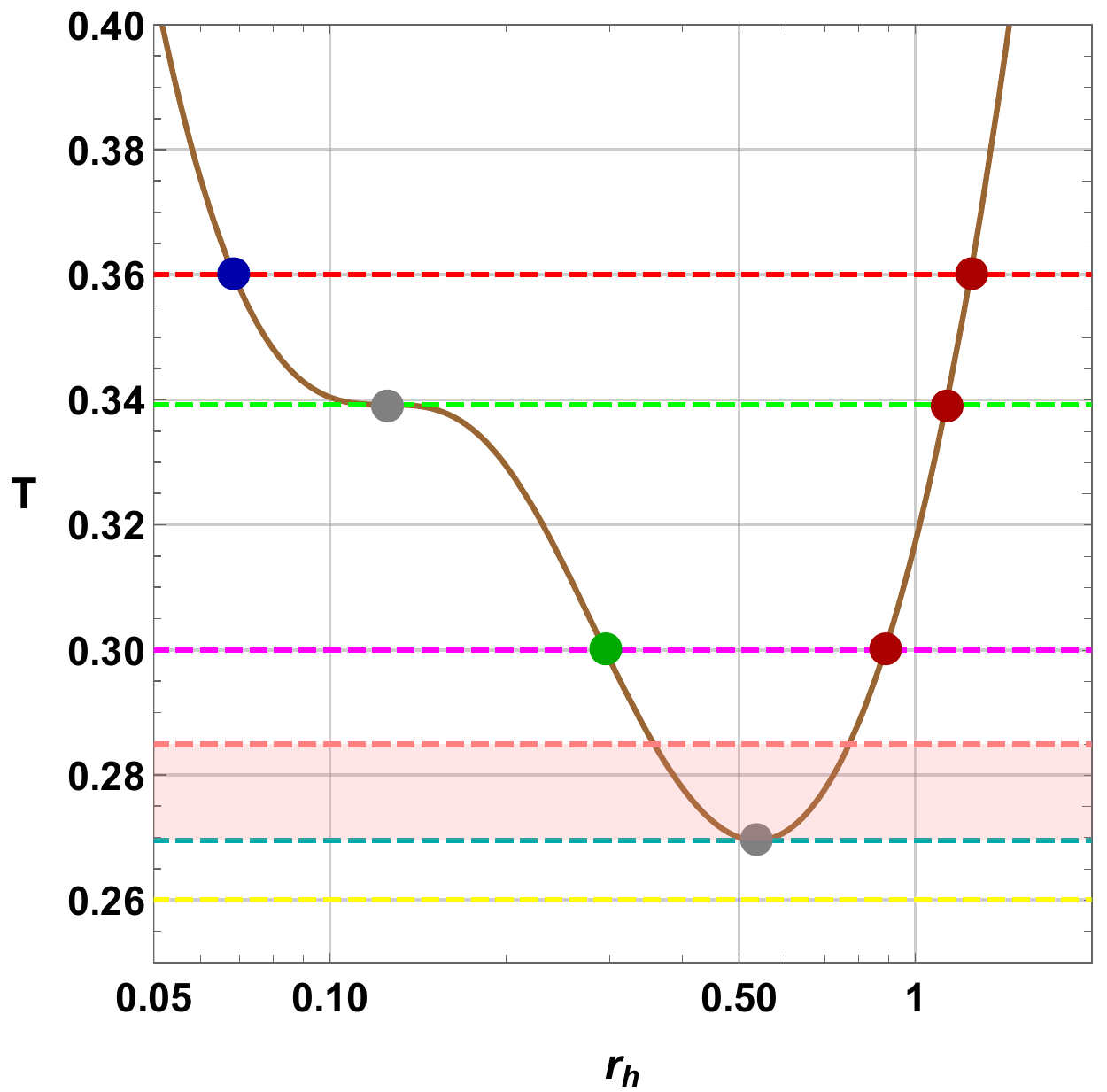}
		\caption{}
		\label{f4_1}
	\end{subfigure}
	\hspace{1pt}	
	\begin{subfigure}[h]{0.45\textwidth}
		\centering \includegraphics[scale=.5]{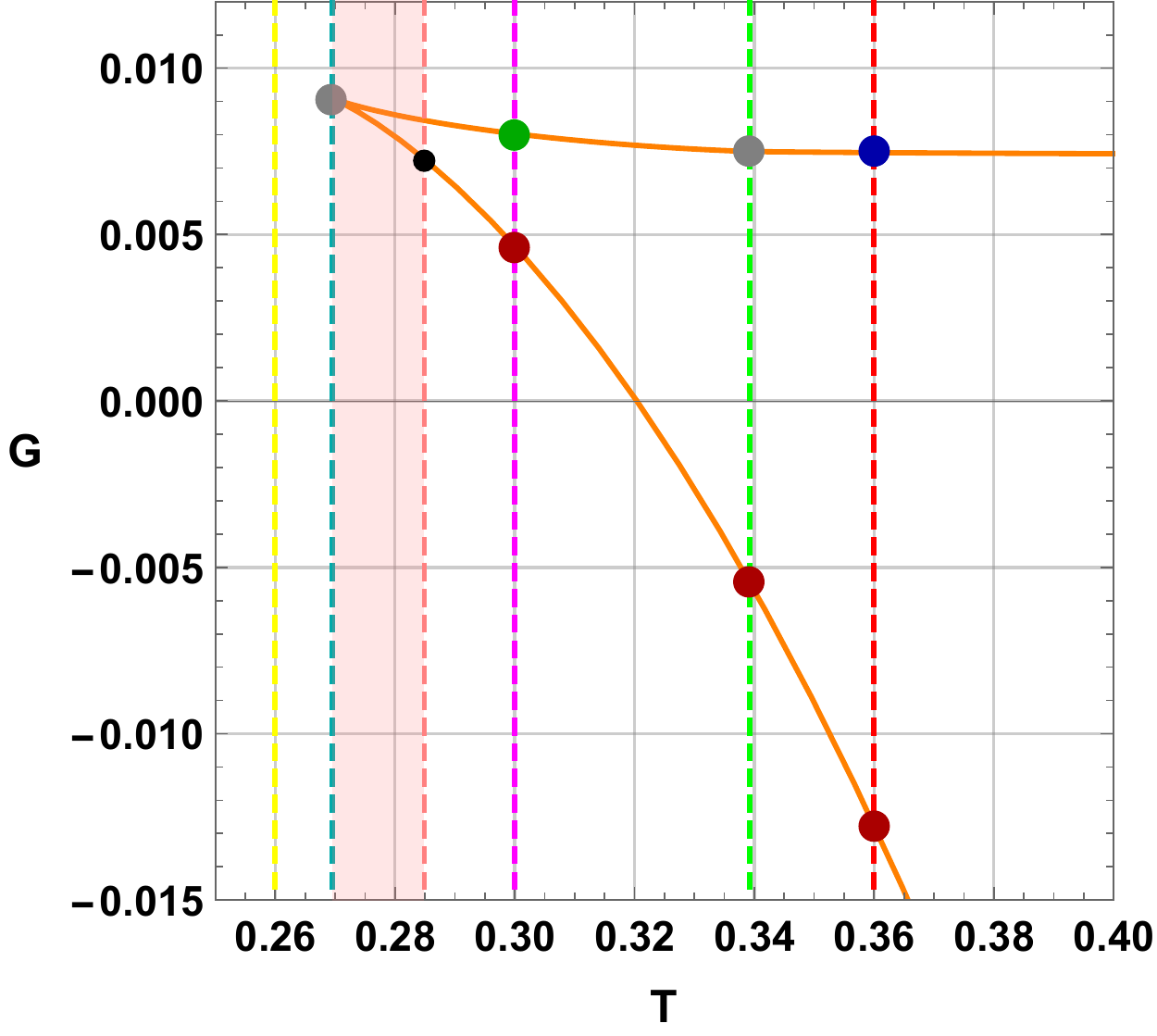}
		\caption{}
		\label{f4_2}		
	\end{subfigure}
	\hspace{1pt}	
	\begin{subfigure}[h]{0.45\textwidth}
		\centering \includegraphics[scale=.5]{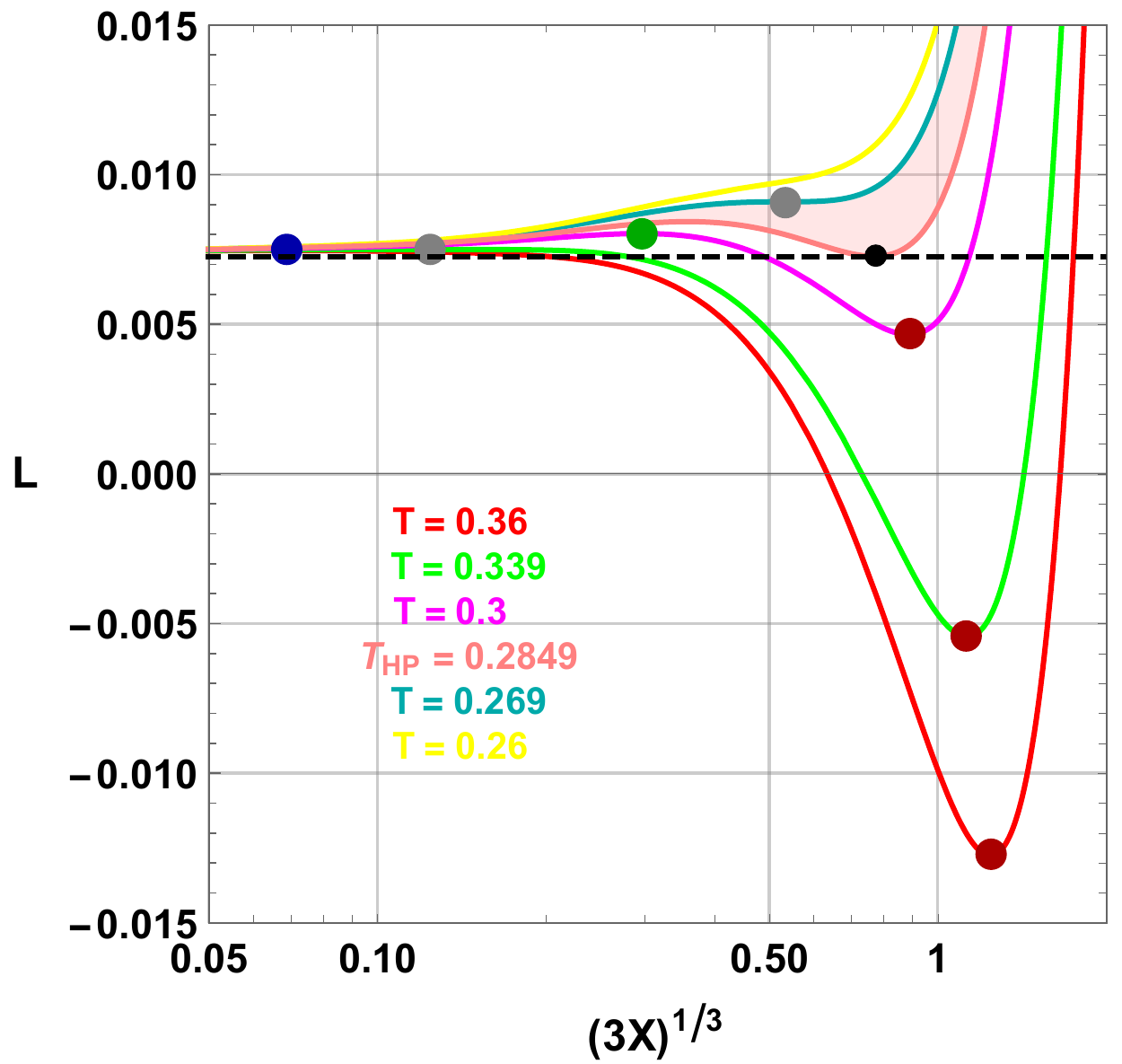}
		\caption{}
		\label{f4_3}	
	\end{subfigure}
	\hspace{1pt}	
	\begin{subfigure}[h]{0.45\textwidth}
		\centering \includegraphics[scale=.5]{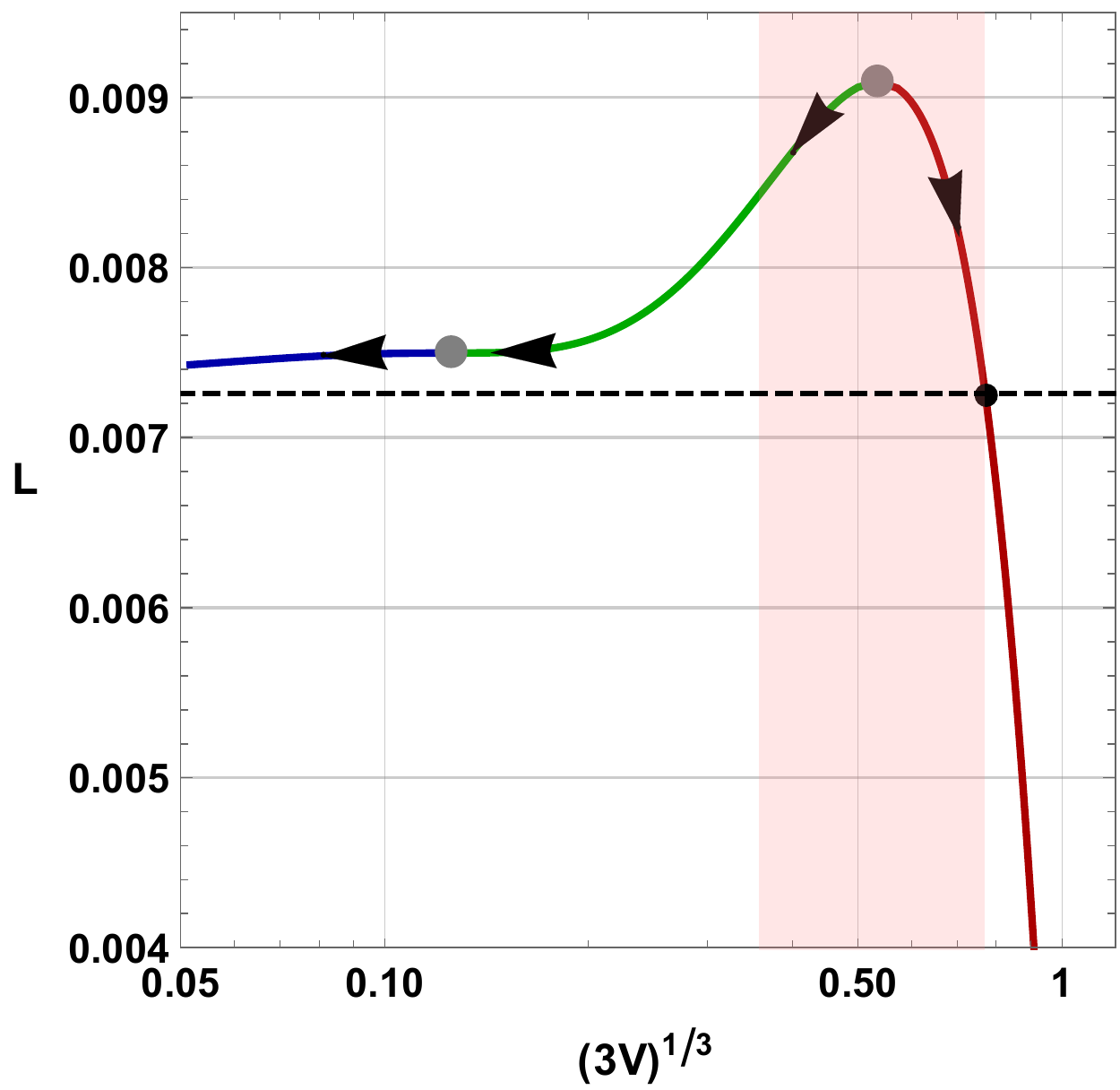}
		\caption{}
		\label{f4_4}	
	\end{subfigure}
	\hspace{1pt}	
	\begin{subfigure}[h]{0.45\textwidth}
		\centering \includegraphics[scale=.5]{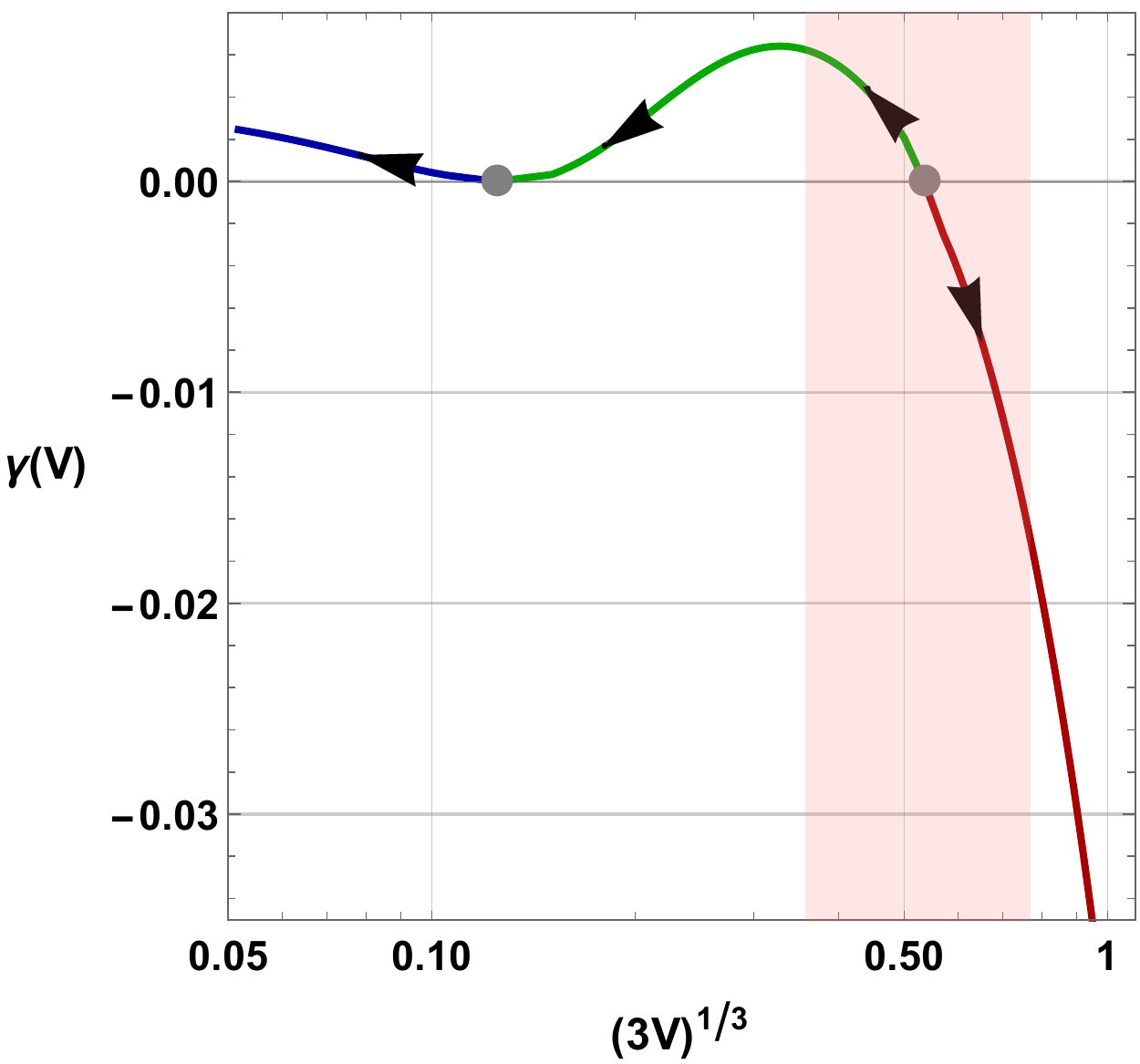}
		\caption{}
		\label{f4_5}
		
	\end{subfigure}
			\hspace{1pt}	
\begin{subfigure}[h]{0.45\textwidth}
	\centering \includegraphics[scale=.5]{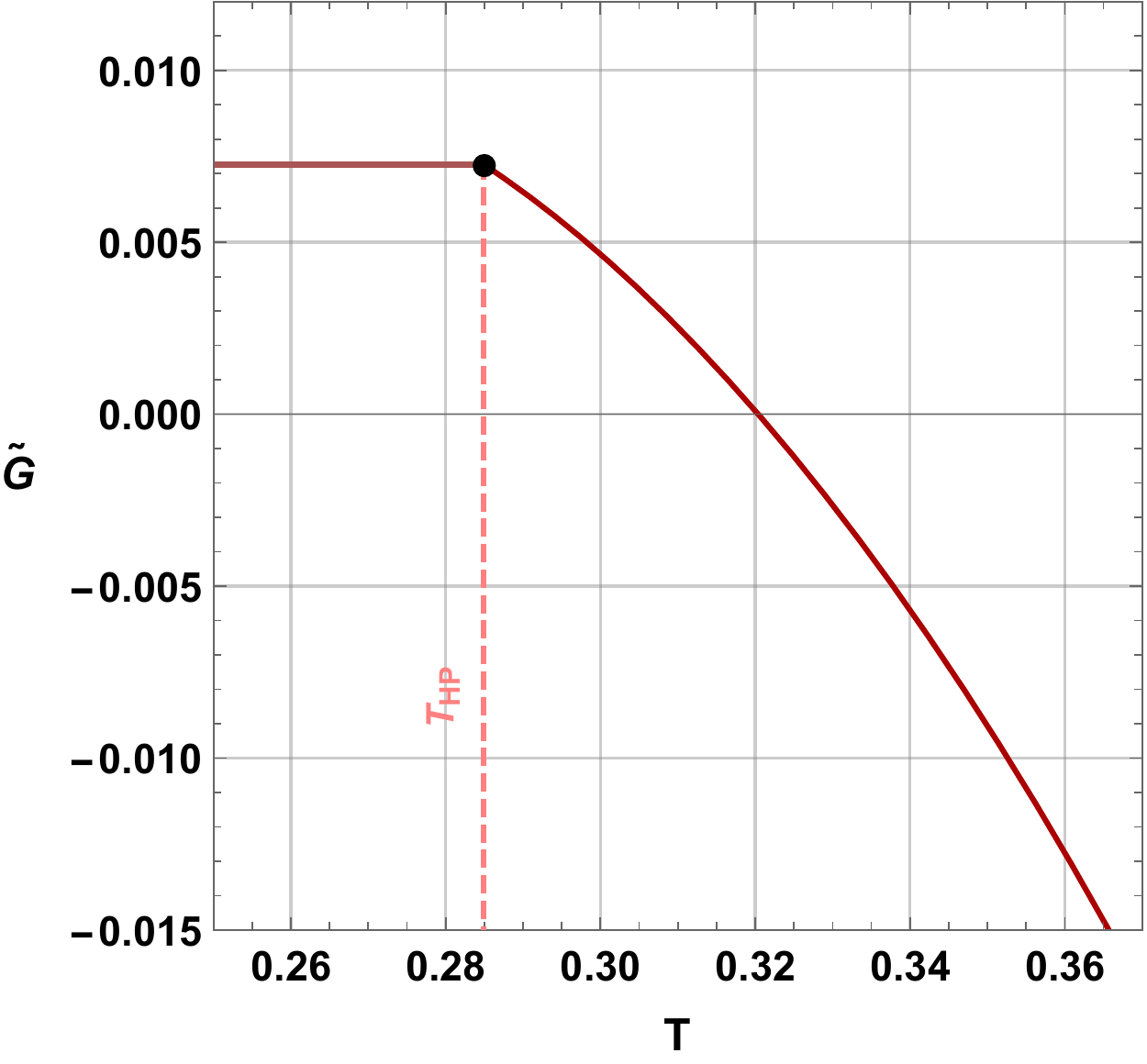}
	\caption{}
	\label{f4_6}
	
\end{subfigure}
	\caption{\footnotesize\it  (a) Temperature versus event horizon radius $r_h$. (b) Gibbs free energy-temperature diagram. (c) Landau function $L$ in terms of the parameter $X$ for different temperatures. (d) Landau function $L$ in terms of the black hole volume $V$. (e)  $\gamma$-function in terms of the black hole volume $V$. (f) On-shell Gibbs free energy $\tilde{G}$ as a function of temperature $T$. The arrows indicate the evolution of the temperature and the pink region indicates where the thermal radiation phase is the global stable phase $(T<T_{HP})$ with $Q = 0.00922$, $l=1$, and $b=3.5$.}
	\label{f4}
\end{figure}
It's remarked that the temperature in terms of the horizon radius shows an inflection point that corresponds to a discontinuity in the first derivative of the Gibbs free energy and an inflection point in the Landau function (green curve). Thus, we have a new black hole phase called intermediate black holes which is an unstable phase like the unstable small black holes as we can see in Fig.\ref{f4_3} where this phase corresponds to a local maximum of Landau function. Moreover, we see in Fig.\ref{f4_4} that the landau function is increasing in terms of black hole volume which is a signature of the instability. We see in Fig.\ref{f4_5} that $\gamma$-function has two fixed points, one is unstable which corresponds to Hawking-Page-like and indicates that the temperature is minimal at this point, while the second fixed point is semi-stable and which separates two unstable phases (small and intermediate black holes) and corresponds to an inflection point in the behavior of the temperature. Finally, Fig.\ref{f4_6} reveals that the black hole is similar to AdS Schwarzschild black holes where only the thermal radiations and large black hole phases are globally stable.

	\item  Now, the charge is set to $Q = 0.01$ in Fig.\ref{f5} %the temperature as a function of horizon radius $r_h$ (Fig.\ref{f5_1}), Gibbs free energy as a function of temperature Fig.\ref{f5_2}, Landau function $L$ in terms of the parameter $X$ for different temperatures (Fig.\ref{f5_3}), Landau function $L$ in terms of the black hole volume $V$ Fig.\ref{f5_4} ,  $\gamma$-function in terms of the black hole volume $V$ (Fig.\ref{f5_5})  and on-shell Gibbs free energy $\tilde{G}$ as a function of temperature (Fig.\ref{f5_6}).
 
\begin{figure}[!ht]
	\centering
	\begin{subfigure}[h]{0.45\textwidth}
		\centering \includegraphics[scale=.5]{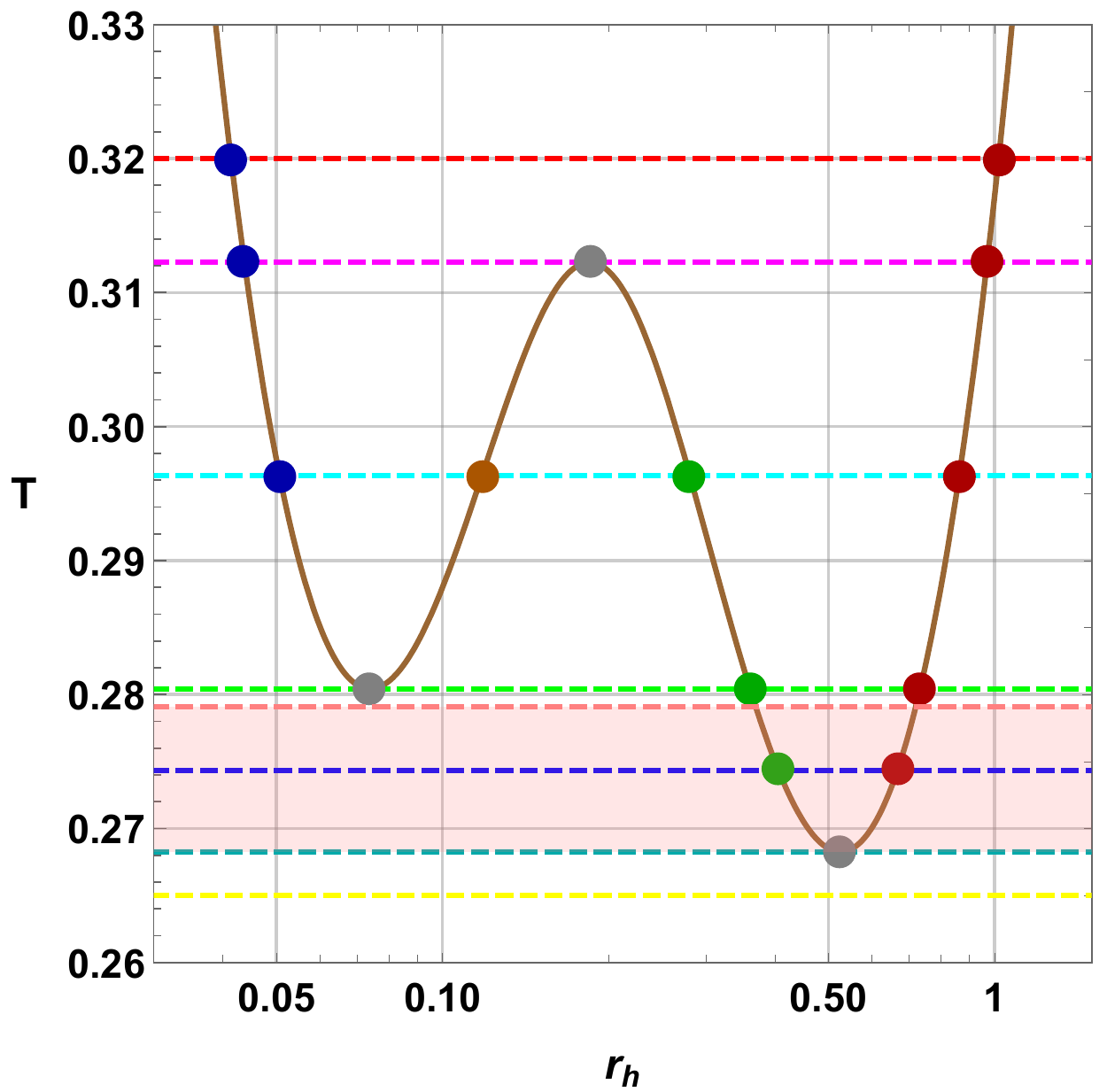}
		\caption{}
		\label{f5_1}
	\end{subfigure}
	\hspace{1pt}	
	\begin{subfigure}[h]{0.45\textwidth}
		\centering \includegraphics[scale=.5]{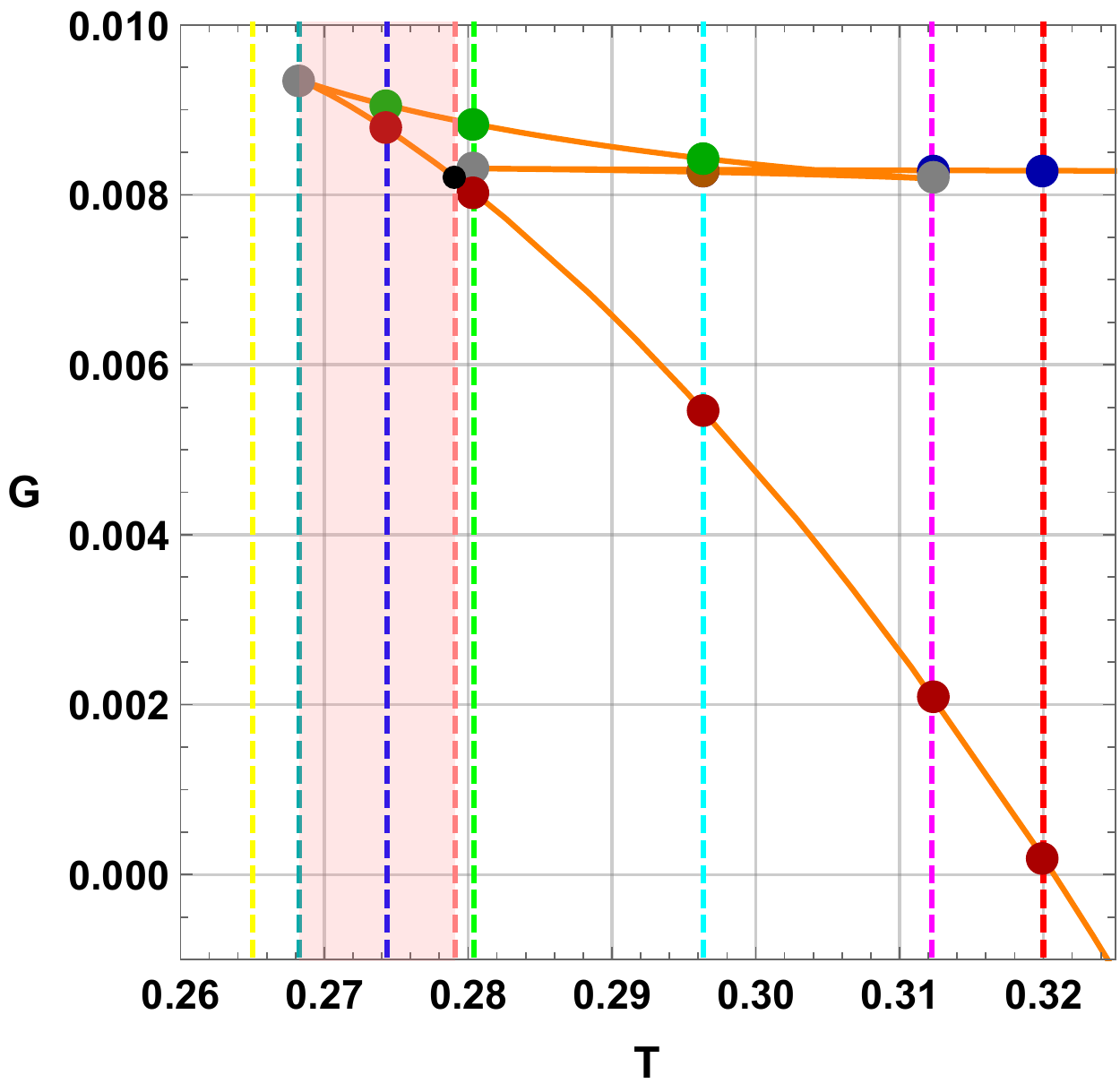}
		\caption{}
		\label{f5_2}		
	\end{subfigure}
	\hspace{1pt}	
	\begin{subfigure}[h]{0.45\textwidth}
		\centering \includegraphics[scale=.5]{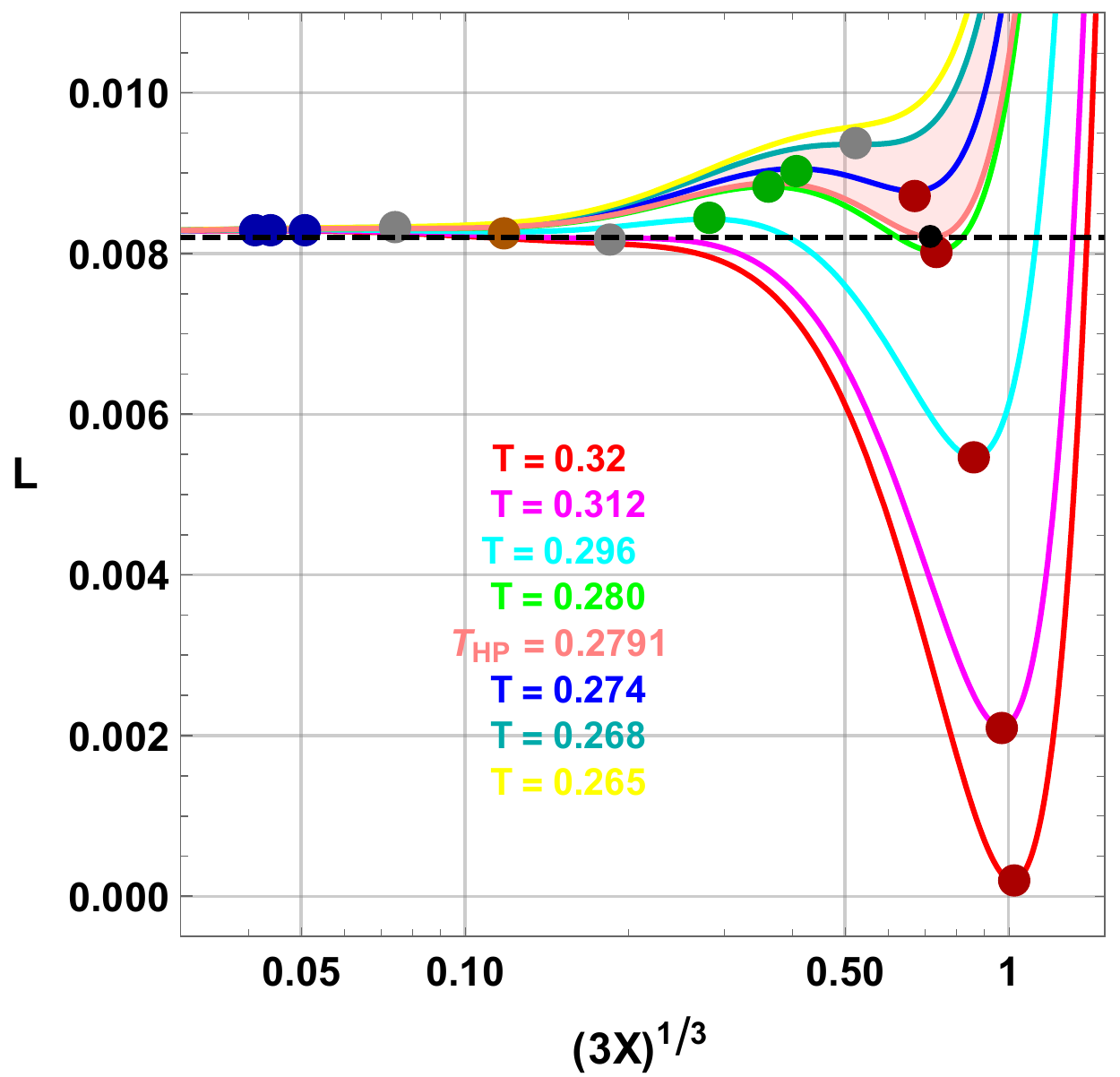}
		\caption{}
		\label{f5_3}	
	\end{subfigure}
	\hspace{1pt}	
	\begin{subfigure}[h]{0.45\textwidth}
		\centering \includegraphics[scale=.5]{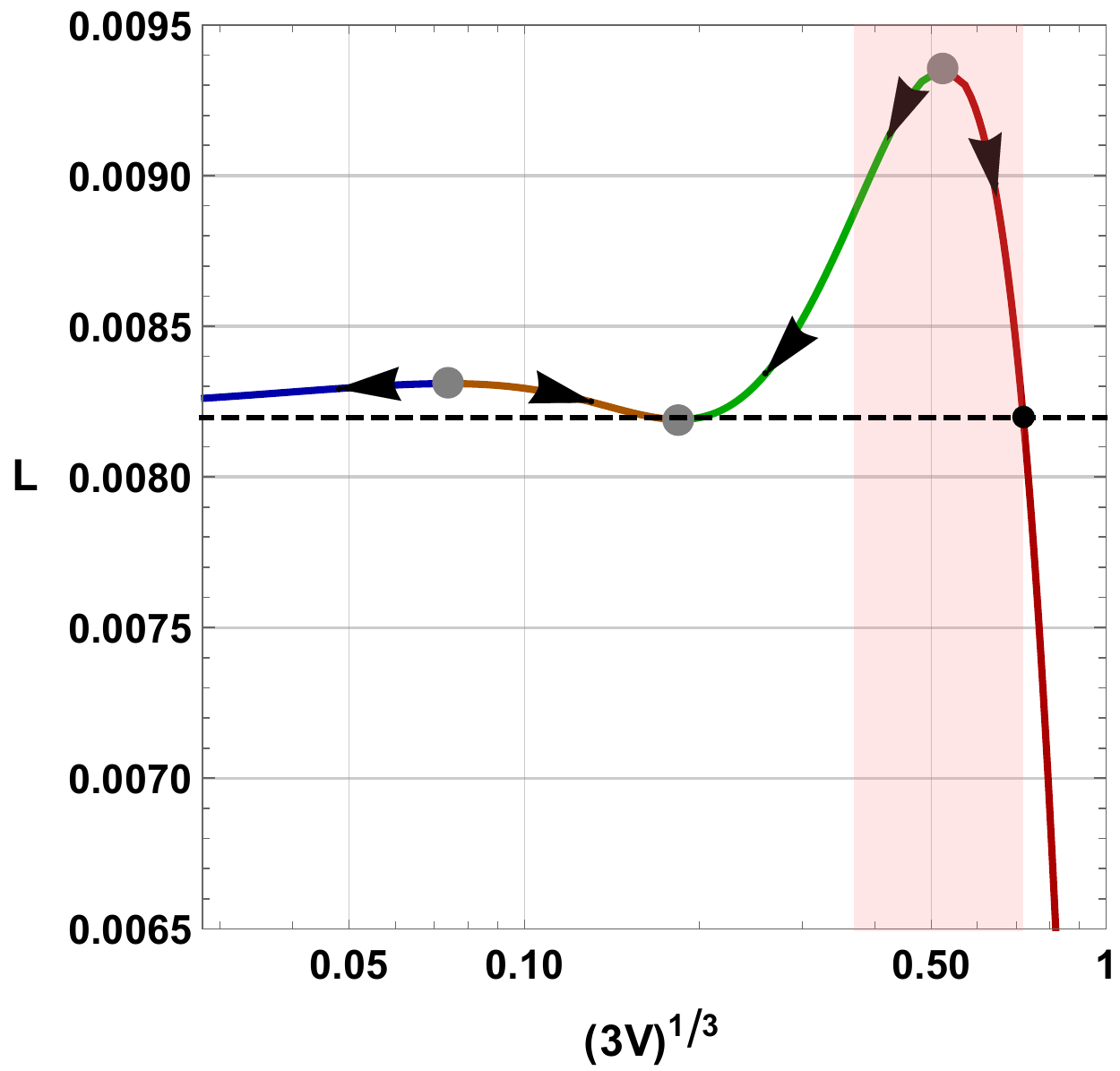}
		\caption{}
		\label{f5_4}	
	\end{subfigure}
	\hspace{1pt}	
	\begin{subfigure}[h]{0.45\textwidth}
		\centering \includegraphics[scale=.5]{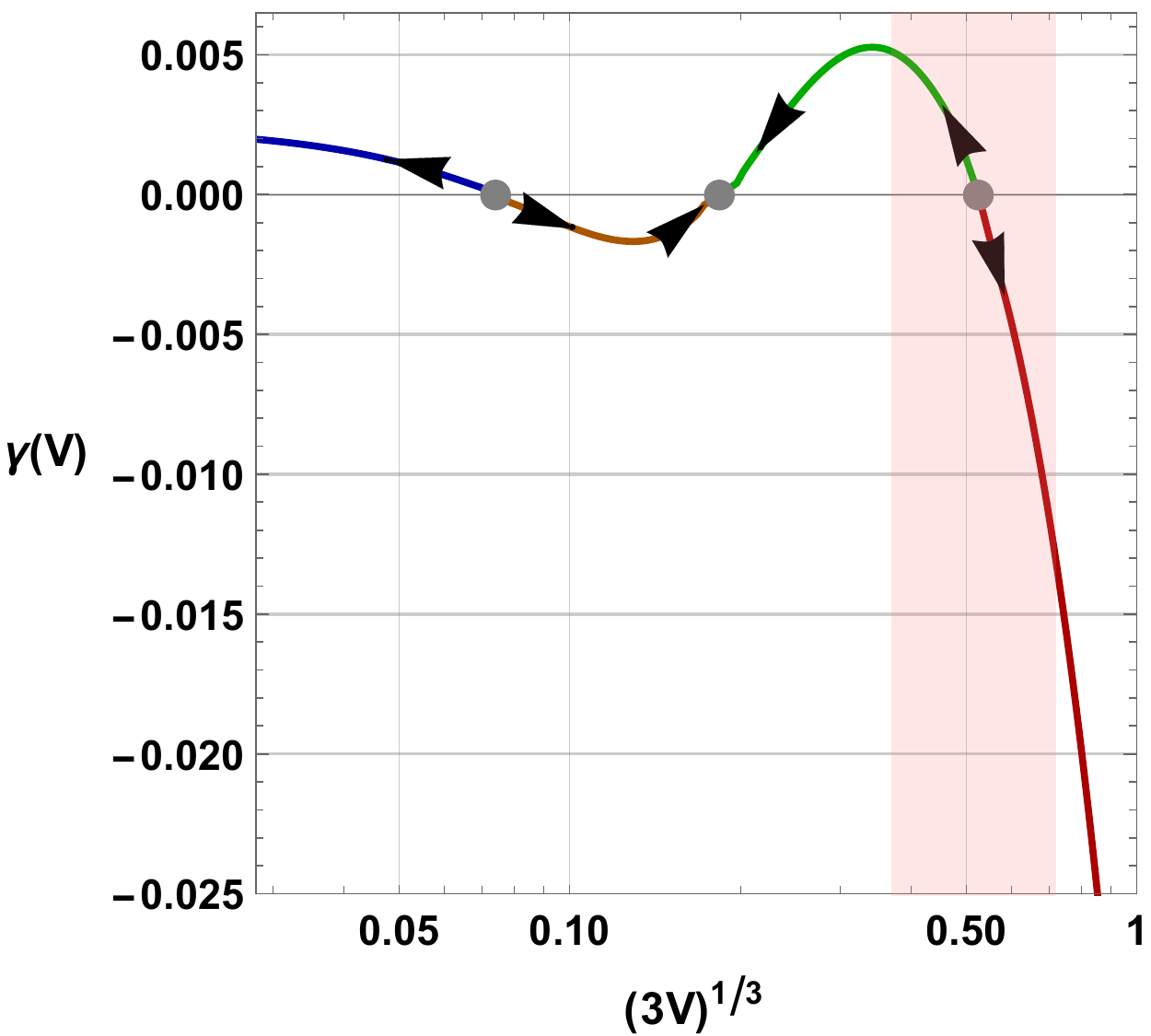}
		\caption{}
		\label{f5_5}
	\end{subfigure}
			\hspace{1pt}	
\begin{subfigure}[h]{0.45\textwidth}
	\centering \includegraphics[scale=.5]{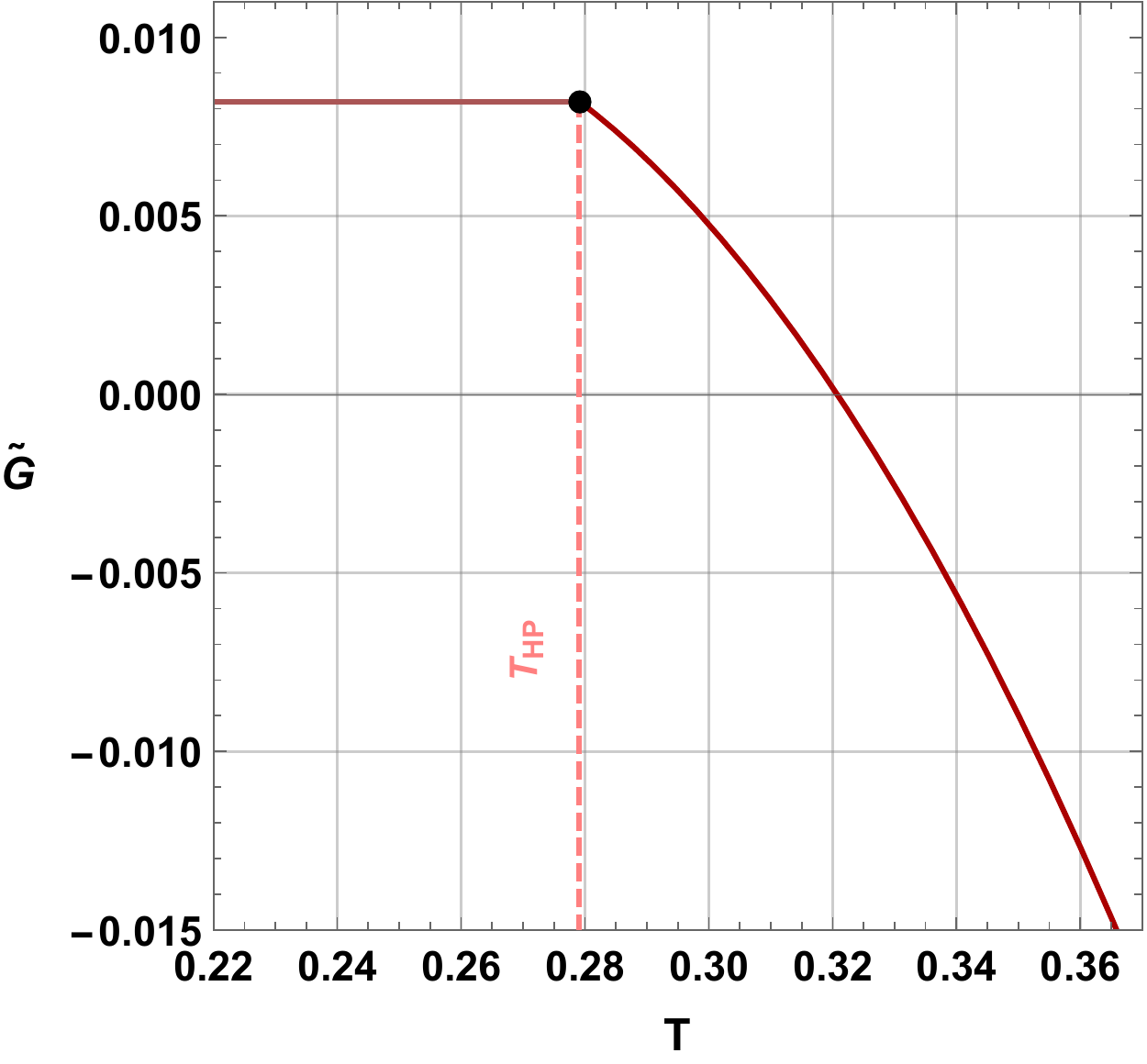}
	\caption{}
	\label{f5_6}
	
\end{subfigure}
	\caption{\footnotesize\it (a) Temperature versus the event horizon radius $r_h$. (b) Gibbs free energy-temperature diagram. (c) Landau function $L$ in terms of the parameter $X$ for different temperatures. (d) Landau function $L$ in terms of the black hole volume $V$. (e)  $\gamma$-function in terms of the black hole volume $V$. (f) On-shell Gibbs free energy $\tilde{G}$ as a function of temperature $T$. The arrows indicate the evolution of the temperature and the pink region indicates where the thermal radiation phase is the global stable phase $(T<T_{HP})$ with $Q = 0.01$, $l=1$, and $b=3.5$.}
	\label{f5}
\end{figure}
 We notice that the temperature curve as a function of the horizon radius presents two minimums (gray points), one is local ($r_h = 0.07$) and the other is global ($r_h = 0.52$). In addition, it has a local maximum at ($r_h=0.18$). These minima and maxima correspond to discontinuities in the first derivative of the Gibbs free energy and inflection points in the Landau function (magenta, orange, and black curves). Thus, we are in the presence of four black hole phases, two stable phases which are large and stable small black holes (dark red and orange points respectively), and two unstable phases which are unstable small and intermediate black holes (dark blue and green points respectively) as illustrated in Fig.\ref{f5_3}, where these phases correspond to local minimums and maximums of the Landau function. Moreover, we see in Fig.\ref{f5_4} that the landau function is decreasing in terms of black hole volume for the stable phases and it is increasing for unstable ones. In addition, one can observe in Fig.\ref{f5_5} that $\gamma$-function has three fixed points, one is unstable (right point) which separates the large black hole phase and the unstable intermediate black hole one which indicates that the temperature is minimal at this point,  the second fixed point is stable (middle point) which disconnects the stable small black holes phase and the unstable intermediate black holes one and corresponds to the temperature maximum, while the third fixed is unstable (left point) which is the limit between the stable small black hole phase and unstable one, it indicates that the temperature has a local minimum at this point. Furthermore, we state that the stable black holes become larger when they get hotter whereas the unstable ones become smaller when they get hotter.  Finally, we see in Fig.\ref{f5_6}, the black hole is similar to AdS Schwarzschild black holes where only the thermal radiations and large black hole phases are globally stable.

\item  The case of $Q = 0.010026$ is depicted  in Fig.\ref{f6} %the temperature as a function of horizon radius $r_h$ (Fig.\ref{f6_1}), Gibbs free energy as a function of temperature (Fig.\ref{f6_2}), Landau function $L$ in terms of the parameter $X$ for different temperatures (Fig.\ref{f6_3}), Landau function $L$ in terms of the black hole volume $V$ Fig.\ref{f6_4} ,  $\gamma$-function in terms of the black hole volume $V$ (Fig.\ref{f6_5})  and on-shell Gibbs free energy $\tilde{G}$ as a function of temperature (Fig.\ref{f6_6}). 

\begin{figure}[!ht]
	\centering
	\begin{subfigure}[h]{0.45\textwidth}
		\centering \includegraphics[scale=.5]{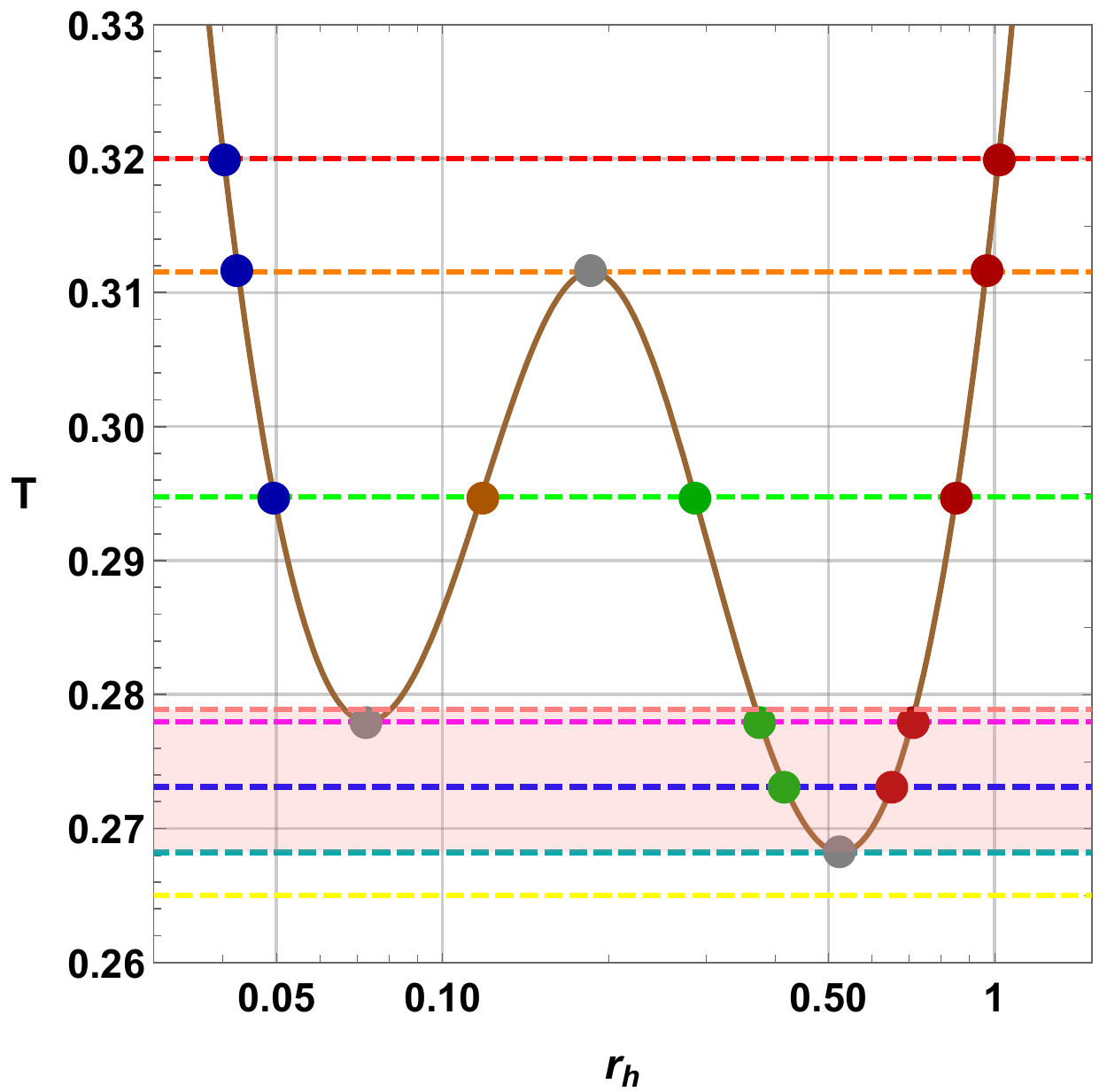}
		\caption{}
		\label{f6_1}
	\end{subfigure}
	\hspace{1pt}	
	\begin{subfigure}[h]{0.45\textwidth}
		\centering \includegraphics[scale=.5]{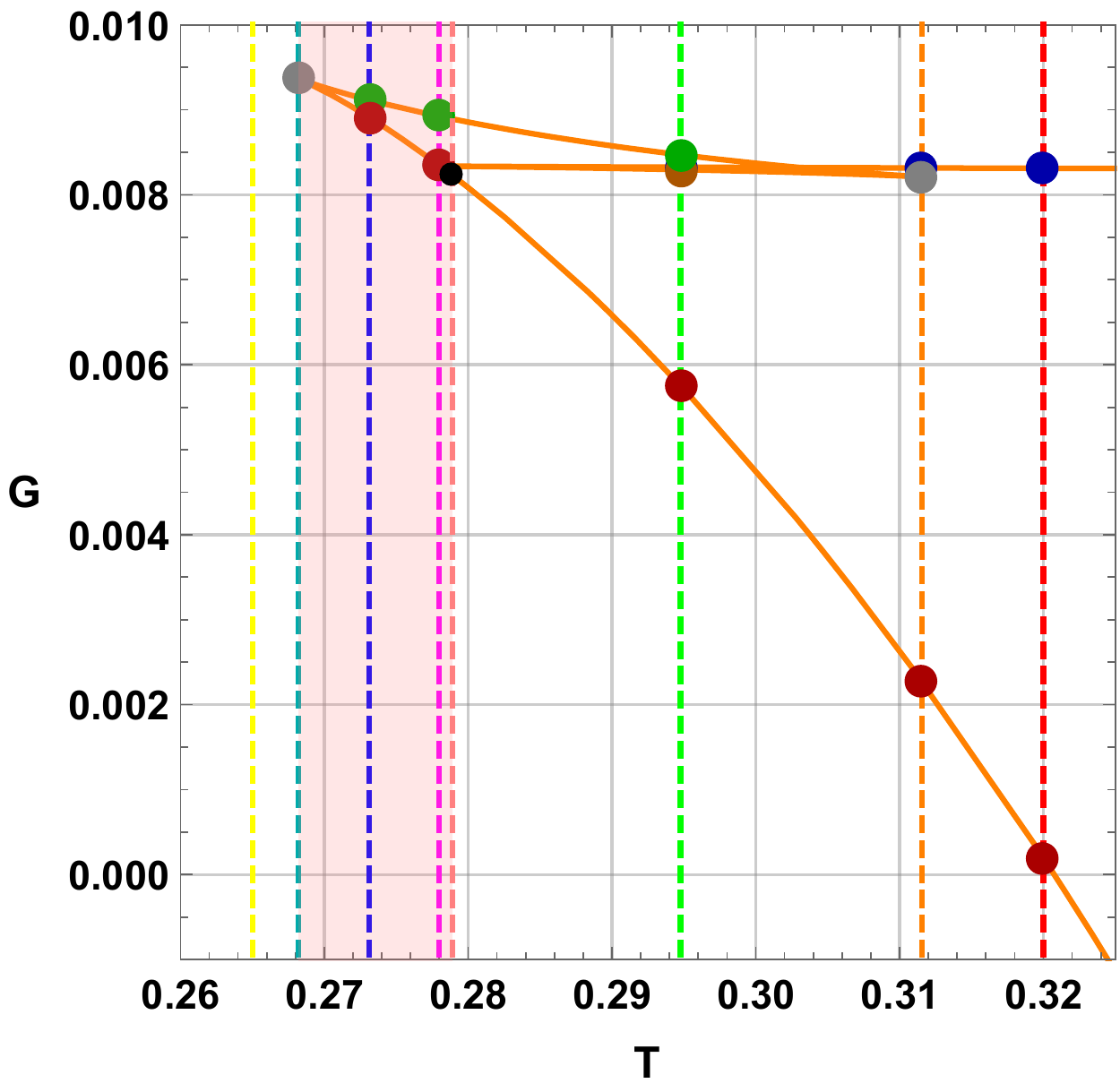}
		\caption{}
		\label{f6_2}		
	\end{subfigure}
	\hspace{1pt}	
	\begin{subfigure}[h]{0.45\textwidth}
		\centering \includegraphics[scale=.5]{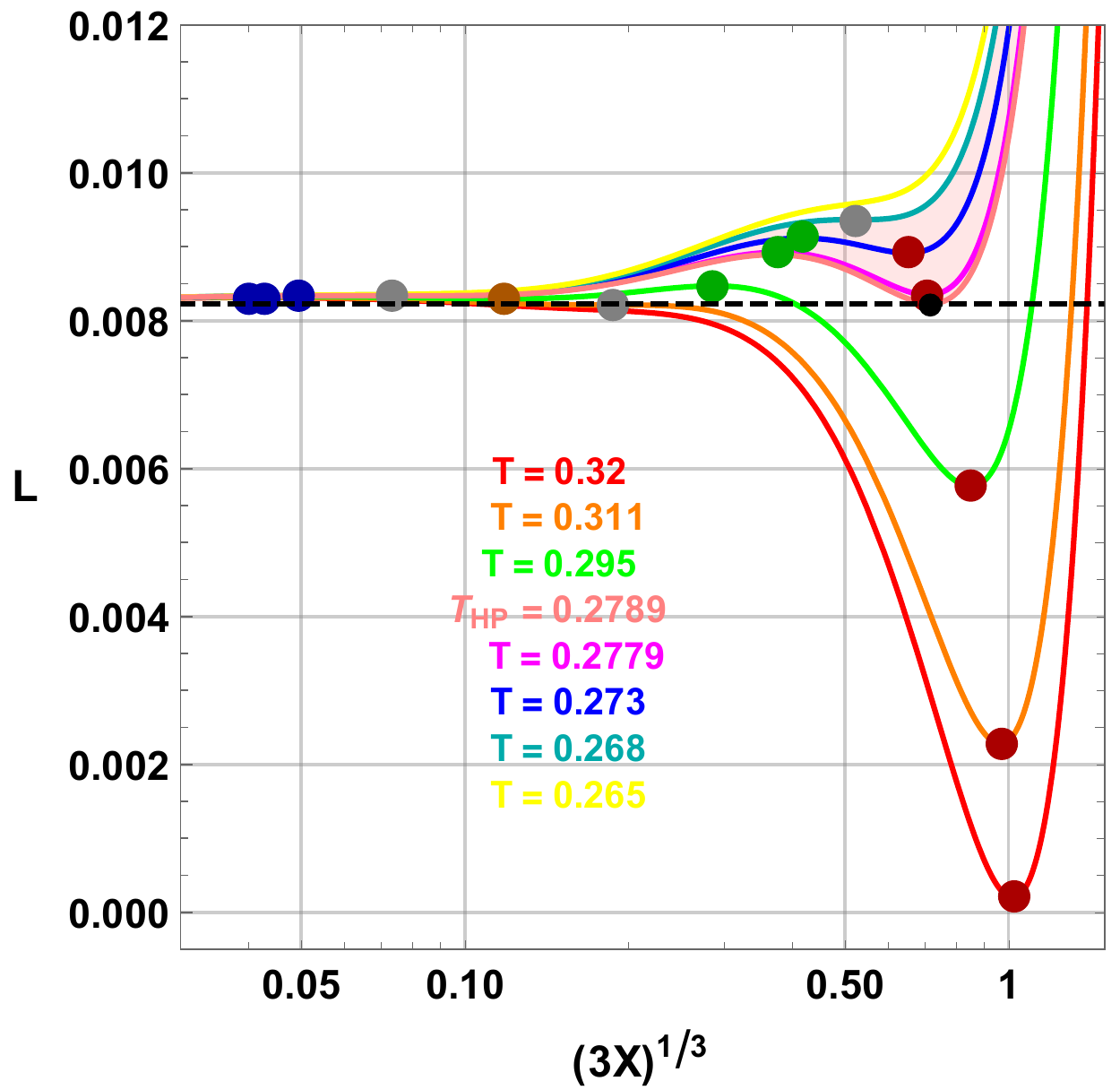}
		\caption{}
		\label{f6_3}	
	\end{subfigure}
	\hspace{1pt}	
	\begin{subfigure}[h]{0.45\textwidth}
		\centering \includegraphics[scale=.5]{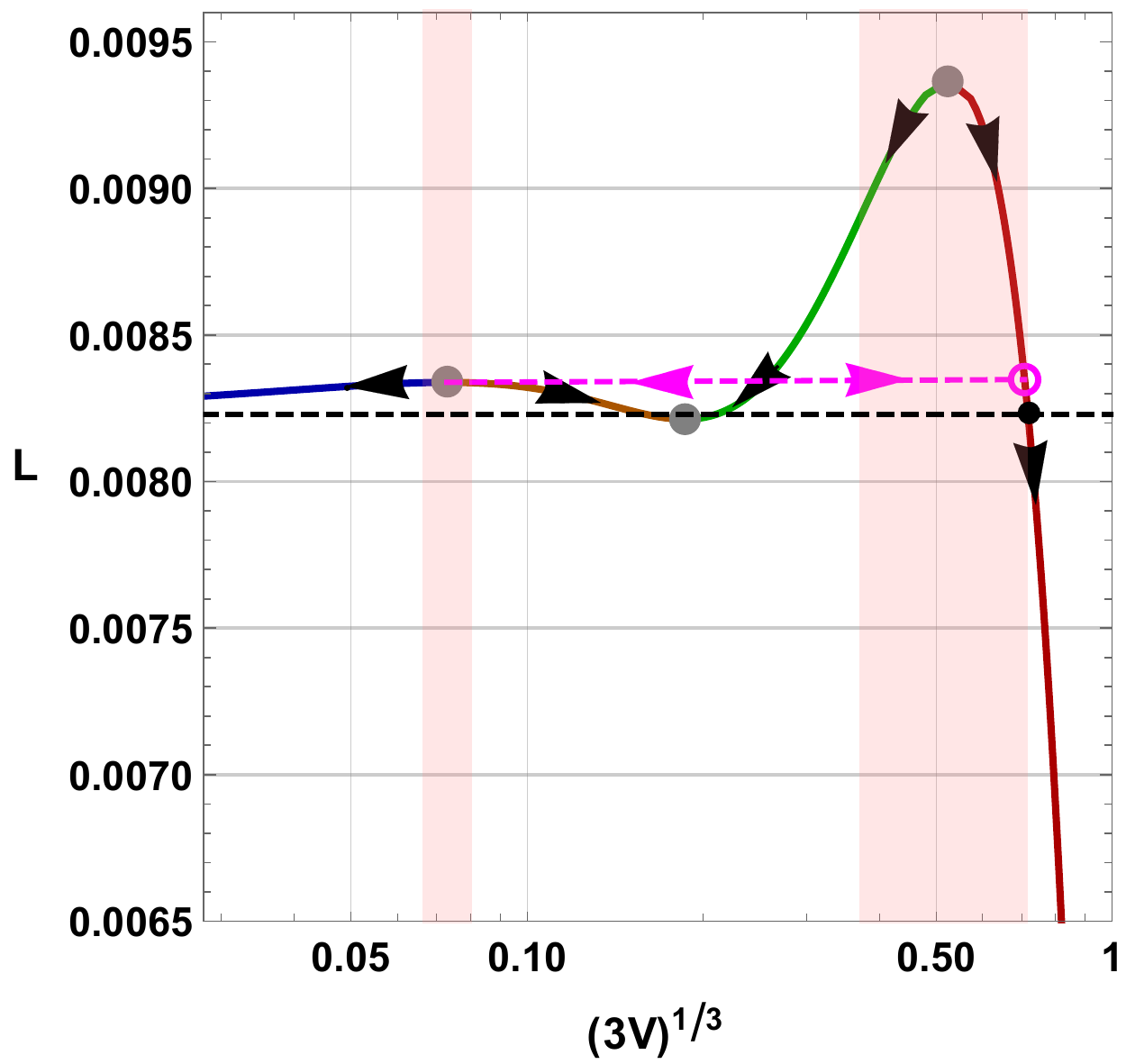}
		\caption{}
		\label{f6_4}	
	\end{subfigure}
	\hspace{1pt}	
	\begin{subfigure}[h]{0.45\textwidth}
		\centering \includegraphics[scale=.5]{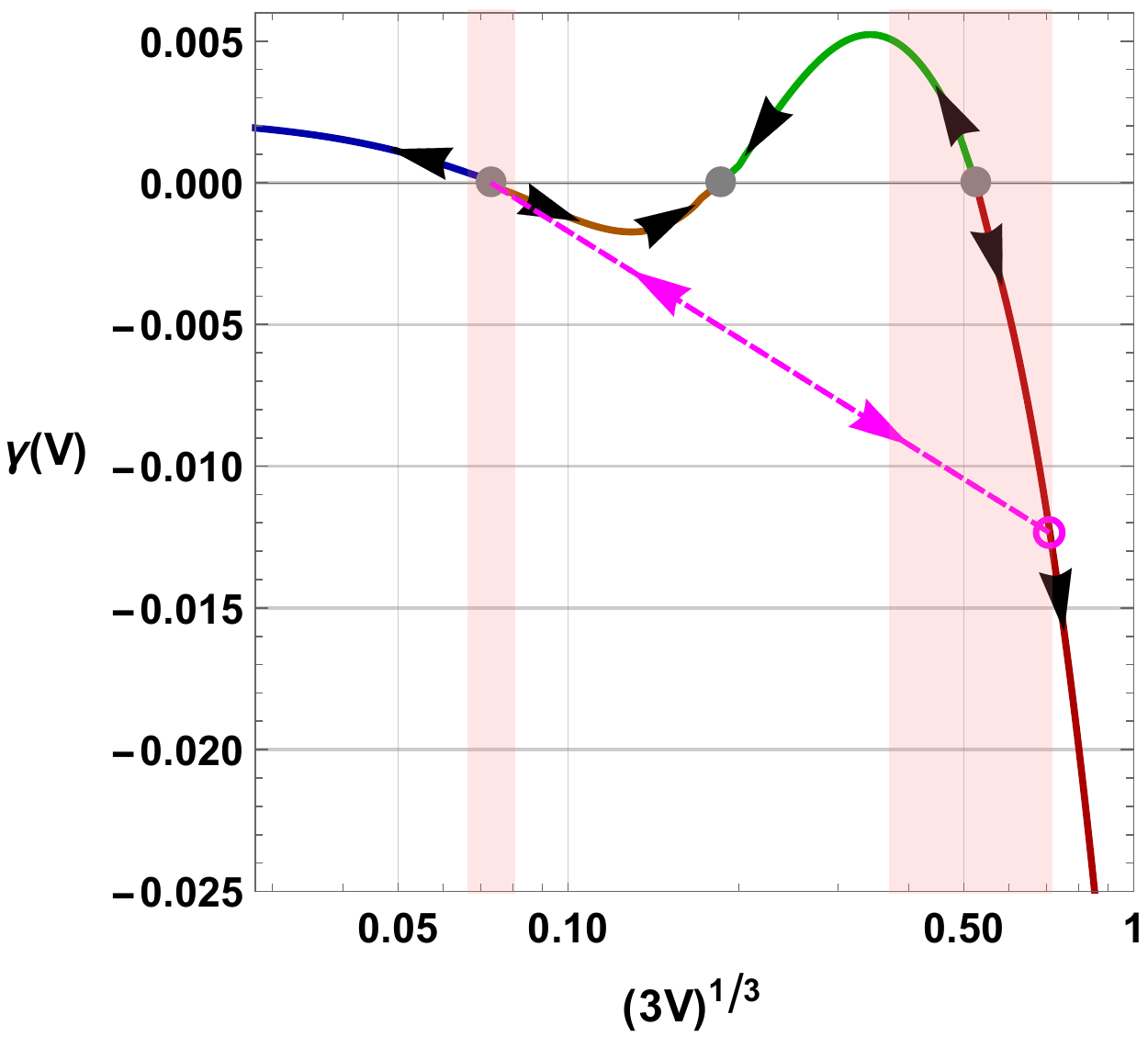}
		\caption{}
		\label{f6_5}
		
	\end{subfigure}
	\hspace{1pt}	
\begin{subfigure}[h]{0.45\textwidth}
	\centering \includegraphics[scale=.5]{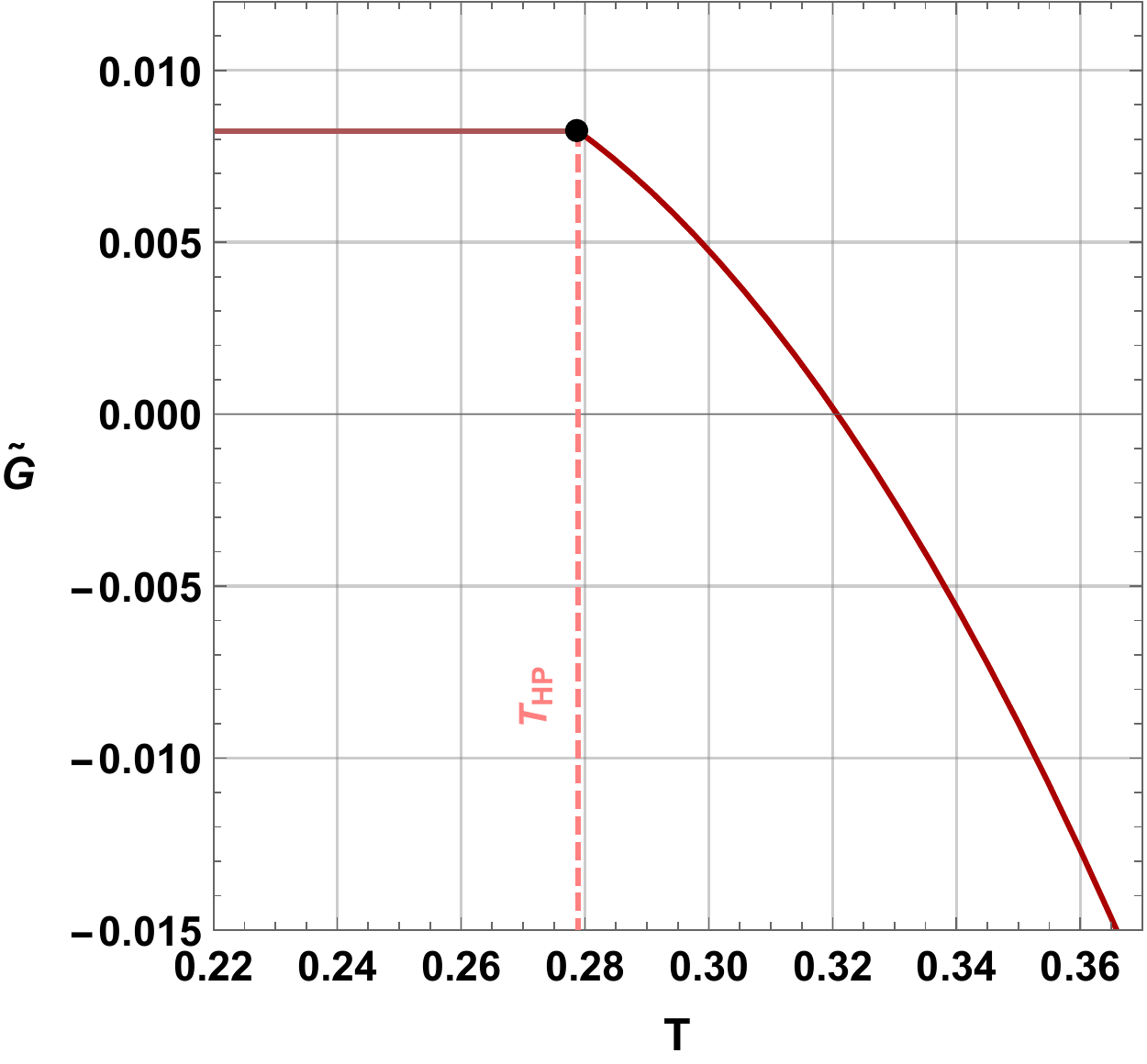}
	\caption{}
	\label{f6_6}
	
\end{subfigure}
	\caption{\footnotesize\it (a) Temperature versus the event horizon radius $r_h$. (b) Gibbs free energy-temperature diagram. (c) Landau function $L$ in terms of the parameter $X$ for different temperatures. (d) Landau function $L$ in terms of the black hole volume $V$. (e)  $\gamma$-function in terms of the black hole volume $V$.  (f) On-shell Gibbs free energy $\tilde{G}$ as a function of temperature $T$. The arrows indicate the evolution of the temperature and the pink region indicates where the thermal radiation phase is the global stable phase $(T<T_{HP})$ with $Q = 0.010026$, $l=1$, and $b=3.5$.}
	\label{f6}
\end{figure}
At such a charge,  the zeroth order phase transition occurs at the intersection point between the stable small black holes branch and the large black holes as depicted in Fig.\ref{f6_2} and which corresponds to an inflection point in Landau function in terms of the black hole volume (magenta curve). We illustrate such a phase transition that occurs at a constant temperature by the  dashed magenta line in Fig.\ref{f6_4} and Fig.\ref{f6_5}. The zeroth order phase transition occurs between the unstable state (gray point) and the large black hole phase. We ought to mention that at this charge begins the appearance of first-order phase transition and reentrant phase transition those we shall see next case.  Finally, we see in Fig.\ref{f6_6}, the black hole reminiscents the AdS Schwarzschild black holes as in previous cases.

\item  Going further in our construction of phase picture, we consider now the case of  $Q =  0.01009$, in Fig.\ref{f7}, in fact, we plot the temperature as a function of horizon radius $r_h$ (Fig.\ref{f7_1}), Gibbs free energy in terms of temperature (Fig.\ref{f7_2}), Landau function $L$ versus $X$ parameter within a variety of temperature (Fig.\ref{f7_3}), Landau function $L$ in terms of the black hole volume $V$ Fig.\ref{f7_4},  $\gamma$-function in terms of the black hole volume $V$ (Fig.\ref{f7_5})  and on-shell Gibbs free energy-temperature $\tilde{G}-T$ diagram (Fig.\ref{f7_6}).

\begin{figure}[!ht]
	\centering
	\begin{subfigure}[h]{0.45\textwidth}
		\centering \includegraphics[scale=.5]{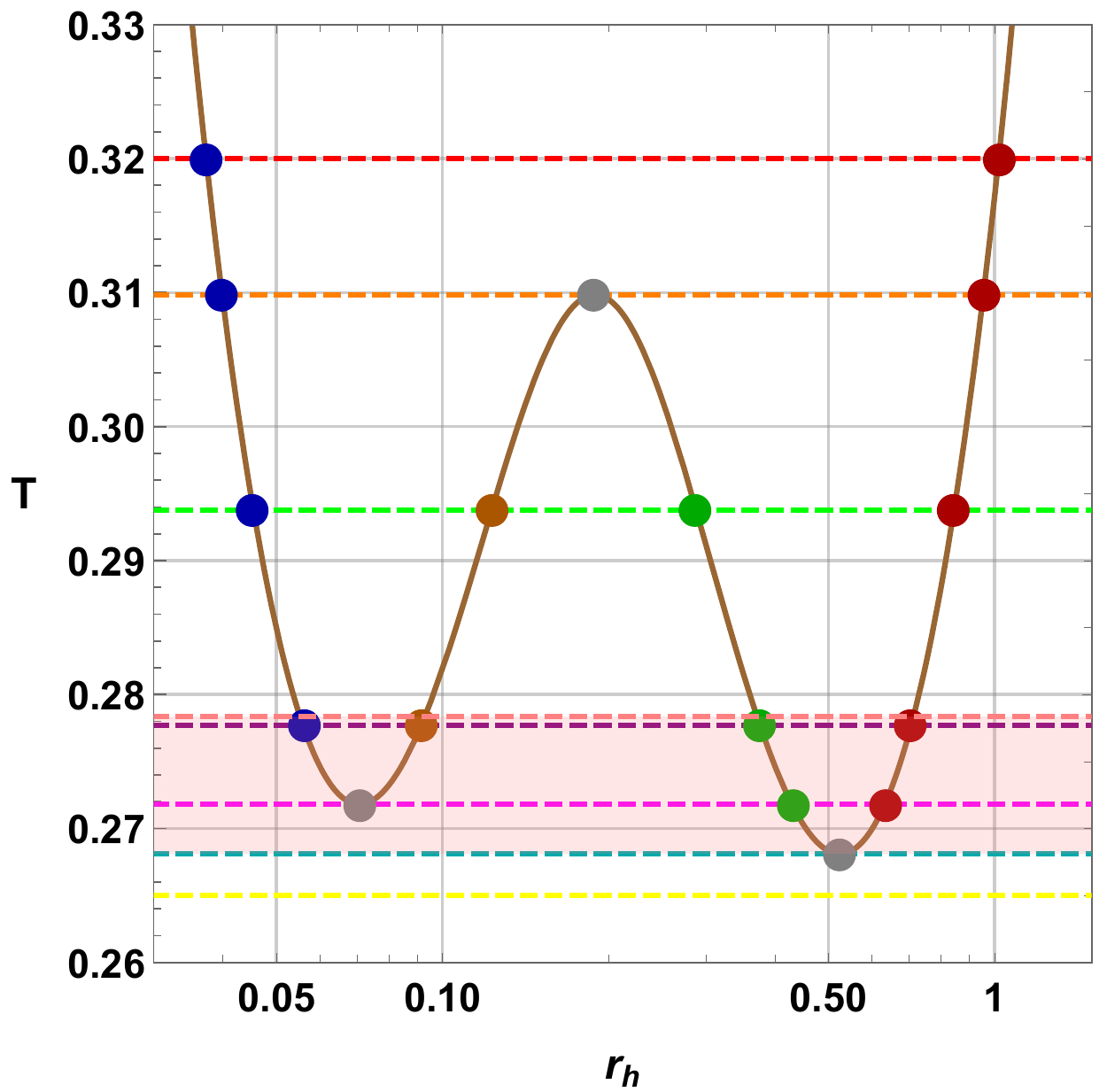}
		\caption{}
		\label{f7_1}
	\end{subfigure}
	\hspace{1pt}	
	\begin{subfigure}[h]{0.45\textwidth}
		\centering \includegraphics[scale=.5]{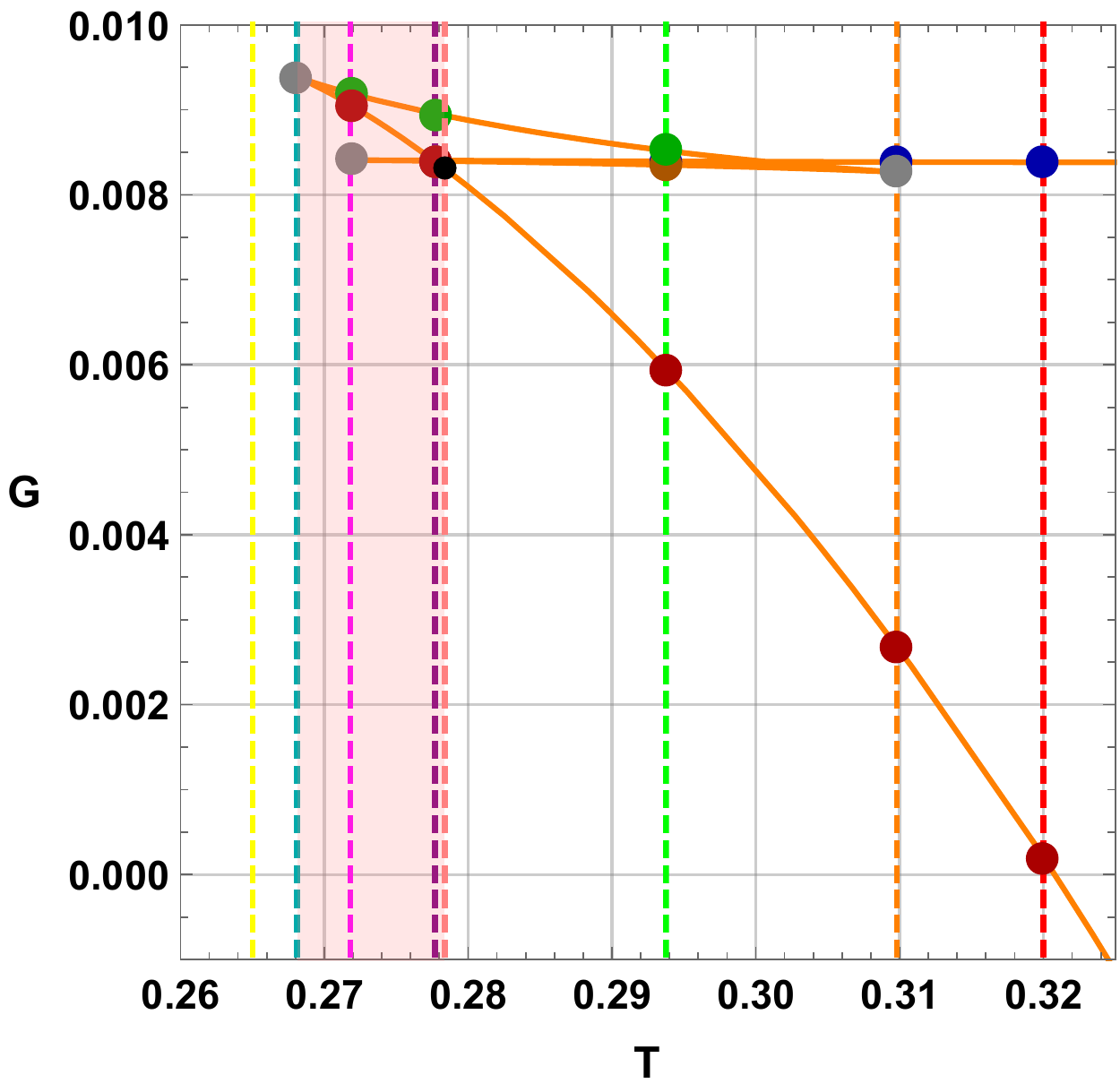}
		\caption{}
		\label{f7_2}		
	\end{subfigure}
	\hspace{1pt}	
	\begin{subfigure}[h]{0.45\textwidth}
		\centering \includegraphics[scale=.5]{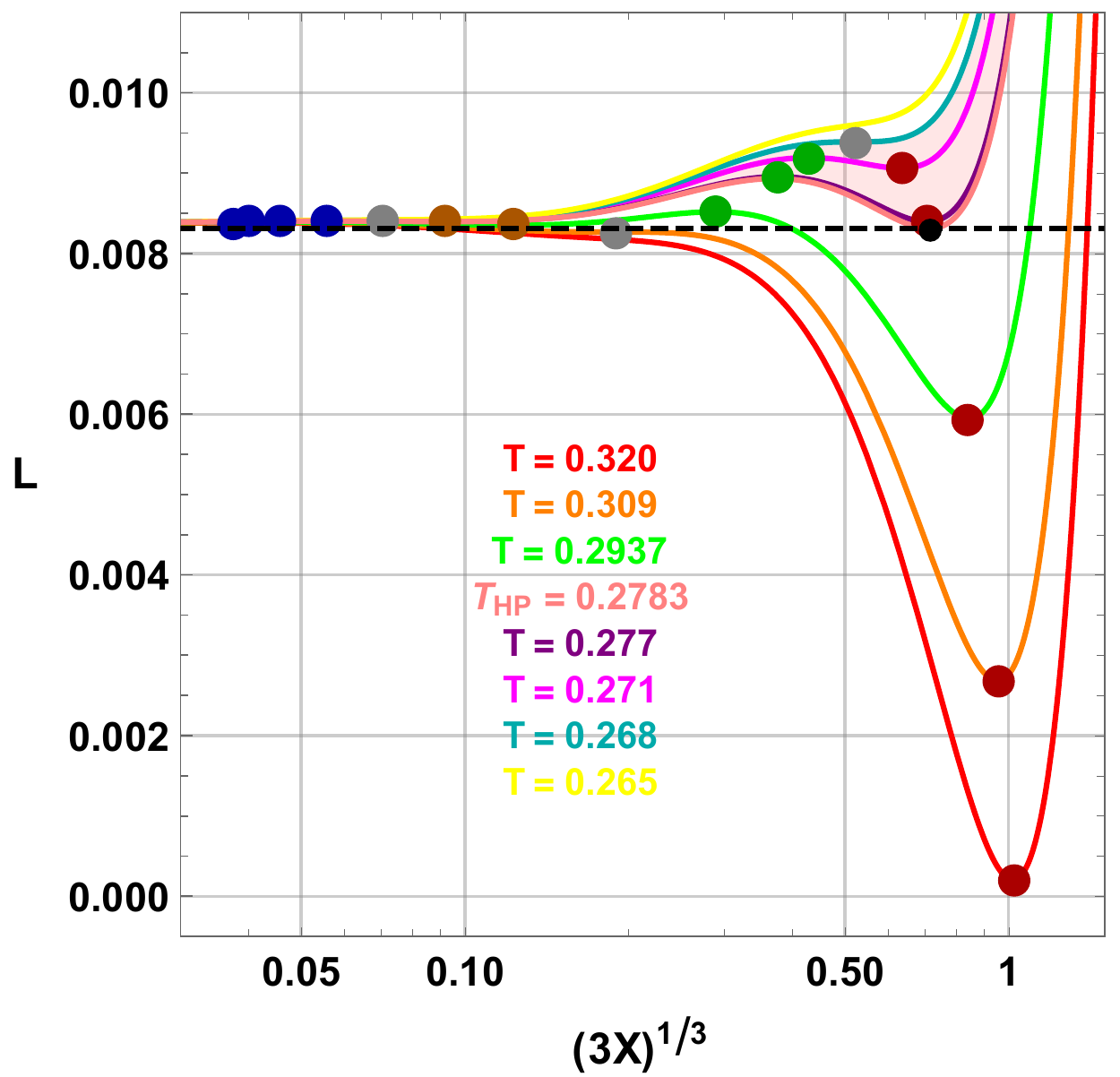}
		\caption{}
		\label{f7_3}	
	\end{subfigure}
	\hspace{1pt}	
	\begin{subfigure}[h]{0.45\textwidth}
		\centering \includegraphics[scale=.5]{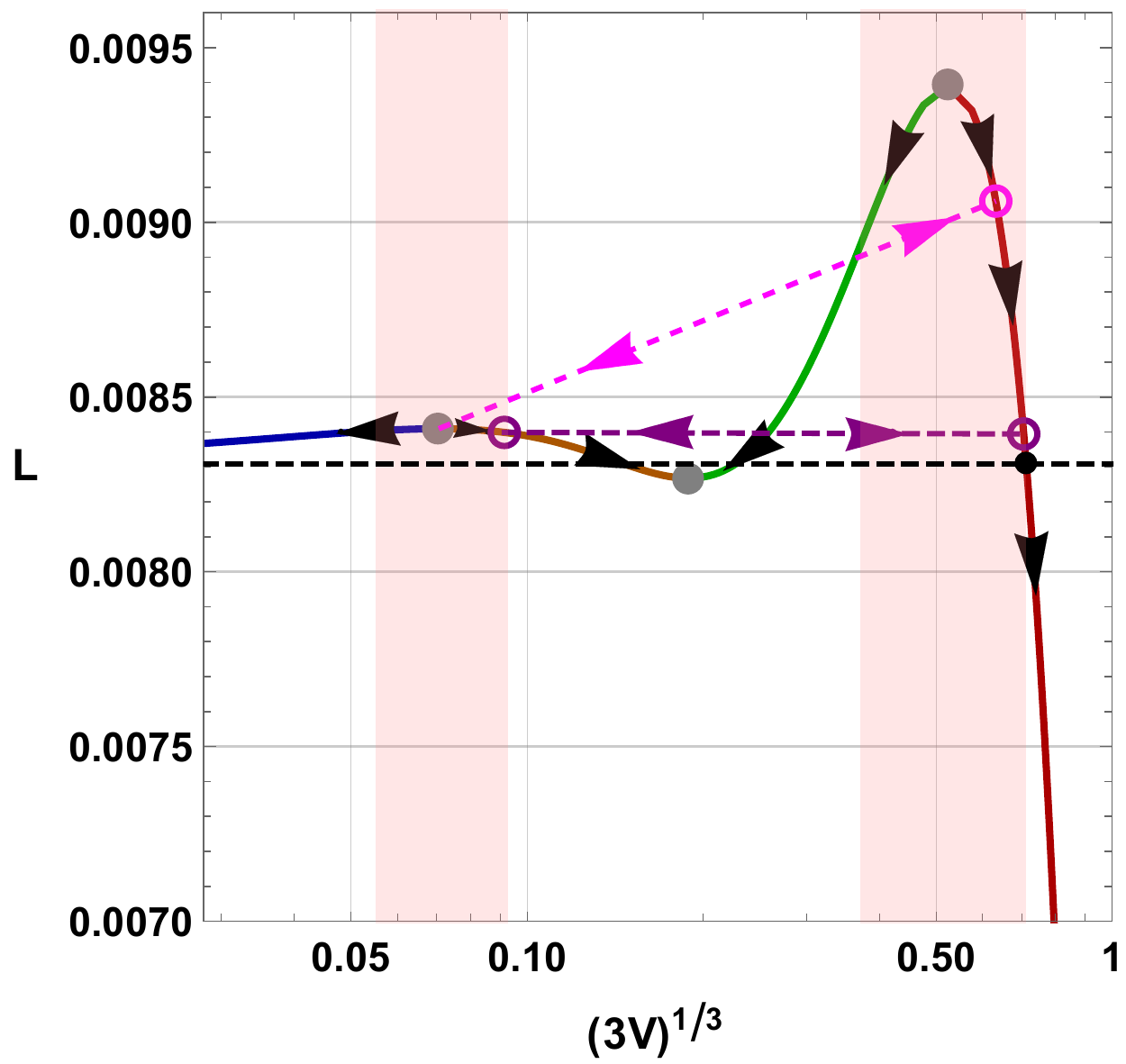}
		\caption{}
		\label{f7_4}	
	\end{subfigure}
	\hspace{1pt}	
	\begin{subfigure}[h]{0.45\textwidth}
		\centering \includegraphics[scale=.5]{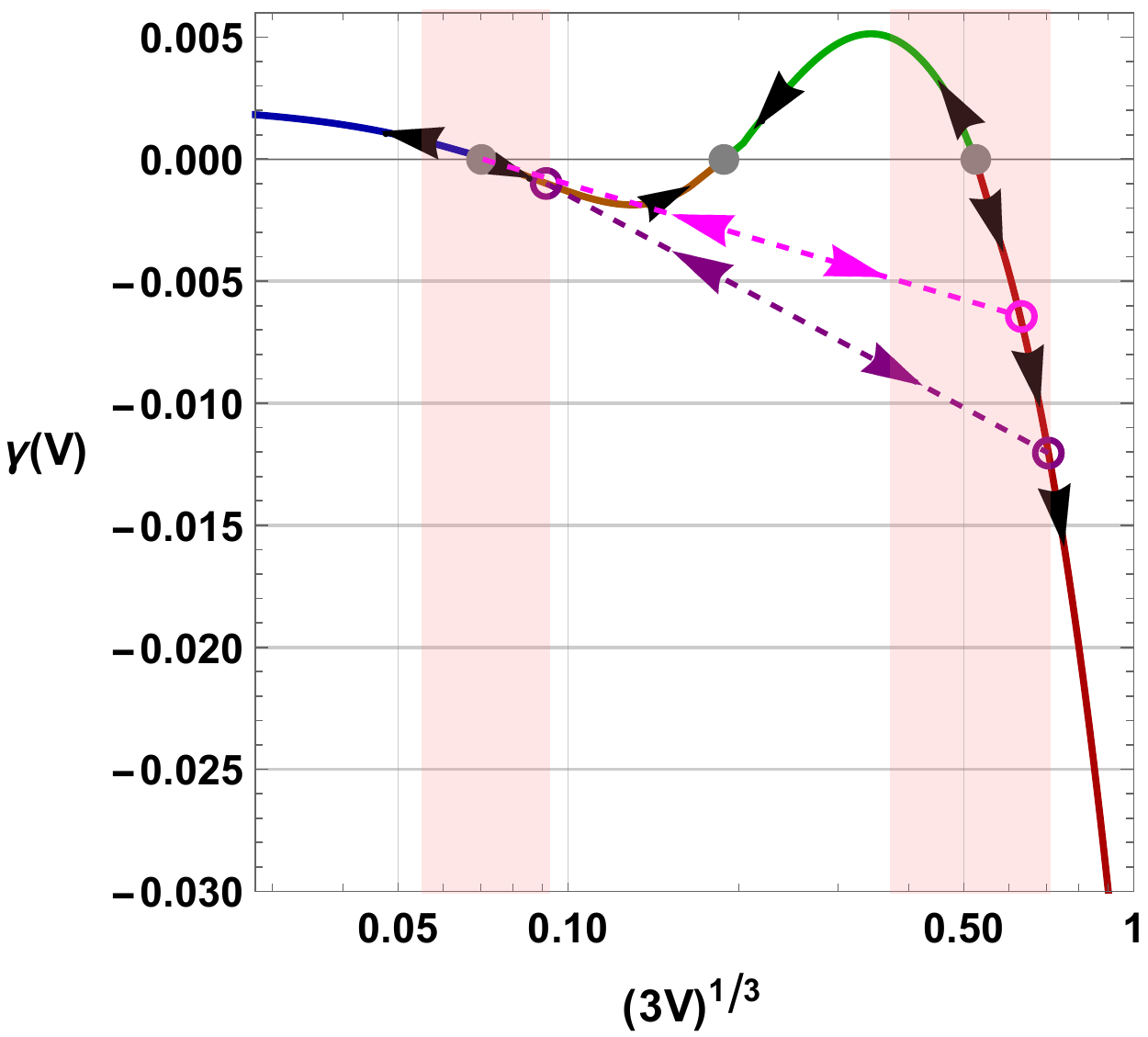}
		\caption{}
		\label{f7_5}
		
	\end{subfigure}
	\hspace{1pt}	
\begin{subfigure}[h]{0.45\textwidth}
	\centering \includegraphics[scale=.5]{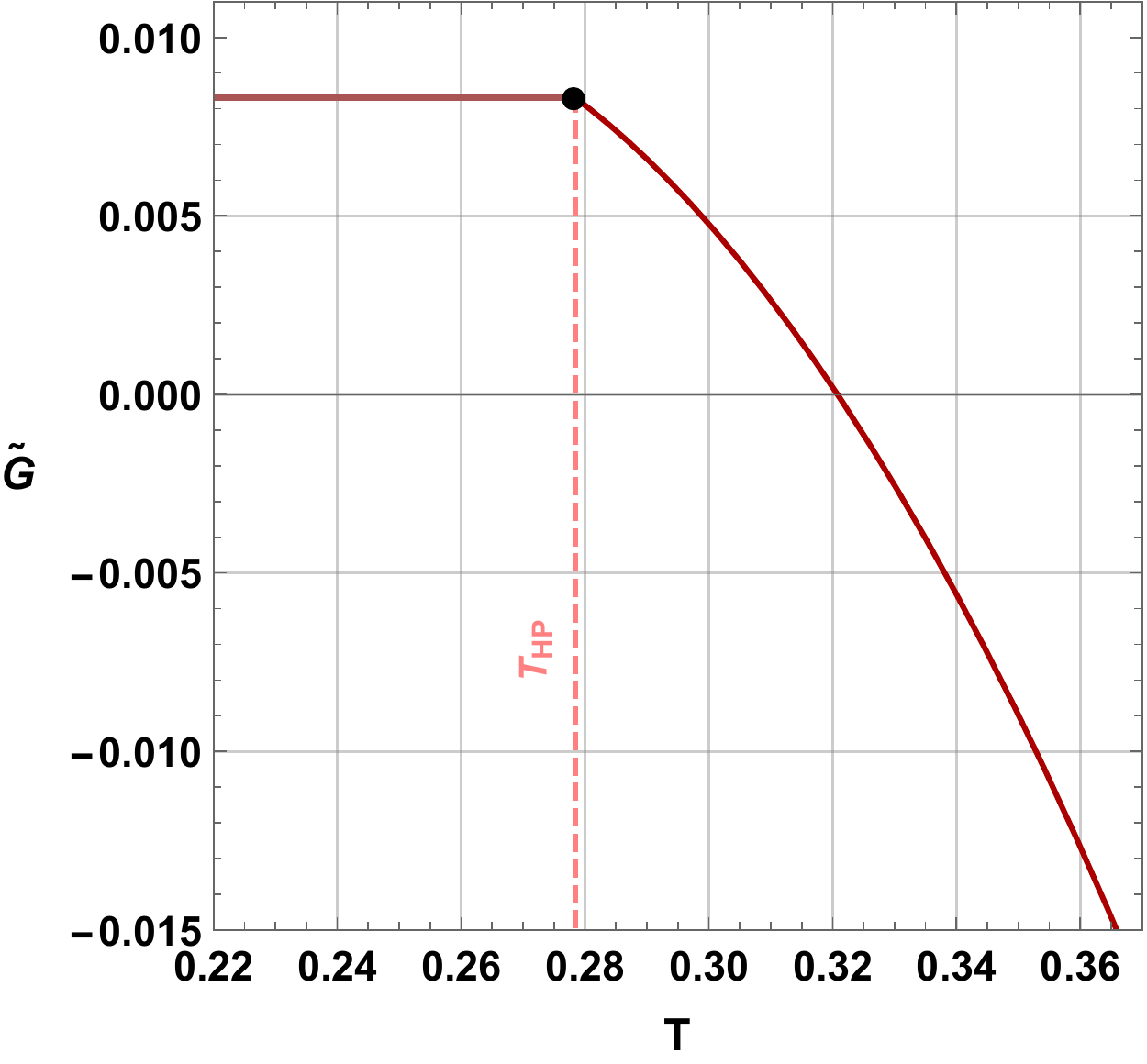}
	\caption{}
	\label{f7_6}
	
\end{subfigure}
	\caption{\footnotesize\it (a) Temperature versus the event horizon radius $r_h$. (b) Gibbs free energy-temperature diagram. (c) Landau function $L$ in terms of the parameter $X$ for different temperatures. (d) Landau function $L$ in terms of the black hole volume $V$. (e)  $\gamma$-function in terms of the black hole volume $V$.  (f) On-shell Gibbs free energy $\tilde{G}$ as a function of temperature $T$. The arrows indicate the evolution of the temperature and the pink region indicates where the thermal radiation phase is the global stable phase $(T<T_{HP})$ with $Q =  0.01009$, $l=1$, and $b=3.5$.}
	\label{f7}
\end{figure}
 In this case, we observe a reentrant phase transition LBH$\to$SBH$\to$LBH. Indeed, we notice in Fig.\ref{f7_2} that we have a zeroth order phase transition between large black holes and small black holes (following the magenta dashed line), this phase transition corresponds to a jump in Gibbs free energy. Moreover, we have a first-order phase transition between small black holes and large ones (purple dashed line) characterized by the swallowtail shape in Gibbs free energy curve. In Fig.\ref{f7_4} and Fig.\ref{f7_5}. Additionally, we remark that the large black hole (red curve) can jump to a small black hole (following the magenta dashed line) keeping the temperature constant then it can undergo a first-order transition to a large black hole phase again.  Finally, from Fig.\ref{f7_6}, and again the black hole mimics  Schwarzschild-AdS black hole.

\item  Increasing the charge to reach $Q =0.010128$ in Fig.\ref{f8} %the temperature as a function of horizon radius $r_h$ (Fig.\ref{f8_1}), Gibbs free energy as a function of temperature (Fig.\ref{f8_2}), Landau function $L$ in terms of the parameter $X$ for different temperatures (Fig.\ref{f8_3}), Landau function $L$ in terms of the black hole volume $V$ Fig.\ref{f8_4} ,  $\gamma$-function in terms of the black hole volume $V$ (Fig.\ref{f8_5})  and on-shell Gibbs free energy $\tilde{G}$ as a function of temperature (Fig.\ref{f8_6}). 

\begin{figure}[!ht]
	\centering
	\begin{subfigure}[h]{0.45\textwidth}
		\centering \includegraphics[scale=.5]{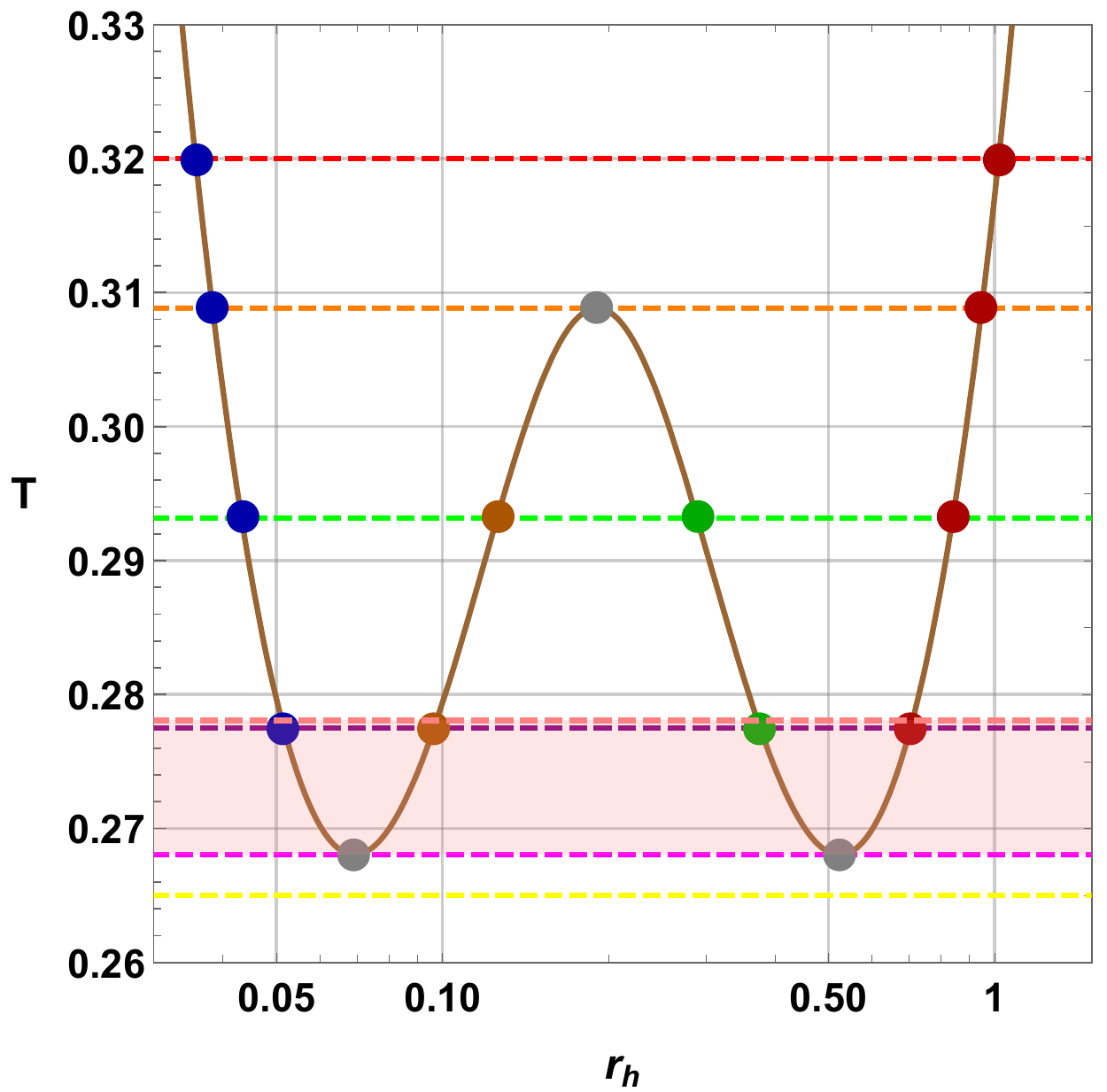}
		\caption{}
		\label{f8_1}
	\end{subfigure}
	\hspace{1pt}	
	\begin{subfigure}[h]{0.45\textwidth}
		\centering \includegraphics[scale=.5]{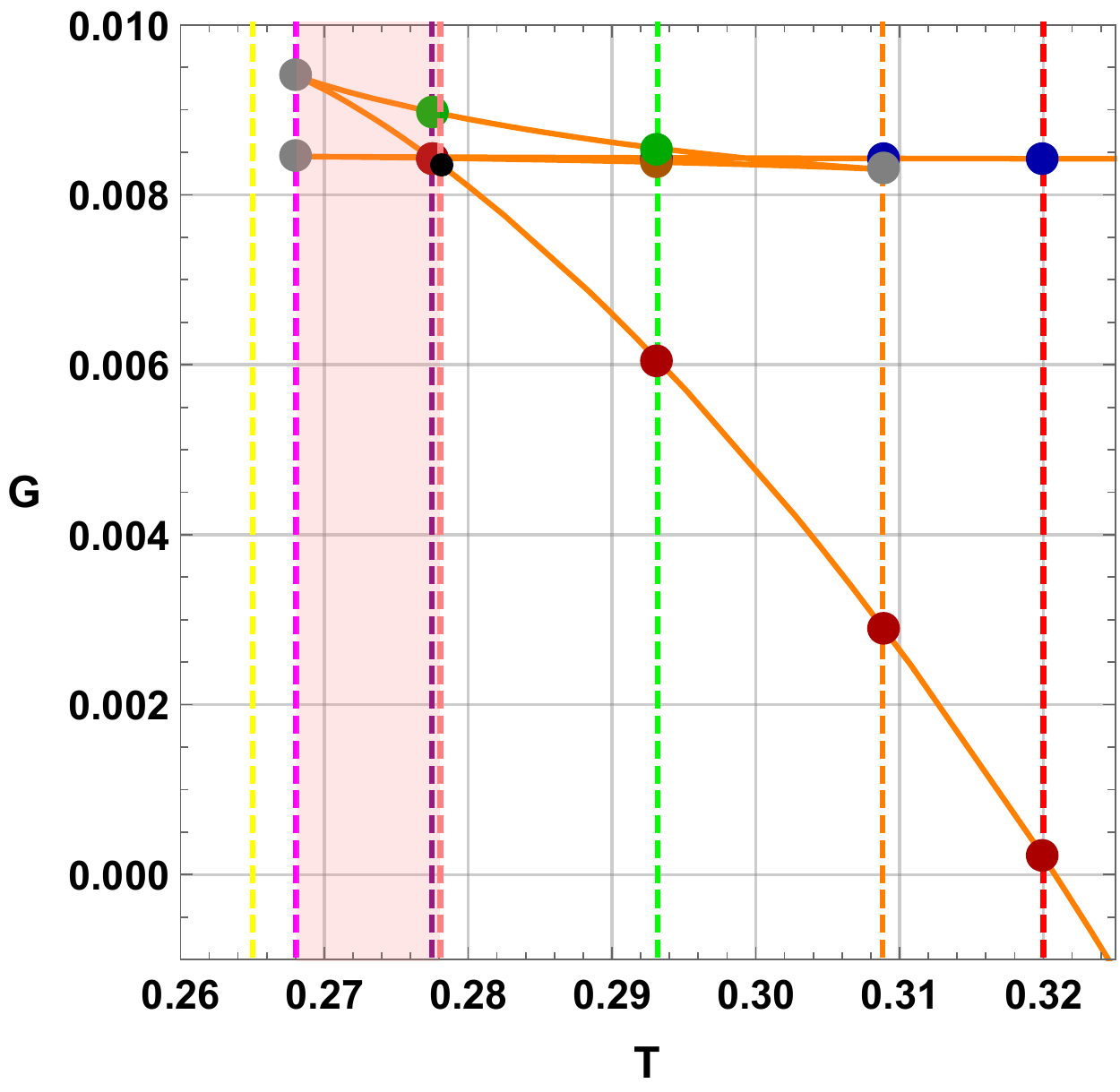}
		\caption{}
		\label{f8_2}		
	\end{subfigure}
	\hspace{1pt}	
	\begin{subfigure}[h]{0.45\textwidth}
		\centering \includegraphics[scale=.5]{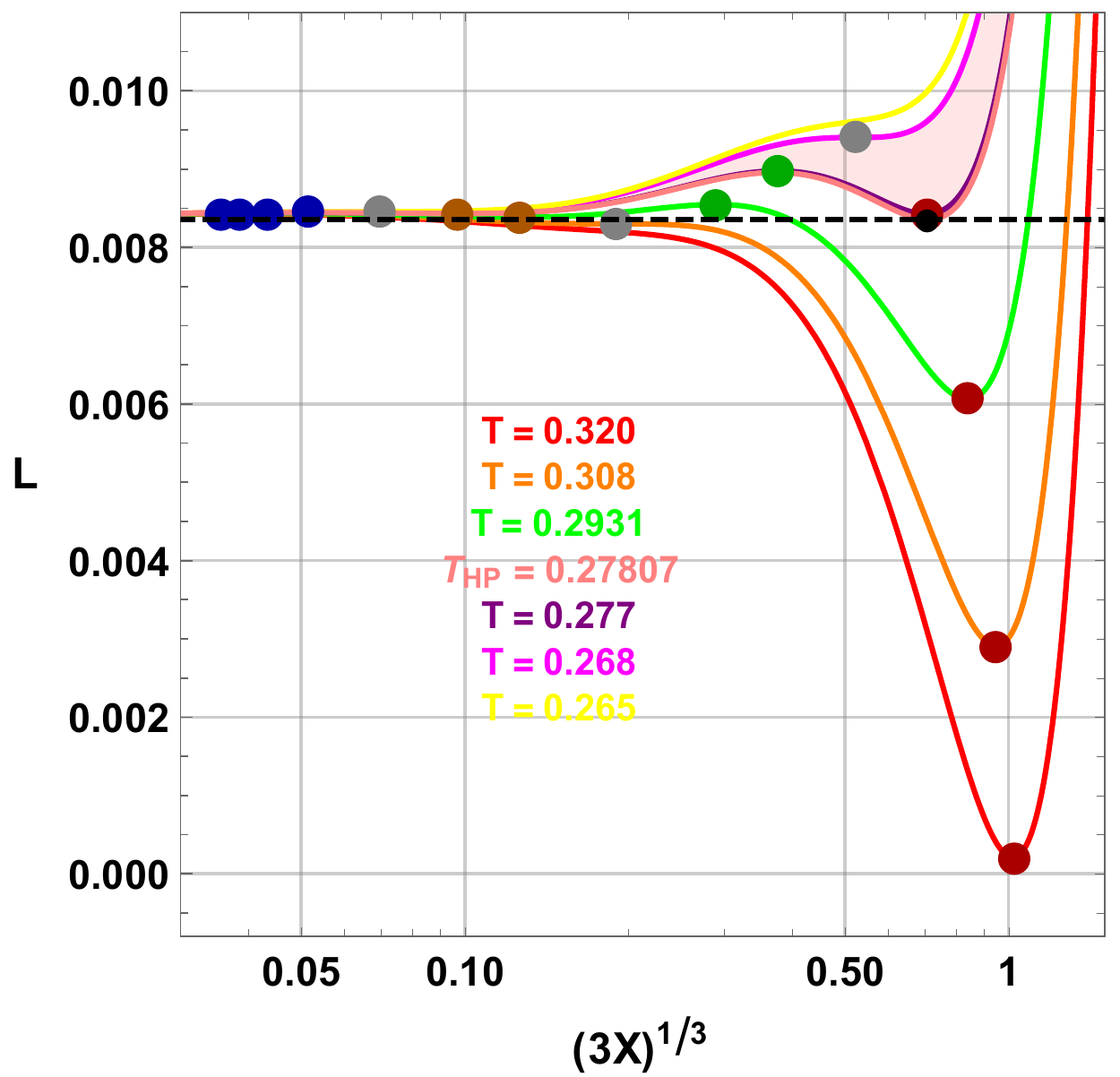}
		\caption{}
		\label{f8_3}	
	\end{subfigure}
	\hspace{1pt}	
	\begin{subfigure}[h]{0.45\textwidth}
		\centering \includegraphics[scale=.5]{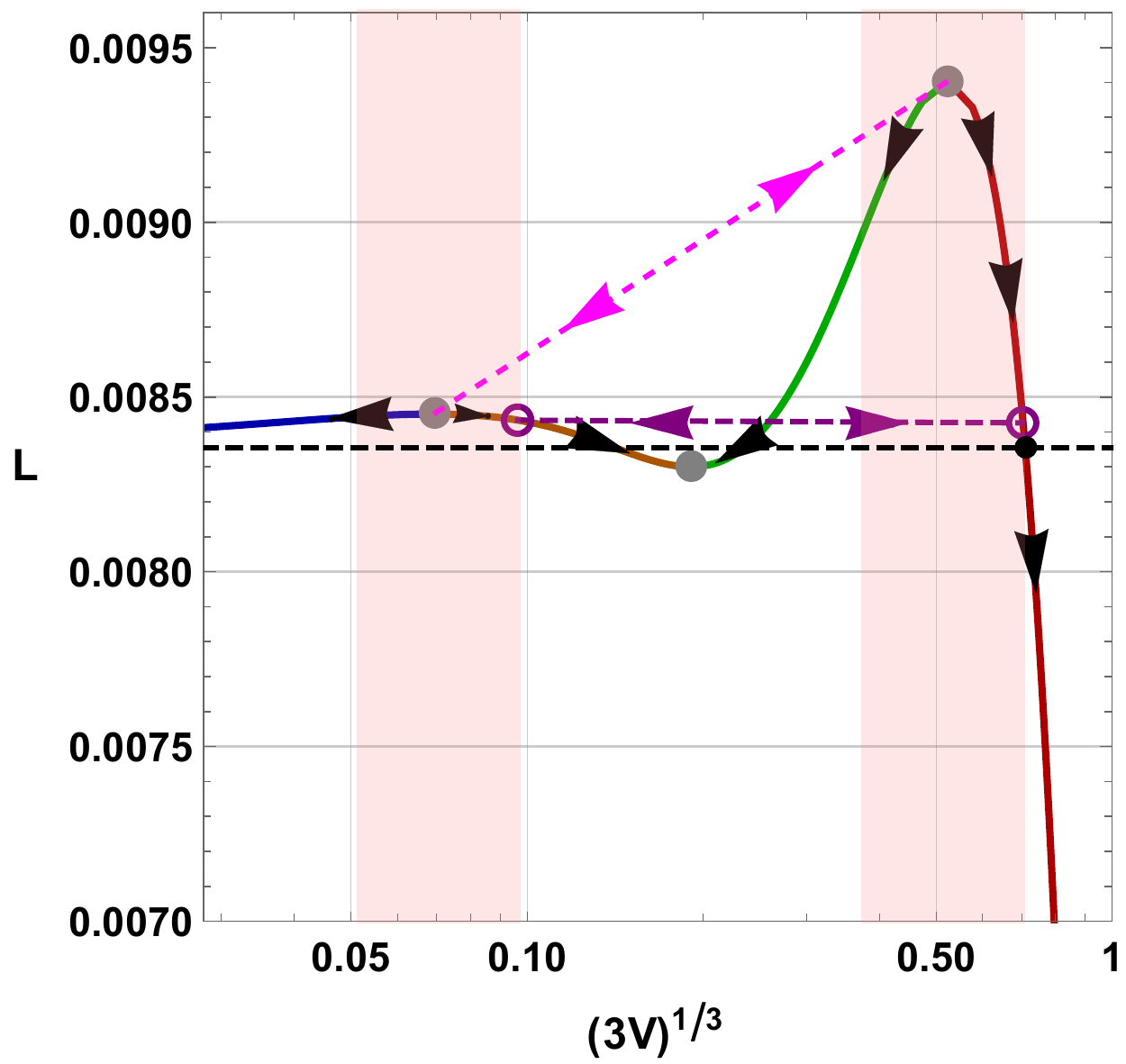}
		\caption{}
		\label{f8_4}	
	\end{subfigure}
	\hspace{1pt}	
	\begin{subfigure}[h]{0.45\textwidth}
		\centering \includegraphics[scale=.5]{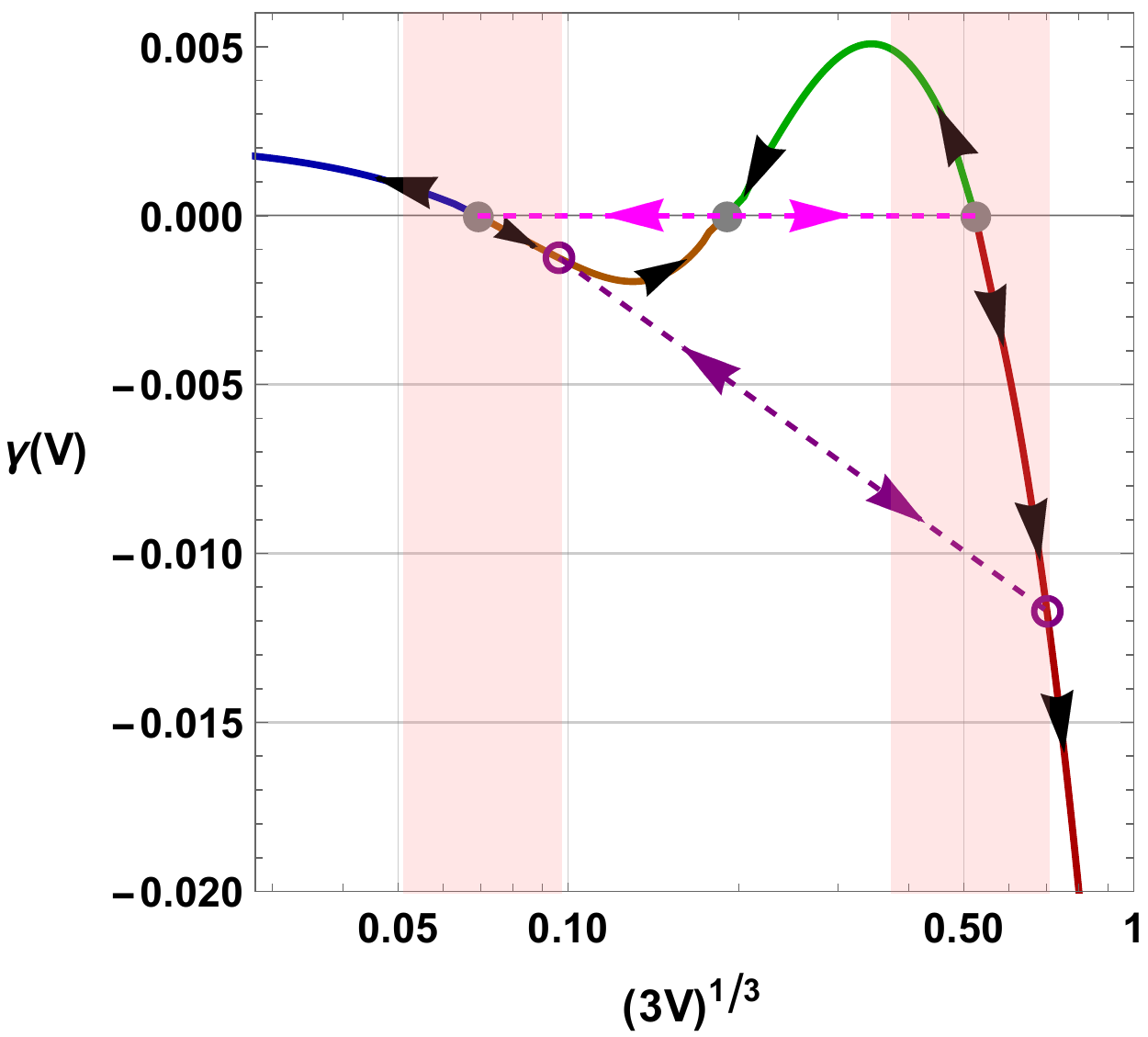}
		\caption{}
		\label{f8_5}
		
	\end{subfigure}
	\hspace{1pt}	
\begin{subfigure}[h]{0.45\textwidth}
	\centering \includegraphics[scale=.5]{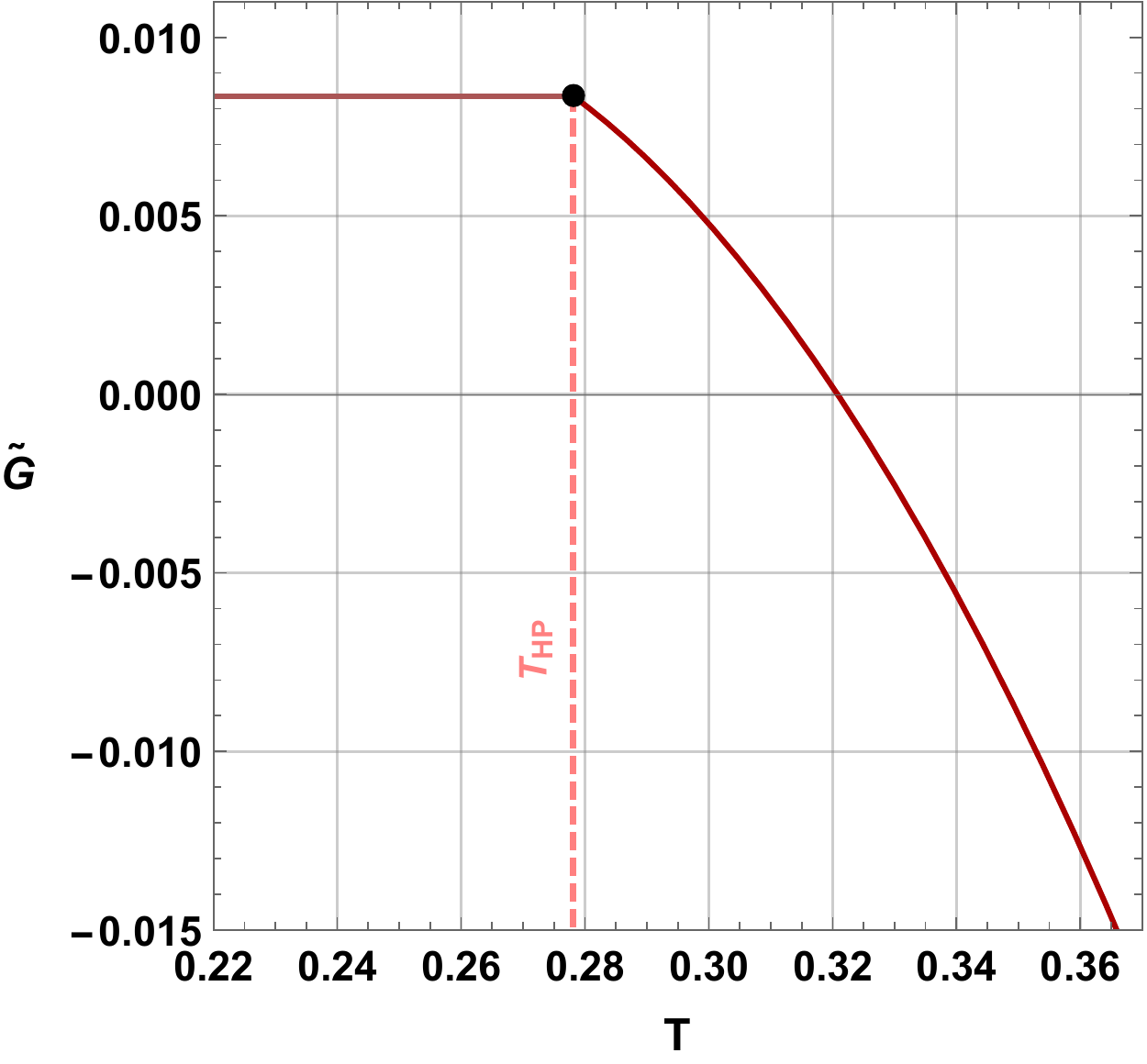}
	\caption{}
	\label{f8_6}
	
\end{subfigure}
	\caption{\footnotesize\it (a) Temperature versus the event horizon radius $r_h$. (b) Gibbs free energy-temperature diagram. (c) Landau function $L$ in terms of the parameter $X$ for different temperatures. (d) Landau function $L$ in terms of the black hole volume $V$. (e)  $\gamma$-function in terms of the black hole volume $V$.  (f) On-shell Gibbs free energy $\tilde{G}$ as a function of temperature $T$. The arrows indicate the evolution of the temperature and the pink region indicates where the thermal radiation phase is the global stable phase $(T<T_{HP})$ with $Q =  0.010128$, $l=1$, and $b=3.5$.}
	\label{f8}
\end{figure}
The system exhibits always a first-order phase transition between small black holes and large ones following the dashed purple line whereas the zeroth-order phase transition is carried out between two unstable points following the magenta line. From this electric charge, the zeroth order phase transition shall disappear as we will see next case. Therefore, the reentrant phase transition that characterized Born-Infeld-AdS black hole will disappear.  Finally, we see in Fig.\ref{f8_6} that the situation remains the same.

\item In Fig.\ref{f9}, we reach $Q =  0.0105$ %the temperature as a function of horizon radius $r_h$ (Fig.\ref{f8_1}), Gibbs free energy as a function of temperature (Fig.\ref{f9_2}), Landau function $L$ in terms of the parameter $X$ for different temperatures (Fig.\ref{f9_3}), Landau function $L$ in terms of the black hole volume $V$ Fig.\ref{f9_4},  $\gamma$-function in terms of the black hole volume $V$ (Fig.\ref{f9_5})  and on-shell Gibbs free energy $\tilde{G}$ as a function of temperature (Fig.\ref{f9_6}). 

\begin{figure}[!ht]
	\centering
	\begin{subfigure}[h]{0.45\textwidth}
		\centering \includegraphics[scale=.5]{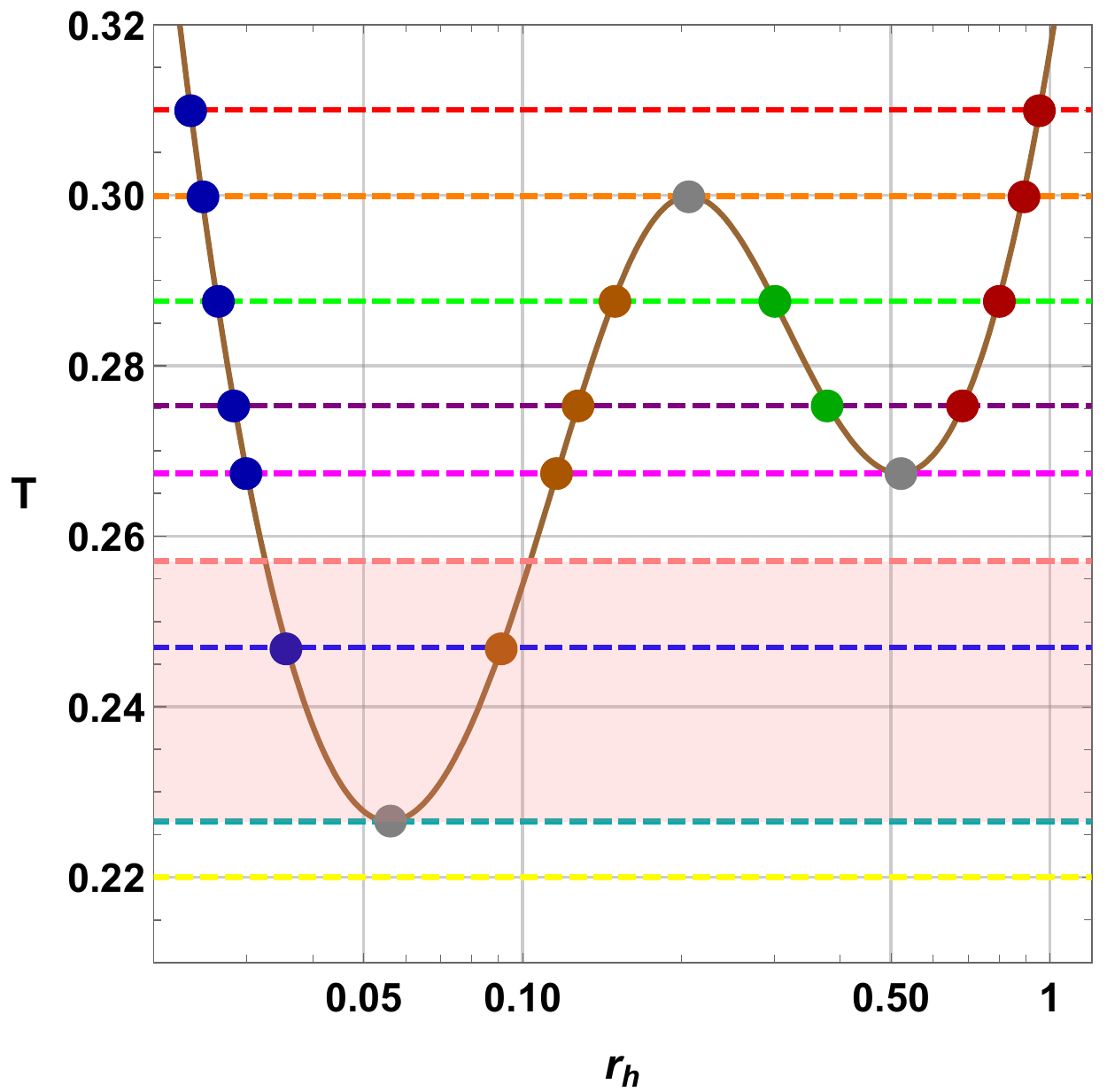}
		\caption{}
		\label{f9_1}
	\end{subfigure}
	\hspace{1pt}	
	\begin{subfigure}[h]{0.45\textwidth}
		\centering \includegraphics[scale=.5]{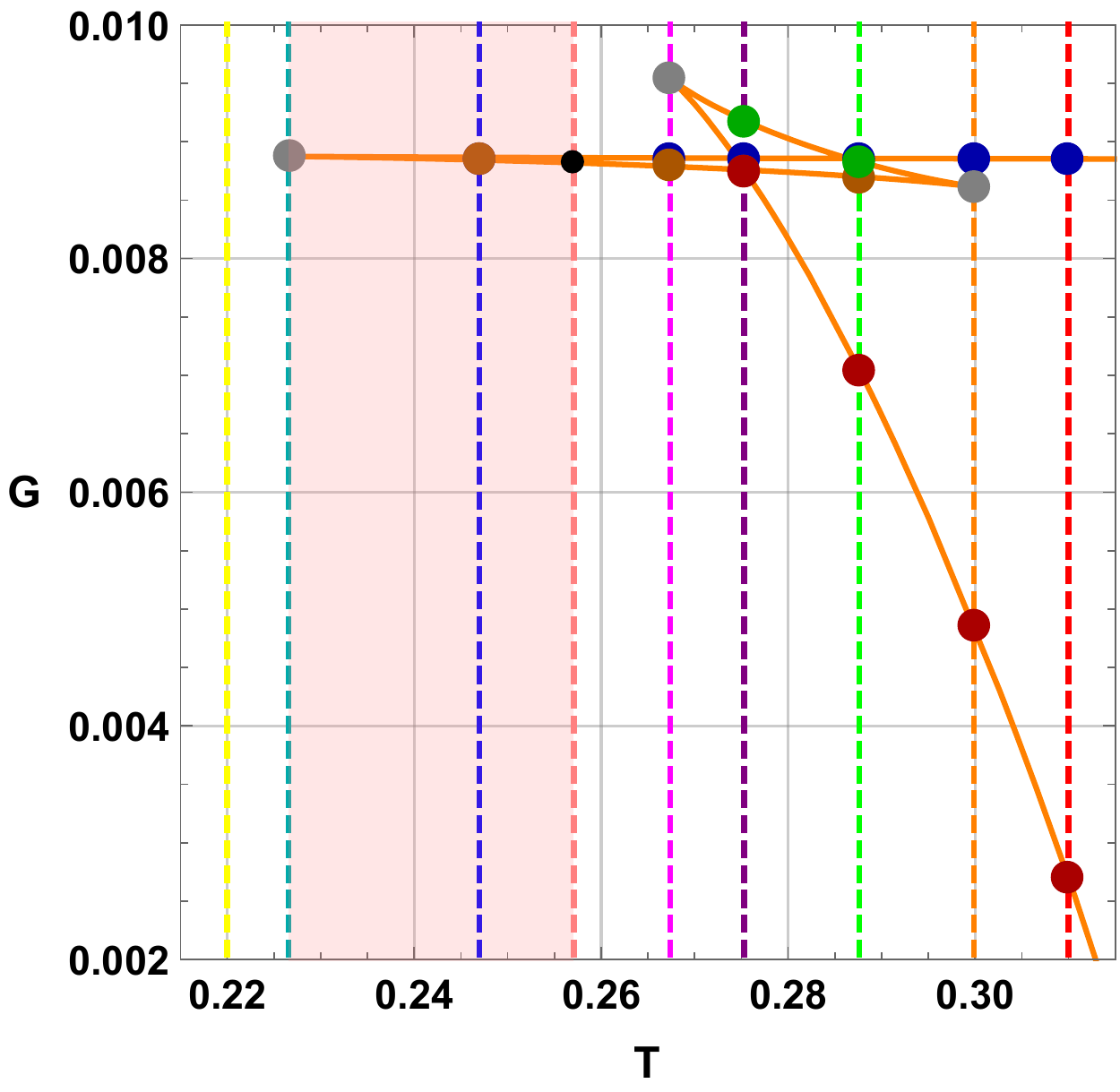}
		\caption{}
		\label{f9_2}		
	\end{subfigure}
	\hspace{1pt}	
	\begin{subfigure}[h]{0.45\textwidth}
		\centering \includegraphics[scale=.5]{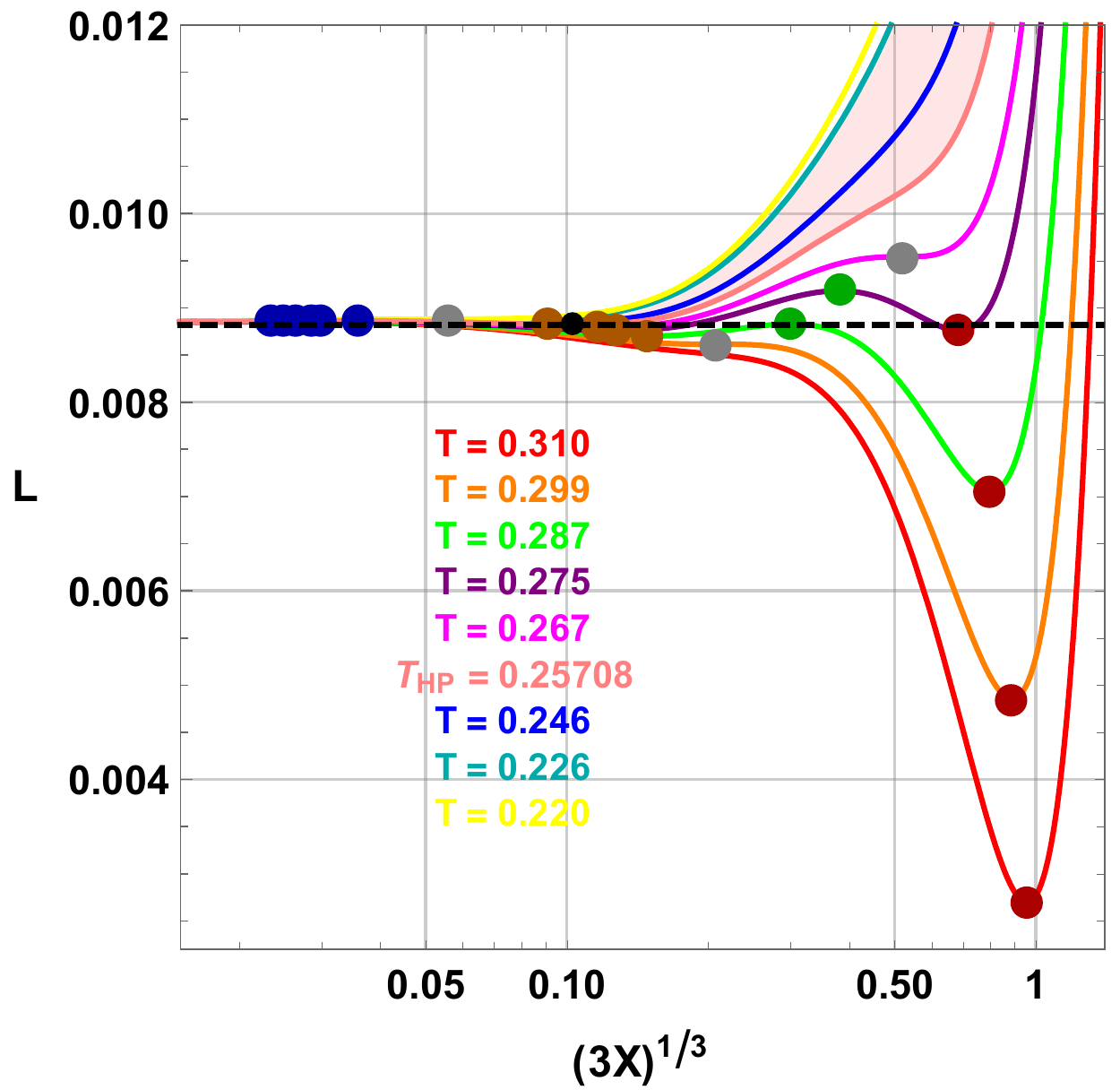}
		\caption{}
		\label{f9_3}	
	\end{subfigure}
	\hspace{1pt}	
	\begin{subfigure}[h]{0.45\textwidth}
		\centering \includegraphics[scale=.5]{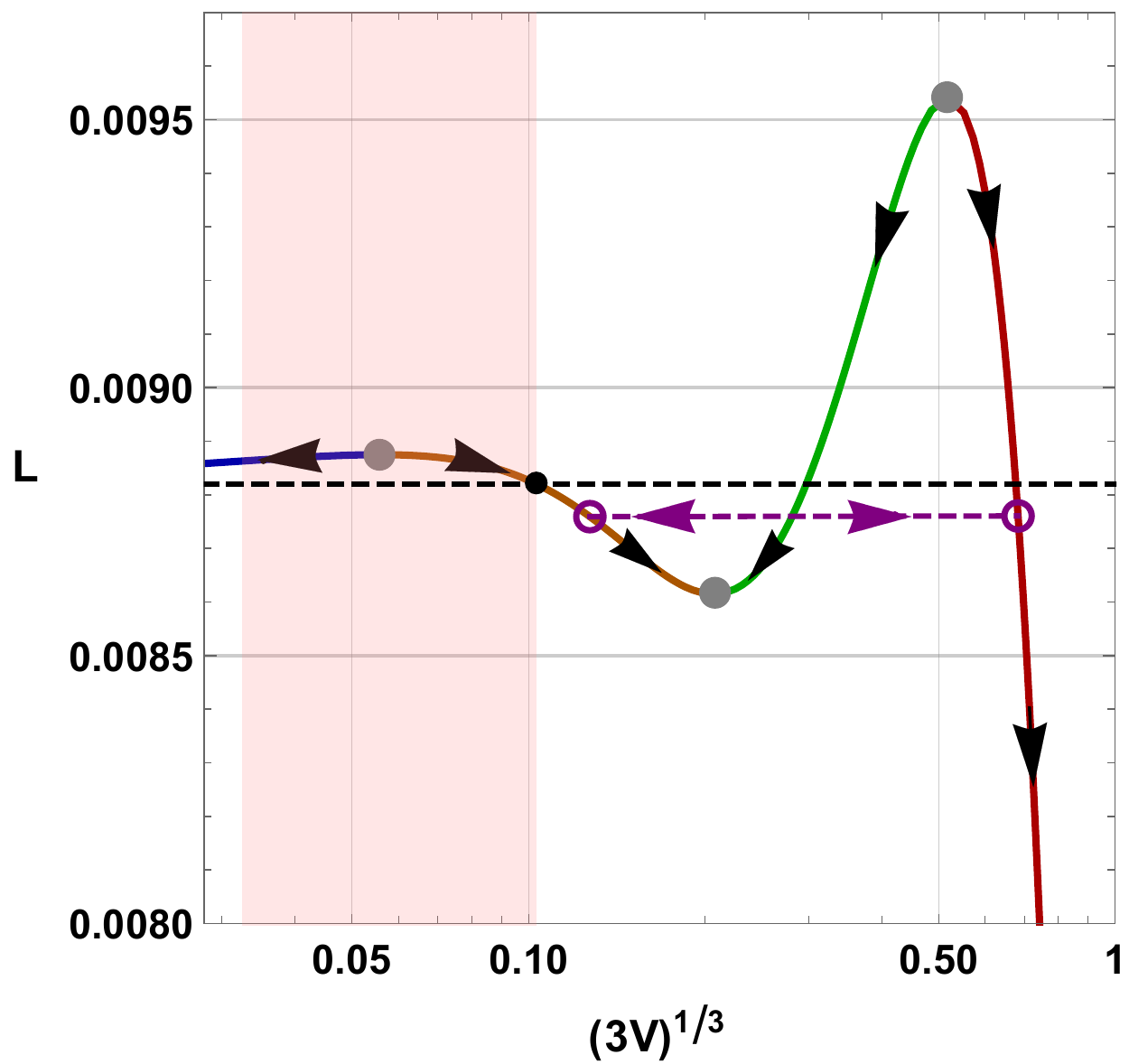}
		\caption{}
		\label{f9_4}	
	\end{subfigure}
	\hspace{1pt}	
	\begin{subfigure}[h]{0.45\textwidth}
		\centering \includegraphics[scale=.5]{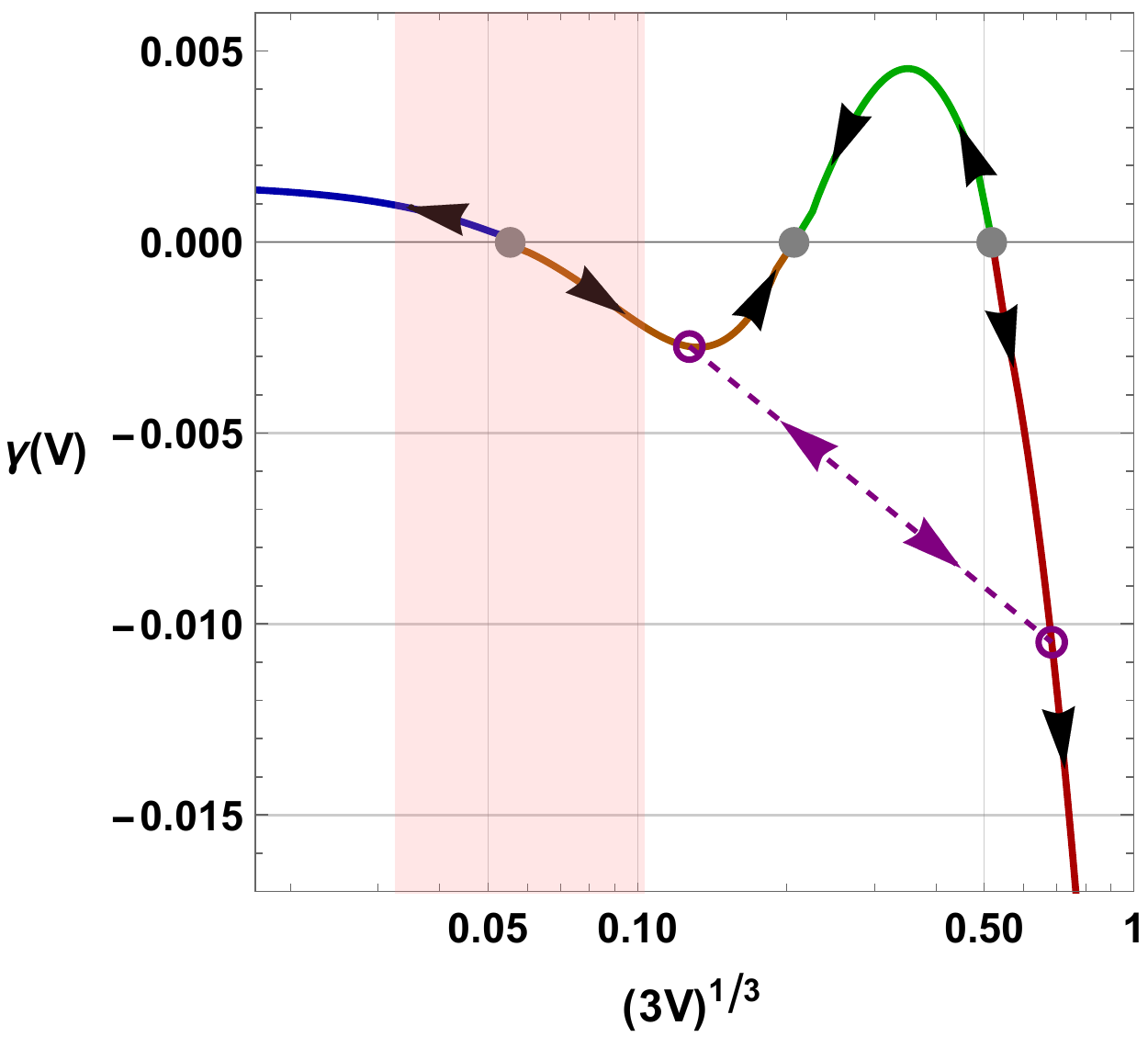}
		\caption{}
		\label{f9_5}
		
	\end{subfigure}
	\hspace{1pt}	
\begin{subfigure}[h]{0.45\textwidth}
	\centering \includegraphics[scale=.5]{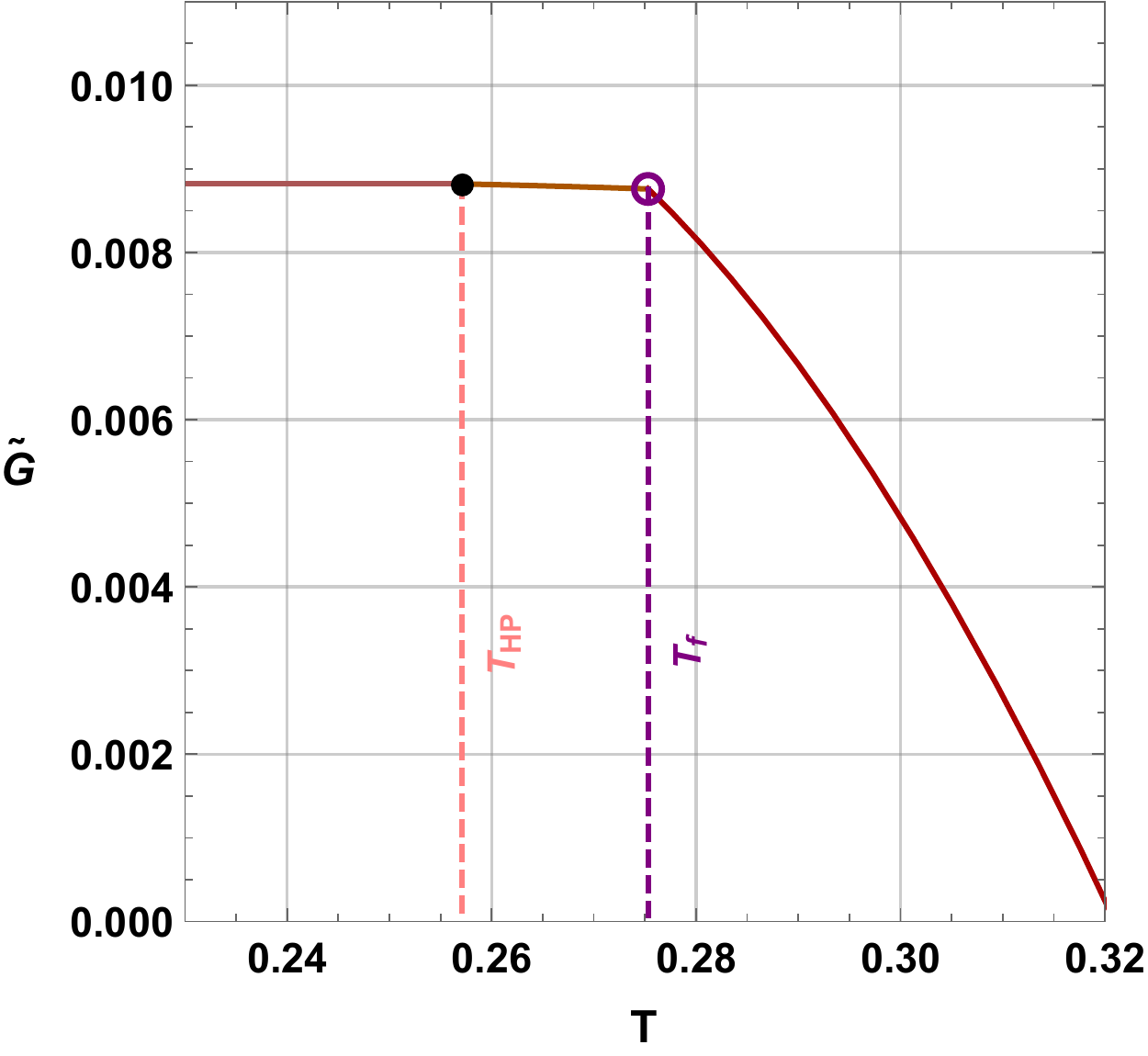}
	\caption{}
	\label{f9_6}
	
\end{subfigure}
	\caption{\footnotesize\it (a) Temperature versus the event horizon radius $r_h$. (b) Gibbs free energy-temperature diagram. (c) Landau function $L$ in terms of the parameter $X$ for different temperatures. (d) Landau function $L$ in terms of the black hole volume $V$. (e)  $\gamma$-function in terms of the black hole volume $V$.  (f) On-shell Gibbs free energy $\tilde{G}$ as a function of temperature $T$. The arrows indicate the evolution of the temperature and the pink region indicates where the thermal radiation phase is the global stable phase $(T<T_{HP})$ with $Q =  0.0105$, $l=1$, and $b=3.5$.}
	\label{f9}
\end{figure}
The key remark is that the zeroth order phase transition has disappeared and consequently there is no reentrant phase transition. The system processes just the first-order phase transition between small black holes and large ones, whereas there is always the Hawking-Page-like phase transition between the unstable small black holes and stable ones and consequently there is no black hole below $T=0.226$.  Finally, we see in Fig.\ref{f9_6} that we have two critical points. the first one corresponds to the Hawking-Page-like transition between thermal radiations and small black holes, while the second one is associated with the first-order phase transition between small and large black holes. Therefore, we have three globally stable phases: thermal radiations, small, and large black holes.

\item  The phase portrait corresponding to $Q =Q_m = 0.0113682$ is depicted in Fig.\ref{f10}% the temperature as a function of horizon radius $r_h$ (Fig.\ref{f10_1}), Gibbs free energy as a function of temperature (Fig.\ref{f10_2}), Landau function $L$ in terms of the parameter $X$ for different temperatures (Fig.\ref{f10_3}), Landau function $L$ in terms of the black hole volume $V$ Fig.\ref{f10_4} ,  $\gamma$-function in terms of the black hole volume $V$ (Fig.\ref{f10_5})  and on-shell Gibbs free energy $\tilde{G}$ as a function of temperature (Fig.\ref{f10_6}).

\begin{figure}[!ht]
	\centering
	\begin{subfigure}[h]{0.45\textwidth}
		\centering \includegraphics[scale=.5]{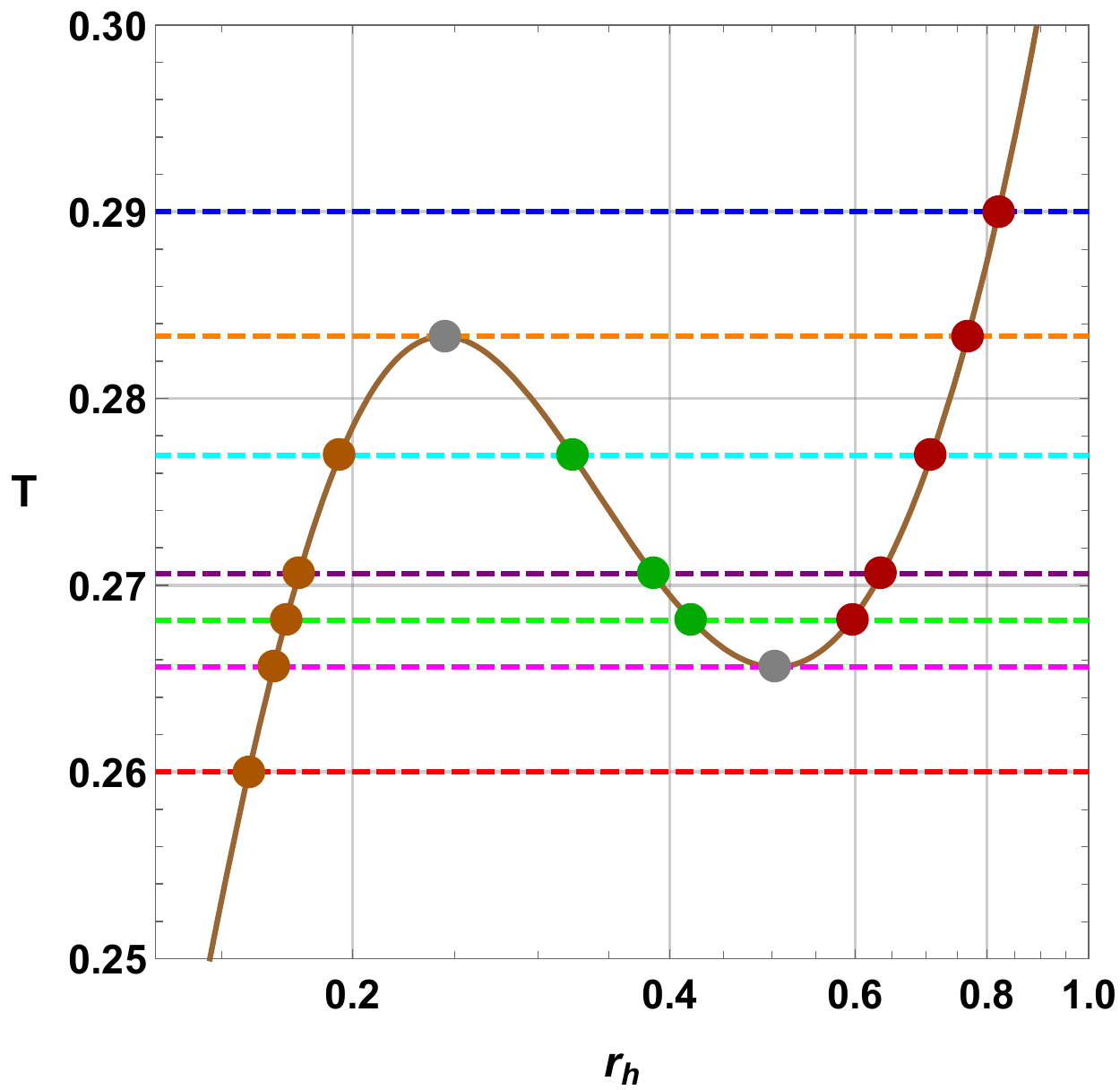}
		\caption{}
		\label{f10_1}
	\end{subfigure}
	\hspace{1pt}	
	\begin{subfigure}[h]{0.45\textwidth}
		\centering \includegraphics[scale=.5]{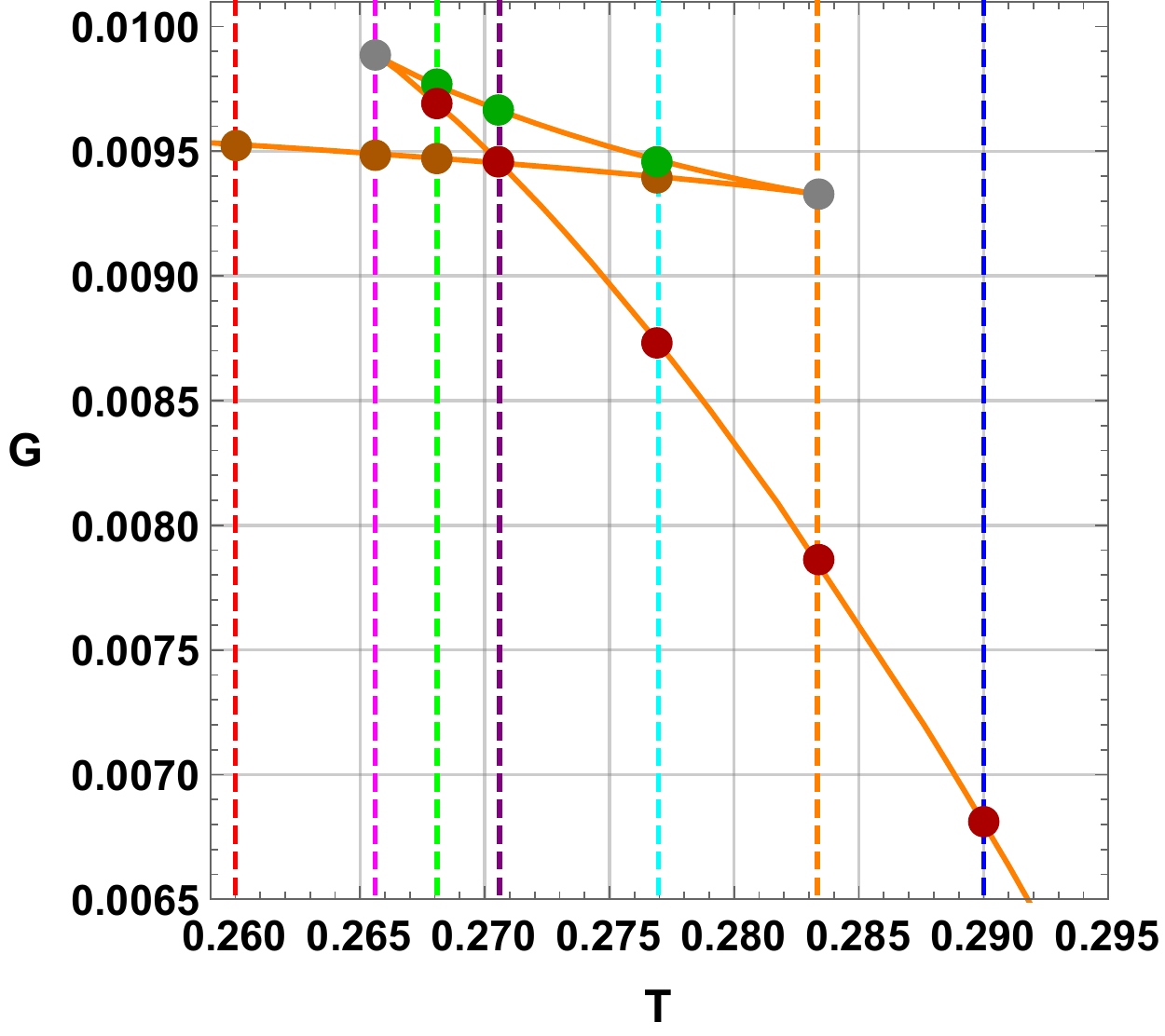}
		\caption{}
		\label{f10_2}		
	\end{subfigure}
	\hspace{1pt}	
	\begin{subfigure}[h]{0.45\textwidth}
		\centering \includegraphics[scale=.5]{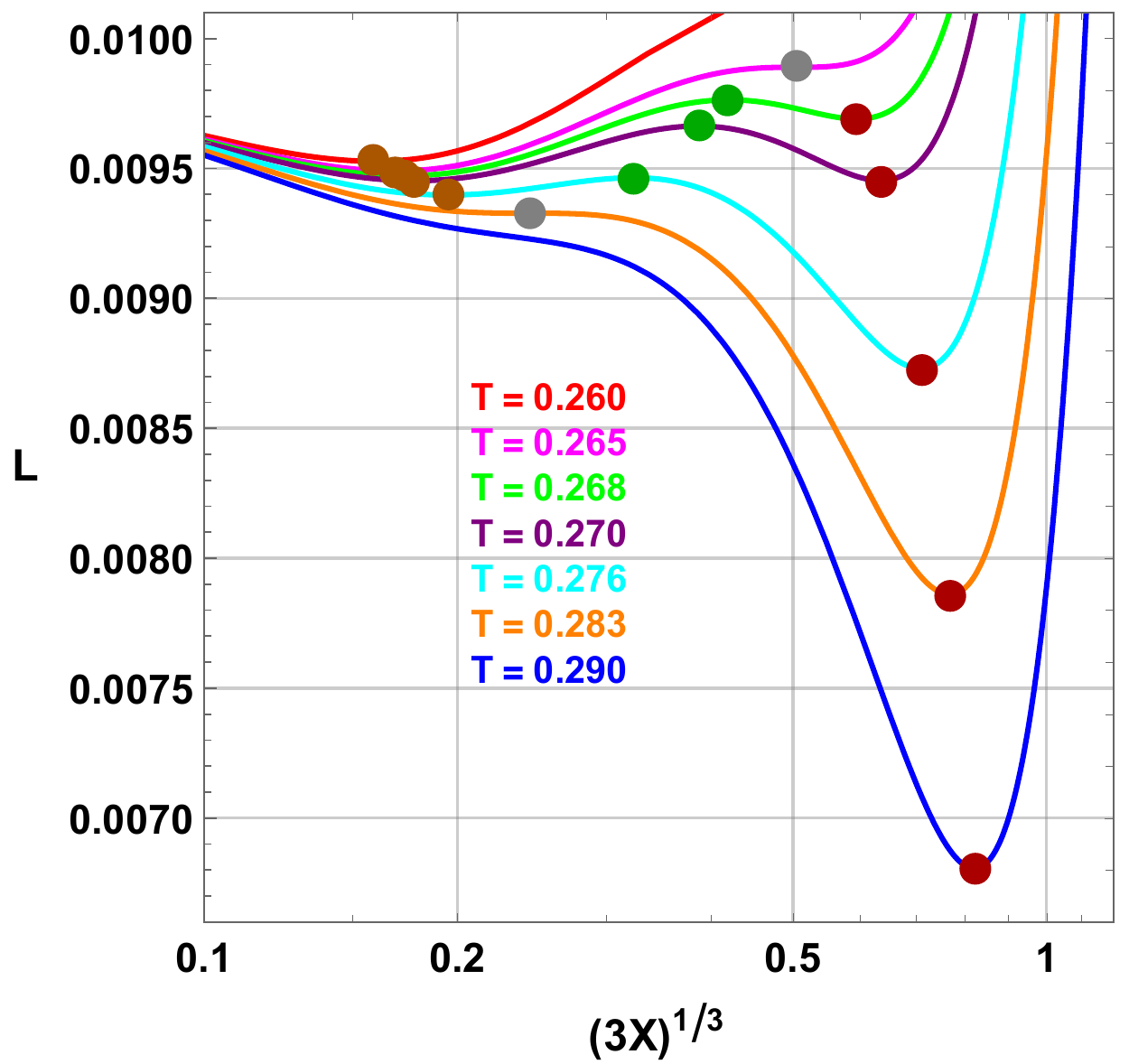}
		\caption{}
		\label{f10_3}	
	\end{subfigure}
	\hspace{1pt}	
	\begin{subfigure}[h]{0.45\textwidth}
		\centering \includegraphics[scale=.5]{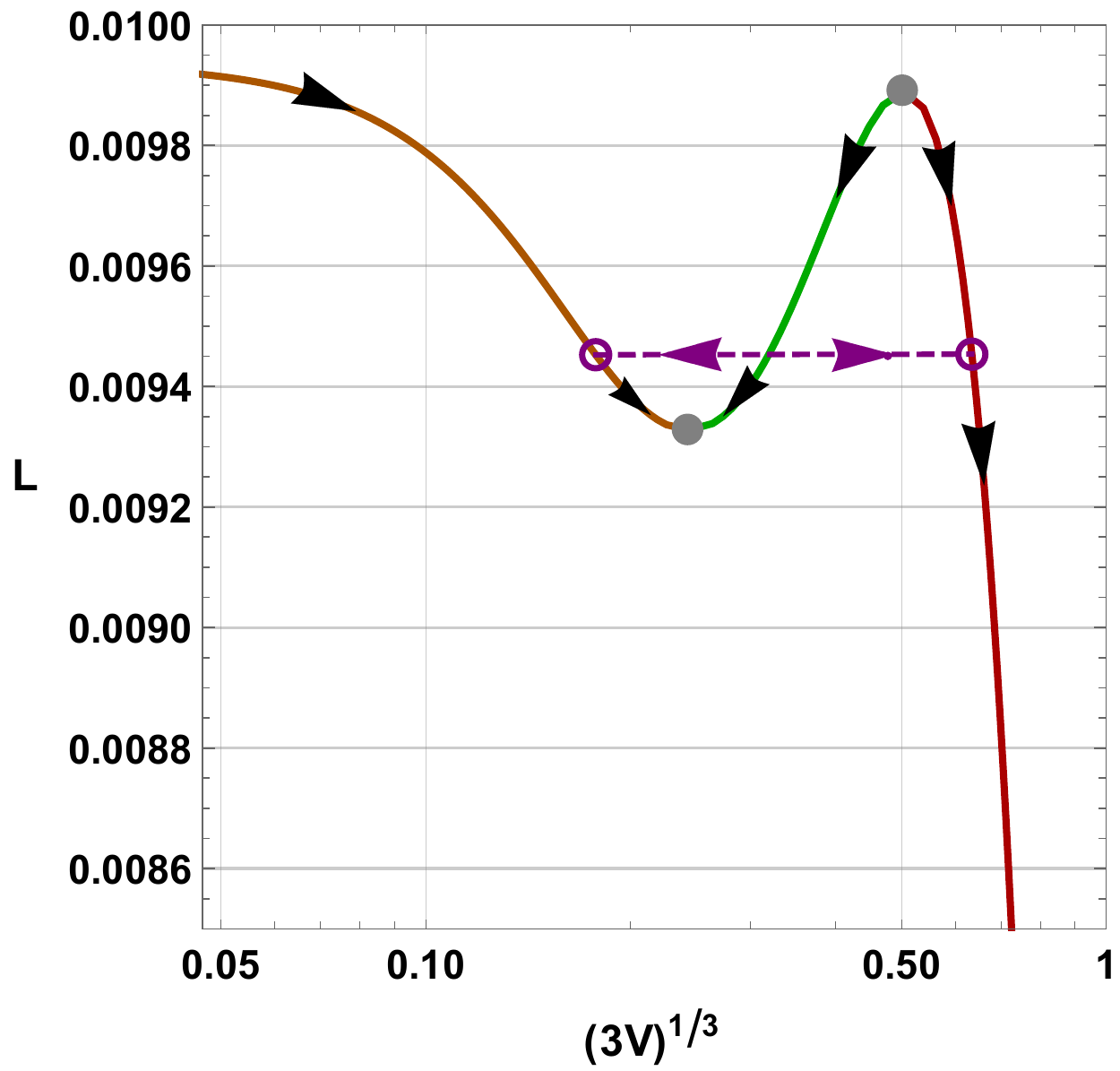}
		\caption{}
		\label{f10_4}	
	\end{subfigure}
	\hspace{1pt}	
	\begin{subfigure}[h]{0.45\textwidth}
		\centering \includegraphics[scale=.5]{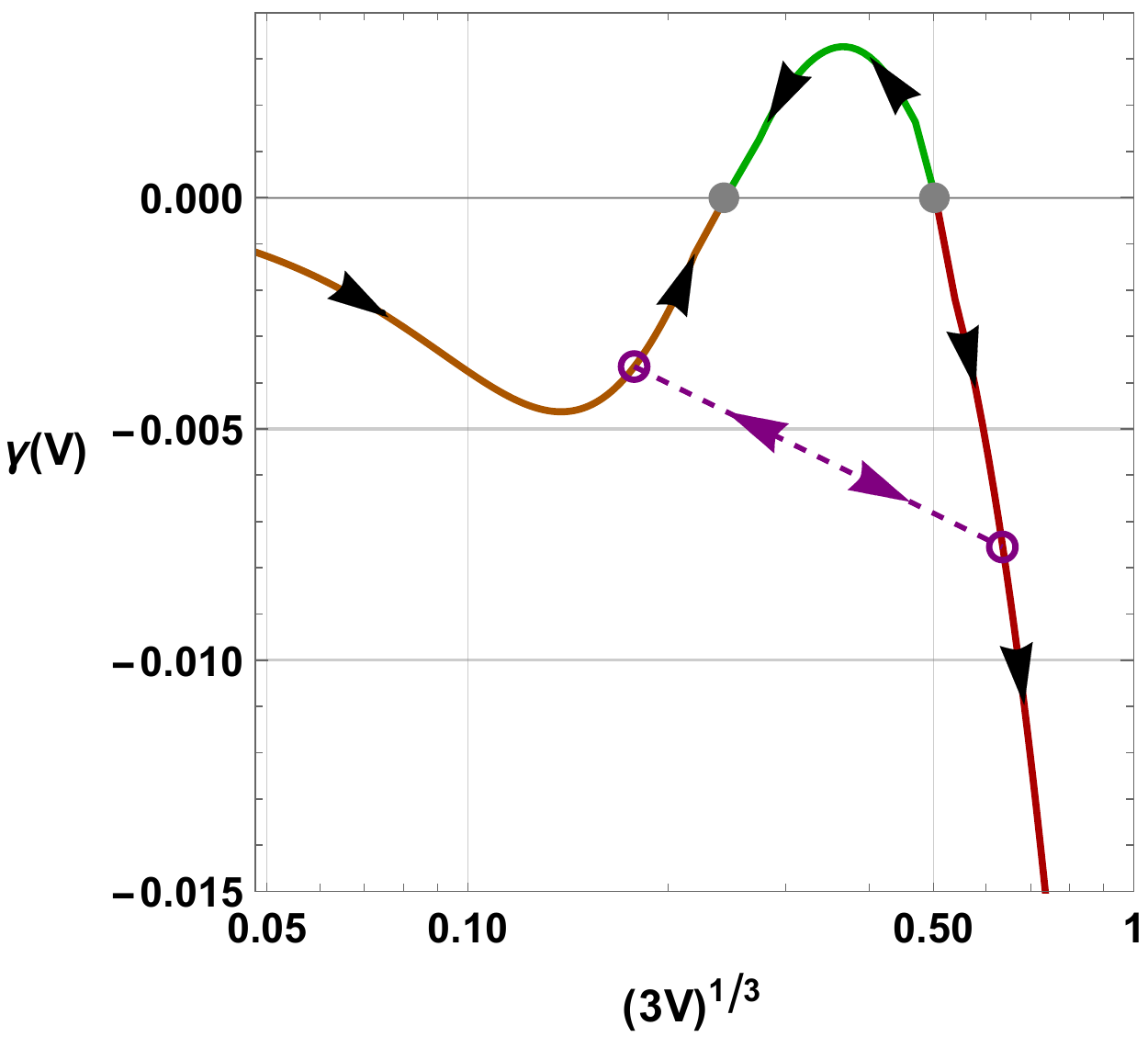}
		\caption{}
		\label{f10_5}
		
	\end{subfigure}
	\hspace{1pt}	
\begin{subfigure}[h]{0.45\textwidth}
	\centering \includegraphics[scale=.5]{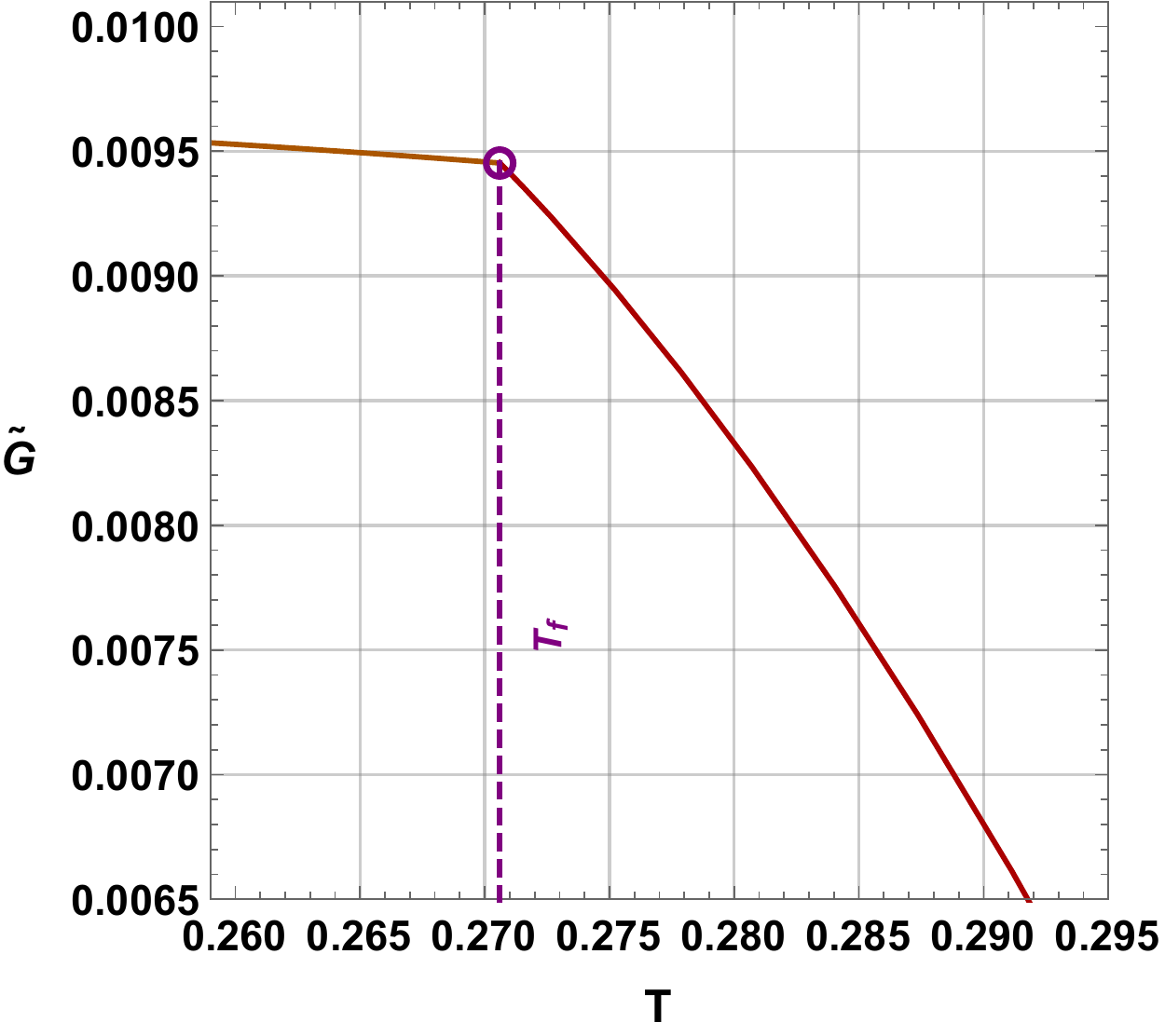}
	\caption{}
	\label{f10_6}
	
\end{subfigure}
	\caption{\footnotesize\it (a) Temperature versus the event horizon radius $r_h$. (b) Gibbs free energy-temperature diagram. (c) Landau function $L$ in terms of the parameter $X$ for different temperatures. (d) Landau function $L$ in terms of the black hole volume $V$. (e)  $\gamma$-function in terms of the black hole volume $V$. (f) On-shell Gibbs free energy $\tilde{G}$ as a function of temperature $T$. The arrows indicate the evolution of the temperature. The arrows indicate the evolution of the temperature with $Q =Q_m = 0.0113682$, $l=1$, and $b=3.5$.}
	\label{f10}
\end{figure}
Obviously, we are in  a Reissner-Nordstrom-like situation, we have a first-order phase transition between small and large black holes (purple arrow) with an unstable intermediate phase.

\item  Fig.\ref{f11} is associated with the case $Q =Q_c = 0.0136024$ %the temperature as a function of horizon radius $r_h$ (Fig.\ref{f11_1}), Gibbs free energy as a function of temperature (Fig.\ref{f11_2}), Landau function $L$ in terms of the parameter $X$ for different temperatures (Fig.\ref{f11_3}), Landau function $L$ in terms of the black hole volume $V$ Fig.\ref{f11_4} ,  $\gamma$-function in terms of the black hole volume $V$ (Fig.\ref{f11_5})  and on-shell Gibbs free energy $\tilde{G}$ as a function of temperature (Fig.\ref{f11_6}). 

\begin{figure}[!ht]
	\centering
	\begin{subfigure}[h]{0.45\textwidth}
		\centering \includegraphics[scale=.5]{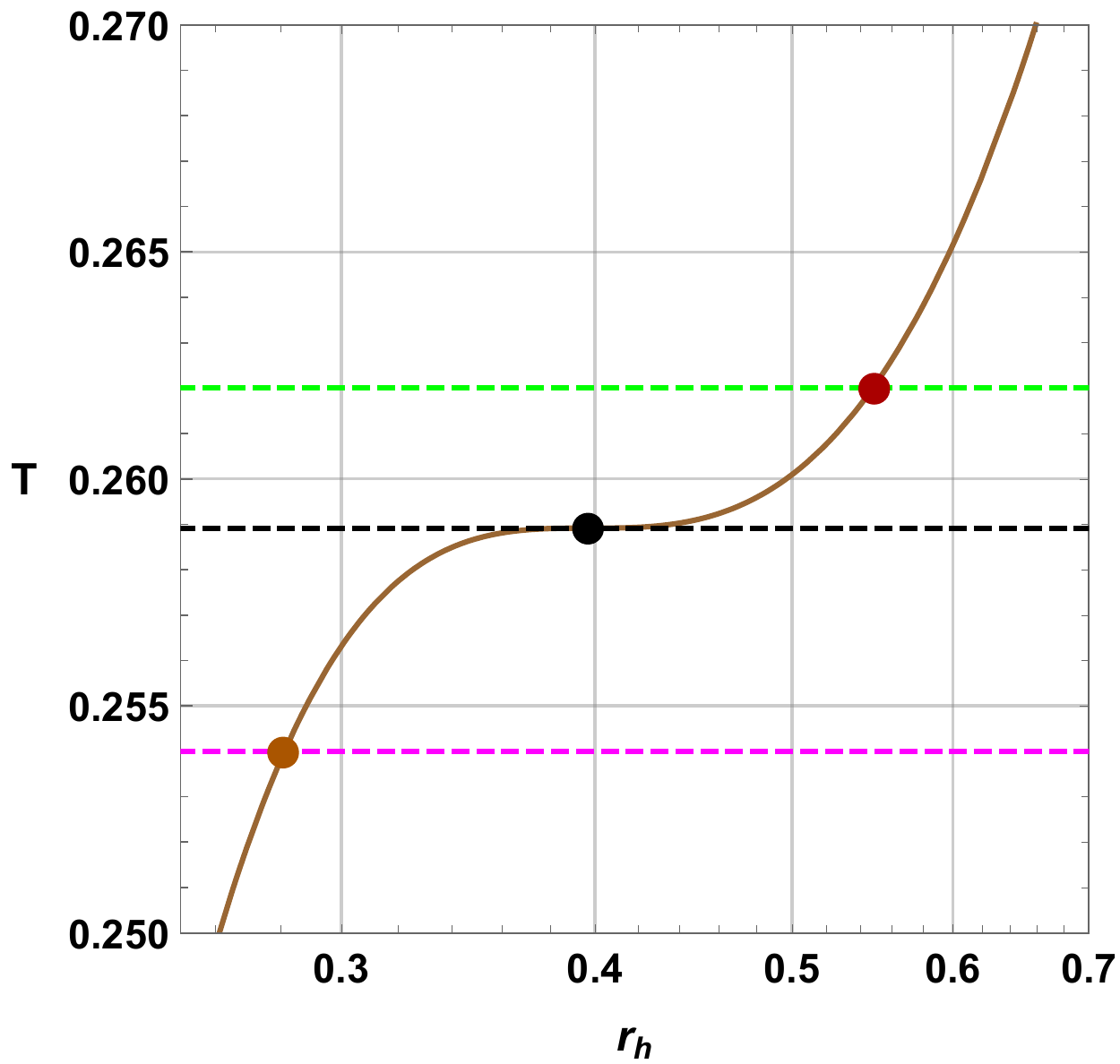}
		\caption{}
		\label{f11_1}
	\end{subfigure}
	\hspace{1pt}	
	\begin{subfigure}[h]{0.45\textwidth}
		\centering \includegraphics[scale=.5]{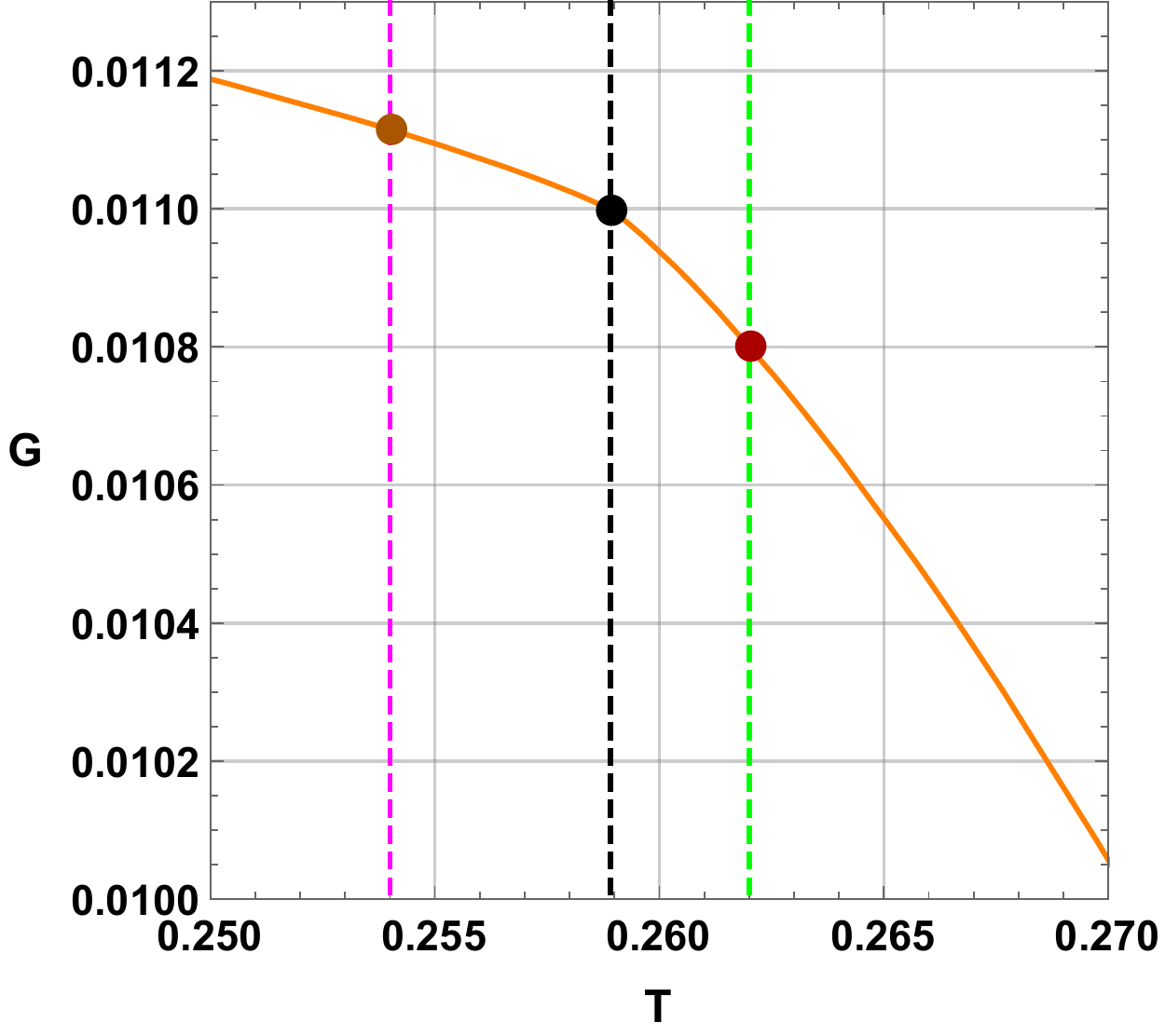}
		\caption{}
		\label{f11_2}		
	\end{subfigure}
	\hspace{1pt}	
	\begin{subfigure}[h]{0.45\textwidth}
		\centering \includegraphics[scale=.5]{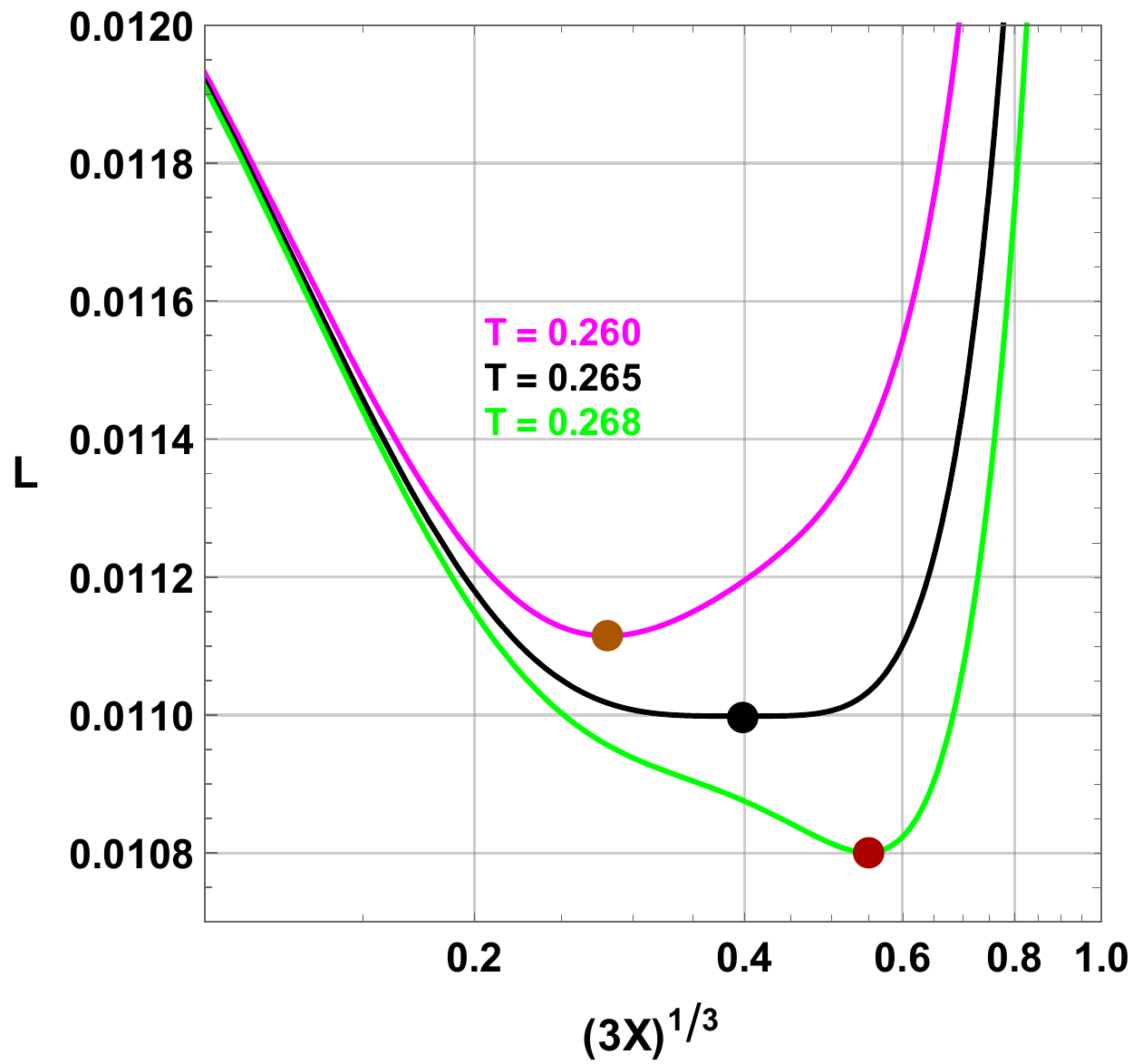}
		\caption{}
		\label{f11_3}	
	\end{subfigure}
	\hspace{1pt}	
	\begin{subfigure}[h]{0.45\textwidth}
		\centering \includegraphics[scale=.5]{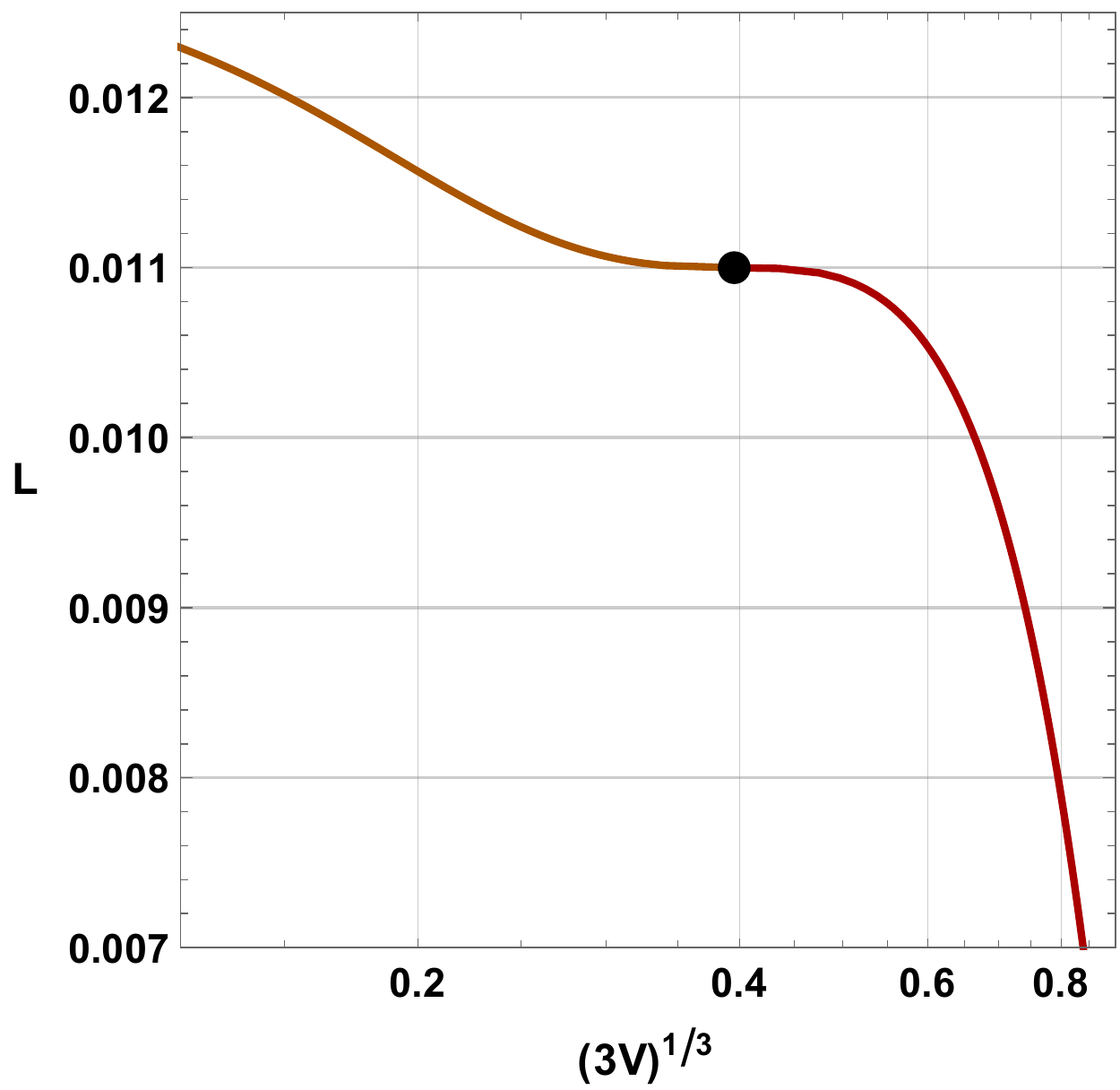}
		\caption{}
		\label{f11_4}	
	\end{subfigure}
	\hspace{1pt}	
	\begin{subfigure}[h]{0.45\textwidth}
		\centering \includegraphics[scale=.5]{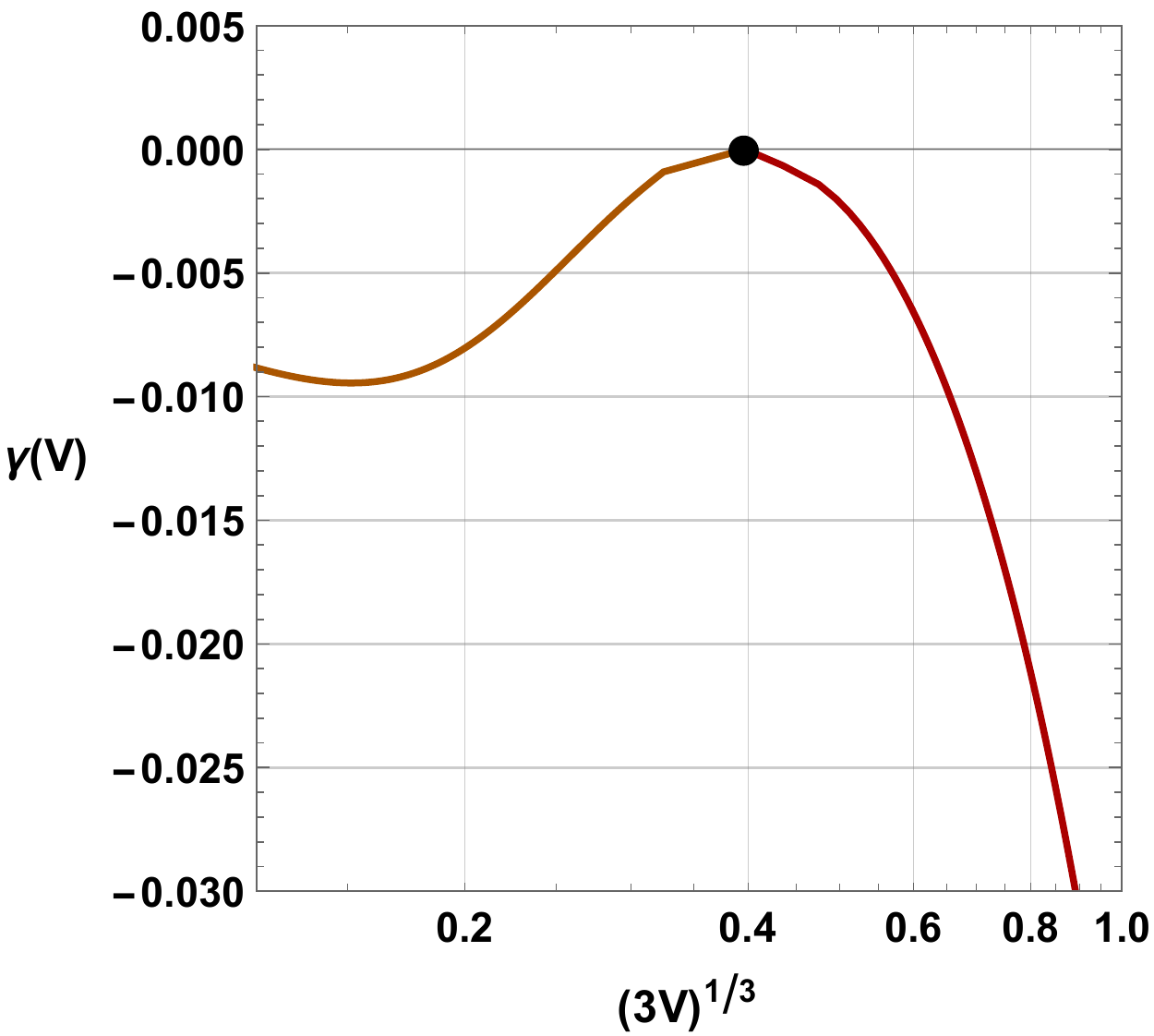}
		\caption{}
		\label{f11_5}
		
	\end{subfigure}
	\hspace{1pt}	
\begin{subfigure}[h]{0.45\textwidth}
	\centering \includegraphics[scale=.5]{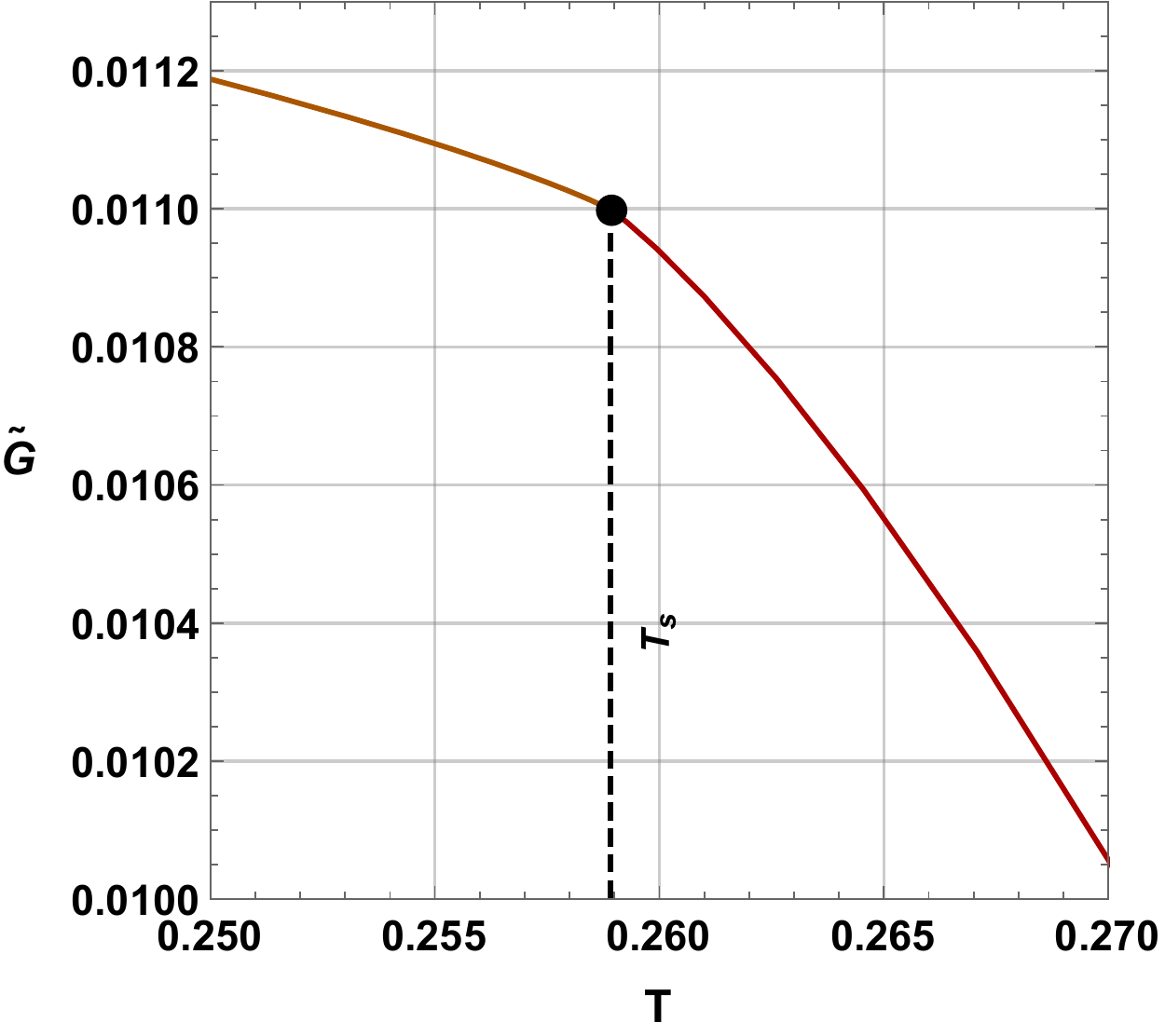}
	\caption{}
	\label{f11_6}
	
\end{subfigure}
	\caption{\footnotesize\it (a) Temperature versus the event horizon radius $r_h$. (b) Gibbs free energy-temperature diagram. (c) Landau function $L$ in terms of the parameter $X$ for different temperatures. (d) Landau function $L$ in terms of the black hole volume $V$. (e)  $\gamma$-function in terms of the black hole volume $V$. (f) On-shell Gibbs free energy $\tilde{G}$ as a function of temperature $T$. The arrows indicate the evolution of the temperature. The arrows indicate the evolution of the temperature with $Q =Q_c = 0.0136024$, $l=1$, and $b=3.5$.}
	\label{f11}
\end{figure}
The panels unveil a critical behavior where (block dot) a second phase transition occurs between small and large black holes at exactly $T = 0.265$.
%{\bf HERE}
\item  The case related to $Q = 0.0154$ is illustrated in Fig.\ref{f12}, where we have depicted just the temperature as a function of horizon radius $r_h$ (Fig.\ref{f12_1}), Gibbs free energy-temperature diagram (Fig.\ref{f12_2}), Landau function $L$ in terms of the parameter $X$ for different temperatures (Fig.\ref{f12_3}), then in terms of the black hole volume $V$ in Fig.\ref{f12_4} panel. The last panel is devoted to $\gamma$-function versus the black hole volume $V$ (Fig.\ref{f12_5}).  

\begin{figure}[!ht]
	\centering
	\begin{subfigure}[h]{0.45\textwidth}
		\centering \includegraphics[scale=.5]{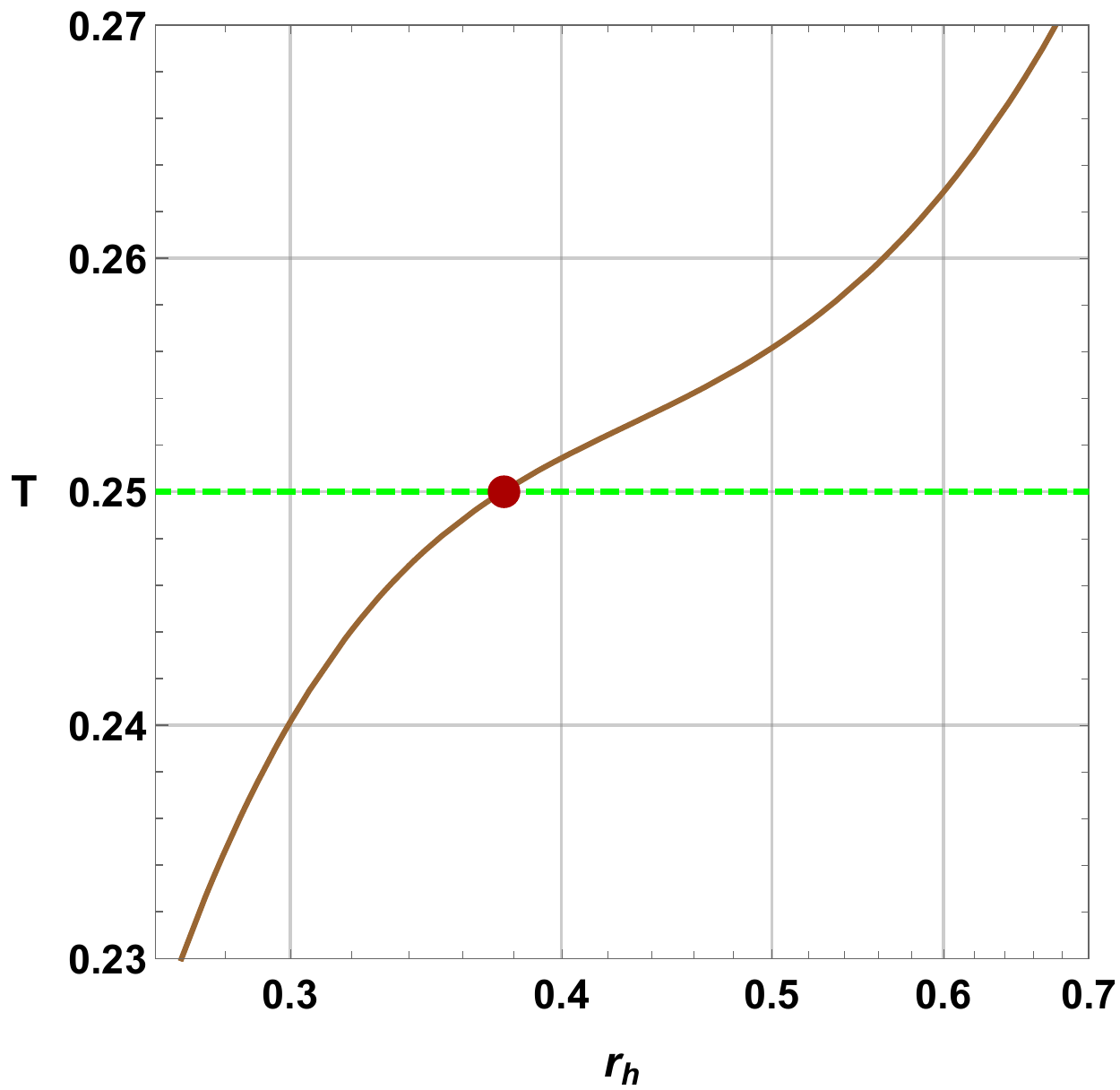}
		\caption{}
		\label{f12_1}
	\end{subfigure}
	\hspace{1pt}	
	\begin{subfigure}[h]{0.45\textwidth}
		\centering \includegraphics[scale=.5]{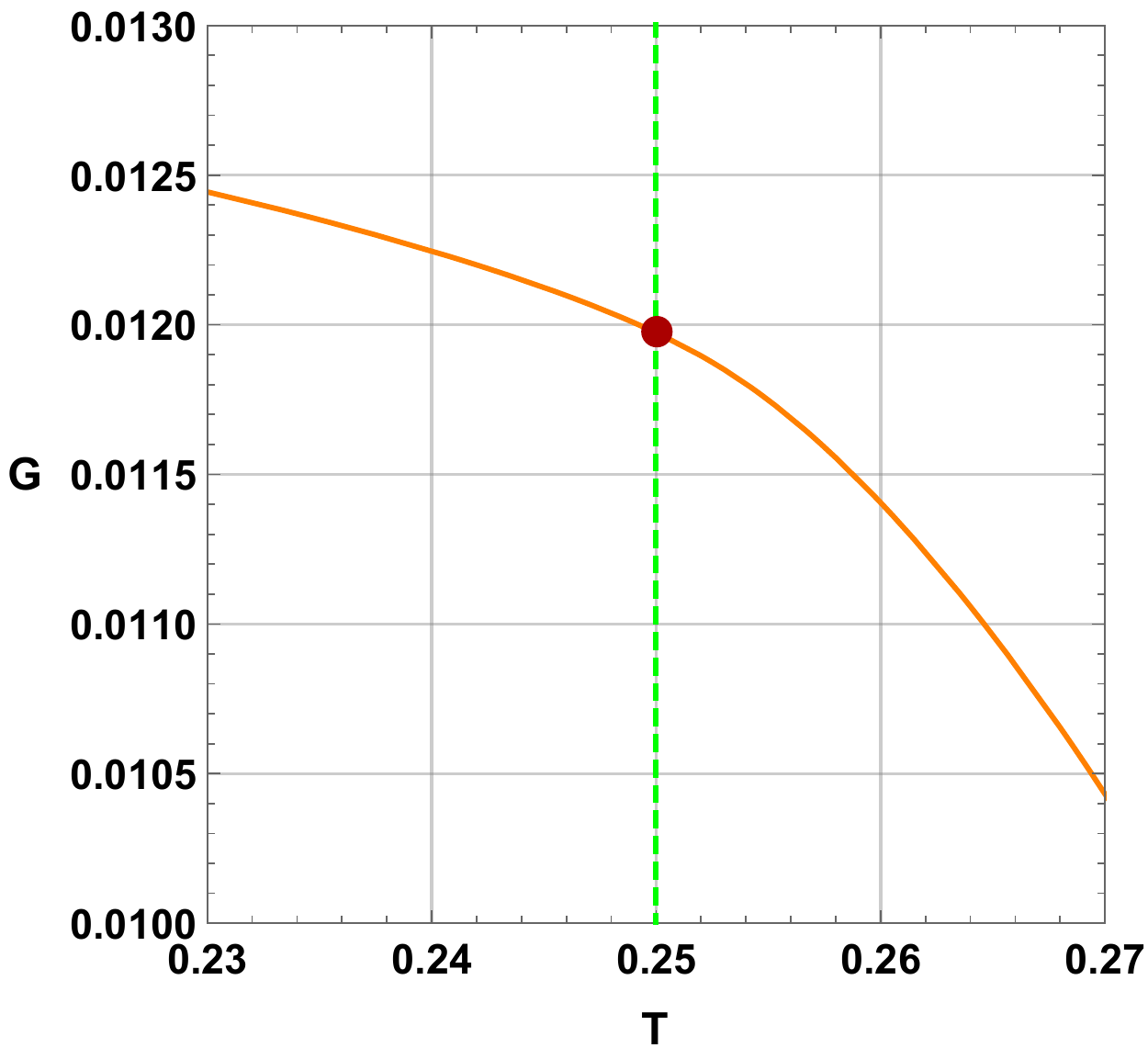}
		\caption{}
		\label{f12_2}		
	\end{subfigure}
	\hspace{1pt}	
	\begin{subfigure}[h]{0.45\textwidth}
		\centering \includegraphics[scale=.5]{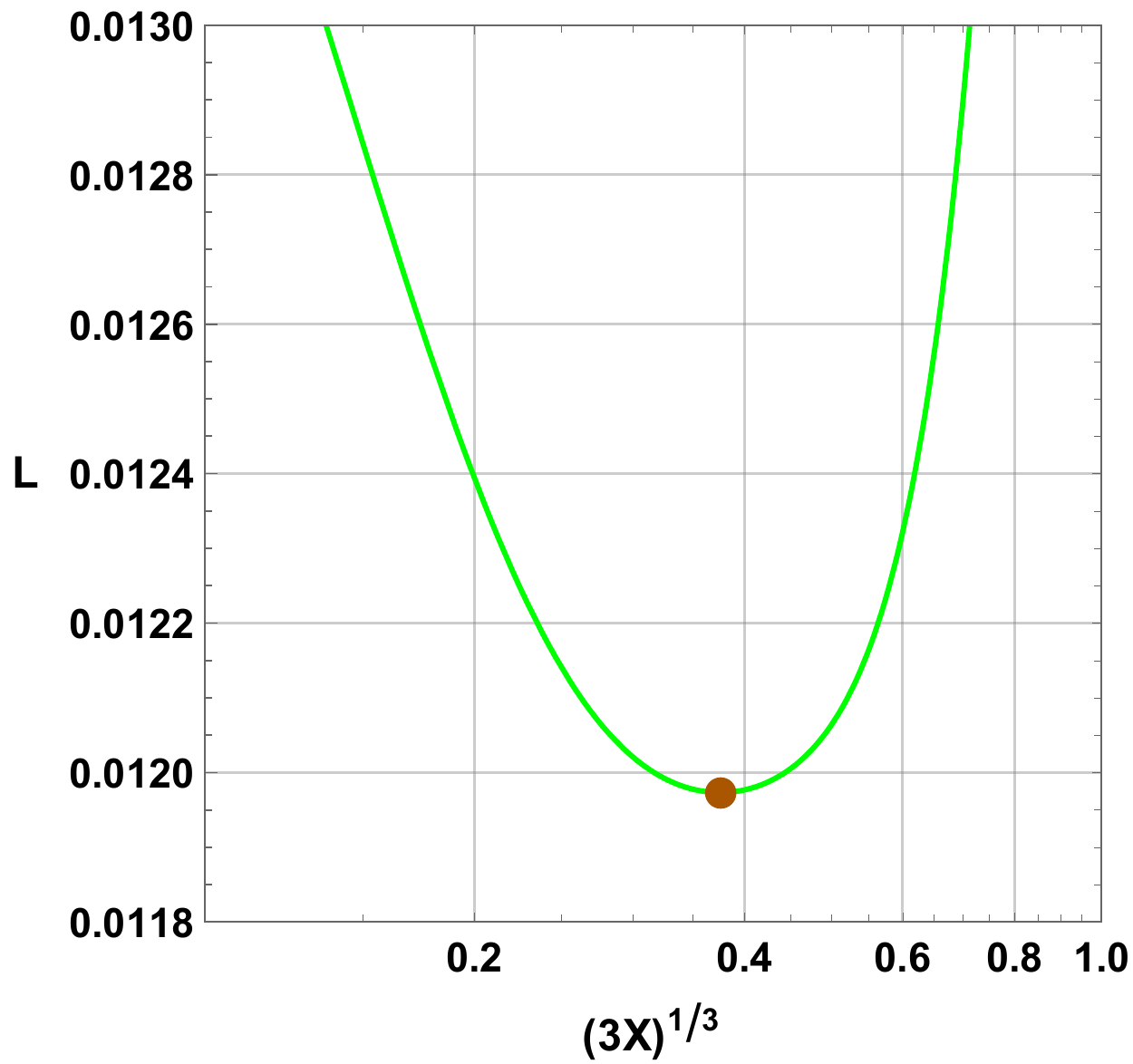}
		\caption{}
		\label{f12_3}	
	\end{subfigure}
	\hspace{1pt}	
	\begin{subfigure}[h]{0.45\textwidth}
		\centering \includegraphics[scale=.5]{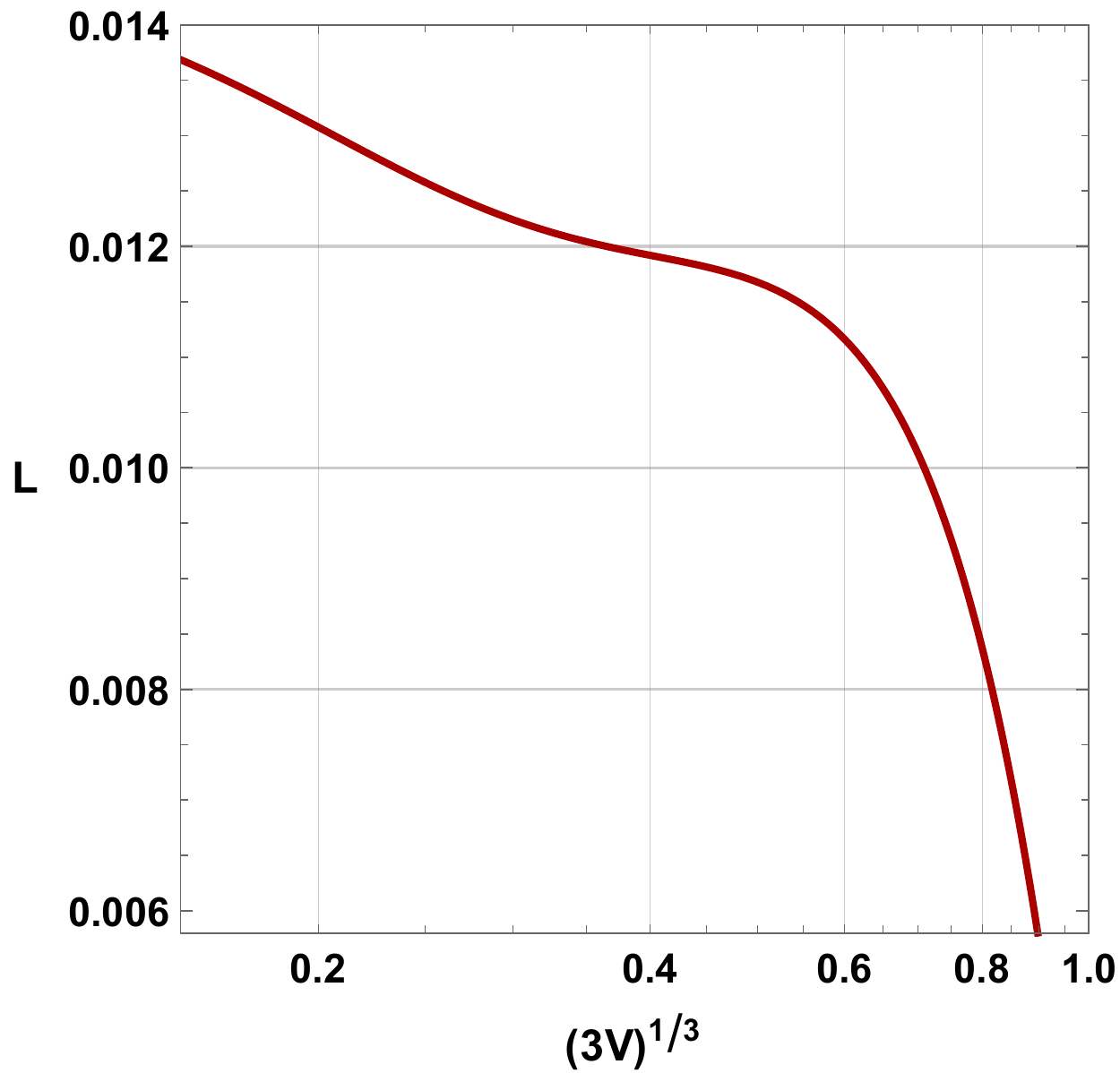}
		\caption{}
		\label{f12_4}	
	\end{subfigure}
	\hspace{1pt}	
	\begin{subfigure}[h]{0.45\textwidth}
		\centering \includegraphics[scale=.5]{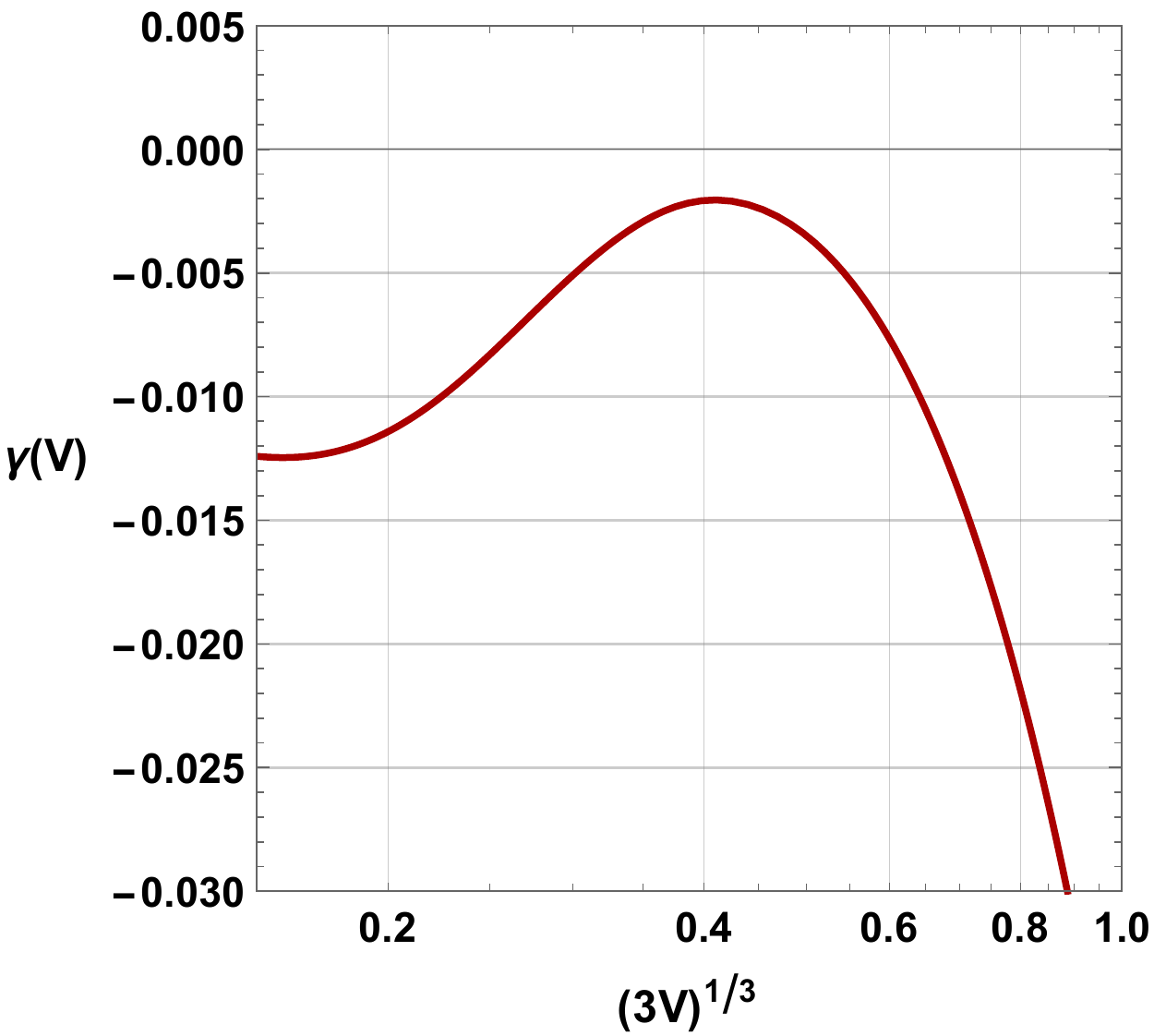}
		\caption{}
		\label{f12_5}
		
	\end{subfigure}
	\caption{\footnotesize\it (a) Temperature versus the event horizon radius $r_h$. (b) Gibbs free energy-temperature diagram. (c) Landau function $L$ in terms of the parameter $X$ for different temperatures. (d) Landau function $L$ in terms of the black hole volume $V$. (e)  $\gamma$-function in terms of the black hole volume $V$. The arrows indicate the evolution of the temperature with $Q =0.015$, $l=1$, and $b=3.5$.}
	\label{f12}
\end{figure}

\end{itemize}
%\textcolor{red}{comment}

 \begin{figure}[!ht]
	\centering
    	\begin{subfigure}[h]{0.47\textwidth}
		\centering \includegraphics[scale=.6]{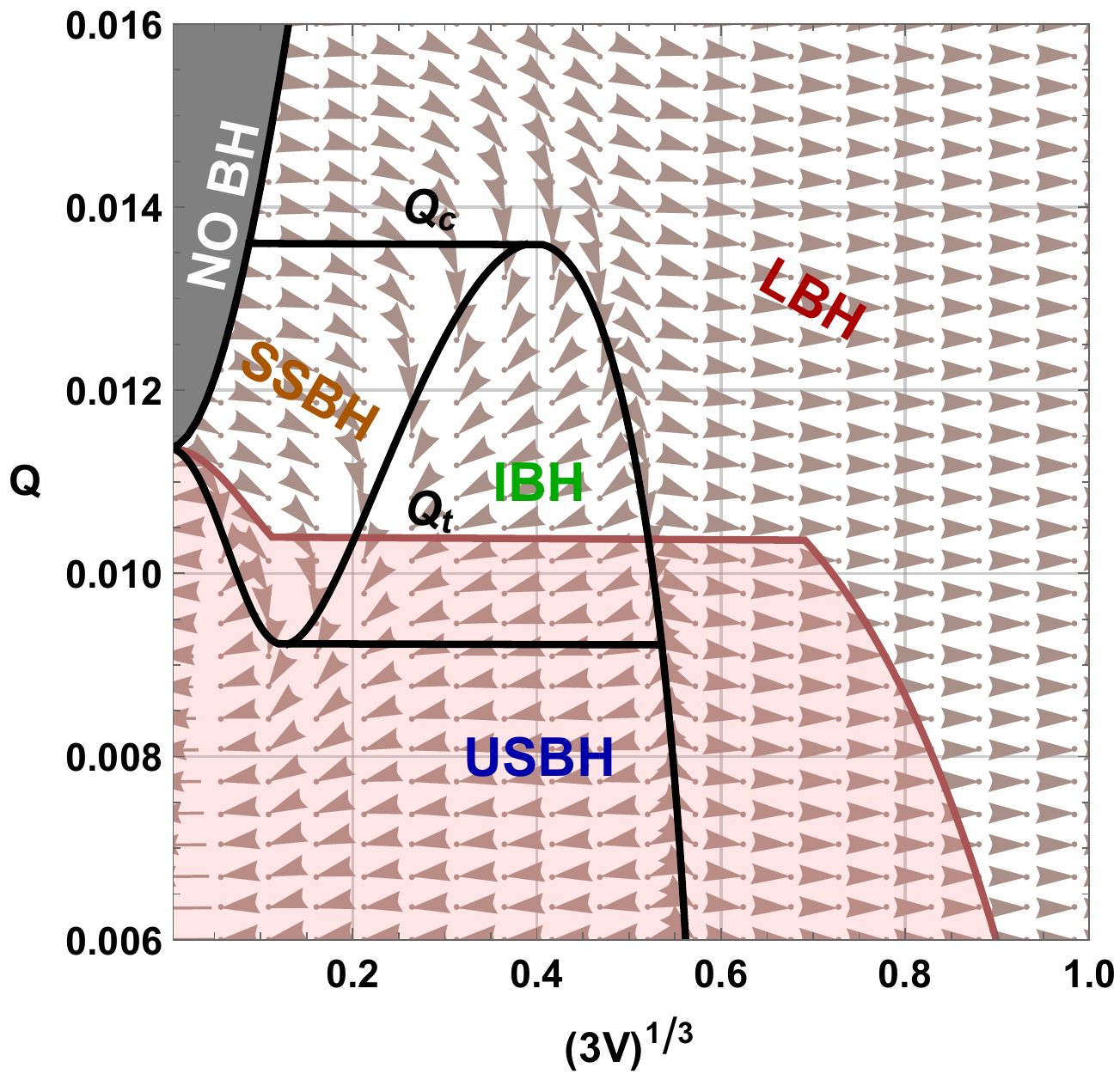}
			\caption{}
			\label{f13_1}
		\end{subfigure}
	\hspace{1pt}
		\begin{subfigure}[h]{0.47\textwidth}
			\centering \includegraphics[scale=.6]{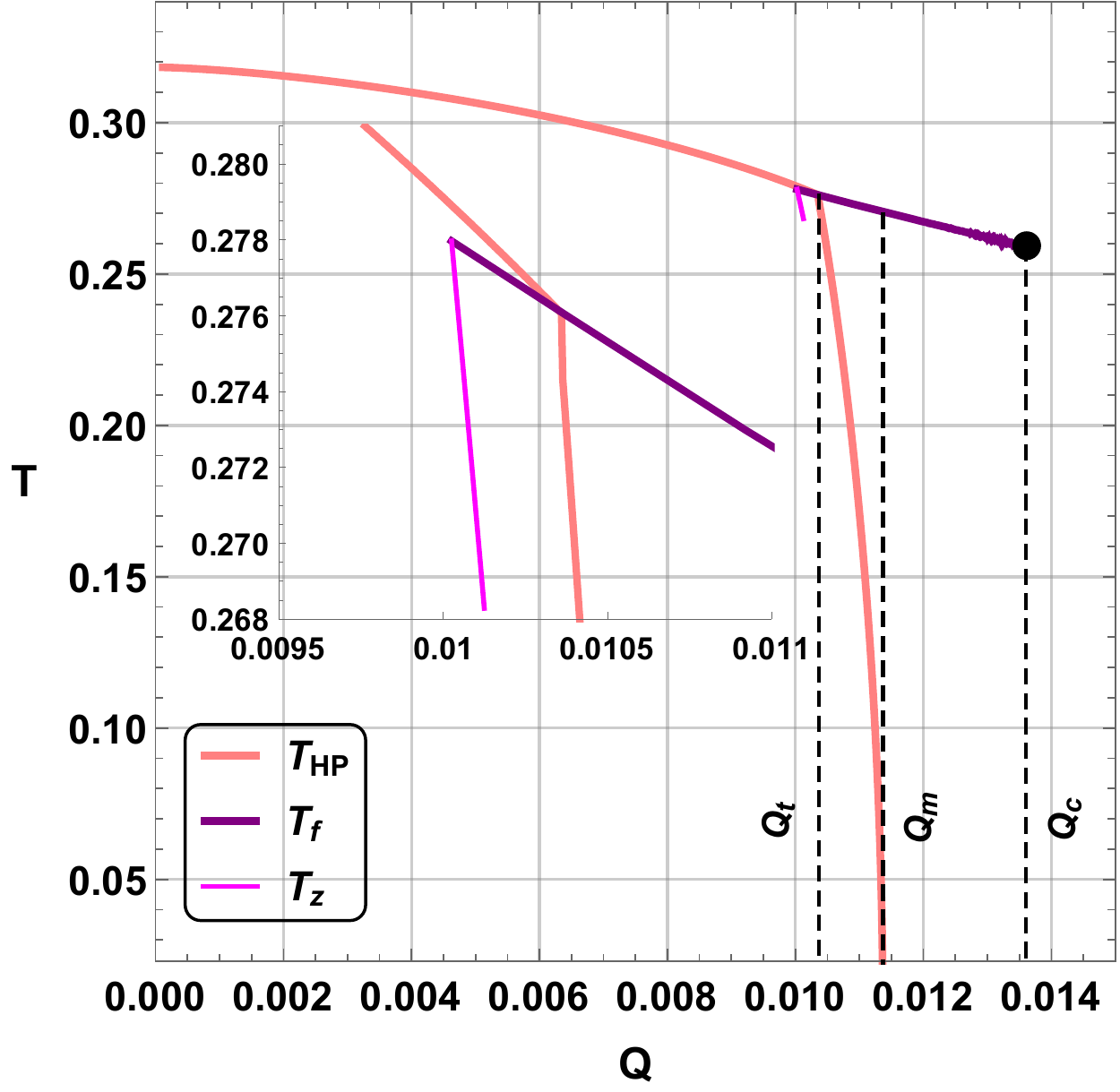}
			\caption{}
			\label{f13_2}
		\end{subfigure}	
	\hspace{1pt}
		\begin{subfigure}[h]{0.47\textheight}
		   \includegraphics[scale=.8]{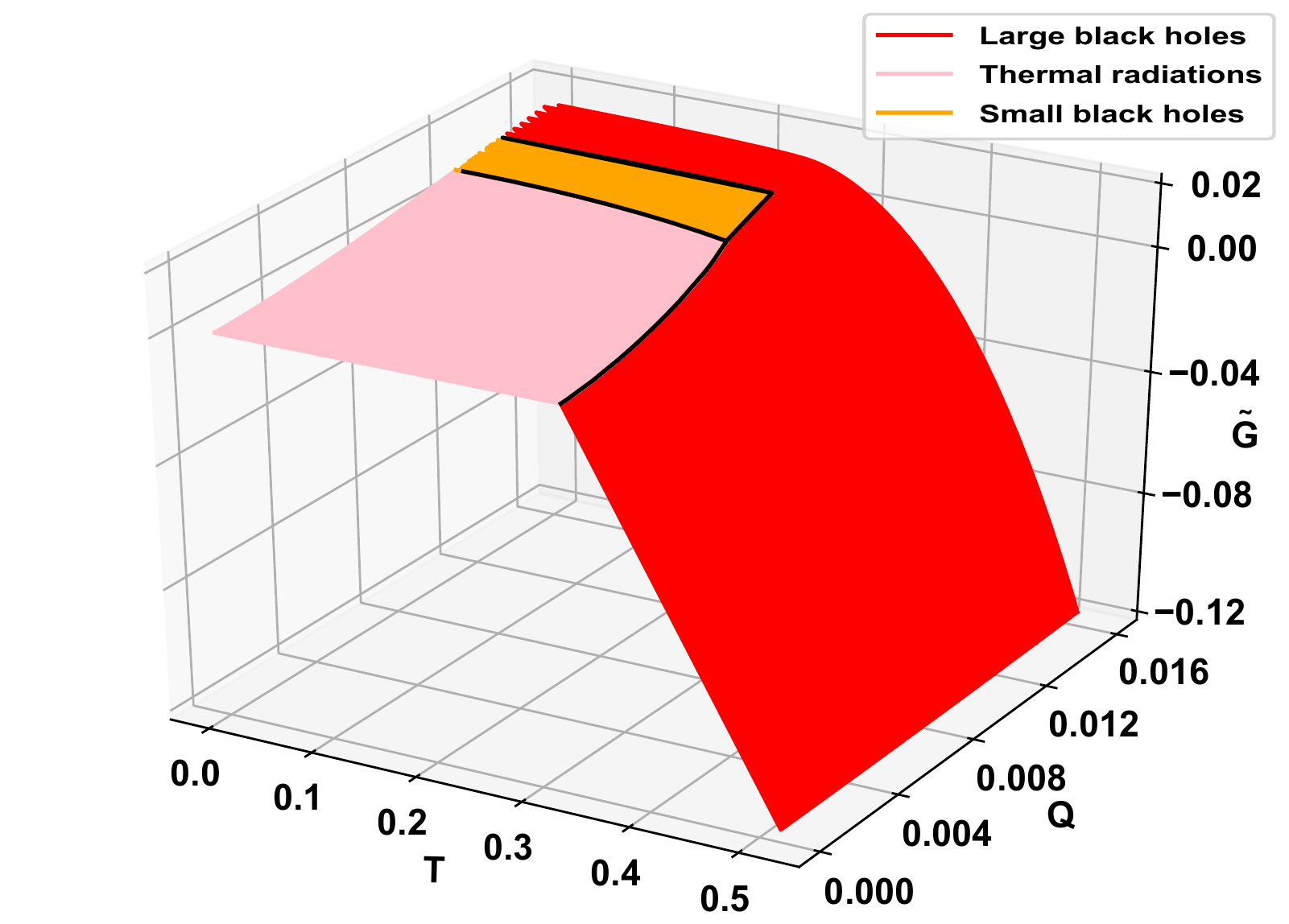}
			\caption{}
			\label{f13_3}		
		\end{subfigure}
	\caption{\footnotesize\it (a) $Q-V$ diagram of Born-Infeld-AdS black hole; the flow indicates the temperature gradient. (b) Critical temperatures as a function of electric charge. (c) On-shell Gibbs free energy as a function of temperature and electric charge. $l=1$ and $b=3.5$.}
	\label{f13}
\end{figure}
  The phase diagram of the Born-Infeld-AdS black hole is depicted in Fig.\ref{f13}. We have displayed in Fig.\ref{f13_1} the $Q-V$ diagram where the flow represented by arrows indicates the temperature gradient. We clearly distinguish four different local phases: stable small black holes (SSBH), unstable small black holes (USBH), intermediate black holes (IBH), and large black holes (LBH), the gray zone represents the zone where the black hole cannot exist and which delimited by the extremal black hole case. The pink zone indicates the region where the thermal radiation is the globally stable phase, that is to say, it corresponds to temperatures below the Hawking-Page temperature. We notice that the locally stable black holes correspond to a right-oriented temperature gradient  which means that the black hole becomes hotter when it gets larger, whereas, the locally unstable black holes correspond to a left-oriented temperature gradient traducing that the black hole becomes hotter when it gets smaller and consequently evaporates. For a more deep probing,  in Fig.\ref{f13_2}, we illustrate the Hawking-Page transition temperature $T_{HP}$, the first order phase transition temperature $T_f$ and the zeroth order phase transition temperature $T_z$ as a function of electric charge $Q$. From such a panel, various remarks can be deduced, first
the zeroth order phase transition temperature $T_z$ is always below the Hawking-Page transition temperature $T_{HP}$, which means that the zeroth order phase transition can not occur. Second, the first order phase transition can occur where $Q>Q_m$. Further, in Fig.\ref{f13_3}, we establish the on-shell Gibbs free energy-temperature $T$-electric charge $Q$ diagram, and from which one can see that we are in the presence of  three globally stable phases: large black holes, thermal radiations, and small black holes.
% \cleardoublepage

 The phase diagram of the Born-Infeld-AdS black hole can be characterized by three electric charge values  as follow  
\begin{itemize}
 \item[$\diamondsuit$] For $Q<Q_t = 0.0103638$, we have just two globally stable phases, thermal radiation, and a large black hole, then consequently there is only one possible phase transition which is the Hawking-Page phase transition. Therefore, there is neither first-order phase transition nor zeroth-order phase transition as it was shown in \cite{Dehyadegari:2017hvd}. Indeed, in \cite{Dehyadegari:2017hvd}, authors have shown that the reentrant phase transition occurs for $0.010026<Q<0.10128$ but they had not proved that the intermediate black holes are globally stable for these values of electric charge. To conclude, for $Q<Q_t$, the Born-Infeld-AdS black hole is Schwrazchild-like, there are only two globally stable phases that are connected by Hawking-Page phase transition. Moreover, there is no extremal (no vanishing temperature) which confirms the results of \cite{PhysRevD.78.084002}.
 
 \item[$\diamondsuit$] For $Q_t<Q<Q_m = 0.0113682$, we have three globally stable phases, thermal radiation (pink zone), small stable black hole (SSBH), and large black hole (LBH) with two possible critical points as we have shown in Fig.\ref{f9}. Indeed, there are two possible phase transitions, the first one is between thermal radiations and small black holes, it's a Hawking-Page-like transition; the second one is the first order phase transition between small and large black holes like that is observed in Reissner-Nordström-AdS black holes \cite{Kubiznak:2012wp}. Moreover, for $Q=Q_t$, we have a triple point where thermal radiation, small, and large black holes coexist together.
 
 \item[$\diamondsuit$] For $Q_m<Q<Q_c = 0.0136024$, the Born-Infeld-AdS black hole is similar to the Reissner-Nordström-AdS black hole case, and in such a situation we got two globally stable phases and a first-order phase transition between small and large black holes. Moreover, we could have an extremal state where the black hole becomes sufficiently small. Therefore, there is no thermal radiation phase, and the Hawking-Page-like transition disappears.
 
  \item[$\diamondsuit$] For $Q>Q_c$, there is only one globally stable phase which is the large black holes phase as in Reissner-Nordström-AdS black hole. Besides, for $Q=Q_c$ the black hole exhibits a second-order phase transition between small and large black holes. 
\end{itemize}

\begin{figure}[!ht]
	\centering
	\begin{subfigure}[h]{0.45\textwidth}
		\centering \includegraphics[scale=.5]{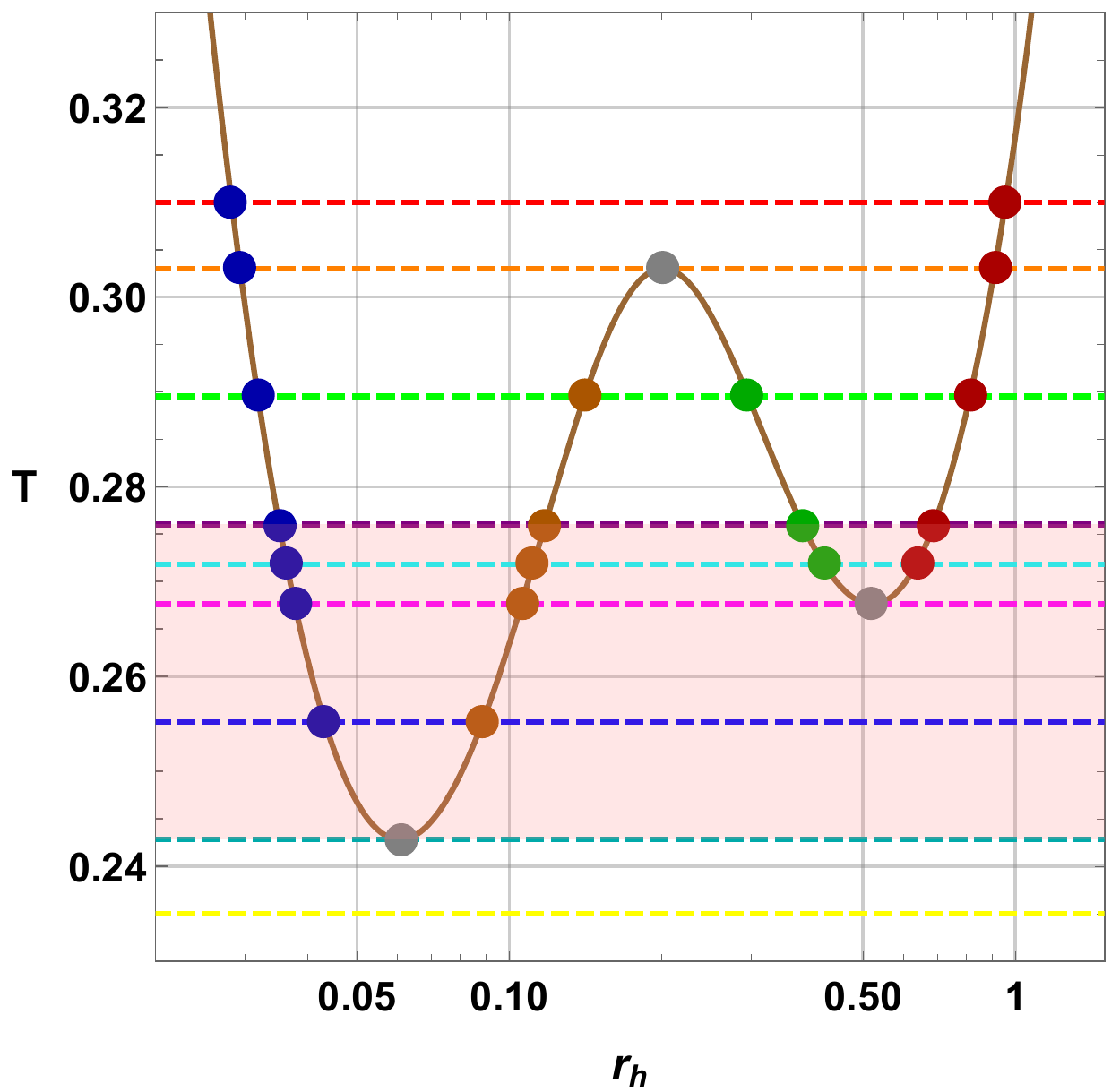}
		\caption{}
		\label{f14_1}
	\end{subfigure}
	\hspace{1pt}	
	\begin{subfigure}[h]{0.45\textwidth}
		\centering \includegraphics[scale=.5]{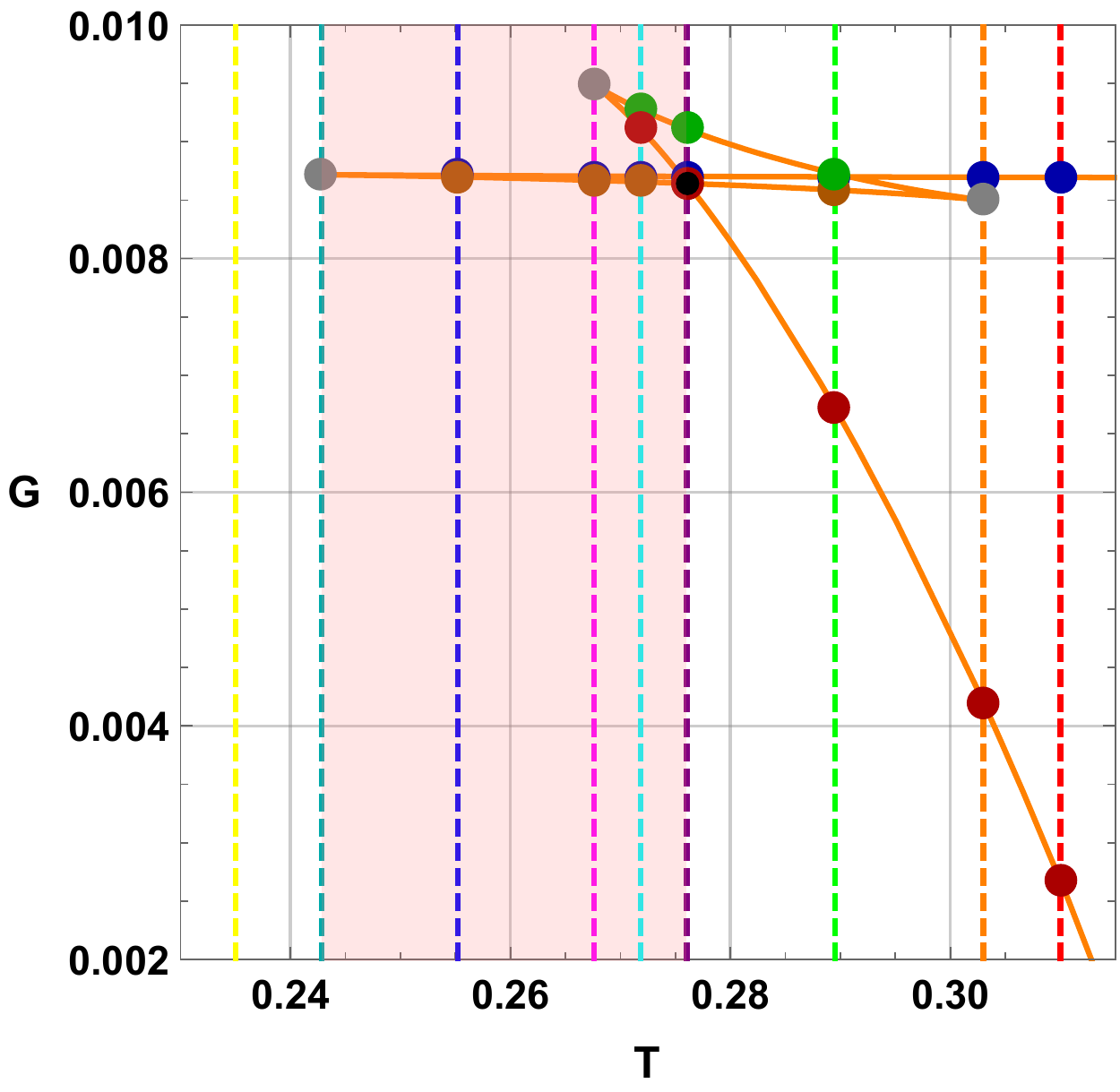}
		\caption{}
		\label{f14_2}		
	\end{subfigure}
	\hspace{1pt}	
	\begin{subfigure}[h]{0.45\textwidth}
		\centering \includegraphics[scale=.5]{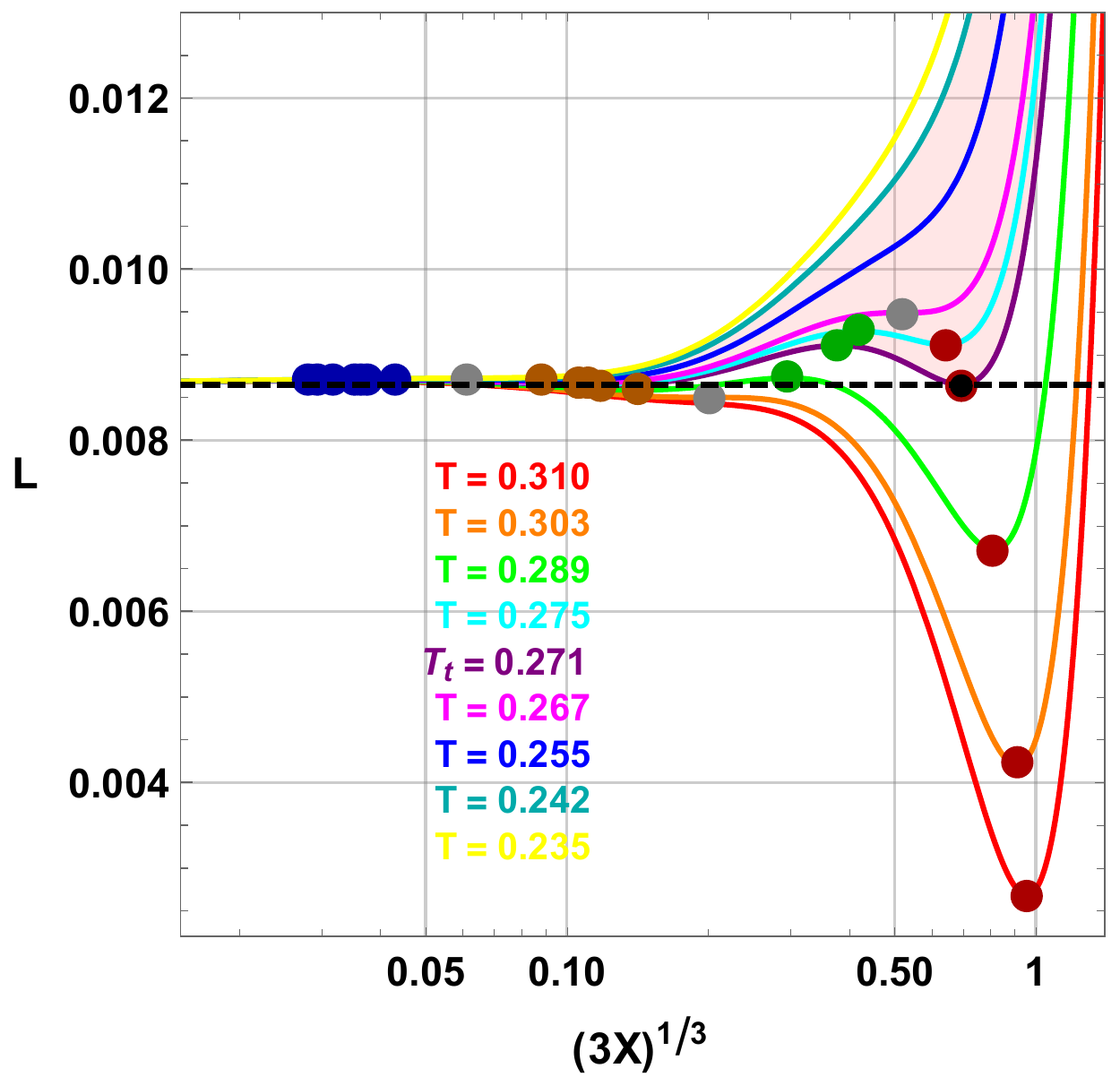}
		\caption{}
		\label{f14_3}	
	\end{subfigure}
	\hspace{1pt}	
	\begin{subfigure}[h]{0.45\textwidth}
		\centering \includegraphics[scale=.5]{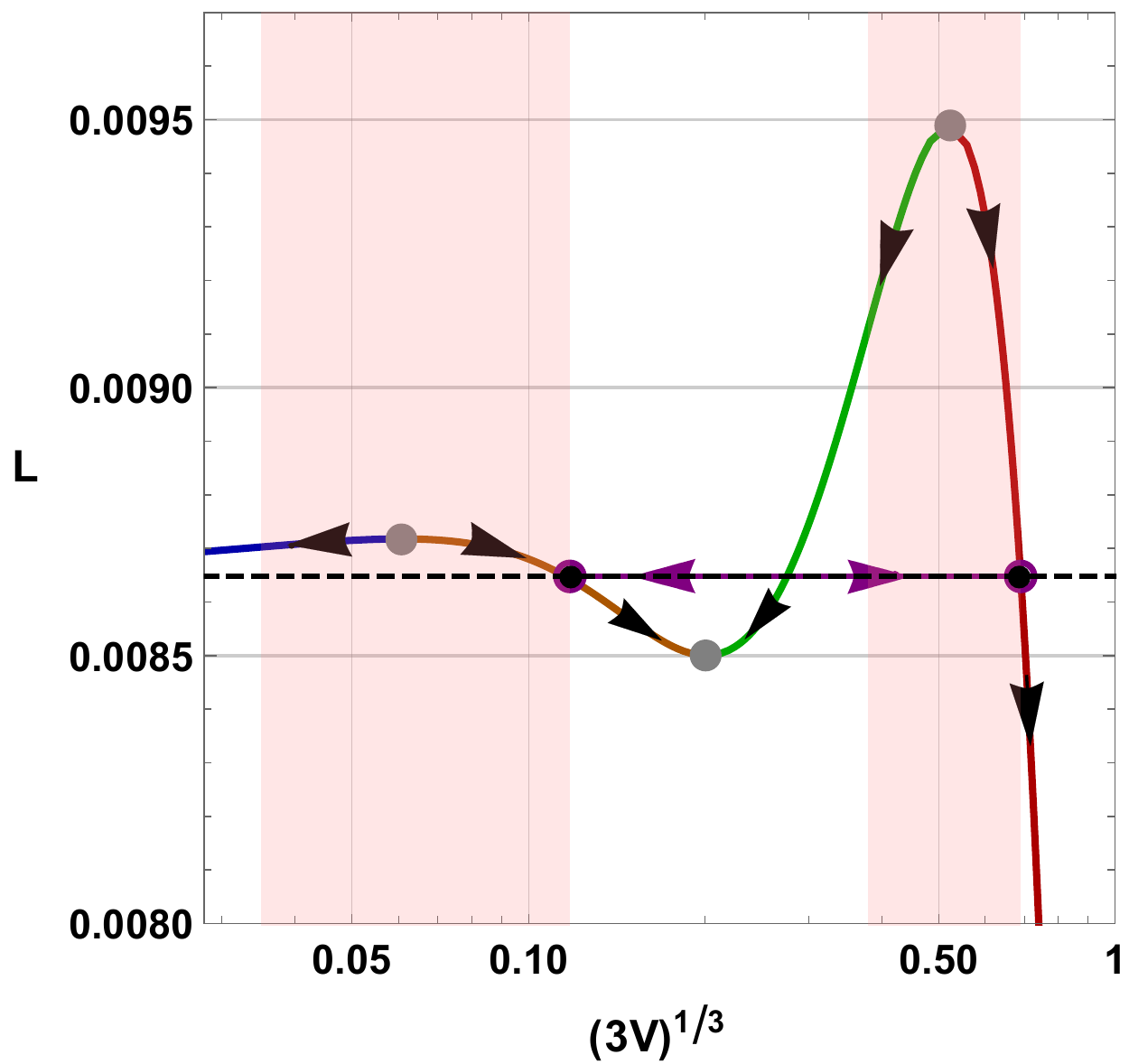}
		\caption{}
		\label{f14_4}	
	\end{subfigure}
	\hspace{1pt}	
	\begin{subfigure}[h]{0.45\textwidth}
		\centering \includegraphics[scale=.5]{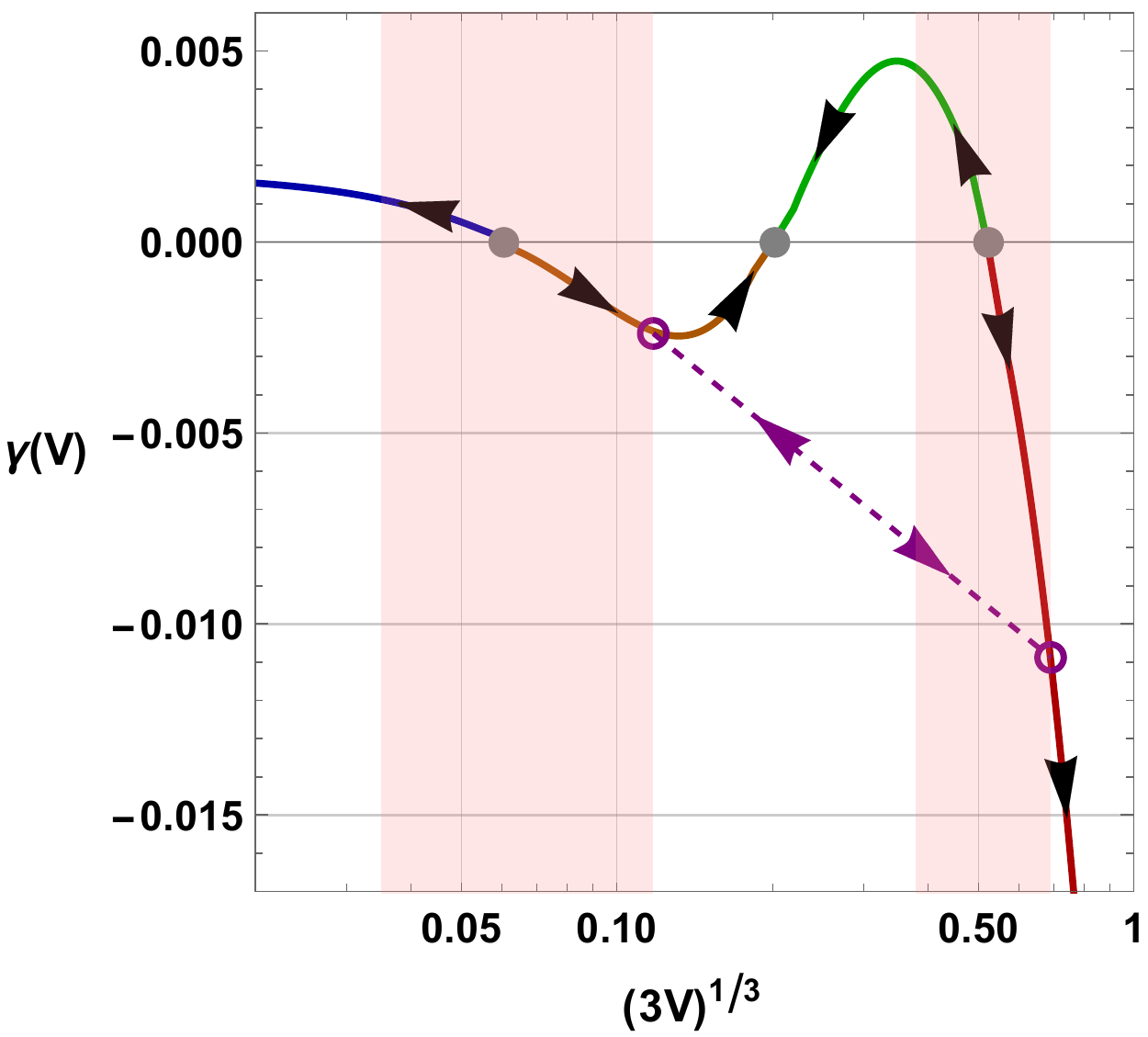}
		\caption{}
		\label{f14_5}
		
	\end{subfigure}
	\hspace{1pt}	
	\begin{subfigure}[h]{0.45\textwidth}
		\centering \includegraphics[scale=.5]{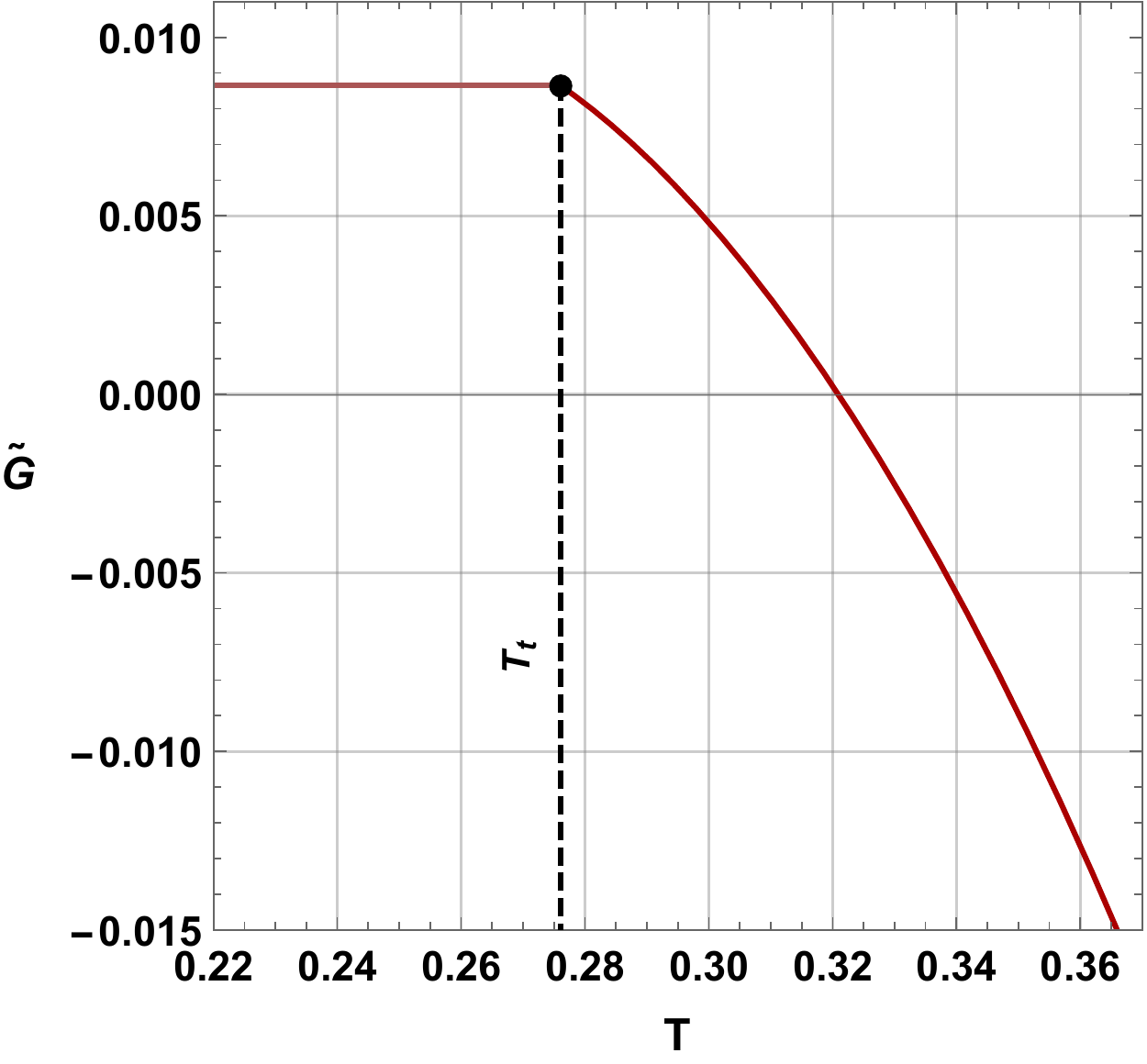}
		\caption{}
		\label{f14_6}
		
	\end{subfigure}
	\caption{\footnotesize (a) Temperature versus the event horizon radius $r_h$. (b) Gibbs free energy-temperature diagram. (c)  Landau function $L$ in terms of the parameter $X$ for different temperatures. (d)  Landau function $L$ in terms of the black hole volume $V$. (e)  $\gamma$-function in terms of the black hole volume $V$. (f) On-shell Gibbs free energy $\tilde{G}$ as a function of temperature $T$. The arrows indicate the evolution of the temperature. The arrows indicate the evolution of the temperature with $Q =Q_t = 0.0103638$, $l=1$, and $b=3.5$.}
	\label{f14}
\end{figure}

Therefore, one concludes that the Born-Infeld-AdS black hole can unveil a triple point where the three globally stable phases can coexist. In Fig.\ref{f14} we illustrate the temperature as a function of horizon radius $r_h$ (Fig.\ref{f14_1}), Gibbs free energy in terms of temperature (Fig.\ref{f14_2}), Landau function $L$ in terms of the parameter $X$ for different temperatures (Fig.\ref{f14_3}), and in terms of the black hole volume $V$ Fig.\ref{f14_4},  $\gamma$-function versus the black hole volume $V$ (Fig.\ref{f14_5}),  and in the last panel (Fig.\ref{f14_6}), the on-shell Gibbs free energy $\tilde{G}$ as a function of temperature. 

\cleardoublepage
Fig.\ref{f14} reveals that Hawking-Page phase transition temperature $T_{HP}$ coincides with first-order phase temperature $T_f$, which means that Hawking-Page transition and first-order phase transition occur simultaneously. Therefore, we have an equilibrium between thermal radiations, small black holes, and large black holes.

Having constructed a complete phases transitions picture of the Born-Infeld-AdS black hole through the Landau formalism, we will turn our attention in the next section to a new thermodynamical tool lent from the dynamical systems context and which nowadays has been used extensively in the literature. Namely the Fokker-Planck equation.

\section{Fokker-Planck equation and its use to determine the probabilistic nature on the free energy landscape}

It was recently developed in \cite{Li:2020khm} a new method for examining the dynamic process of phase transition on the free energy landscape, demonstrating that a black hole can escape from one phase to another due to thermal fluctuations. Then in \cite{Li2020ThermalDP}, the authors investigated the criticality of RN-AdS black holes from the perspective of the free energy landscape, obtaining the probability distribution of states and the time distribution of the first passage kinetic process of black hole state switching. Within this perception, we will probe the  Born-Infeld-AdS black hole phase structure via such a formalism.

As in \cite{Li2020ThermalDP} and for the sake of simplicity, the horizon radius $r_h$ will be noted simply as $r$, thus the Gibbs free energy which is written in terms of the order parameter $r$ will be noted $G(r)$. In what follows and in order to unveil the evolutional response of the system under thermal fluctuation, we shall take the probability distribution of these evolving states over time to be a function of the order parameter $r$ and the time $t$. Hence, $\rho(r, t)$ stands for the probability distribution of the spacetime state in the ensemble.  

The explicit Fokker-Planck equation for the probabilistic evolution on the free energy landscape is obtained to be  \cite{doi:10.1021/j100356a007, zwanzig2001nonequilibrium}
\begin{equation}\label{29}	
 \dfrac{\partial \rho(r,t)}{\partial t} = D \dfrac{\partial}{\partial r}\left[ e^{-\beta G(r)} \dfrac{\partial}{\partial r}\left[  e^{\beta G(r)}\rho(r,t)\right]  \right],
\end{equation}
in which $ \beta = 1/k_BT$ denotes the inverse temperature and $D = k_BT /\zeta $ stands for the diffusion coefficient with $k_B$ is the Boltzmann constant while $\zeta$ called dissipation coefficient. For commodity and by preserving the generality, one will set $k_B$ and $\zeta$ equal to the unit in the rest of our analysis.

Depending on the considered question, two types of boundary conditions should be enforced at the computing domain's boundaries in order to solve the Fokker-Planck equation.
For instance, at $r = r_0$, we provide the following boundary conditions
\begin{itemize}
	\item A reflecting boundary condition : 
	\begin{equation}\label{30}	
	\left.  \dfrac{\partial}{\partial r}\left[  e^{\beta G(r)}\rho(r,t)\right]\right|_{r=r_0} = 0 .
	\end{equation}
    \item Then, an absorbing boundary condition:
    \begin{equation}\label{31}	
    	\rho(r_0,t) = 0 .
    \end{equation}
\end{itemize}
Within the canonical ensemble, we investigate the time evolution of the probability of state distribution.
During the evolution phase, the reflecting border condition will maintain probability conservation and  the initial condition is selected as Gaussian packet\footnote{In the numerical calculation processes, a Gaussian wave packet is a good approximation of the $\delta$ distribution and it's extensively used.}
     \begin{equation}\label{32}	
 	\rho(r,0) = \dfrac{1}{\sqrt{\pi}a} e^{-\left(r-r_i\right) ^2/a^2 } , \quad\text{ with }\quad a \leq 0.01.
 \end{equation}
%with the parameter $a \leq 0.01$. 

By taking the partial derivative $\frac{\partial \rho(r,t)}{\partial t}=0$ null in the Fokker-Planck equation, one can achieve the final stationary distribution as $\rho(r,t_{\infty}) \propto \exp (-G(r)/T) $.
%The final stationary distributions are determined by $\rho(r,t_{\infty}) \propto \exp (-G(r)/T) $ , which can be obtained easily from the Fokker-Planck equation by setting $\frac{\partial \rho(r,t)}{\partial t}=0$. 
This is in accordance with the Boltzmann link between free energy and equilibrium probability distribution. In fact, as a result of the long-time evolution, the stationary distribution reaches the equilibrium probability. Therefore, the maximum of the final stationary distribution is then used to define the thermodynamic stable state \cite{Li:2020nsy}.

We propose investigating the time-dependent characteristics of black hole probability distributions in extended phase space at various electrical charges and temperatures.

\begin{itemize}
	\item For $Q = 0.01009$: In \cite{Dehyadegari:2017hvd}, authors affirmed that there is a reentrant phase transition between the small and large black hole, nevertheless, we shall show that there is no reentrant phase transition and only thermal radiations and large black holes phases are globally stable and most probable states. We plot \footnote{$y$-axis is rescaled with a logarithmic scale such that $y\longrightarrow 10^{50(y-1)}\left( y \longleftarrow \dfrac{\log(y)}{50}+1\right) $} in Fig.\ref{f15} the probability distribution $\rho(r,t)$ governed by Fokker-Planck equation for different temperatures with $Q  = 0.01009$, $l=1$ and $b=3.5$. For $T = T_z = 0.2718$, 
 
 \begin{figure}[!ht]
		\centering
			\text{$\quad \quad \quad \quad \quad \quad \quad \quad \quad \quad \quad \quad \quad \quad \quad \quad T = T_z = 0.2718 \quad \quad \quad \quad \quad \quad \quad \quad \quad \quad \quad \quad \quad \quad \quad \quad \quad \quad $}
		\begin{subfigure}[h]{0.48\textwidth}
			\centering \includegraphics[scale=.52]{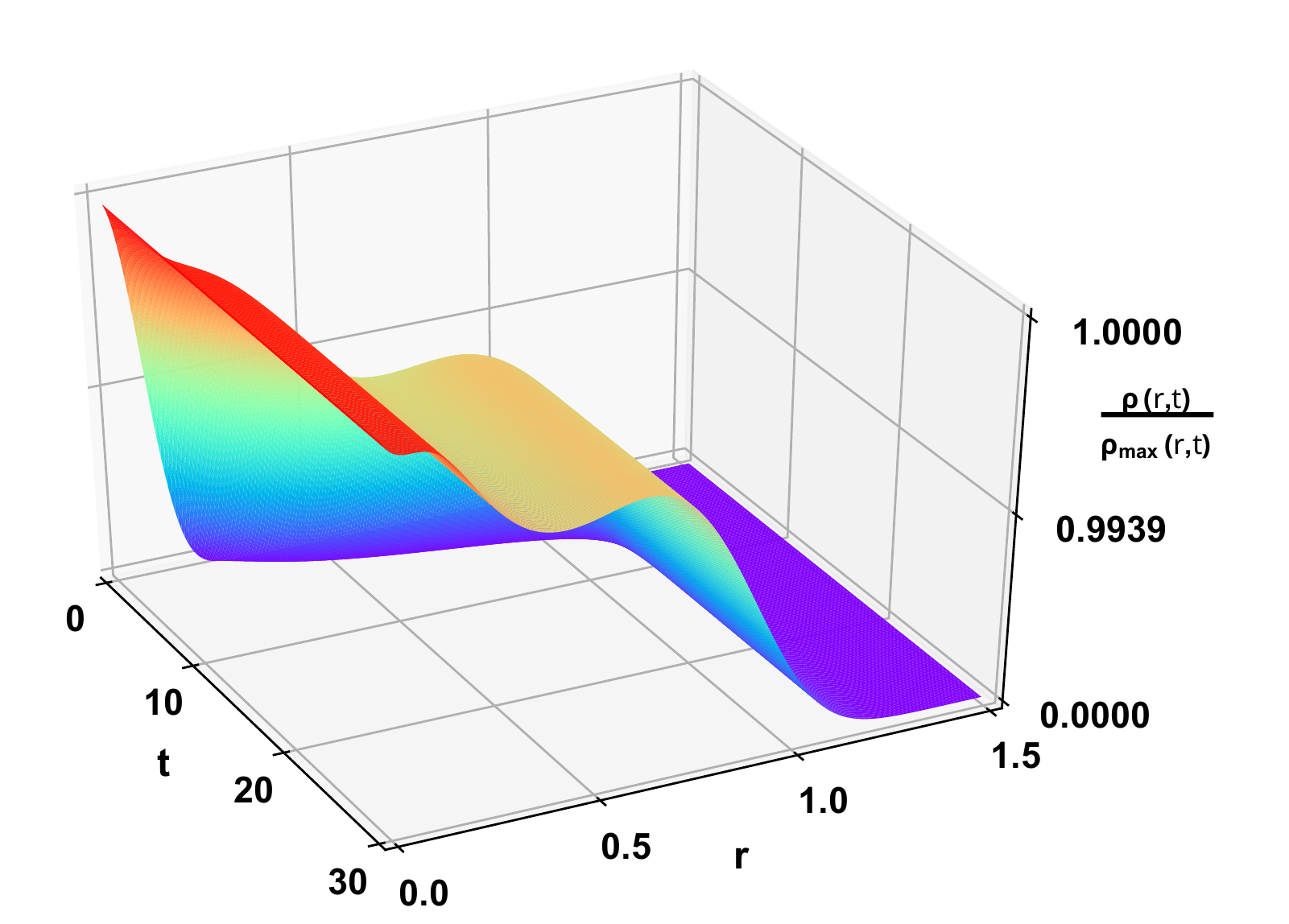}
			\caption{}
			\label{f15_1}
		\end{subfigure}
		\hspace{1pt}	
		\begin{subfigure}[h]{0.5\textwidth}
			\centering \includegraphics[scale=.5]{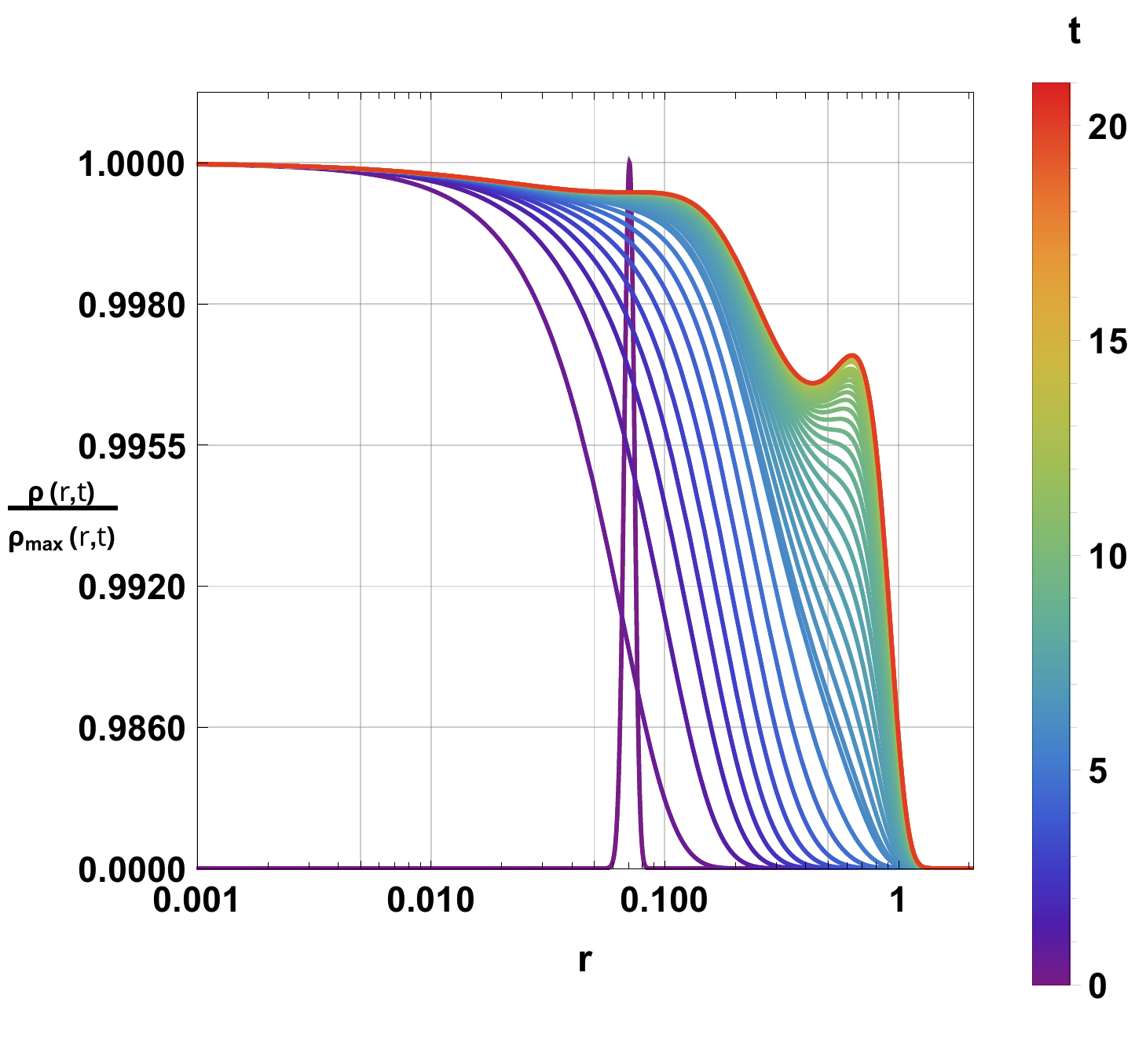}
			\caption{}
			\label{f15_2}		
		\end{subfigure}
		\hspace{1pt}
		\text{$\quad \quad \quad \quad \quad \quad \quad \quad \quad \quad \quad \quad \quad \quad \quad \quad T = T_{HP} = 0.2783 \quad \quad \quad \quad \quad \quad \quad \quad \quad \quad \quad \quad \quad \quad \quad \quad \quad \quad $}	
		\begin{subfigure}[h]{0.48\textwidth}
			\centering \includegraphics[scale=.52]{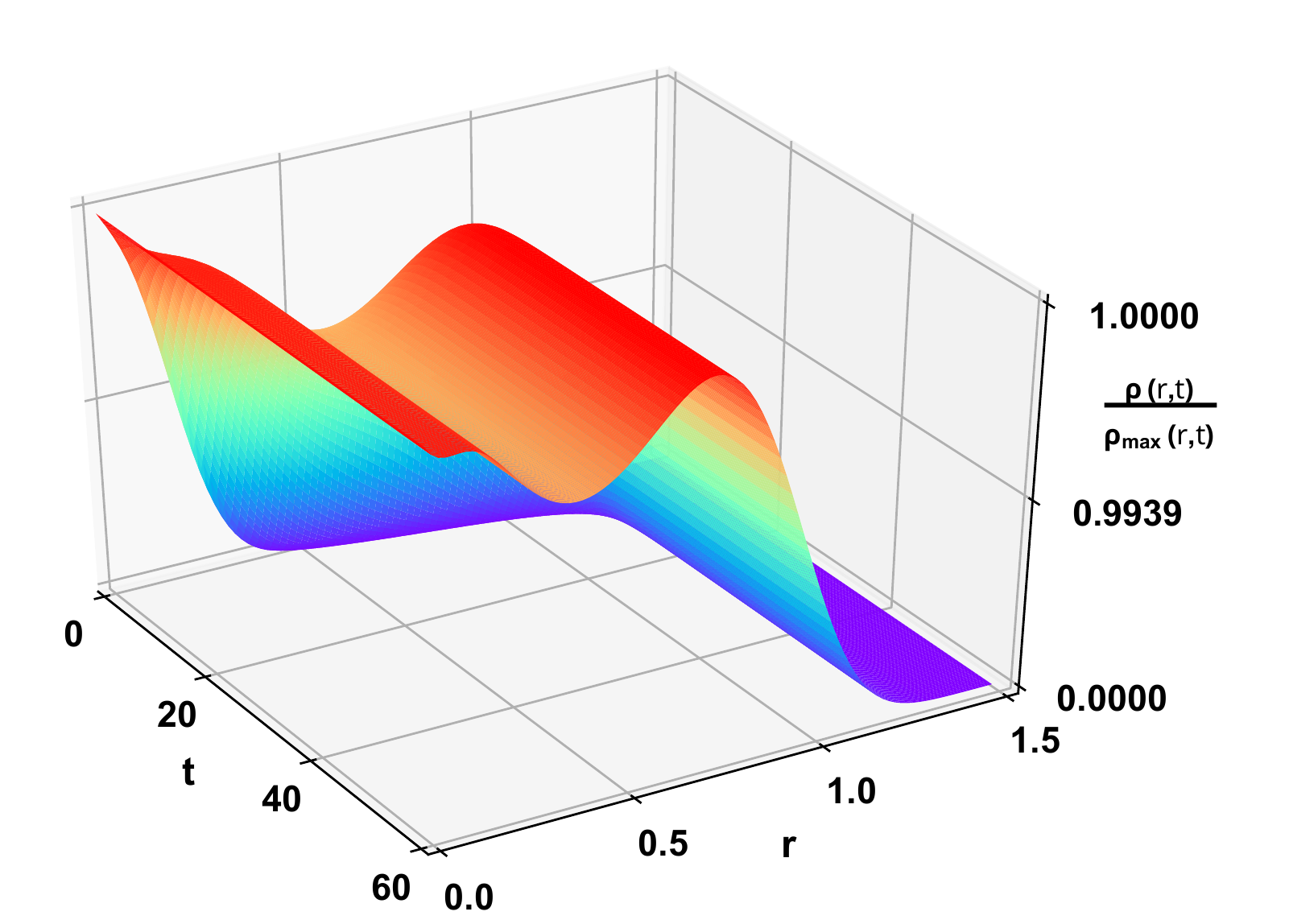}
			\caption{}
			\label{f15_3}	
		\end{subfigure}
		\hspace{1pt}	
		\begin{subfigure}[h]{0.5\textwidth}
			\centering \includegraphics[scale=.5]{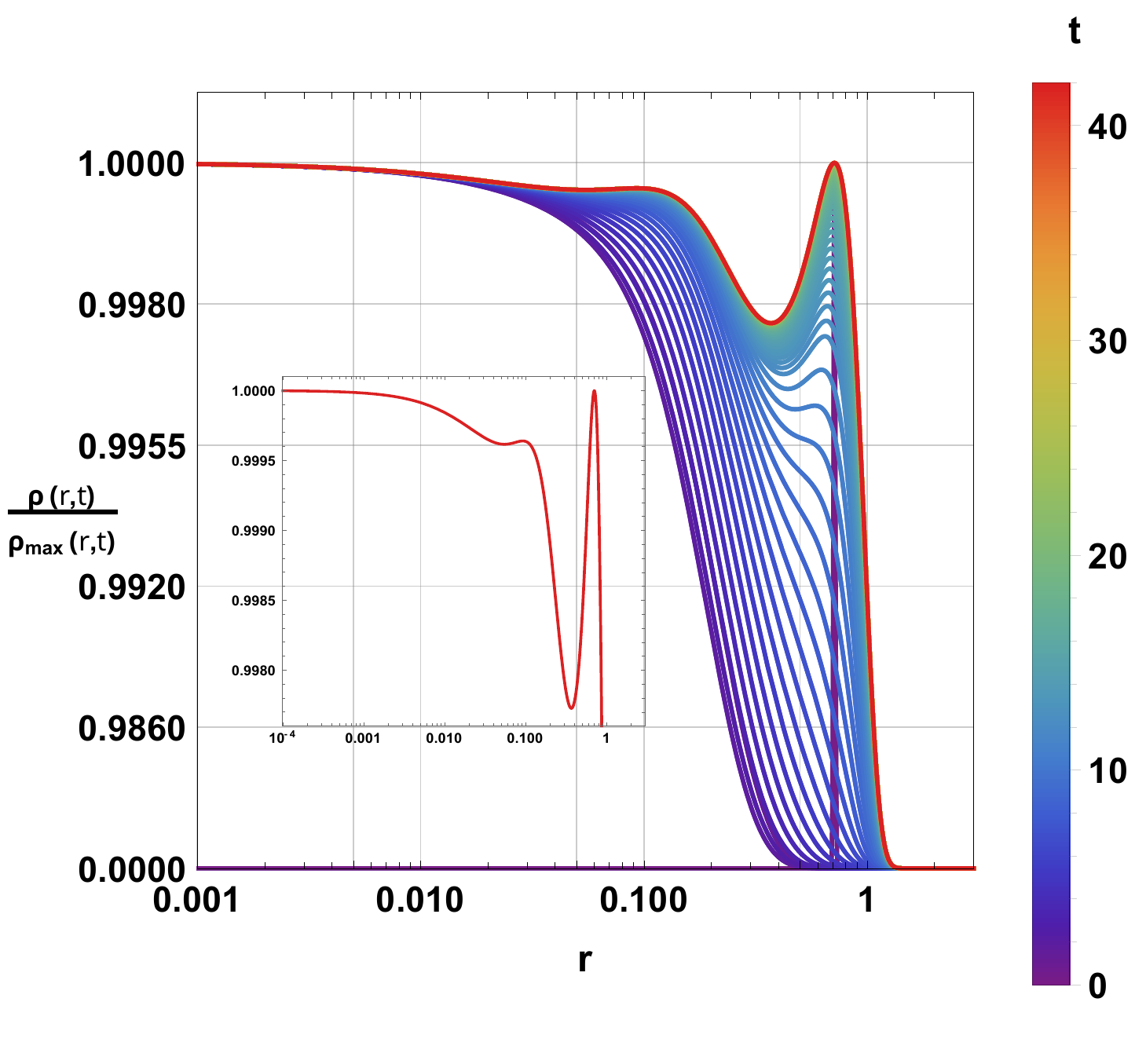}
			\caption{}
			\label{f15_4}	
		\end{subfigure}
		\hspace{1pt}
		\text{$\quad \quad \quad \quad \quad \quad \quad \quad \quad \quad \quad \quad \quad \quad \quad \quad T = 0.2937 \quad \quad \quad \quad \quad \quad \quad \quad \quad \quad \quad \quad \quad \quad \quad \quad \quad \quad $}	
		\begin{subfigure}[h]{0.48\textwidth}
			\centering \includegraphics[scale=.52]{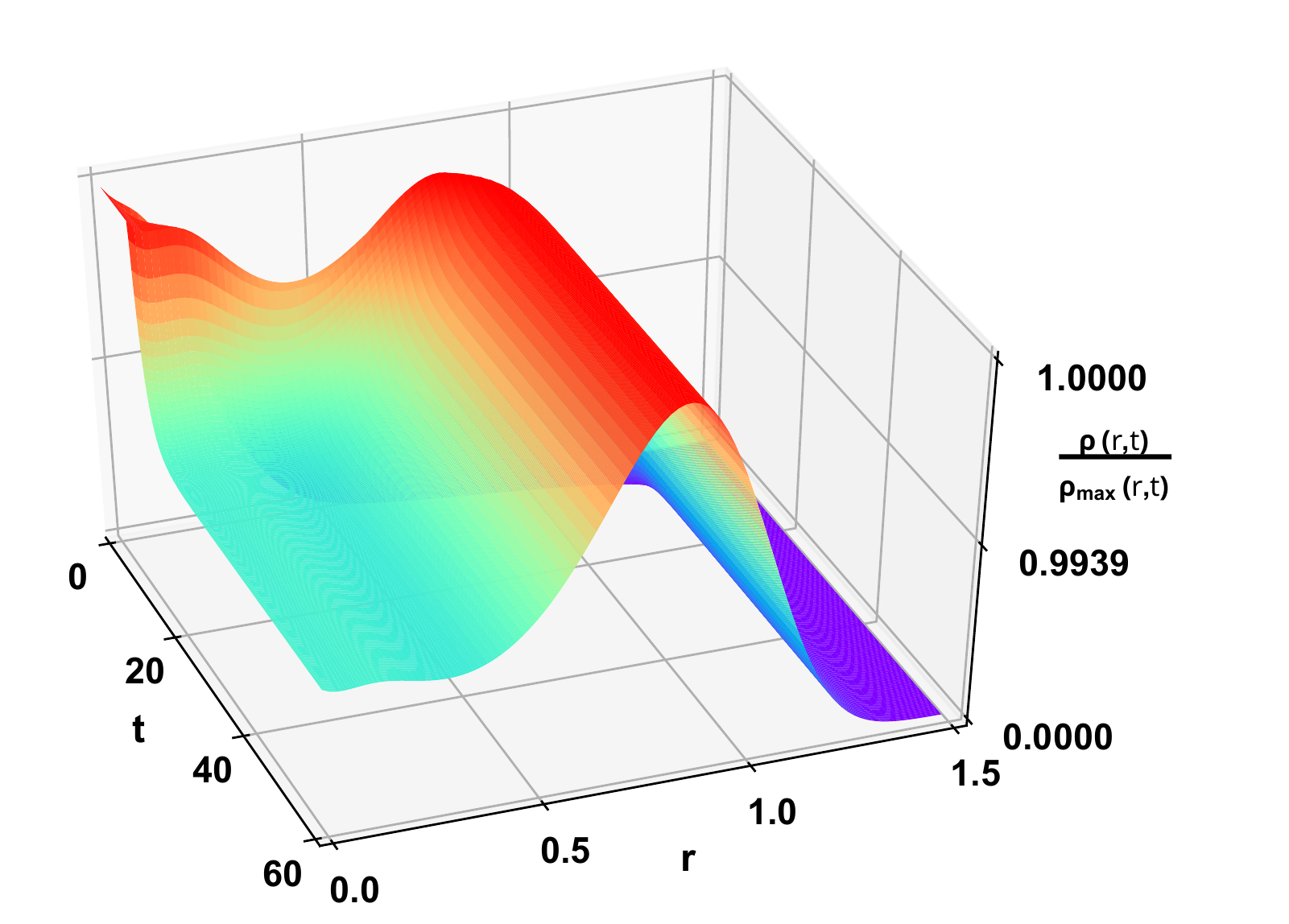}
			\caption{}
			\label{f15_5}
			
		\end{subfigure}
		\hspace{1pt}	
		\begin{subfigure}[h]{0.5\textwidth}
			\centering \includegraphics[scale=.5]{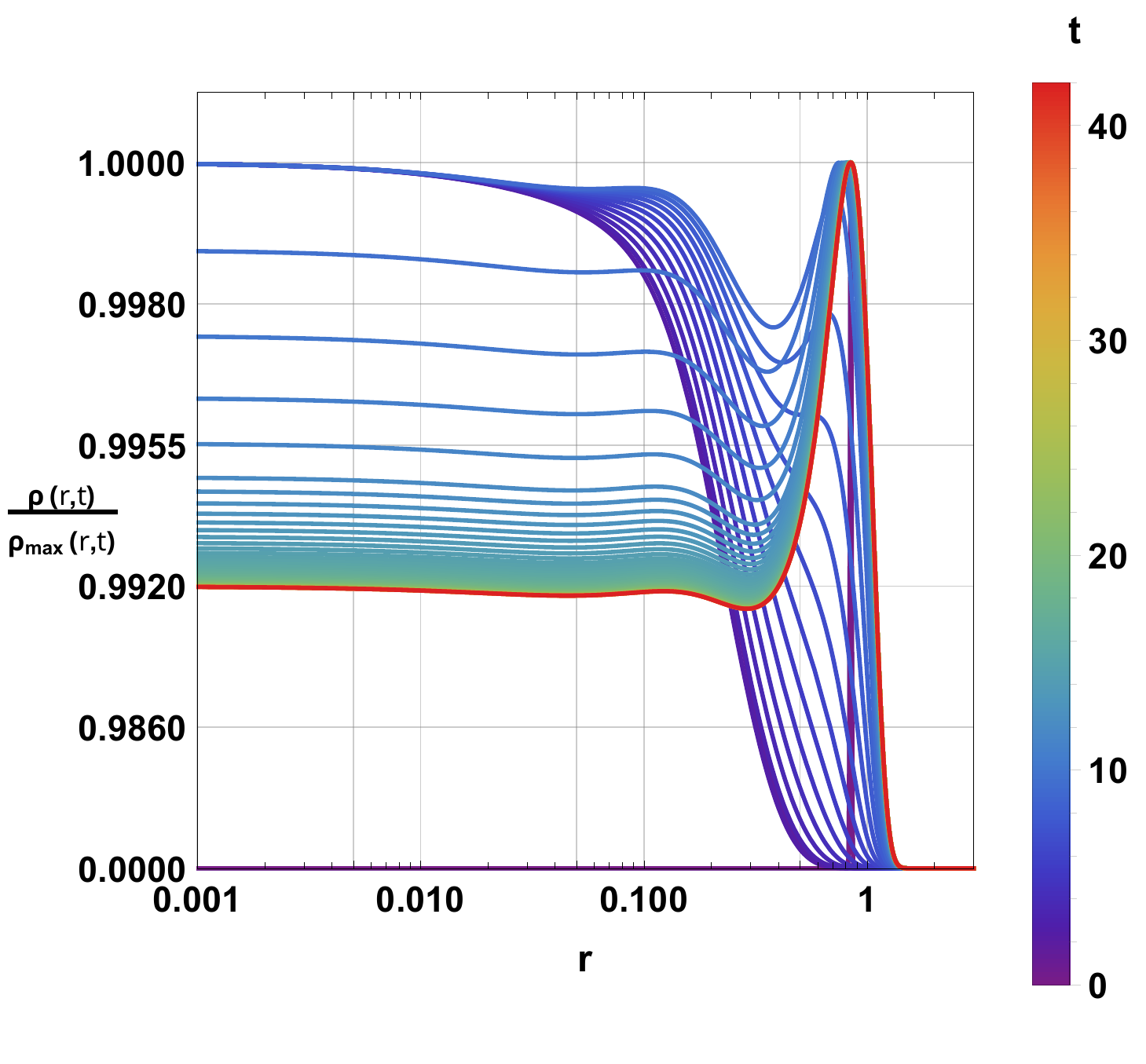}
			\caption{}
			\label{f15_6}
			
		\end{subfigure}
		\caption{\footnotesize\it Probability distribution $\rho(r,t)$ governed by Fokker-Planck equation for different temperatures with $Q  = 0.01009$, $l=1$ and $b=3.5$.}
		\label{f15}
	\end{figure}
 where it is supposed to occur a reentrant phase transition (zeroth order phase transition), we observe that the probability $\rho(r,t)$ which is initially centered around the small black hole state, $r_i = 0.0707$, leaks quickly to thermal radiations state, $r=0$, which is the only globally stable state where the probability is maximal. This result is another proof that there is no reentrant phase transition because $\rho(r_{l},t_\infty)< \rho(r_{s},t_\infty)< \rho(r=0,t_\infty)$, where $r_l = 0.6332$ and $r_s = 0.0707$ are the large and the small black holes horizon radii respectively, and the thermal radiations is the most stable phase. Afterward, in the second line of the figure, we consider $T = T_{HP} = 0.2783$, which corresponds to the Hawking-Page transition, one can notice that the probability $\rho(r,t)$ which is initially centered around the large black hole state, $r_i = 0.7138$, leaks quickly to thermal radiations state, i.e $r=0$, then it comes back to form another peak around the large black hole state, and by the end, we have $\rho(r = 0,t) = \rho(r_l,t)$ traducing the coexistence of the large black hole and the thermal radiations where the probability is maximal.  In the bottom panels where we have taken $T = 0.2937$,  one can remark that the probability $\rho(r,t)$ which is initially centered around the large black hole state, $r_i = 0.8448$, leaks quickly to thermal radiations state ($r=0$), then as $t$ increases, the thermal radiations phase probability decreases wheres the large black holes phase probability increases forming a peak around $r_l$. Therefore, the large black holes phase is the most probable and stable phase.

	\item For the charge value $Q = Q_t = 0.0103638$, associated with Fig.\ref{f14} in which we have shown the existence of a triple point, we suggest confirming this funding through the probability distribution. To this end, we depict in Fig.\ref{f16} the probability distribution $\rho(r,t)$ ruled by the Fokker-Planck equation for different temperatures with $Q = Q_t  = 0.01009$, $l=1$ and $b=3.5$. For $T = 0.25$, 
 
 	\begin{figure}[!ht]
	\centering
	\text{$\quad \quad \quad \quad \quad \quad \quad \quad \quad \quad \quad \quad \quad \quad \quad \quad T = 0.25 \quad \quad \quad \quad \quad \quad \quad \quad \quad \quad \quad \quad \quad \quad \quad \quad \quad \quad $}
	\begin{subfigure}[h]{0.48\textwidth}
		\centering \includegraphics[scale=.52]{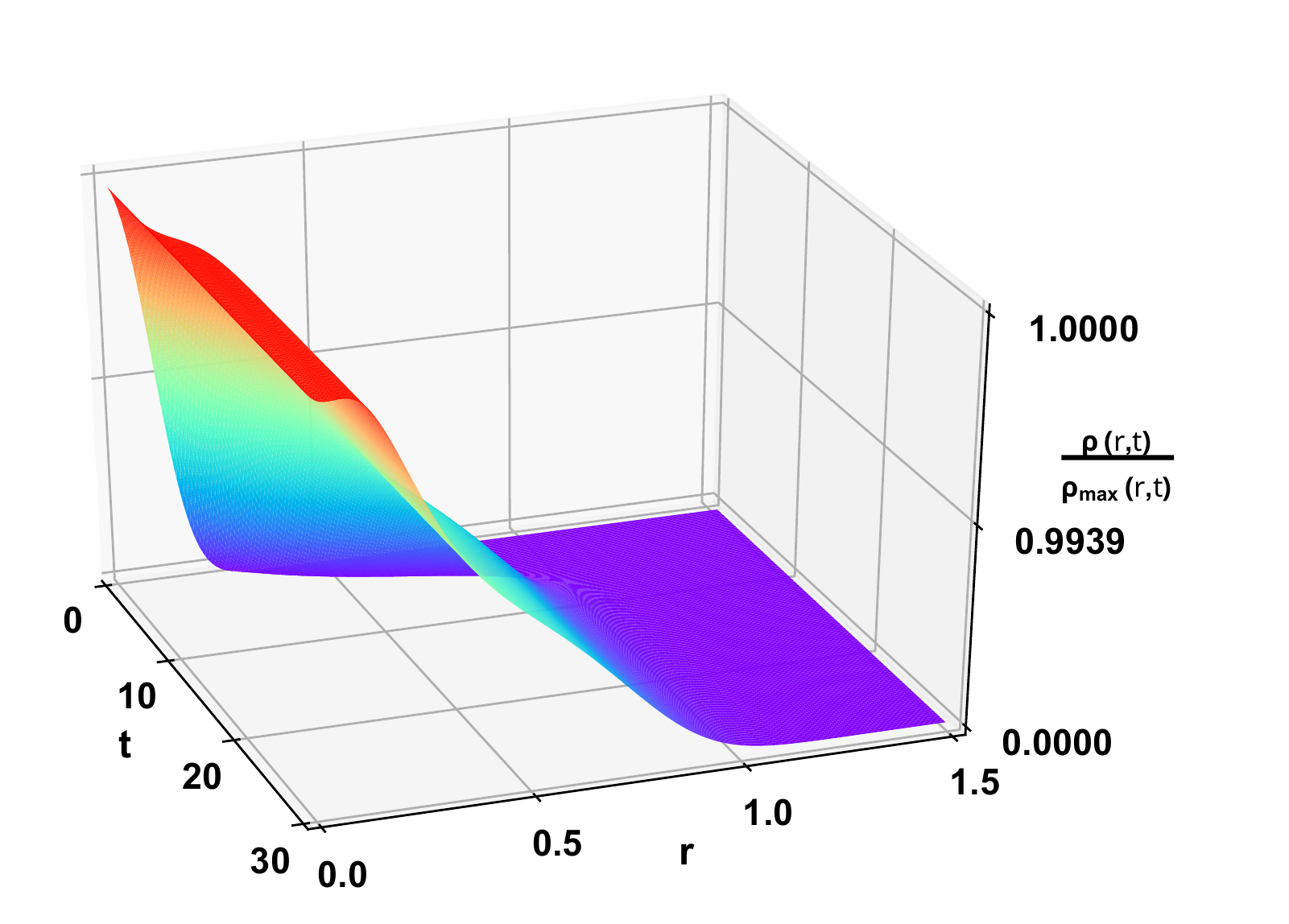}
		\caption{}
		\label{f16_1}
	\end{subfigure}
	\hspace{1pt}	
	\begin{subfigure}[h]{0.5\textwidth}
		\centering \includegraphics[scale=.5]{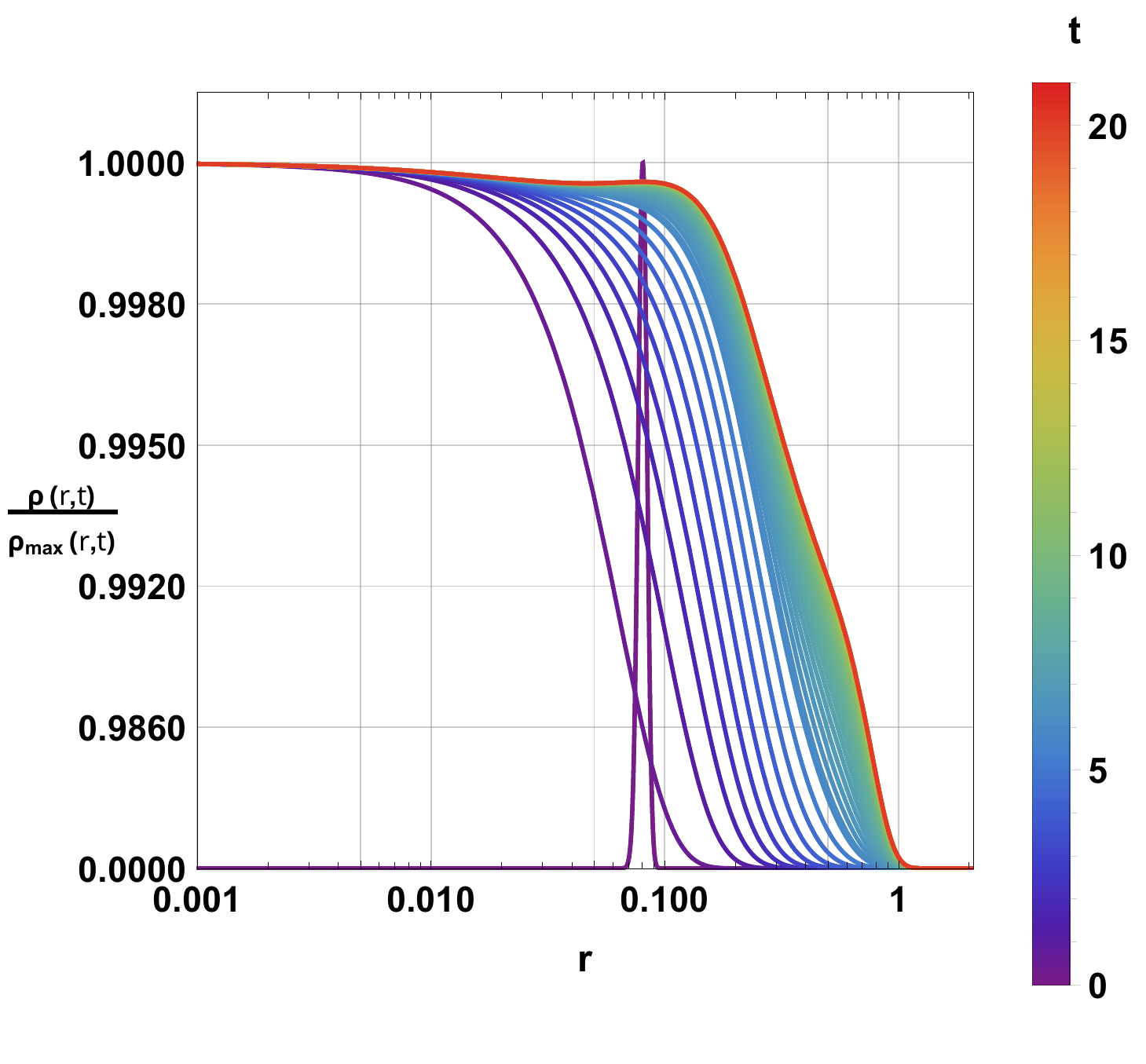}
		\caption{}
		\label{f16_2}		
	\end{subfigure}
	\hspace{1pt}
	\text{$\quad \quad \quad \quad \quad \quad \quad \quad \quad \quad \quad \quad \quad \quad \quad \quad T= T_t = 0.276 \quad \quad \quad \quad \quad \quad \quad \quad \quad \quad \quad \quad \quad \quad \quad \quad \quad \quad $}	
	\begin{subfigure}[h]{0.48\textwidth}
		\centering \includegraphics[scale=.52]{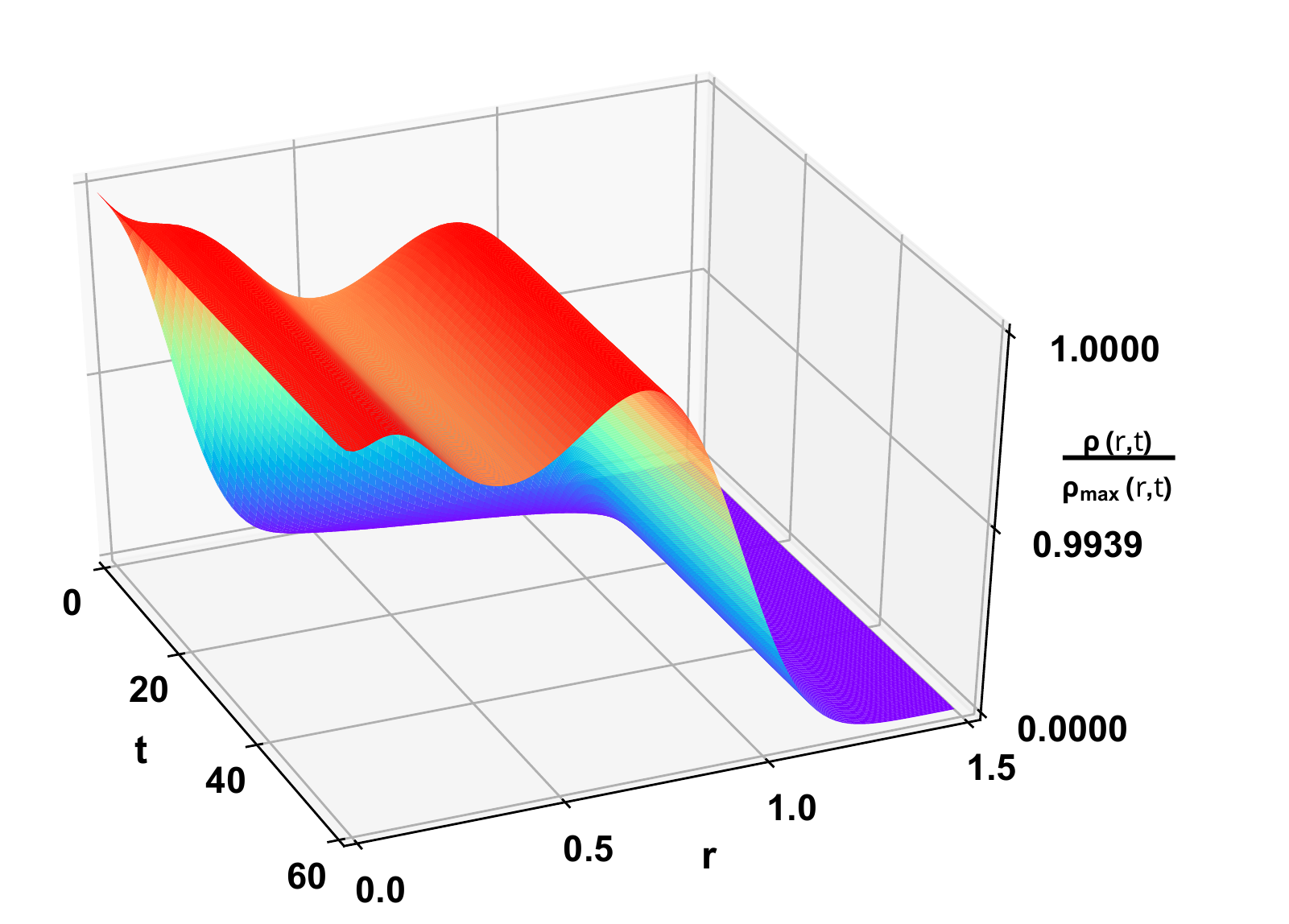}
		\caption{}
		\label{f16_3}	
	\end{subfigure}
	\hspace{1pt}	
	\begin{subfigure}[h]{0.5\textwidth}
		\centering \includegraphics[scale=.5]{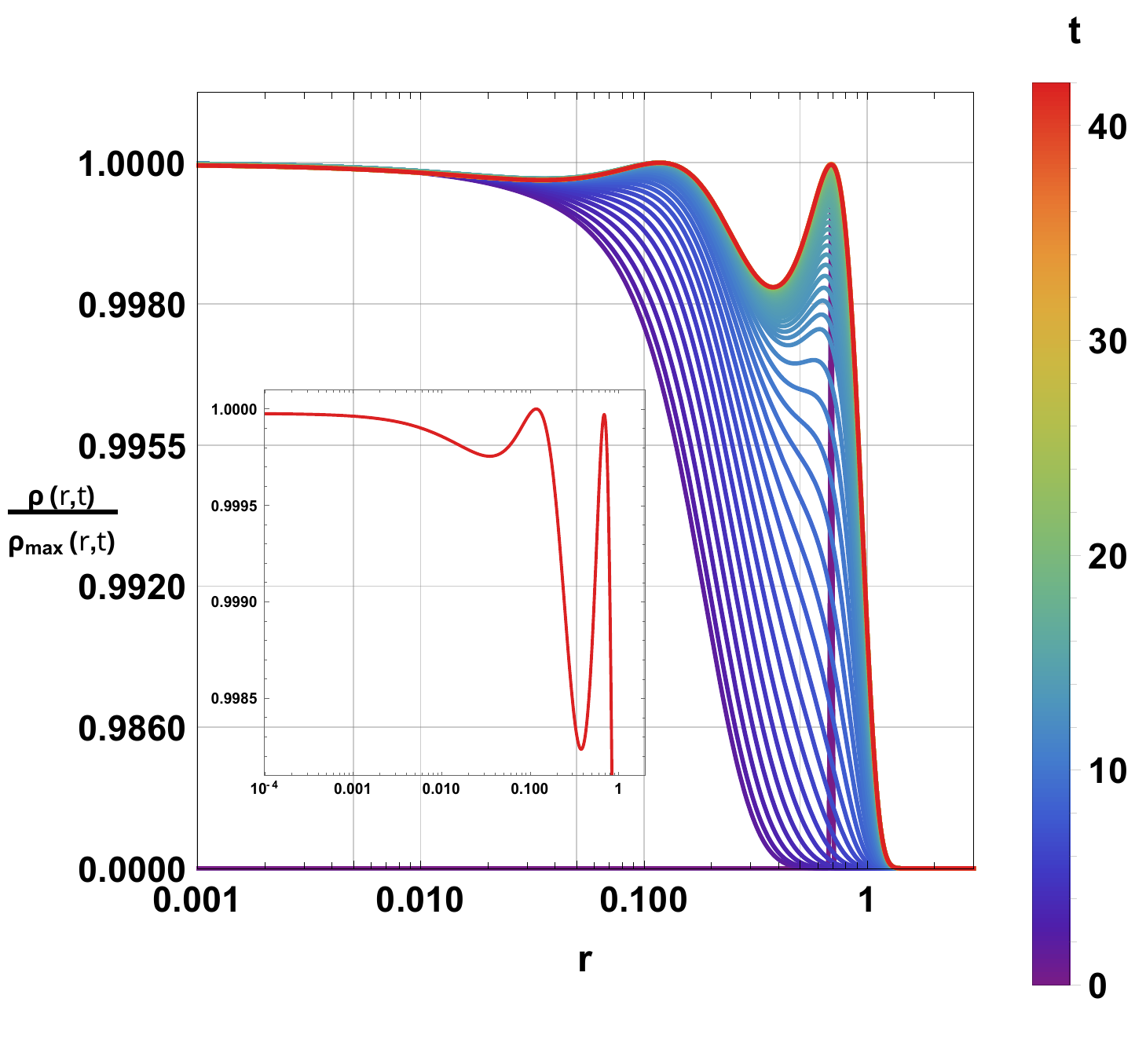}
		\caption{}
		\label{f16_4}	
	\end{subfigure}
	\hspace{1pt}
	\text{$\quad \quad \quad \quad \quad \quad \quad \quad \quad \quad \quad \quad \quad \quad \quad \quad T = 0.29 \quad \quad \quad \quad \quad \quad \quad \quad \quad \quad \quad \quad \quad \quad \quad \quad \quad \quad $}	
	\begin{subfigure}[d]{0.48\textwidth}
		\centering \includegraphics[scale=.52]{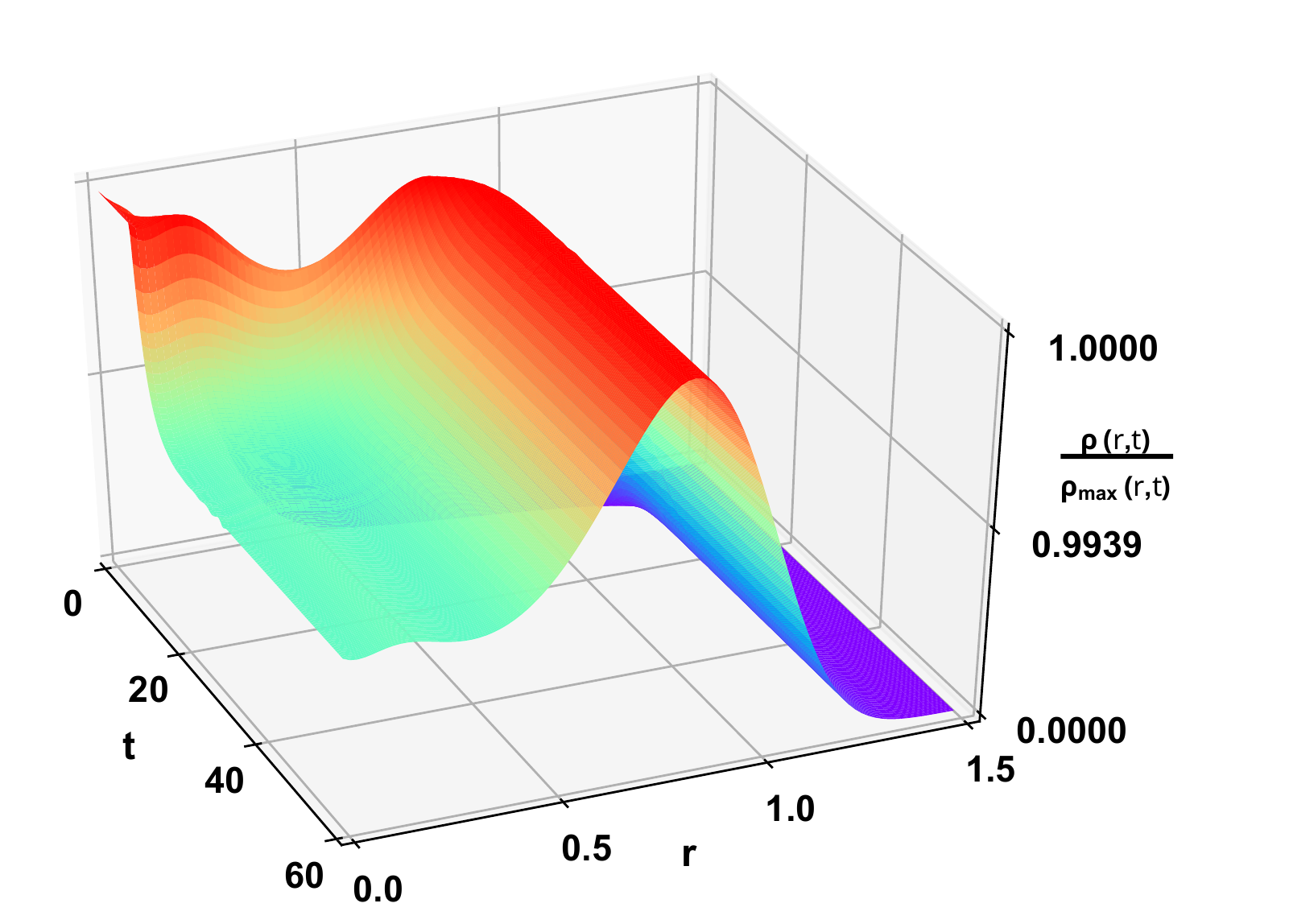}
		\caption{}
		\label{f16_5}
		
	\end{subfigure}
	\hspace{1pt}	
	\begin{subfigure}[h]{0.5\textwidth}
		\centering \includegraphics[scale=.5]{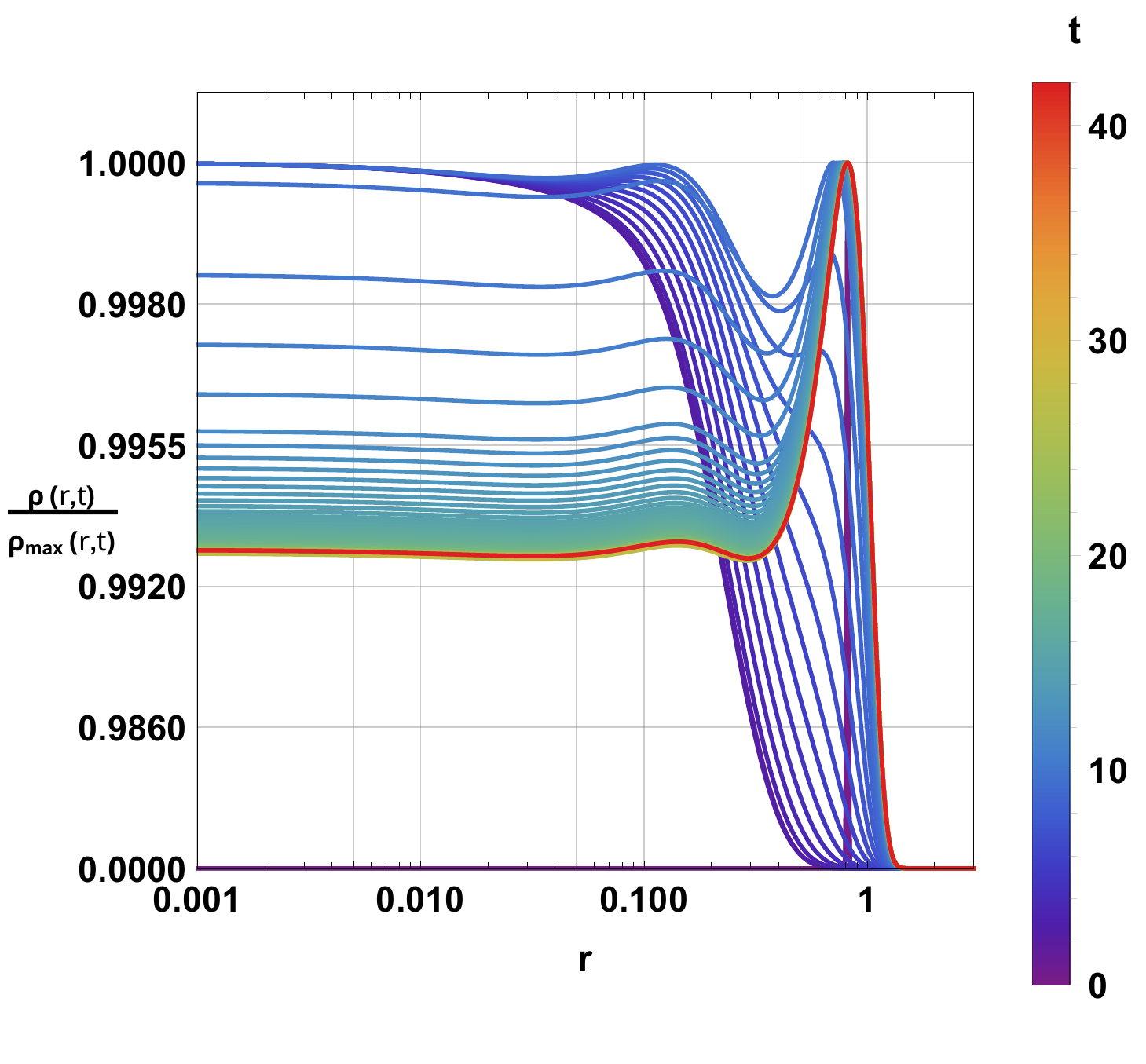}
		\caption{}
		\label{f16_6}
		
	\end{subfigure}
	\caption{\footnotesize\it Probability distribution $\rho(r,t)$ governed by Fokker-Planck equation for different temperatures with $Q =Q_t = 0.0103638$, $l=1$ and $b=3.5$.}
	\label{f16}
\end{figure}
 It's remarked that the probability $\rho(r,t)$ initially centered around the small black hole state, $r_i = 0.0805$, leaks immediately to thermal radiation state, $r=0$ associated with the only globally stable state where the probability is maximal. Indeed, $\rho(r_{s},t_\infty)< \rho(r=0,t_\infty)$, where $r_s = 0.0805$ is the small black hole horizon radius, and then the thermal radiations is the most stable phase  confirming our previous result that the thermal radiation is the only globally stable phase. Increasing the temperature to $T = T_t = 0.276$,  we notice that the probability $\rho(r,t)$ which is initially centered around the large black hole state, $r_i = 0.6909$, leaks quickly to thermal radiations state, $r=0$, but after then it comes back to form two peaks around the small and the large black holes horizon radii. Thus, we are in presence of a triple point, i.e a location where the thermal radiations, small black holes, and large black holes coexist together. Indeed, $\rho(r_{s},t_\infty) = \rho(r_l,t_\infty)= \rho(r=0,t_\infty)$  where $r_s = 0.1177$ and $r_l = 0.6909$, the three phases are equiprobable and globally stable. Now reaching $T = 0.29$,  we see that the probability $\rho(r,t)$ initially centered around the large black hole state, $r_i = 0.8172$, leaks rapidly  to thermal radiation and small black hole states forming the same shape as in triple point case, but after then $\rho(r =0,t)$ and $\rho(r_s,t)$ decrease with the time whereas $\rho(r_l,t)$ increases to reach its maximum. Therefore, $\rho(r = 0,t_\infty) < \rho(r_s,t_\infty) < \rho(r_l,t_\infty)$  where $r_s = 0.1418$ and $r_l = 0.8172$, and the large black holes phase is the most probable state and then the only globally stable phase.

\item Now, with $Q = 0.0105$ that corresponds to Fig.\ref{f9} and where we have unveiled the existence of two critical points at such a value of charge, that is to say, two-phase transitions occur, the first one is between thermal radiation and small black holes phases which is a Hawking-Page-like transition, while the second one is a first-order phase transition between small and black holes phases. We illustrate in Fig.\ref{f17} the probability distribution $\rho(r,t)$  derived from the Fokker-Planck equation for different temperatures around thermal radiations-small black holes transition with $Q = 0.0105$, $l=1$ and $b=3.5$. For $T  = 0.227$,

\begin{figure}[!ht]
	\centering
	\hspace{1pt}
	\text{$\quad \quad \quad \quad \quad \quad \quad \quad \quad \quad \quad \quad \quad \quad \quad \quad T = 0.227 \quad \quad \quad \quad \quad \quad \quad \quad \quad \quad \quad \quad \quad \quad \quad \quad \quad \quad $}
	\begin{subfigure}[h]{0.48\textwidth}
		\centering \includegraphics[scale=.52]{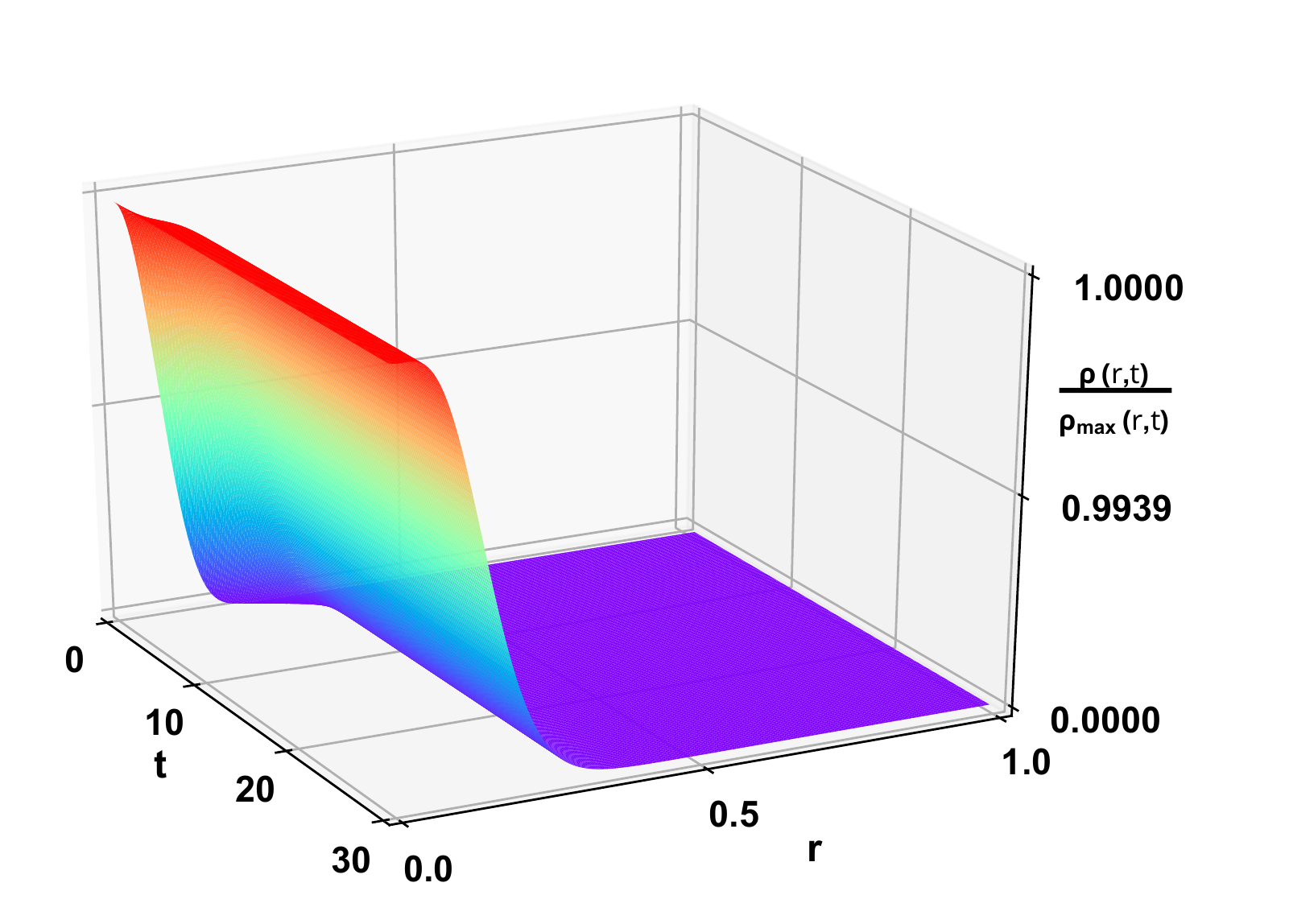}
		\caption{}
		\label{f17_1}
	\end{subfigure}
	\hspace{1pt}	
	\begin{subfigure}[h]{0.5\textwidth}
		\centering \includegraphics[scale=.5]{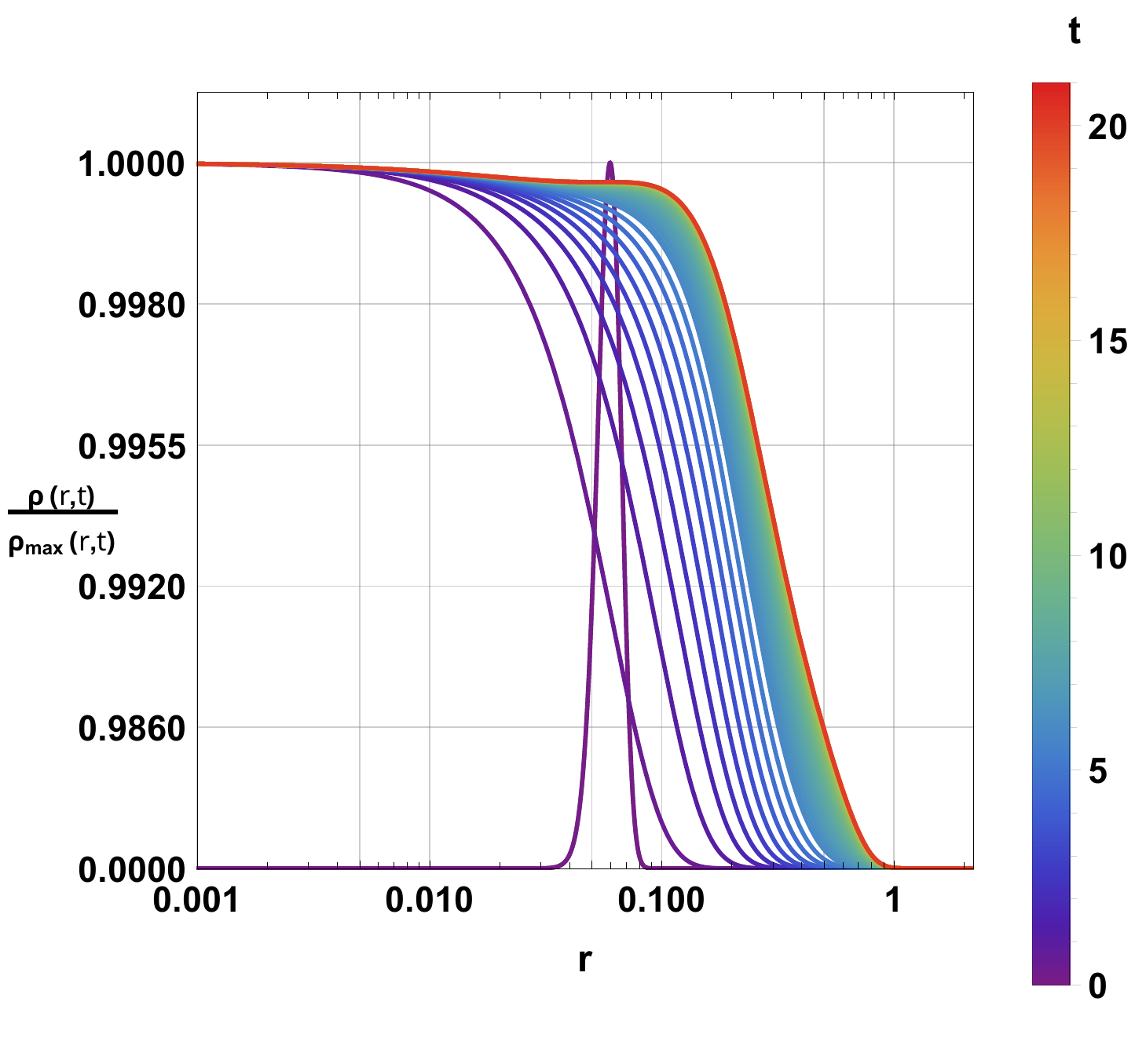}
		\caption{}
		\label{f17_2}		
	\end{subfigure}
	\hspace{1pt}
	\text{$\quad \quad \quad \quad \quad \quad \quad \quad \quad \quad \quad \quad \quad \quad \quad \quad T = T_{HP} = 0.257 \quad \quad \quad \quad \quad \quad \quad \quad \quad \quad \quad \quad \quad \quad \quad \quad \quad \quad $}	
	\begin{subfigure}[h]{0.48\textwidth}
		\centering \includegraphics[scale=.52]{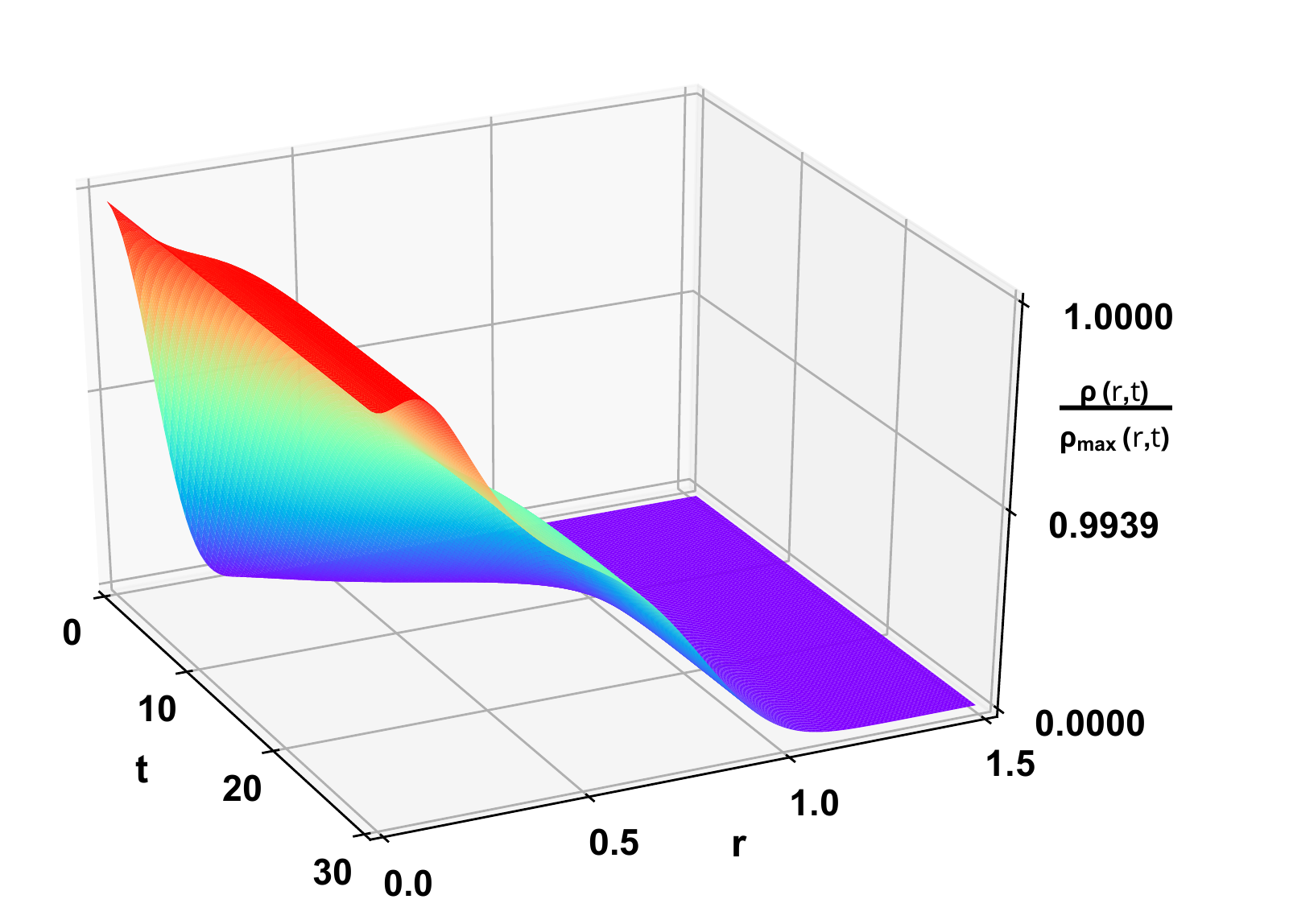}
		\caption{}
		\label{f17_3}	
	\end{subfigure}
	\hspace{1pt}	
	\begin{subfigure}[h]{0.5\textwidth}
		\centering \includegraphics[scale=.5]{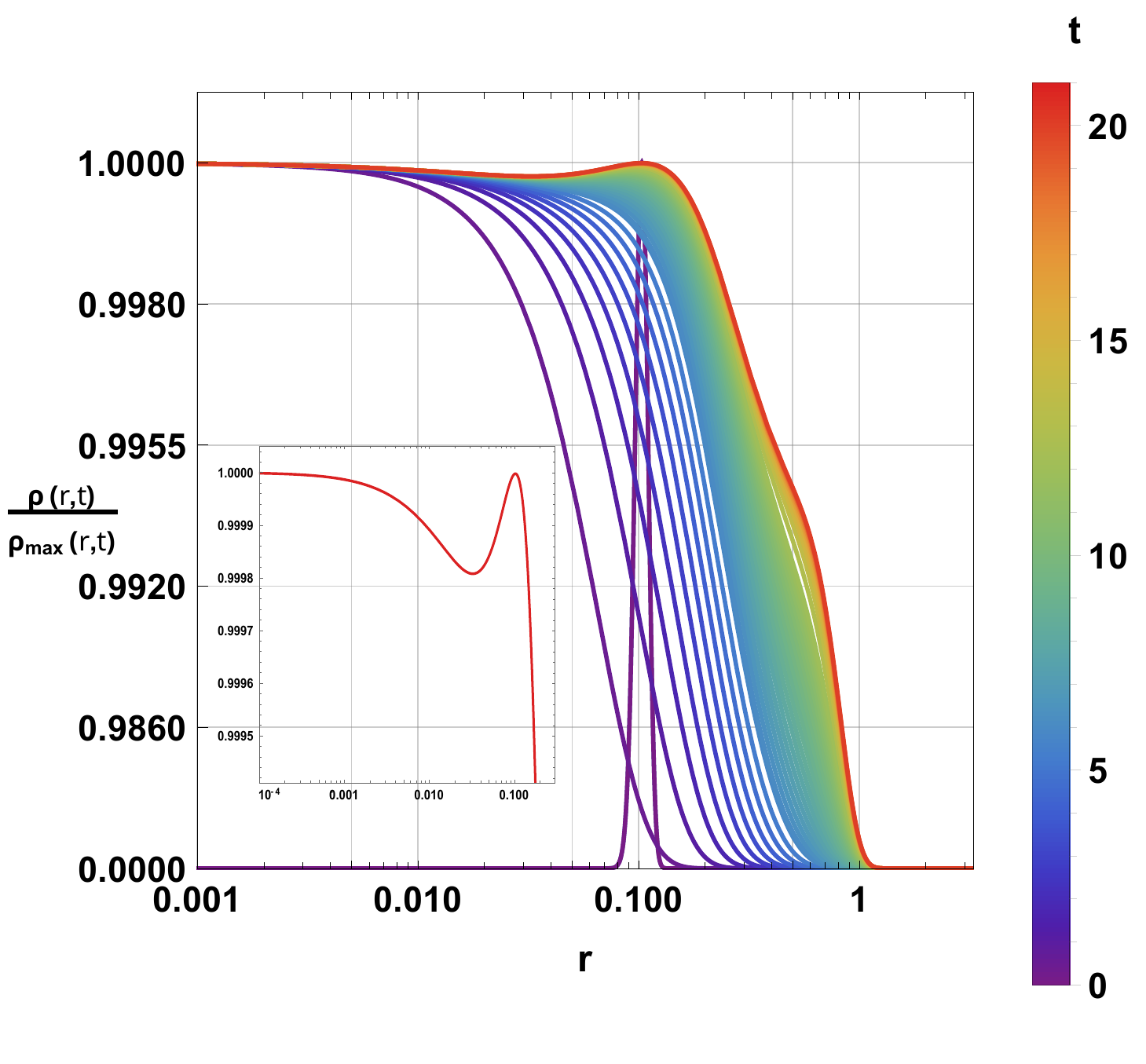}
		\caption{}
		\label{f17_4}	
	\end{subfigure}
	\hspace{1pt}
	\text{$\quad \quad \quad \quad \quad \quad \quad \quad \quad \quad \quad \quad \quad \quad \quad \quad T = 0.272 \quad \quad \quad \quad \quad \quad \quad \quad \quad \quad \quad \quad \quad \quad \quad \quad \quad \quad $}
	\hspace{10pt}	
	\begin{subfigure}[d]{0.48\textwidth}
		\centering \includegraphics[scale=.52]{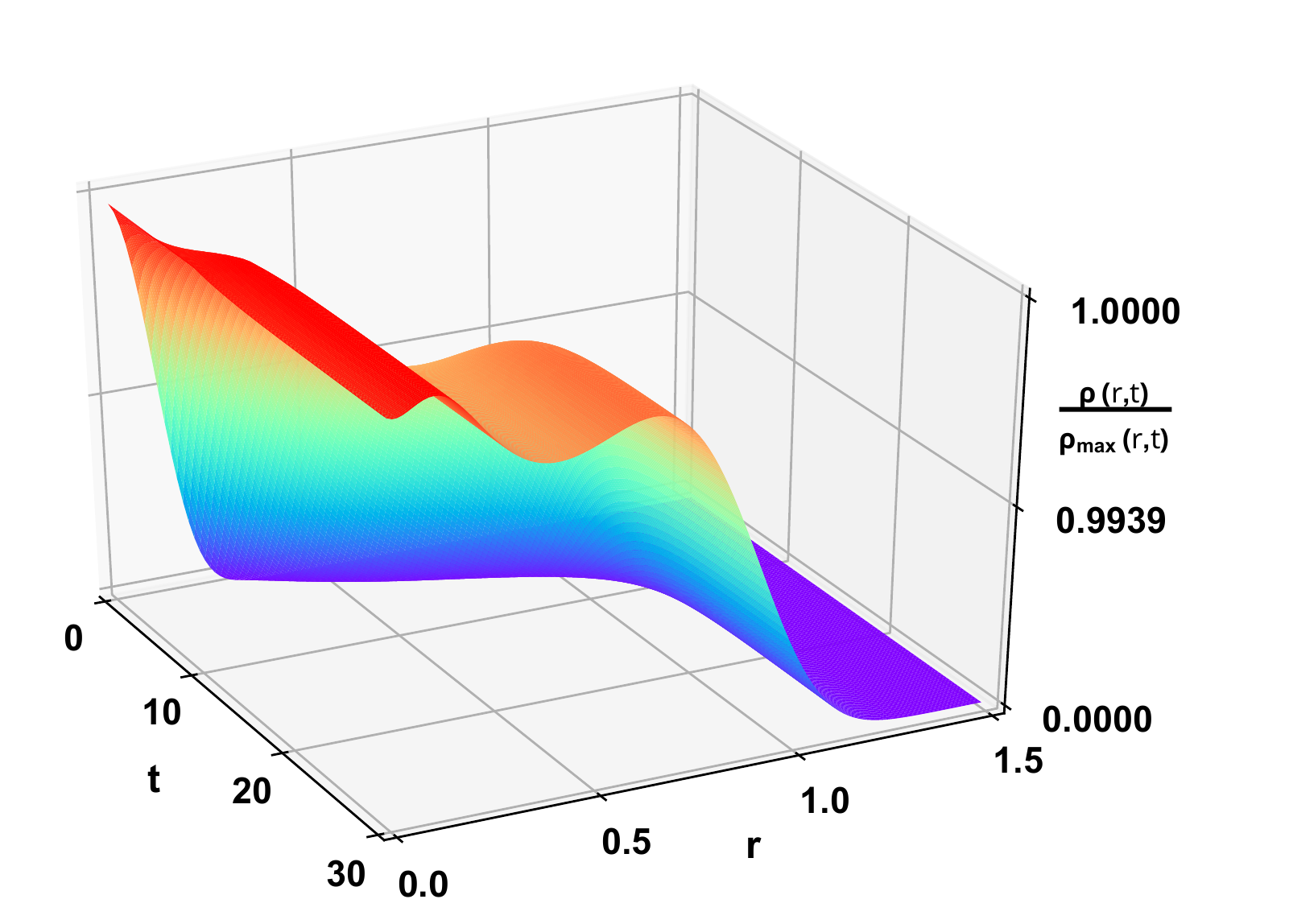}
		\caption{}
		\label{f17_5}
		
	\end{subfigure}
	\hspace{1pt}	
	\begin{subfigure}[h]{0.5\textwidth}
		\centering \includegraphics[scale=.5]{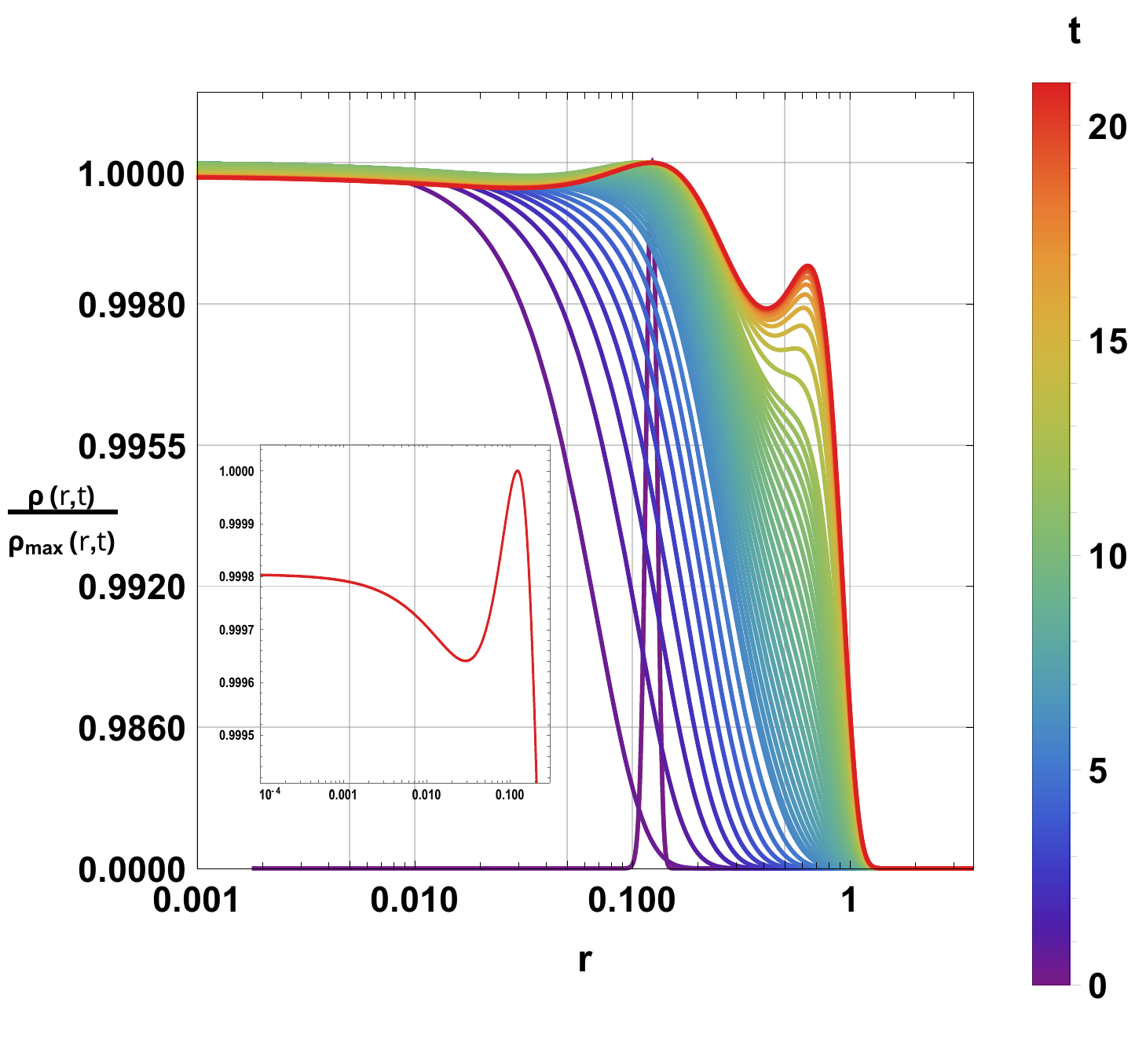}
		\caption{}
		\label{f17_6}
		
	\end{subfigure}
	\caption{\footnotesize\it  Probability distribution $\rho(r,t)$ governed by Fokker-Planck equation for different temperatures around thermal radiations-small black holes transition with $Q = 0.0105$, $l=1$ and $b=3.5$}
	\label{f17}
\end{figure}
Such a figure reveals serval remarks, one can first notice that the probability $\rho(r,t)$ which  initially centered around the small black hole state, $r_i = 0.0599$, leaks fast to the thermal radiation state, $r=0$, which is the only globally stable state where the probability is maximal. Indeed, $\rho(r_{s},t_\infty)< \rho(r=0,t_\infty)$, where $r_s = 0.0599$, and then the thermal radiations is the most stable phase which confirms our previous result that the thermal radiation is the only globally stable phase. Second, for $T = T_{HP} = 0.25708$, which corresponds to the Hawking-Page-like transition between thermal radiation and small black holes phases, we see that the probability $\rho(r,t)$ which is initially centered around the small black hole state, $r_i = 0.1034$, leaks quickly to thermal radiations state, $r=0$, then it comes back to form another peak around the small black hole state, and by the end, one achieves $\rho(r = 0,t) = \rho(r_s,t)$, at $r_s = 0.1034 $, and then the coexistence of the small black hole and the thermal radiations at the maximum of the probability. The third situation associated with $T = 0.272$ reveals that the probability $\rho(r,t)$ which is initially centered around the small black hole state, $r_i = 0.1226$, leaks quickly to thermal radiations state i.e $r=0$, then as $t$ increases, the thermal radiations phase probability decreases whereas the small black holes phase probability increases forming a maximal peak around $r_s$. Therefore, the small black holes phase is the most probable and the most stable phase. 

 Within Fig.\ref{f18}, we depict the probability distribution $\rho(r,t)$  for different temperatures around small-large black holes transition with the same parameters as in Fig.\ref{f17}.

	\begin{figure}[!ht]
	\centering
	\text{$\quad \quad \quad \quad \quad \quad \quad \quad \quad \quad \quad \quad \quad \quad \quad \quad T = T_f = 0.2753 \quad \quad \quad \quad \quad \quad \quad \quad \quad \quad \quad \quad \quad \quad \quad \quad \quad \quad $}
	\begin{subfigure}[h]{0.48\textwidth}
		\centering \includegraphics[scale=.5]{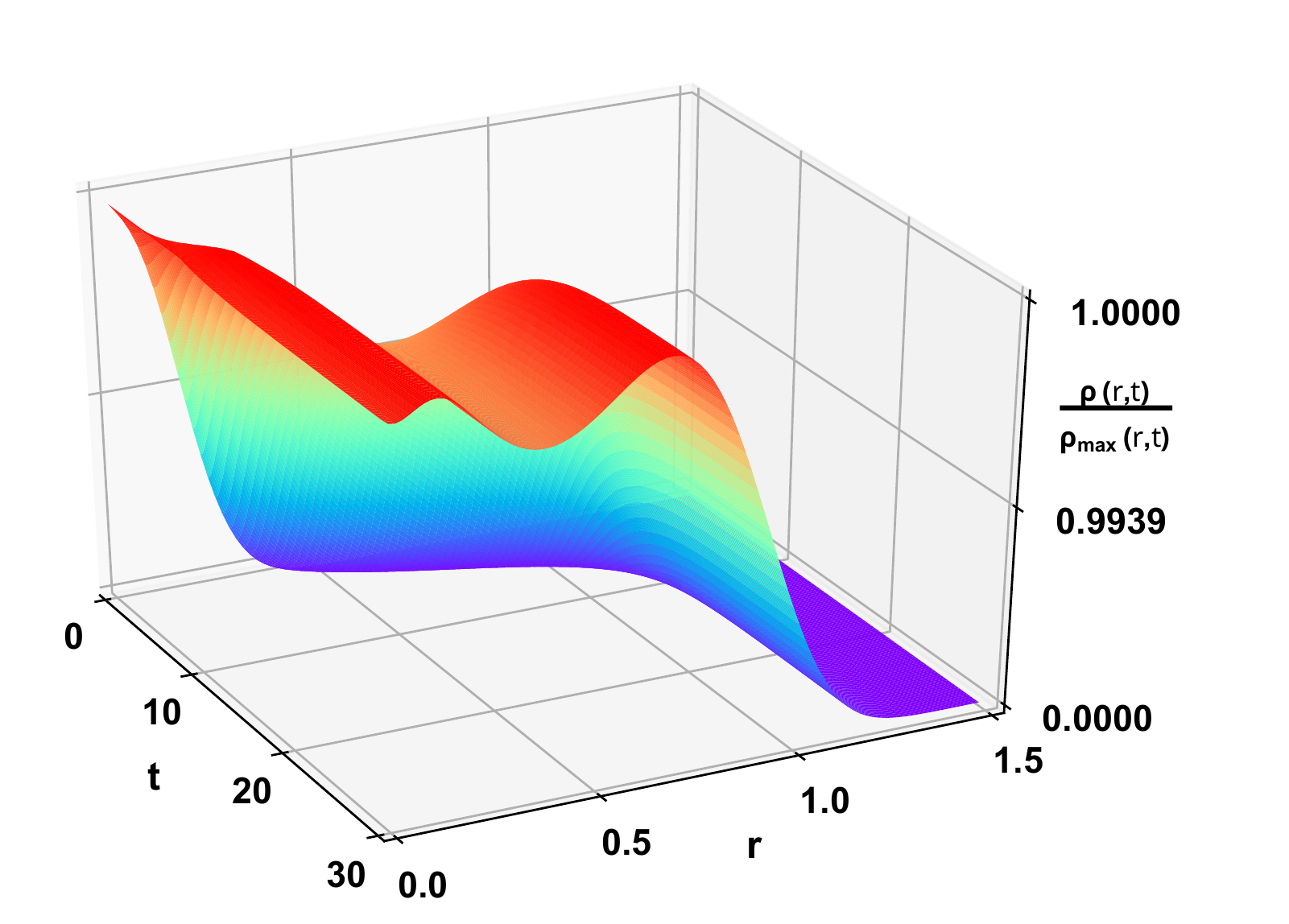}
		\caption{}
		\label{f18_1}
	\end{subfigure}
	\hspace{1pt}	
	\begin{subfigure}[h]{0.5\textwidth}
		\centering \includegraphics[scale=.5]{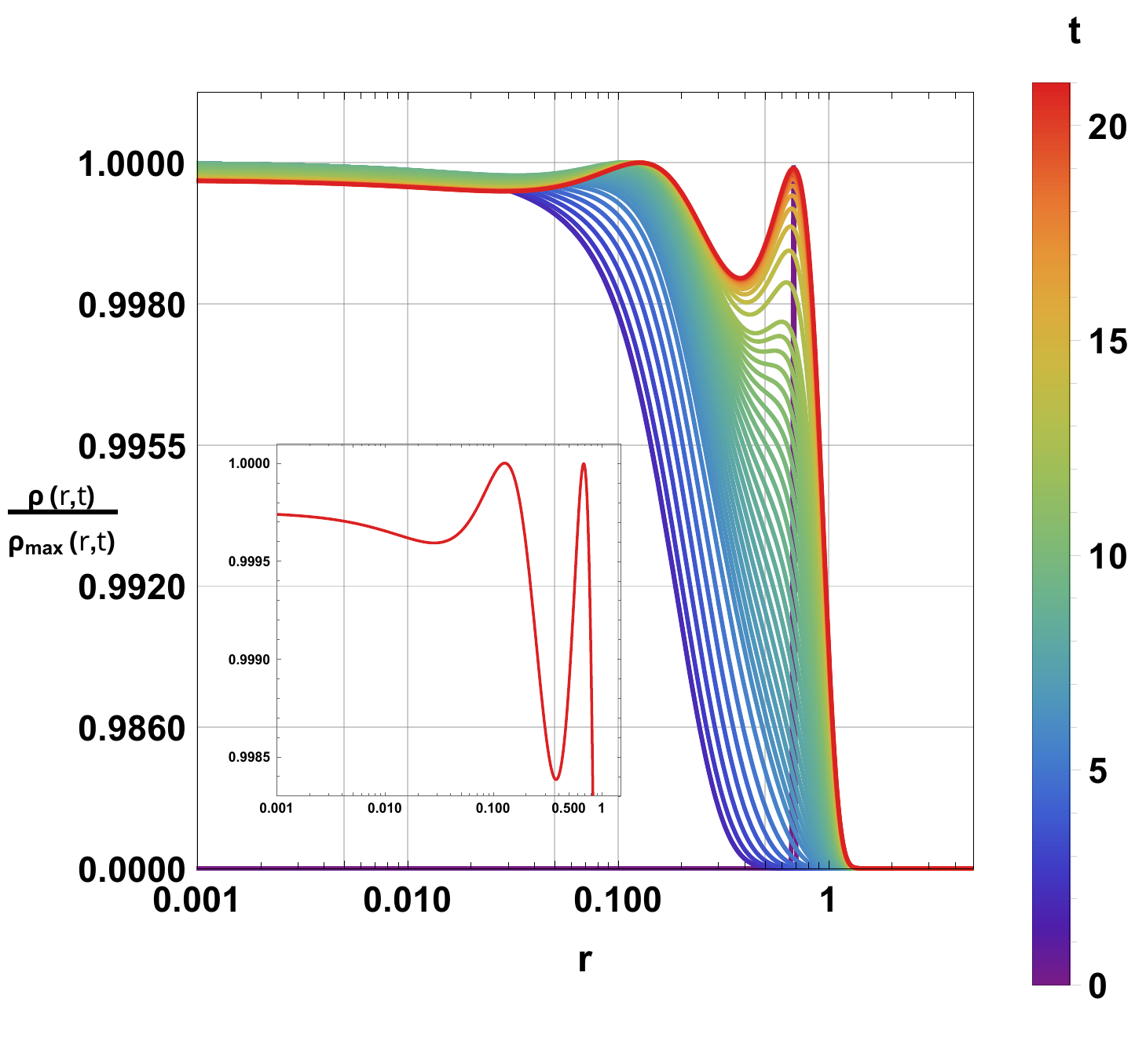}
		\caption{}
		\label{f18_2}		
	\end{subfigure}
	\hspace{1pt}
	\text{$\quad \quad \quad \quad \quad \quad \quad \quad \quad \quad \quad \quad \quad \quad \quad \quad T = 0.29 \quad \quad \quad \quad \quad \quad \quad \quad \quad \quad \quad \quad \quad \quad \quad \quad \quad \quad $}	
	\begin{subfigure}[h]{0.48\textwidth}
		\centering \includegraphics[scale=.52]{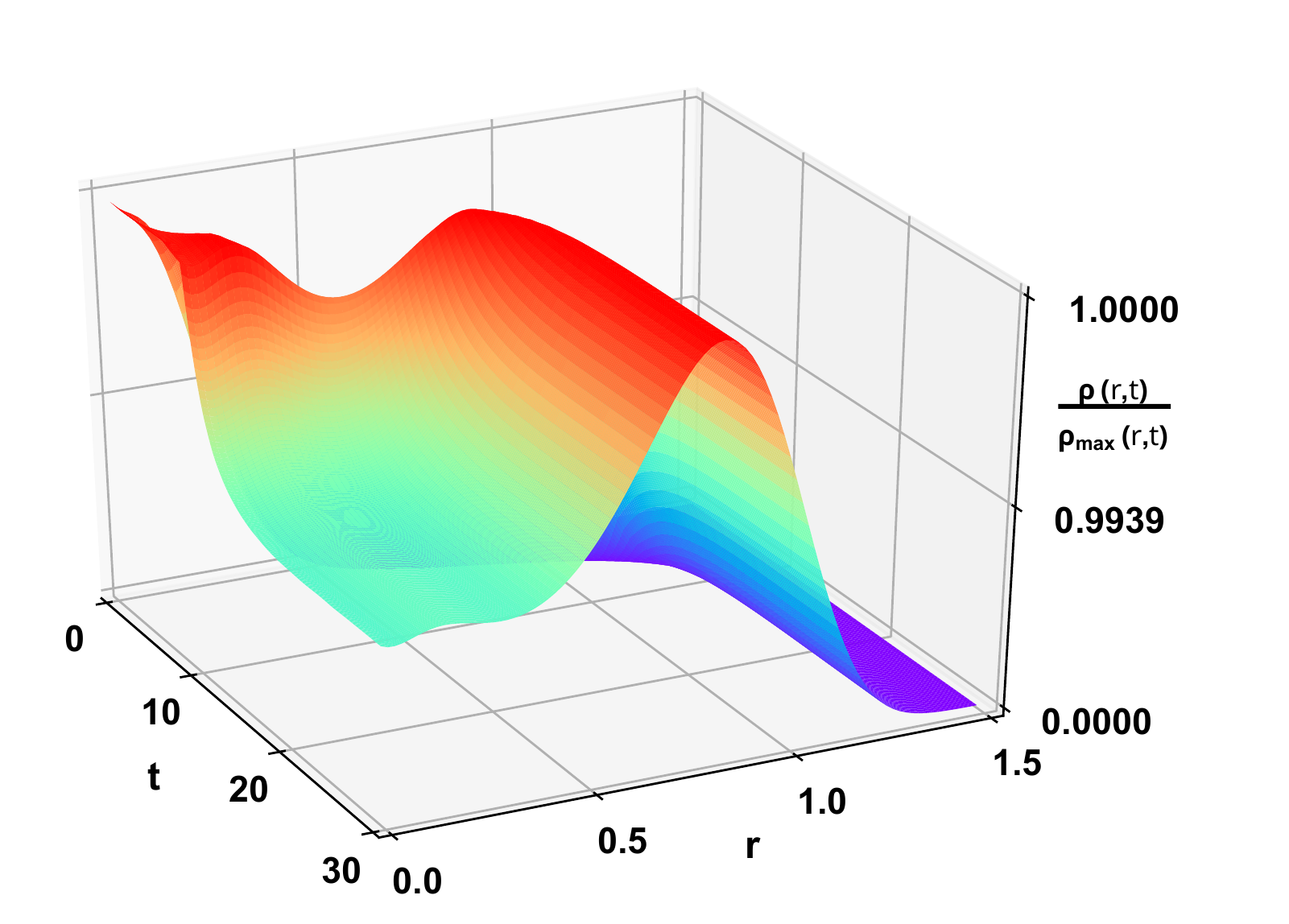}
		\caption{}
		\label{f18_3}	
	\end{subfigure}
	\hspace{1pt}	
	\begin{subfigure}[h]{0.5\textwidth}
		\centering \includegraphics[scale=.5]{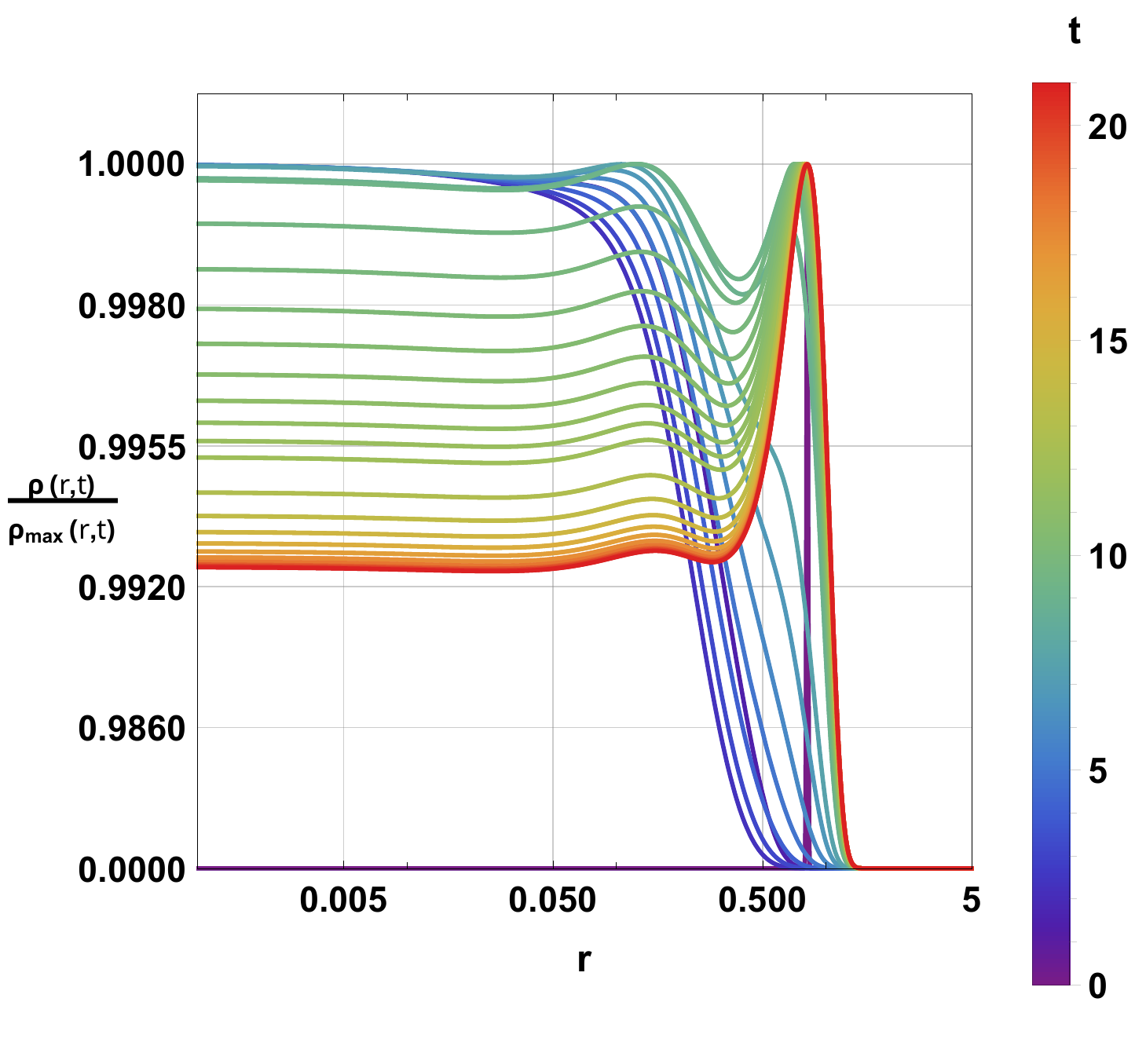}
		\caption{}
		\label{f18_4}	
	\end{subfigure}

	\caption{\footnotesize\it Probability distribution $\rho(r,t)$ governed by Fokker-Planck equation for different temperatures around small-large black holes transition with $Q = 0.0105$, $l=1$ and $b=3.5$}
	\label{f18}
\end{figure}
 
 For $T = T_f = 0.2753$  corresponding to the first-order phase transition between small and large black holes phases, we observe that the probability $\rho(r,t)$ which is initially centered around the large black hole state, $r_i = 0.6838$, leaks rapidly to thermal radiations state, $r=0$, then it comes back to form two peaks around the small and the large black holes horizon radii. Thus, we have a critical point where the small black hole and the large black hole coexist together. Indeed, $\rho(r_{s},t_\infty) = \rho(r_l,t_\infty)$  where $r_s = 0.1273$ and $r_l = 0.6838$, the two phases are both equiprobable and globally stable. Once the system reaches $T = 0.29$,  we notice that the probability $\rho(r,t)$ which is initially centered around the large black hole state, $r_i = 0.8177$, leaks now quickly to thermal radiations and small black hole states forming the same shape as in critical point case, but after then $\rho(r_s,t)$ decreases with the time whereas $\rho(r_l,t)$ increases to reach its maximum. Therefore, $ \rho(r_s,t_\infty) < \rho(r_l,t_\infty)$  where $r_s = 0.1542$ and $r_l = 0.8177$, and the large black holes phase is the most probable state and then the only globally stable phase.

\end{itemize}

\begin{figure}[!ht]
	\centering
	\begin{subfigure}[h]{0.48\textwidth}
		\centering \includegraphics[scale=.62]{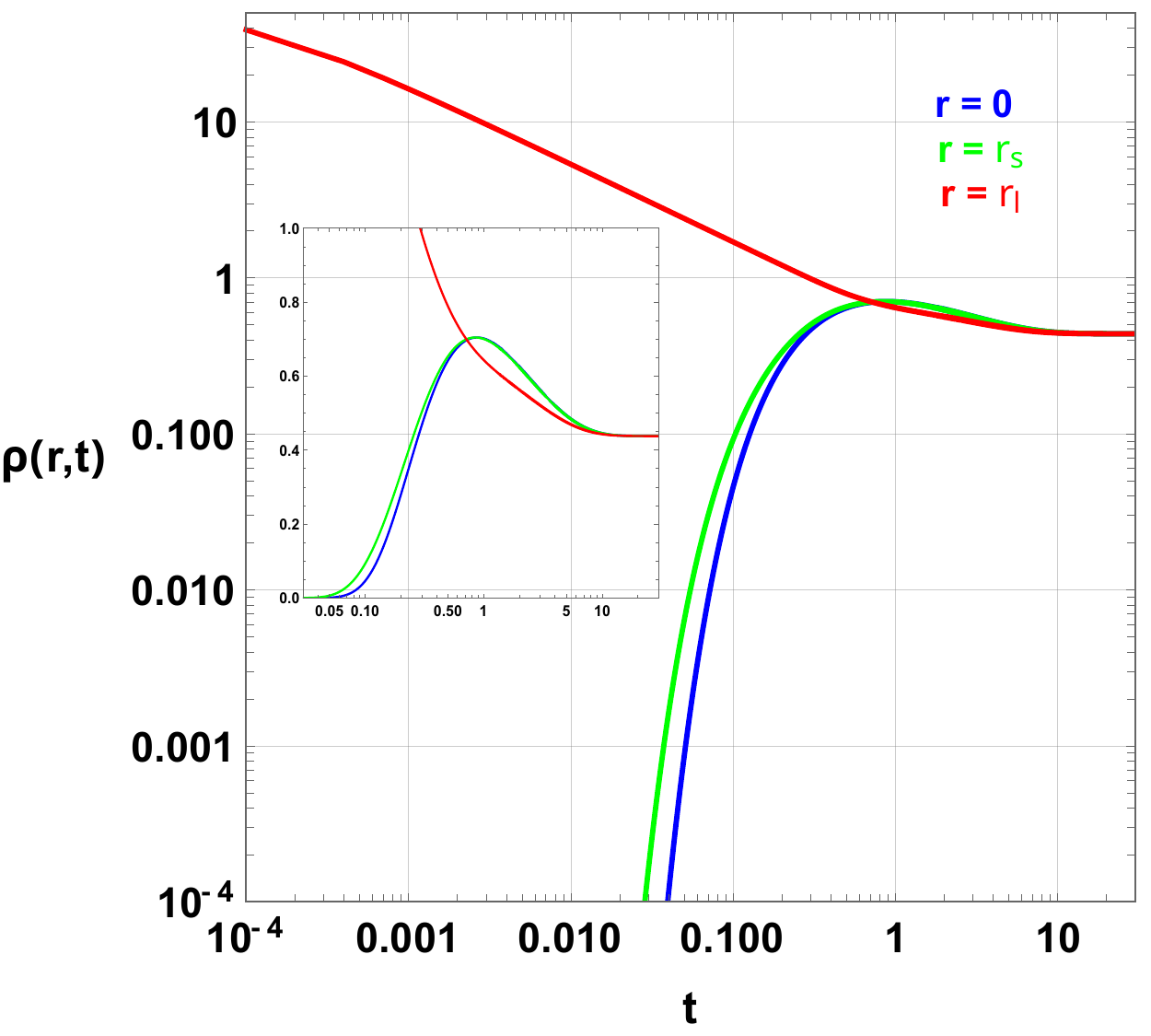}
		\caption{}
		\label{f19_1}
	\end{subfigure}
	\hspace{1pt}	
	\begin{subfigure}[h]{0.5\textwidth}
		\centering \includegraphics[scale=.62]{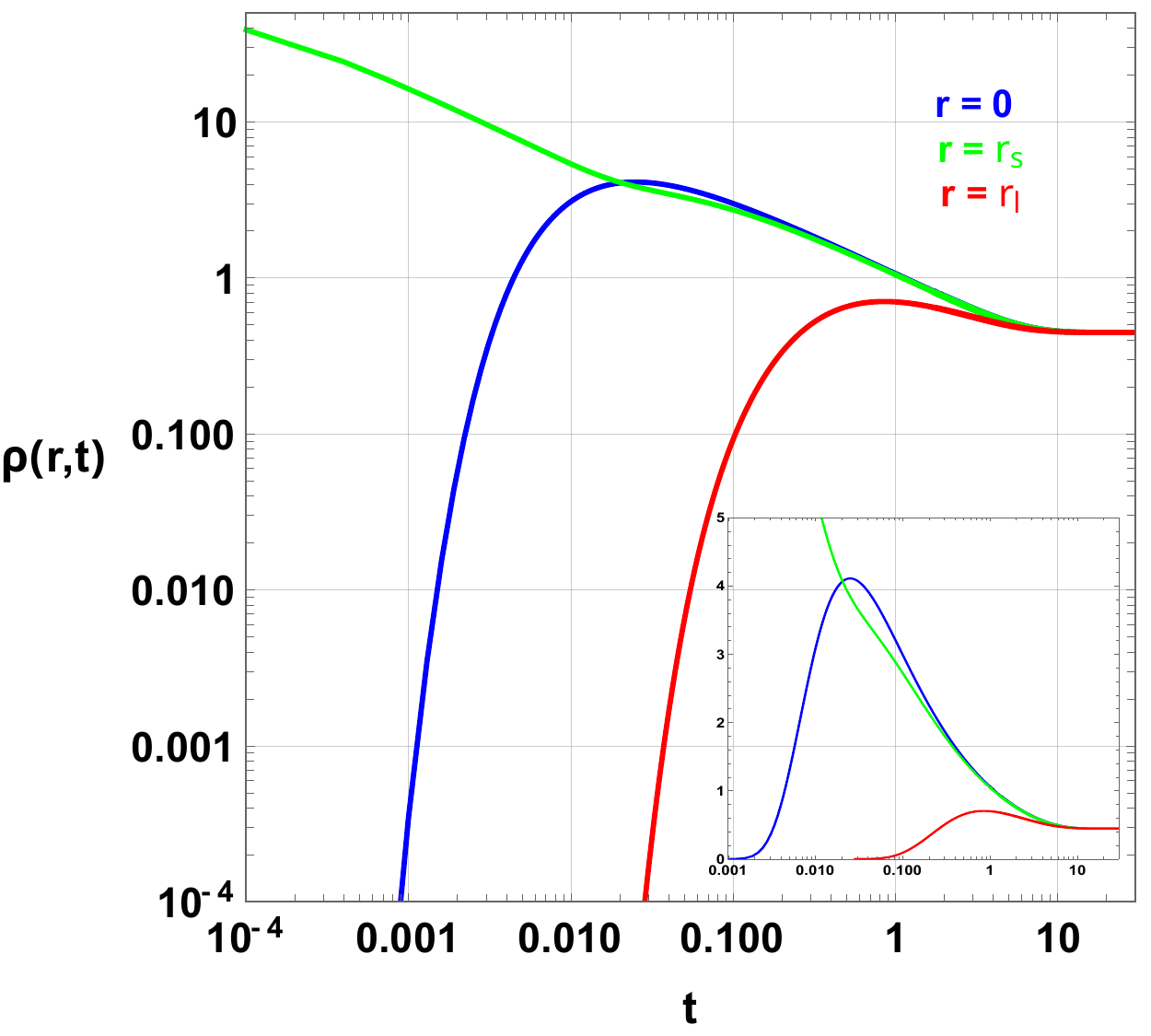}
		\caption{}
		\label{f19_2}		
	\end{subfigure}
	
	\caption{\footnotesize\it Behaviors of the probability $\rho(r,t)$ at the triple point, when the initial Gaussian wave packet is peaked at the (a) large  and (b) small black hole states, with $Q = Q_t = 0.0103638$, $T=T_t = 0.276$, $l=1$ and $b=3.5$.}
	\label{f19}
\end{figure}

%{\bf HERE}
Now we focus on what happens in the triple point where we the coexistence of three globally stable phases: thermal radiation at $r = 0$, the small black hole at $r_s$, and the large black hole at $r_l$. To a clear illustration,  we plot in Fig.\ref{f19} the evolution of $\rho(r_k,t)$  \footnote{$r_k = 0, r_s, r_l$} when the initial Gaussian wave packet peaks at large or small black hole states. 

We observe that the initially large value$\rho(r_{l,s},t = 0)$ in each case rapidly decays to a stationary value, while the other two states (initially zero) grow toward this value. 

\begin{figure}[!ht]
	\begin{center}
		\centering
		\includegraphics[scale=.7]{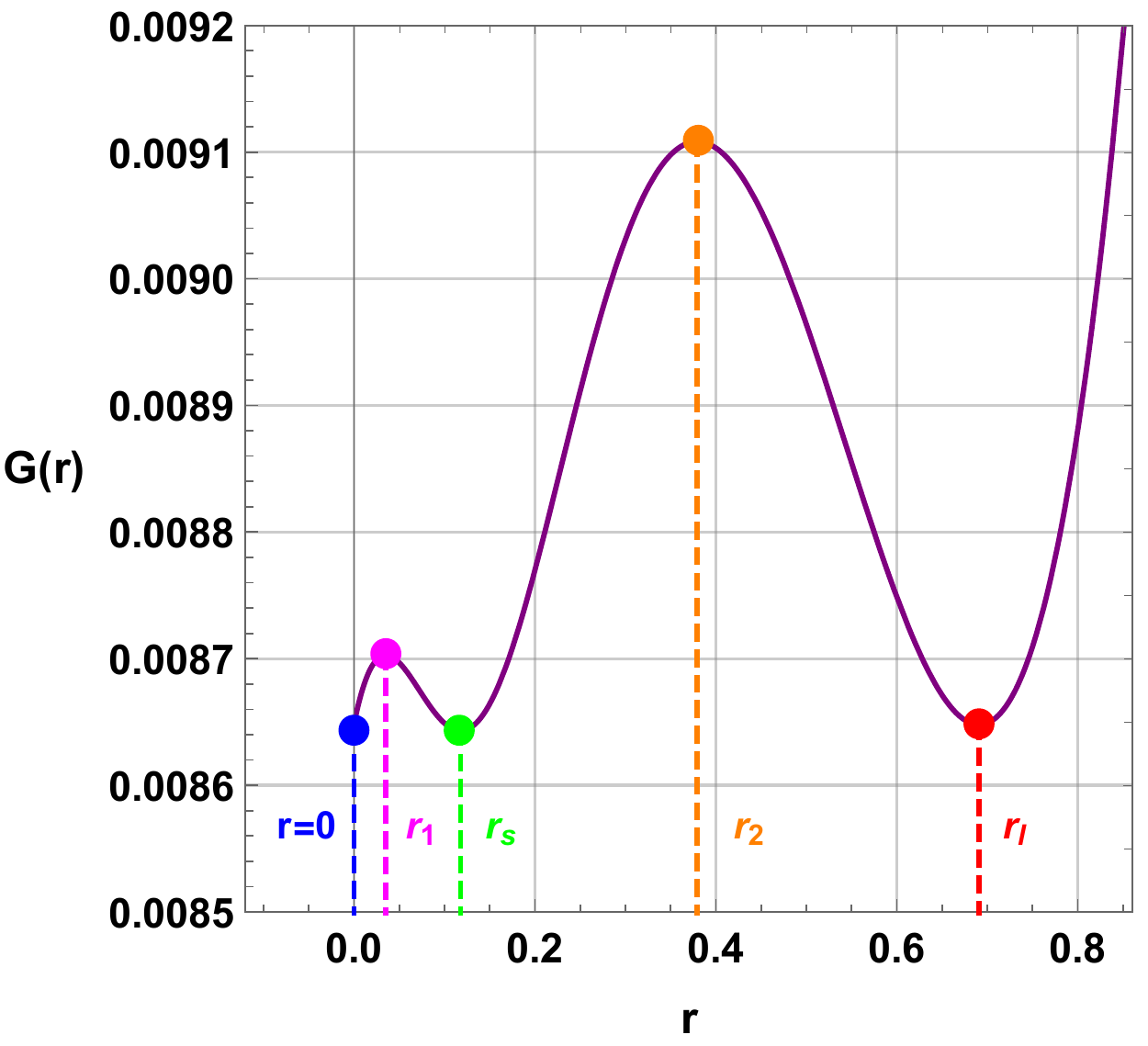} 
	\end{center}
	\caption{\footnotesize\it Gibbs free energy via landscape at the triple point with $Q = Q_t = 0.0103638$, $T=T_t = 0.276$, $l=1$ and $b=3.5$.}
	\label{f20}
\end{figure}

Henceforth, considering the first case shown Fig.\ref{f19_1}, the initial Gaussian wave packet $\rho(r,t = 0)$ peaks at the large black hole phase with
$\rho(r_l,t = 0) = 56.4189$, while $\rho(r = 0,t = 0) \simeq \rho(r_s,t = 0) \simeq 0$. As $t$ increases $\rho(r_l,t)$ decreases (red curve), and $\rho(r=0,t)$ and $\rho(r_s,t)$ increase (blue and green curves respectively) as expected, with $\rho(r=0,t) < \rho(r_s,t)$ because the initial state must surmount two barriers to reach thermal radiations state as we can see in Fig.\ref{f20}.
Indeed, we see in Fig.\ref{f20} where we have displayed the Gibbs free energy via landscape at the triple point that we have three wells (blue, green, and red dots) with the same depth separated by two barriers (magenta and orange dots) and then a large black hole located at $r_l$ should surmount the two barriers to reach $r = 0$. The transition rate from the large to small black hole
state is higher than the total rate from the small to the large black hole and thermal state states. Once $t > 20$, $\rho(r=0,t) = \rho(r_s,t) = \rho(r_l,t) = 0.4377$ , where the final stationary state is achieved. During the evolution, we notice that $\rho(r_s,t)$ increases to a maximum of $0.7043$ at $t = 0.8544$ and then decreases with time to its stationary value of $0.4377$. This behavior occurs
because the small black hole state can transit to both thermal radiation and large black hole states, while the system will stay longer at the small black hole and the thermal radiation states because the barrier between them is relatively small  than the large black hole state. As the small
black hole and thermal radiation states become more populated, they will transit back to the large black hole state. Because $\rho(r_s,t)$ and $\rho(r=0,t)$ can be greater than $\rho(r_l,t)$, we refer to this behavior as a strong oscillatory phenomenon \cite{Wei:2021bwy}.

Considering second case shown Fig.\ref{f19_2}, the initial Gaussian wave packet $\rho(r,t = 0)$ peaks at the small black hole phase with $\rho(r_s,t = 0) = 56.4189$, while $\rho(r = 0,t = 0) \simeq \rho(r_l,t = 0) \simeq 0$. As expected, $\rho(r_s,t = 0)$ decreases whereas $\rho(r=0,t = 0)$
and $\rho(r_l,t = 0)$ both increase. Once $t > 20$, $\rho(r=0,t) = \rho(r_s,t) = \rho(r_l,t) = 0.4377$, where the final stationary state is achieved as in the previous case. During the evolution, we observe that the probability leaks from the small black hole state to both thermal radiations and large black
hole states, with $\rho(r=0,t)$ increasing faster than $\rho(r_l,t)$ because of the lower barrier height shown in Fig.\ref{f20}. Moreover, we see that $\rho(r=0,t)$ increases quickly to a maximum of $4.1097$ at $t = 0.0245$ and then decreases with time to its stationary value of $0.4377$. In addition, for $t > 0.0205 $, we have $\rho(r=0,t) > \rho(r_s,t) > \rho(r_l,t)$, which means that $\rho(r=0,t)$ is dominant among the three probabilities, indicating that the system has a large probability of staying at the thermal radiations state. Since $\rho(r=0,t)$ dominates, we refer to this behavior (similar to that observed in the previous case) as a strong oscillatory phenomenon. During the evolution of $\rho(r_l,t)$, we observe that it increases to a maximum of $0.7043$ at $t=0.8551$ and then decreases with time to its stationary value of $0.4377$. Since $\rho(r_l,t)$ is always smaller than $\rho(r_s,t)$, such a behavior is known as a weak oscillatory phenomenon \cite{Wei:2021bwy}.
\cleardoublepage

\newpage
\section{Kinetics: First passage process}

In a triple point setting where thermal radiation, large, and small black hole states are coexisting, we will explore the kinetics of the first passage event from one black hole state to another.
The time required for the black hole state to reach an unstable black hole phase, as represented by the free energy peak, is characterized as the first passage time.
The mean first passage time denotes the average timescale for a stochastic event to occur for the first time  \cite{Li:2020nsy}. From Fig.\ref{f20}, we have three cases: thermal radiations  to the small black holes (case 1), small black holes to thermal radiations and large black holes (case 2), and large black holes to small black holes (case 3). 

The distribution of first passage times is defined as $F_p(t)$, where $\Sigma (t)$ is the probability that the state of the black hole hasn't performed a first passage by time $t$. $F_p(t)$ and $\Sigma(t)$ distributions are connected by 
\begin{equation}\label{33}	
	F_p(t) = - \mathcal{A} \dfrac{d \Sigma(t)}{d t},
\end{equation}
herein, $\mathcal{A}$ is nothing than a normalization constant such that $\int_{0}^{+ \infty} F_p(t) dt = 1 $. 
The definition of $\Sigma(t)$ is the probability of a black hole being present in the system at time $t$ \cite{Li:2020nsy}. Hence, we get
\begin{equation}\label{34}	
	\Sigma_{1}(t) = \int_{0}^{r_1} \rho(r,t) dr,
\end{equation}
\begin{equation}\label{35}	
	\Sigma_{2}(t) = \int_{r_1}^{r_2} \rho(r,t) dr,
\end{equation}
\begin{equation}\label{36}	
	\Sigma_{3}(t) = \int_{r_2}^{+\infty} \rho(r,t) dr.
\end{equation}
We apply reflective boundary conditions at $r = 0$ and $r = +\infty$ (sufficiently big $r$) and absorbing boundary conditions at $r_1$ and $r_2$ peaks. We assumed that the time required to transition from the intermediate transition state of the black hole to the large black hole state is substantially shorter than the first passage time.
Furthermore, if a black hole state makes the first passage via thermal fluctuation, the black hole state exits the system.
In this case, the probability distribution's normalization will not be kept.

Using Eqs.\eqref{33}, \eqref{34}, \eqref{35}, \eqref{36} and the Fokker-Planck equation Eq.\eqref{29},
one can express the first passage rate $F_p(t)$ for
the three cases as \footnote{We have used $\lim\limits_{r \rightarrow +\infty} \rho(r,t) = 0$}
\begin{equation}\label{37}	
	F_{p1}(t) =  \mathcal{A}_1 \left( - \left. \dfrac{\partial \rho(r,t)}{\partial r}\right| _{r = r_1} +\dfrac{1}{T} \left. \rho(r,t) \dfrac{\partial G(r)}{\partial r}\right| _{r = 0}\right)  ,
\end{equation}
\begin{equation}\label{38}	
	F_{p2}(t) =  \mathcal{A}_2 \left( - \left. \dfrac{\partial \rho(r,t)}{\partial r}\right| _{r = r_2} +\ \left.  \dfrac{\partial \rho(r,t)}{\partial r}\right| _{r = r_1}\right)  ,
\end{equation}
\begin{equation}\label{39}	
	F_{p3}(t) = \mathcal{A}_3 \left. \dfrac{\partial \rho(r,t)}{\partial r}\right| _{r = r_2}  .
\end{equation}

We may calculate the mean first passage time and its relative fluctuation using the time distributions. The average first passage time is given by 
\begin{equation}\label{40}	
	\left\langle t\right\rangle = \int_{0}^{+ \infty} t F_p(t) dt ,
\end{equation}
while the relative fluctuation is obtained to be 
\begin{equation}\label{41}	
	\mathfrak{f} = \dfrac{\left\langle t^2\right\rangle - \left\langle t\right\rangle^2}{\left\langle t\right\rangle^2}.
\end{equation}

\begin{figure}[!ht]
	\centering
	\begin{subfigure}[h]{0.45\textwidth}
		\centering \includegraphics[scale=.62]{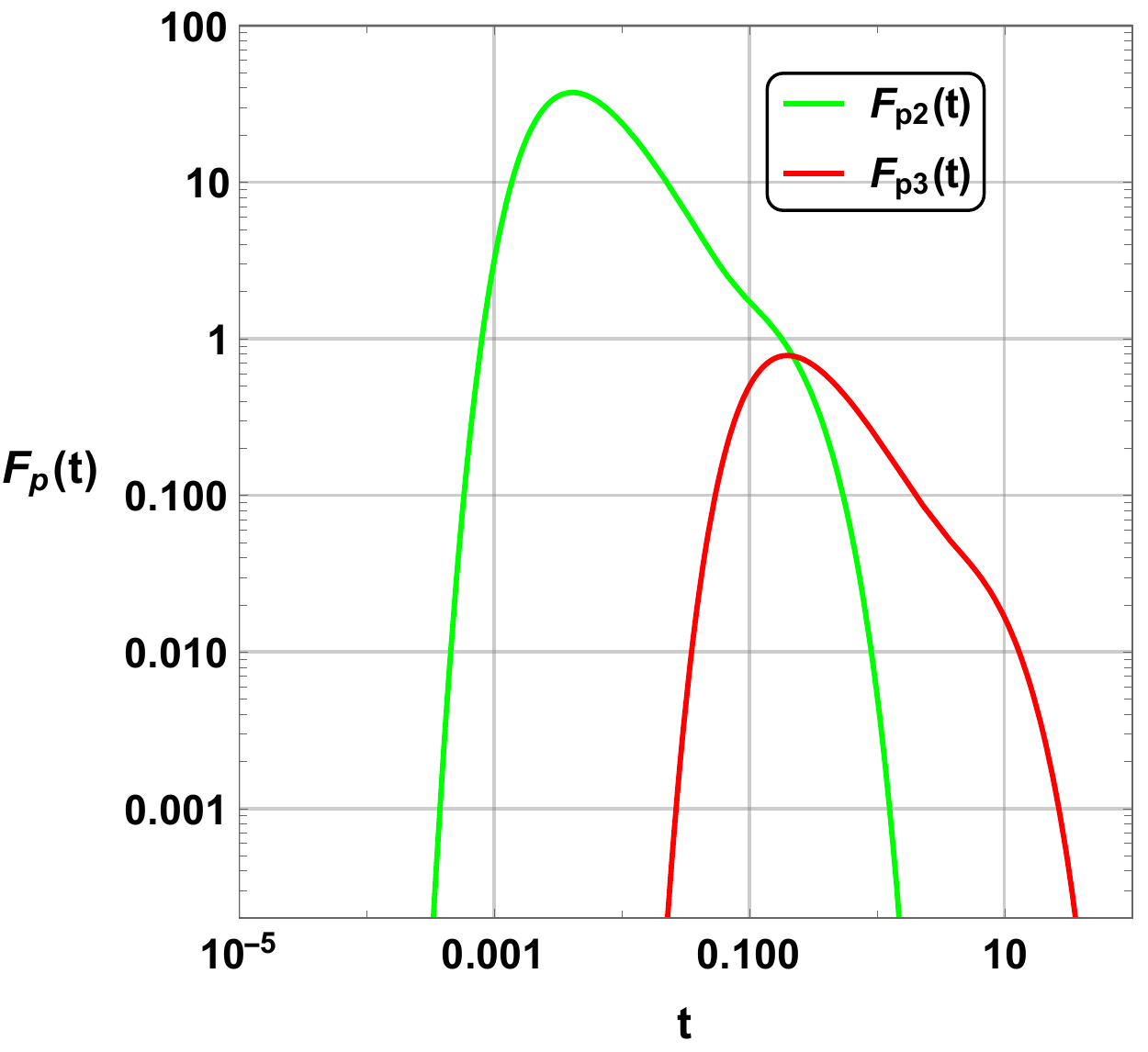}
		\caption{}
		\label{f21_1}
	\end{subfigure}
	\hspace{1pt}	
	\begin{subfigure}[h]{0.45\textwidth}
		\centering \includegraphics[scale=.6]{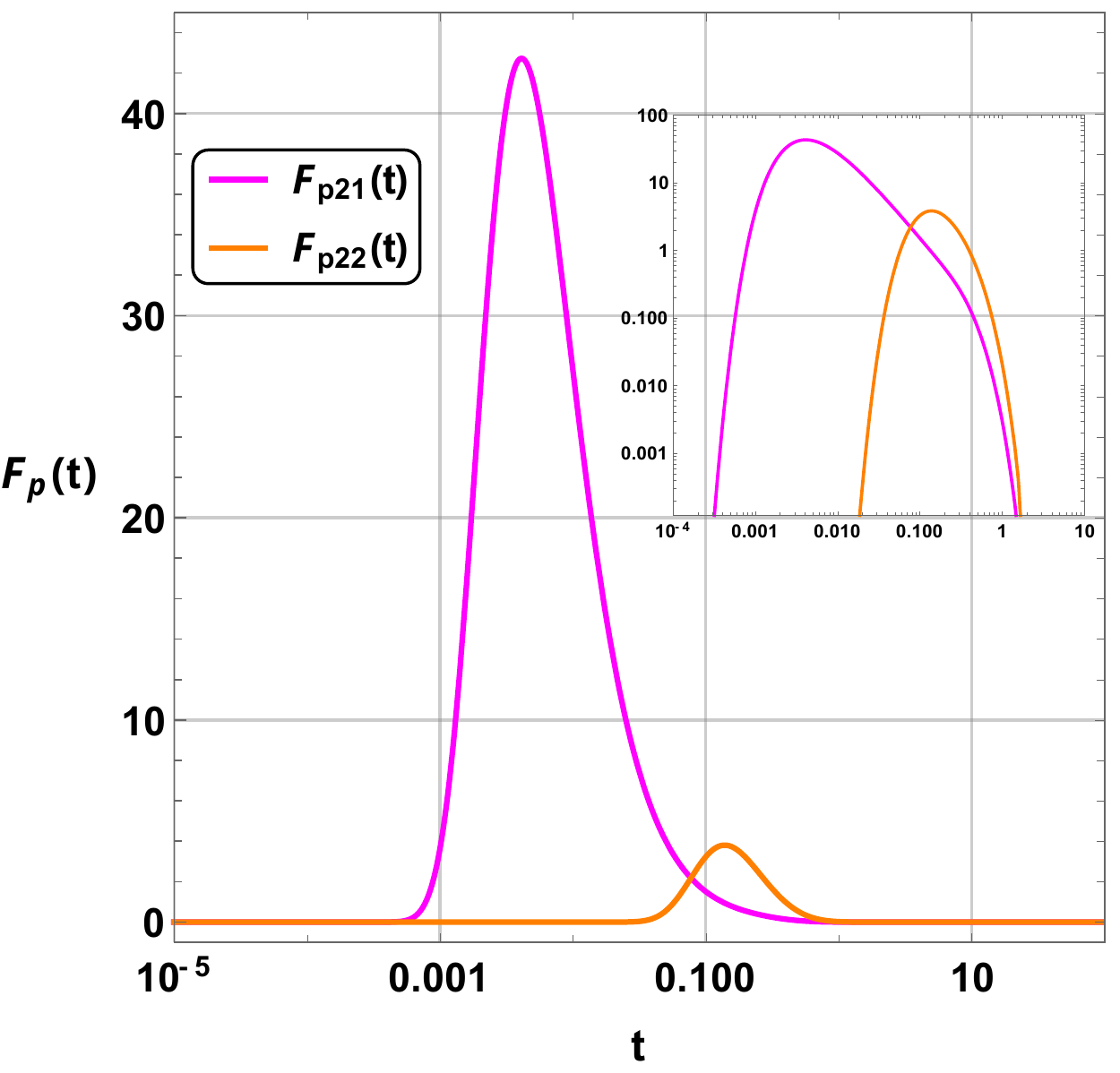}
		\caption{}
		\label{f21_2}		
	\end{subfigure}
	\hspace{1pt}	
\begin{subfigure}[h]{0.45\textwidth}
	\centering \includegraphics[scale=.6]{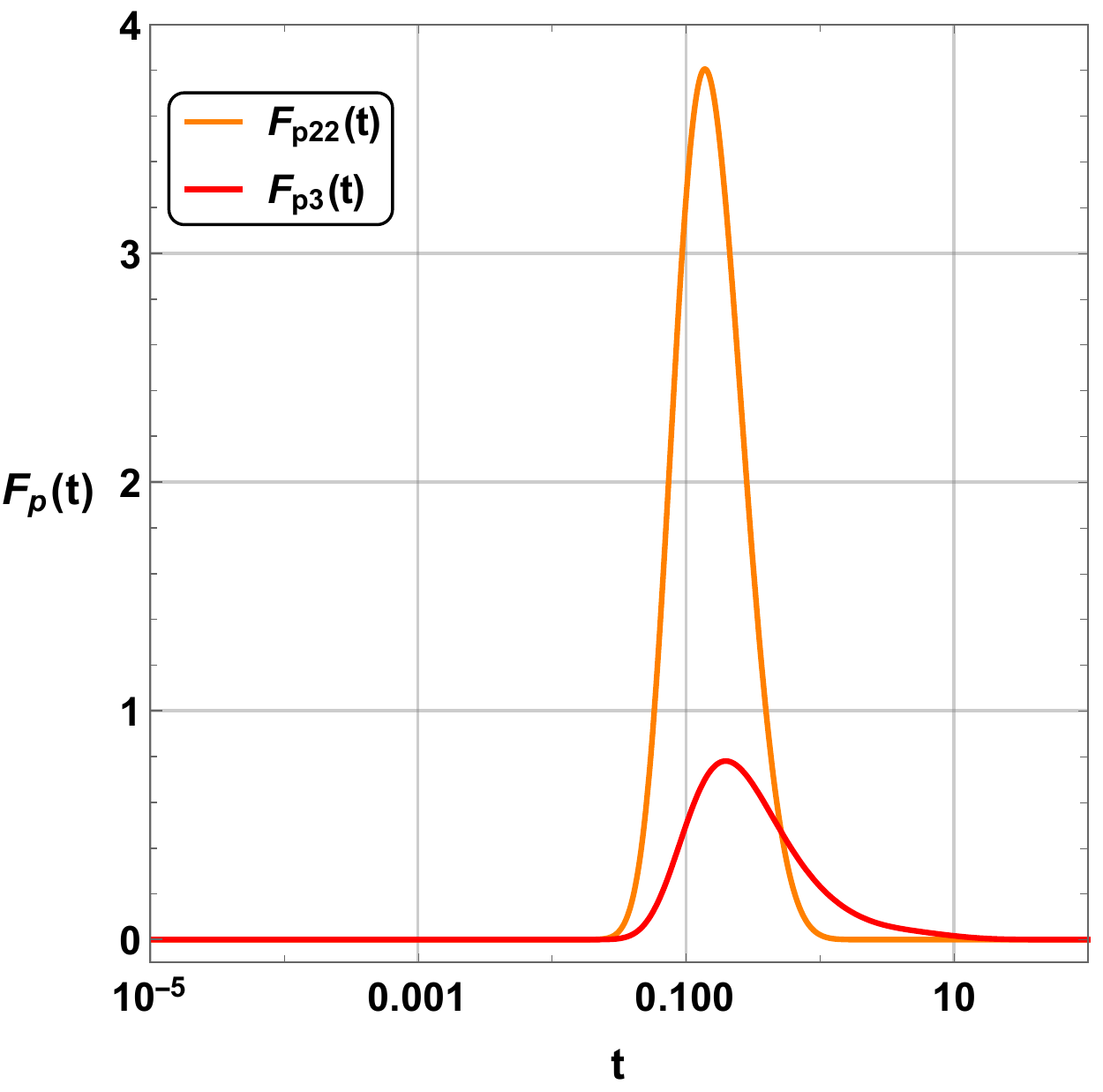}
	\caption{}
	\label{f21_3}		
\end{subfigure}
	
	\caption{\footnotesize\it (a) First passage time distribution for cases 2 and 3. (b)
		First and second parts of the first passage time for $F_{p2}(t)$. (c) First passage time distribution for case 3 and the second part of the first passage time for $F_{p2}(t)$.}
	\label{f21}
\end{figure}

We suggest studying just the second and the third cases because the initial state in the first case coincides with the boundary $r = 0$, and we have the numerical instabilities\footnote{due to the limits of performance associated with our calculator and the smallness of the studied domain}. The numerical results for $F_{p2}(t)$ and $F_{p3}(t)$ are plotted in Fig.\ref{f21_1}.
For each case, there is a single peak at $t_2 = 0.0041$  and $t_3 = 0.1986$, which can be interpreted as the length of time at which the system remains in its initial state for each before first transiting to another state. For the second case (green curve), the system transits to other states (thermal radiation and large black hole) after a short time, whereas for the third case (red curve), the system remains in the large black hole state longer. The mean first passage time for the second case is $\left\langle t_2\right\rangle = 0.0856$ with a relative fluctuation $\mathfrak{f}_2 = 2.3474$, whereas, for the third case, $\left\langle t_3\right\rangle = 3.3835$ with a relative fluctuation $\mathfrak{f}_3 = 1.9227$. 

In order to understand the behavior observed in Fig.\ref{f19}, we depict the two parts of $F_{p2}(t)$ in Fig.\ref{f21_2} where $F_{p21}(t)$ (magenta curve) is the distribution of first passage time from small black hole to thermal radiation, and $F_{p22}(t)$ (orange curve) is the distribution of first passage time from small black hole to the large black hole. Obviously, each distribution shows a single peak at $t_{21} = t_2 = 0.0041$  and $t_{22} = 0.1387$. The mean first passage time for each situation is $\left\langle t_{21}\right\rangle = 0.0612$ with a relative fluctuation $\mathfrak{f}_{21} = 2.9721$, and $\left\langle t_{22}\right\rangle = 0.2550$ with a relative fluctuation $\mathfrak{f}_{22} = 0.4081$. 

Moreover, we remark that $\left\langle t_{21}\right\rangle< \left\langle t_{22}\right\rangle$, which means that when the initial state peaks at the small black hole phase, the system transits suddenly to thermal radiations phase as we have seen in Fig.\ref{f19_2} where $\rho(r=0,t)$ increases very quickly and passes over $\rho(r_s,t)$, whereas  $\rho(r_l,t)$ increases slowly before to reach its maximum with $\rho(r_l,t)$ is always smaller than $\rho(r_s,t)$ because the transition to large black hole phase takes more time than the transition to thermal radiations which explains the weak oscillations observed in this case. Indeed, the barrier height between the small black hole and thermal radiation is smaller than that between small and large black holes. When the initial state peaks at the large black hole phase, the transition takes quite more time ($\left\langle t_{3}\right\rangle< \left\langle t_{2}\right\rangle$) to transit to the small black hole phase, and after then the system can transit quickly to thermal radiations phase. Moreover, we see in Fig.\ref{f21_3} that $F_{p22}(t) \neq F_{p3}(t) $, which means that the transition between small and large black holes is not symmetric with $\left\langle t_{3}\right\rangle< \left\langle t_{22}\right\rangle$. That is to say, a large black hole takes more time to transit to a small black hole even though the height of the barrier is the same. Nevertheless, the relative fluctuation of first passage time is very important in the transition from a large black hole to a small black hole ($\mathfrak{f}_{3}>\mathfrak{f}_{22}$), and that what could explain the strong oscillatory behavior observed in Fig.\ref{f19_1}.

\cleardoublepage

\section*{Conclusion}
The connection of gravity, thermodynamics, and quantum field theory is an essential tool to probe the quantum nature of black holes. The thermodynamic analysis and the phase transition behavior during the black hole evaporation might be a crucial areas of investigation to determine the microscopic description of the black hole thermodynamics in particular, and the spacetime structure in general. One possible way to investigate such microscopic structure is probed through the dynamic and kinetic evolution of the black hole during the thermal phase transitions. This paper is devoted to inquiring into the small black hole and the large black hole phases as well as the reentrant phase transition for the Born-Infeld-AdS black holes from the perspectives of the free energy landscape via the Landau free energy formalism and the stochastic processes through a certain probability distribution using the Fokker-Planck equation.\\

We start with a brief discussion of the Born-Infeld-AdS black hole solutions, the related thermodynamic quantities, and their thermal properties. We then moved on to compute the critical points characterizing different black hole configurations during the phase transitions. We systematically analyze the profile of the heat capacity at a constant electric charge, $Q$, and study its stable and unstable branches. Depending upon the values of the charge parameter, we categorically discussed the phase transition behavior. For $Q<Q_m$, $Q_m$ denotes some marginal charge expressed in terms of the Born-Infeld parameter, we have Schwarzschild-AdS-like behavior, whereas for $Q\geq Q_m$ we have the characteristics behavior of the heat capacity mimicking that of the Reissner-Nordstr$\Ddot{o}$m-AdS black holes. The condition $Q_m\leq Q<Q_c$, where $Q_c$ is the critical point, exhibits a first-order phase transition where a small-sized black hole transited into the large black holes. At the critical value $Q=Q_c$ of the charge parameter, a second-order phase transition occurred between the small and large black holes. For other values of the charge parameter greater than $Q_c$, we observed that the black holes are locally stable, thereby indicating the positive heat capacity.\\

Next, we focused our attention to explore the thermal phase transition behavior of the Born-Infeld AdS black holes by considering a general prescription of the Landau-free energy functional of the VdW fluids. The Landau functional provides us with a phenomenological description of the VdW fluids when it undergoes a second-order phase transition. The analysis is quite interesting for our present analysis of the Born-Infeld AdS black holes, for that we have to take into account the parameter space $\lbrace X,T,P, Q\rbrace$ describing the state of the system pertaining to a set of physical conditions to be imposed. Such an analysis actually leads us to connect the thermal behavior among the different states of the VdW-like fluids and the Born-Infeld AdS black holes phases. We computed for such purposes the convexity of the Landau functional to determine its extreme points and the corresponding stable and unstable phases of the thermal black hole systems. We plotted the temperature, $T$ vs. the horizon radius, $r_h$, the Gibbs free energy, $G$ as a function of temperature, $T$, the Landau functional as a function of the state variable $X$ as well as the volumes, $V$ for different values of the charge parameter $Q$. As a further investigation of the thermal systems, we also plotted the on-shell Gibbs free energy $\Tilde{G}$ as a function of the temperature. When the charge parameter $Q=0.005<Q_0$, we have the small and large black hole phases. Note that $Q_0$ corresponds to the value of the charge parameter where the temperature profile where $T-r_h$ plot has an inflection point. Therefore, $Q=0.00922=Q_0$ indicated that the first derivative of the Gibbs free energy is discontinuous and thereby indicating a corresponding discontinuity of the Landau functional. \\

Additionally, the dynamics of the AdS black holes and their kinematical descriptions are well addressed through the thermal fluctuations of the system under study during their evolution process. Since the fluctuating variables need probability distribution during the thermal phase transitions they also need time dependence. Such a description is guided by the probability distribution of the spacetime state in the ensemble designated as $\rho(r,t)$. The thermal phase transitions are mastered through a stochastic process and are a function of the order parameter and the dynamics during its phases are determined through the fluctuating macroscopic variables, the process during the whole evolution is probed by the Fokker-Planck equation. Since the small to large black hole phase transitions or the reentrant phase transition faced a  sudden change in the size of the event horizon, and at the critical points the change become vanishing, we referred to the event horizon as an order parameter. We analyzed the stability and the black hole thermal phase transitions of the Born-Infeld-AdS spacetime through heat capacity and Gibbs free energy. Rather than confront the phase transitions during the dynamical evolution, the usual Gibbs free energy is raised to the status of the generalized off-shell Gibbs free energy as a function of the order parameter. Analytical solutions for the order parameter are not possible but a numerical analysis found its lower bound. Having the off-shell Gibbs free energy for different black hole configurations we depicted them for different values of the temperatures. In the next paragraph, we discuss in detail the time-dependent behaviors of the probability distribution function of the Born-Infeld-AdS black holes in the extended phase space for various charge and temperature values. \\

During the small-large black hole phases, the switching process becomes faster than it existed earlier and the final stationary states should be of Boltzmann type. Such parables of the small-large black hole phases existed in the literature for a wide variety of AdS black hole systems. However, for the AdS black holes in Born-Infeld  gravity, the notions of the reentrant phase transitions and the leaking of the small to radiation or the large to radiation phases was an exciting investigation. However, in the process of the dynamical evolution, the charge parameter $Q=0.01009$, we ruled out the possible existence of the reentrant phase transition behavior as claimed in the previous literature. The systems instead of going through a reentrant phase transition, it drained down to the pure thermal state and the large black hole phase reflecting the most probable states. At slightly larger values of $Q=Q_t=0.0103638$, the system co-existed as a triple point comprising of the small, large, and thermal phases as was confirmed by the probability distribution. Further slight increase in the charge value, the phase transitions were distilled into two distinct phases, namely, a Hawking-Page-like phase transition between small to radiation, and another first-order phase transition among the small and the large black holes. Lately, we determined the first passage time for the kinetic evolution of different black hole states, particularly for the triple point setup. We observed a finite peak in the distribution of the first passage time for three different cases, (1) pure thermal radiation to the small black hole, (2) small black hole pure thermal radiation and large black hole state, and (3) large black hole to small black hole phases. Such processes occur within a short interval of first passage time. \\

The study of the frictional effects on the kinetic and dynamics processes is a must-investigated research area that we demonstrated a little bit in our current work by calculating the mean first passage time and the corresponding fluctuations. The friction in a dynamical system is accounted for by the microscopic description, thereby speculating ideas about the macroscopic behavior of the AdS black holes (e.g., in terms of the order parameter). A future direction to this work might be worth studying the microscopic degrees of freedom and the interactions of the black hole molecules during the kinetic turnover.

\section*{Acknowledgment}
The research of M. S. A. is supported by the National Postdoctoral Fellowship of the Science and Engineering Research Board (SERB), Department of Science and Technology (DST), Government of India, File No., PDF/2021/003491.

\appendix
\section{Numerical resolution of Focker-Planck equation}

\subsection{Introduction}
We propose two numerical schemes to resolve the Focker-Planck equation in the context of black hole thermodynamics where the temperature $T$ is kept constant in Gibbs $G(r)$ free energy is a function of radial parameter $r$. Our aim is to find out the probability distribution $\rho(r, t)$.  The first method we can use it with NDSolve function of Mathematica whereas the second needs an explicit implementation.

\subsection{First Method}

The Focker-Planck equation reads
\begin{equation}\label{an1}	
	\dfrac{\partial \rho(r,t)}{\partial t} = D \dfrac{\partial}{\partial r}\left[ e^{-\beta G(r)} \dfrac{\partial}{\partial r}\left[  e^{\beta G(r)}\rho(r,t)\right]  \right],
\end{equation}
where $D=T$ and $\beta = 1/T$. Then
\begin{equation}\label{an2}	
	\dfrac{\partial \rho(r,t)}{\partial t} = T \dfrac{\partial}{\partial r}\left[ e^{- \frac{G(r)}{T}} \dfrac{\partial}{\partial r}\left[  e^{\frac{G(r)}{T}}\rho(r,t)\right]  \right].
\end{equation}
We put
\begin{equation}\label{an3}	
	\rho(r,t)=u(r,t)e^{-\frac{G(r)}{T} },
\end{equation}
and we substitute it in Eq.\ref{an2} :
\begin{equation}\label{4}	
	\dfrac{\partial}{\partial t}\left[ u(r,t)e^{-\frac{G(r)}{T} }\right]  = T \dfrac{\partial}{\partial r}\left[ e^{- \frac{G(r)}{T}} \dfrac{\partial}{\partial r}\left[  e^{\frac{G(r)}{T}}u(r,t)e^{-\frac{G(r)}{T} }\right]  \right]
\end{equation}
\begin{equation}\label{an5}	
	\Longrightarrow  e^{-\frac{G(r)}{T}} \dfrac{\partial  u(r,t)}{\partial t} = T \dfrac{\partial}{\partial r}\left[ e^{- \frac{G(r)}{T}} \dfrac{\partial u(r,t)}{\partial r}  \right]
\end{equation}
\begin{equation}\label{an6}	
	\Longrightarrow   \dfrac{\partial  u(r,t)}{\partial t} = T \dfrac{\partial^2 u(r,t)}{\partial r^2} - \dfrac{\partial G(r)}{\partial r} \dfrac{\partial u(r,t)}{\partial r}  ,
\end{equation}
which is similar to the transport equation in one dimension : 
\begin{equation}\label{an7}	
	\Longrightarrow   \dfrac{\partial  u(x,t)}{\partial t} = D \dfrac{\partial^2 u(x,t)}{\partial x^2} - v(x) \dfrac{\partial u(x,t)}{\partial x} .
\end{equation}
Reflecting boundaries condition reads : 
	\begin{equation}\label{an8}	
	\left.  \dfrac{\partial}{\partial r}\left[  e^{\frac{G(r)}{T}}\rho(r,t)\right]\right|_{Boundaries} = 0 
\end{equation} 
\begin{equation}\label{an9}	
	\Longrightarrow  \left. \dfrac{\partial  u(r,t)}{\partial r} \right|_{Boundaries} = 0 .
\end{equation}

Using the boundaries condition Eq.\ref{an8}, the stationary solution ($t \to +\infty$) of Focker-Planck equation reads : 
\begin{equation}\label{an10}	
	 \dfrac{\partial  \rho(r,t)}{\partial t}  = 0 	\Longrightarrow \rho(r) = Ce^{-\frac{G(r)}{T}},
\end{equation}
where $C$ is an integration constant. Then, the stationary solution ($t \to +\infty$) of Eq.\eqref{an6} is 
\begin{equation}\label{an11}	
	u(r) = C.
\end{equation}
The NDSolve function of Mathematica can resolve Eq.\eqref{an6} with the initial condition 
\begin{equation}\label{an12}	
	u(r,0) = \rho(r,0) e^{\frac{G(r)}{T}},
\end{equation}
and the boundaries condition Eq.\eqref{an9}.

We can implement this method explicitly  using the numerical scheme below.

\subsubsection{Discretization}

We put
\begin{align}\label{an13}
		\Delta x &= x_{i+1} -x_{i},  &\Delta t &= t_{j+1} -t_{j},\\
		u(r_i,t_j) &= u_i^j, &\left. \dfrac{\partial G(r)}{\partial r}\right| _{r=r_i} &= g_i,\\
		1 \leq & i \leq N+1, 	&1 \leq &j \leq M+1, 
\end{align}

\subsubsection{Finite differences / Implicit method / Backward time-central space / Crank-Niclson method}

Eq.\eqref{an6} $\Longrightarrow$
\begin{multline}\label{an16}	
	\dfrac{u_i^{j+1}-u_i^j}{\Delta t} = \dfrac{T}{2}\left( \dfrac{u_{i-1}^j-2u_i^j+u_{i+1}^j}{\Delta x ^2} + \dfrac{u_{i-1}^{j+1}-2u_i^{j+1}+u_{i+1}^{j+1}}{\Delta x ^2}\right) \\ - \dfrac{g_i}{2}\left( \dfrac{u_{i+1}^j-u_{i-1}^j}{2 \Delta x } + \dfrac{u_{i+1}^{j+1}-u_{i-1}^{j+1}}{2 \Delta x}\right)
\end{multline}

\begin{multline}\label{an17}	
	\Longrightarrow -\left( \dfrac{T \Delta t}{\Delta x^2} + \dfrac{g_i \Delta t}{2 \Delta x}\right) u^{j+1}_{i-1} + 2\left( 1+\dfrac{T \Delta t}{\Delta x^2}\right) u^{j+1}_{i}  -\left( \dfrac{T \Delta t}{\Delta x^2} - \dfrac{g_i \Delta t}{2 \Delta x}\right) u^{j+1}_{i+1}  = \\  \left( \dfrac{T \Delta t}{\Delta x^2} + \dfrac{g_i \Delta t}{2 \Delta x}\right) u^{j}_{i-1} + 2\left( 1-\dfrac{T \Delta t}{\Delta x^2}\right) u^{j}_{i} +\left( \dfrac{T \Delta t}{\Delta x^2} - \dfrac{g_i \Delta t}{2 \Delta x}\right) u^{j}_{i+1}.
\end{multline}

We put
\begin{equation}\label{an18}	
\dfrac{T \Delta t}{\Delta x^2} = a \quad\quad \text{and}\quad \quad b_i = \dfrac{g_i \Delta t}{2 \Delta x},
\end{equation}
then 
\begin{multline}\label{an19}	
-\left( a + b_i\right) u^{j+1}_{i-1} + 2\left( 1+a\right) u^{j+1}_{i}  -\left( a - b_i\right) u^{j+1}_{i+1}  =   \left( a + b_i\right) u^{j}_{i-1} + 2\left( 1-a\right) u^{j}_{i} +\left(a - b_i\right) u^{j}_{i+1}.
\end{multline}
For $i = 1$ : Eq.\eqref{an9} $\Longrightarrow$
\begin{equation}\label{an20}	
	\dfrac{u_2^{j+1}-u_1^{j+1}}{\Delta x} = 0 \quad \Longrightarrow -u_1^{j+1}+u_2^{j+1} =0
\end{equation}
For $i = N+1$ : Eq.\eqref{an9} $\Longrightarrow$
\begin{equation}\label{an21}	
	\dfrac{u_{N+1}^{j+1}-u_N^{j+1}}{\Delta x} = 0 \quad \Longrightarrow -u_N^{j+1}+u_{N+1}^{j+1} =0.
\end{equation}
Therefore, Eqs.\eqref{an19},\eqref{an20} and \eqref{an21} read
\begin{equation}\label{an22}	
	A u^{j+1} = B u^{j} \quad \Longrightarrow u^{j+1} = A^{-1}B u^{j}  ,
\end{equation}
with 
\begin{equation}\label{an23}
	A = 
	\begin{pmatrix}
		-1 & 1 & 0 & \cdots & \cdots & \cdots & 0 \\
		-(a+b_2) & 2(1+a) & -(a-b_2) & 0 & \cdots & \cdots & 0 \\
		0 & -(a+b_3) & 2(1+a) & -(a-b_3) & 0 & \cdots & 0 \\
		\vdots  & \ddots  & \ddots  & \ddots  & \ddots  & \ddots & \vdots  \\
		\vdots  & \ddots  & \ddots  & \ddots  & \ddots  & \ddots & \vdots  \\
		0 & \cdots & \cdots & 0 & -(a+b_N) & 2(1+a) & -(a-b_N) \\
		0 & \cdots & \cdots & \cdots & 0 & -1 & 1
	\end{pmatrix}
,
\end{equation}
\begin{equation}\label{an24}
	B = 
	\begin{pmatrix}
		0 & 0 & 0 & \cdots & \cdots & \cdots & 0 \\
		(a+b_2) & 2(1-a) & (a-b_2) & 0 & \cdots & \cdots & 0 \\
		0 & (a+b_3) & 2(1-a) & (a-b_3) & 0 & \cdots & 0 \\
		\vdots  & \ddots  & \ddots  & \ddots  & \ddots  & \ddots & \vdots  \\
		\vdots  & \ddots  & \ddots  & \ddots  & \ddots  & \ddots & \vdots  \\
		0 & \cdots & \cdots & 0 & (a+b_N) & 2(1-a) & (a-b_N) \\
		0 & \cdots & \cdots & \cdots & 0 & 0 & 0
	\end{pmatrix}
\end{equation}
and
\begin{equation}\label{an25}
	u^j = 
	\begin{pmatrix}
		u_1^j  \\
		u_2^j \\
		\vdots  \\
		\vdots  \\
		u_{N}^j \\
		u_{N+1}^j
	\end{pmatrix}
.
\end{equation}
\subsection{Second Method}
Following \cite{CHANG19701, PALLESCHI1990378, https://doi.org/10.48550/arxiv.2006.11038}, we consider the general form of Focker-Planck equation 
\begin{equation}\label{an26}
\dfrac{\partial u(x,t)}{\partial t} = \dfrac{1}{A(x)} \dfrac{\partial }{\partial x}\left[ B(x,t) u(x,t) + C(x,t)\dfrac{\partial u(x,t)}{\partial x}\right] . 
\end{equation}
Using this general form, the Focker-Planck equation in black hole thermodynamics can be read
\begin{equation}\label{an27}
	\dfrac{\partial \rho(r,t)}{\partial t} =  \dfrac{\partial }{\partial r}\left[ \dfrac{\partial G(r)}{\partial r} \rho(r,t) + T\dfrac{\partial \rho(r,t)}{\partial r}\right].  
\end{equation}
We put 
\begin{equation}\label{an28}
	F(r,t)=   \dfrac{\partial G(r)}{\partial r} \rho(r,t) + T\dfrac{\partial \rho(r,t)}{\partial r},  
\end{equation}
then
\begin{equation}\label{an29}
	\dfrac{\partial \rho(r,t)}{\partial t} =  \dfrac{\partial F(r,t) }{\partial r}.  
\end{equation}
We put
\begin{align}\label{an30}
	\Delta x &= x_{i+1} -x_{i},  &\Delta t &= t_{j+1} -t_{j},\\
	\rho(r_i,t_j) &= \rho_i^j, &\left. \dfrac{\partial G(r)}{\partial r}\right| _{r=r_i} &= g_i,\\
	\omega_i & = \Delta x \dfrac{g_{i+1/2}}{T}, 	&\delta_i &=\dfrac{1}{\omega_i}-\dfrac{1}{e^{\omega_i}-1},\\
	0 \leq & i \leq N, 	&0 \leq &j \leq M, 
\end{align}
with $x_{i+1/2} =(x_{i+1} +x_{i})/2 $.

Eq.\eqref{an29} $\Longrightarrow$
\begin{equation}\label{an34}
	\dfrac{\rho_{i}^{j+1}-\rho_{i}^{j}}{\Delta t} = \dfrac{F_{i+1/2}^{j+1}-F_{i-1/2}^{j+1}}{\Delta x}.
\end{equation}
Eq.\eqref{an28} $\Longrightarrow$
\begin{equation}\label{an35}
	F_{i+1/2}^{j+1} = \left[ (1-\delta_i)g_{i+1/2}+\dfrac{T}{\Delta x}\right]\rho_{i+1}^{j+1} -\left( \dfrac{T}{\Delta x}-\delta_i g_{i+1/2}\right) \rho_{i}^{j+1}
\end{equation}

\begin{multline}\label{an36}	
	\Longrightarrow \rho_{i}^{j+1} - \rho_{i}^{j} = \dfrac{\Delta t}{\Delta x} \left\lbrace \left[ (1-\delta_i) g_{i+1/2}+\dfrac{T}{\Delta x}\right] \rho_{i+1}^{j+1} \right. \\
	\left. -\left[ \dfrac{2 T}{\Delta x} + (1-\delta_{i-1})g_{i-1/2}-\delta_{i} g_{i+1/2}\right] \rho_{i}^{j+1}  + \left[ \dfrac{T}{\Delta x} - \delta_{i-1} g_{i-1/2}\right] \rho_{i-1}^{j+1} \right\rbrace 
\end{multline}

\begin{multline}\label{an37}	
	\Longrightarrow \rho_{i}^{j} = - \dfrac{\Delta t}{\Delta x} \left[ \dfrac{T}{\Delta x}-\delta_{i-1}g_{i-1/2}\right] \rho_{i-1}^{j+1} \\ +\left\lbrace  1+ \dfrac{\Delta t}{\Delta x}\left[ \dfrac{2 T}{\Delta x}+(1-\delta_{i-1})g_{i-1/2}-\delta_i g_{i+1/2}\right] \right\rbrace \rho_{i}^{j+1}\\
	-\dfrac{\Delta t}{\Delta x}\left[ (1-\delta_i)g_{i+1/2} + \dfrac{T}{\Delta x}\right] \rho_{i+1}^{j+1} 
\end{multline}
\begin{equation}\label{an38}	
	\Longrightarrow \rho_{i}^{j} = a_i \rho_{i-1}^{j+1} + b_i \rho_{i}^{j+1}+c_i \rho_{i+1}^{j+1} ,
\end{equation}
with
\begin{equation}\label{an39}	
\begin{split}
	a_i & = - \dfrac{\Delta t}{\Delta x} \left[ \dfrac{T}{\Delta x}-\delta_{i-1}g_{i-1/2}\right] \\
	b_i &=  1+ \dfrac{\Delta t}{\Delta x}\left[ \dfrac{2 T}{\Delta x}+(1-\delta_{i-1})g_{i-1/2}-\delta_i g_{i+1/2}\right]\\
	c_i &= -\dfrac{\Delta t}{\Delta x}\left[ (1-\delta_i)g_{i+1/2} + \dfrac{T}{\Delta x}\right].
\end{split}
\end{equation}

Using Eq.\eqref{an29}, reflecting boundaries condition read 
\begin{equation}\label{an40}	
\left. \dfrac{\partial F(r,t) }{\partial r} \right|  _{Boundaries} = 0.
\end{equation}

For  $i=0$ : Eq.\eqref{an34} $\Longrightarrow$
\begin{equation}\label{an41}	
	\rho_{0}^{j} = \rho_{0}^{j+1} -\dfrac{\Delta t}{\Delta x}F_{1/2}^{j+1},
\end{equation}
because $F_{-1/2}^{j+1} = 0$. Then
\begin{equation}\label{an42}	
	\rho_{0}^{j} = \rho_{0}^{j+1} -\dfrac{\Delta t}{\Delta x}\left[ (1-\delta_0)g_{1/2}+\dfrac{T}{\Delta x}\right]\rho_{1}^{j+1} +\dfrac{\Delta t}{\Delta x} \left( \dfrac{T}{\Delta x}-\delta_0 g_{1/2}\right) \rho_{0}^{j+1}
\end{equation}
\begin{equation}\label{an43}	
\Longrightarrow	\rho_{0}^{j} = \left[ 1+\dfrac{\Delta t}{\Delta x} \left( \dfrac{T}{\Delta x}-\delta_0 g_{1/2}\right)\right]  \rho_{0}^{j+1}
	 -\dfrac{\Delta t}{\Delta x}\left[ (1-\delta_0)g_{1/2}+\dfrac{T}{\Delta x}\right]\rho_{1}^{j+1} 
\end{equation}
\begin{equation}\label{an44}	
	\Longrightarrow	\rho_{0}^{j} = b_0  \rho_{0}^{j+1}
+ a_0\rho_{1}^{j+1} ,
\end{equation}
with 
\begin{equation}\label{an45}	
	\begin{split}
		b_0 &=  1+\dfrac{\Delta t}{\Delta x} \left( \dfrac{T}{\Delta x}-\delta_0 g_{1/2}\right)\\
		c_0 &= -\dfrac{\Delta t}{\Delta x}\left[ (1-\delta_0)g_{1/2}+\dfrac{T}{\Delta x}\right].
	\end{split}
\end{equation}

For  $i=N$ : Eq.\eqref{an34} $\Longrightarrow$
\begin{equation}\label{an46}	
	\rho_{N}^{j} = \rho_{N}^{j+1} +\dfrac{\Delta t}{\Delta x}F_{N-1/2}^{j+1},
\end{equation}
because $F_{N+1/2}^{j+1} = 0$. Then
\begin{equation}\label{an47}	
	\rho_{N}^{j} = \rho_{N}^{j+1} +\dfrac{\Delta t}{\Delta x}\left[ (1-\delta_{N-1})g_{N-1/2}+\dfrac{T}{\Delta x}\right]\rho_{N}^{j+1} -\dfrac{\Delta t}{\Delta x} \left( \dfrac{T}{\Delta x}-\delta_{N-1} g_{N-1/2}\right) \rho_{N-1}^{j+1}
\end{equation}
\begin{equation}\label{an48}	
	\Longrightarrow	\rho_{N}^{j} =-\left( \dfrac{T}{\Delta x}-\delta_{N-1}g_{N-1/2}\right) \rho_{N-1}^{j+1}
	 +\left\lbrace 1 + \dfrac{\Delta t}{\Delta x} \left[ \left( 1-\delta_{N-1}\right) g_{N-1/2}+ \dfrac{T}{\Delta x}\right] \right\rbrace \rho_{N}^{j+1}
\end{equation}
\begin{equation}\label{an49}	
	\Longrightarrow	\rho_{N}^{j} = a_N  \rho_{N-1}^{j+1}
	+ b_N\rho_{N}^{j+1} ,
\end{equation}
with 
\begin{equation}\label{an50}	
	\begin{split}
		a_N &= -\left( \dfrac{T}{\Delta x}-\delta_{N-1}g_{N-1/2}\right)\\
		b_N &= 1 + \dfrac{\Delta t}{\Delta x} \left[ \left( 1-\delta_{N-1}\right) g_{N-1/2}+ \dfrac{T}{\Delta x}\right].
	\end{split}
\end{equation}

Therefore, Eqs.\eqref{an38},\eqref{an44} and \eqref{an49} read
\begin{equation}\label{51}	
	\rho^{j} = M \rho^{j+1} \quad \Longrightarrow \rho^{j+1} = M^{-1} \rho^{j}.
\end{equation}
Hence
\begin{equation}\label{an52}	
 \rho^{j} = M^{-j} \rho^{0}.
\end{equation}
with 
\begin{equation}\label{an53}
	M = 
	\begin{pmatrix}
		b_0 & c_0 & 0 & \cdots & \cdots & \cdots & 0 \\
		a_1 & b_1 & c_1 & 0 & \cdots & \cdots & 0 \\
		0 & a_2 & b_2 & c_2 & 0 & \cdots & 0 \\
		\vdots  & \ddots  & \ddots  & \ddots  & \ddots  & \ddots & \vdots  \\
		\vdots  & \ddots  & \ddots  & \ddots  & \ddots  & \ddots & \vdots  \\
		0 & \cdots & \cdots & 0 & a_{N-1} & b_{N-1} & c_{N-1} \\
		0 & \cdots & \cdots & \cdots & 0 & a_N & b_N
	\end{pmatrix}
	,
\end{equation}

and
\begin{equation}\label{an25}
	\rho^j = 
	\begin{pmatrix}
		\rho_0^j  \\
		\rho_1^j \\
		\vdots  \\
		\vdots  \\
		\rho_{N-1}^j \\
		\rho_{N}^j
	\end{pmatrix}
.
\end{equation}

\newpage
\bibliographystyle{unsrt}
\bibliography{BI.bib}

\end{document}